\documentclass[12pt,phd,a4paper,oneside, helvetic]{article}
\usepackage[left=25mm, right=20mm]{geometry}
\DeclareMathAlphabet{\pazocal}{OMS}{zplm}{m}{n}
\newcommand{\Lb}{\pazocal{X}}
\usepackage{dsfont}
\usepackage{subfig}
\usepackage{amssymb}
\usepackage{url}
\usepackage{float}
\usepackage{longtable}
\usepackage{pdflscape}
\usepackage[utf8]{inputenc}
\usepackage[english]{babel}
\usepackage{natbib}
\usepackage{booktabs}
\usepackage[flushleft]{threeparttable} 
\usepackage{hyperref}
 \hypersetup{
    colorlinks=true,
    linkcolor=blue,
    filecolor=magenta,      
    urlcolor=blue,
    citecolor=blue,
}

\usepackage{multirow}
\usepackage{tablefootnote}
\usepackage{commath}
\usepackage[skins,breakable]{tcolorbox}
\usepackage{pifont} 
\newcommand{\cmark}{\ding{51}}%
\newcommand{\xmark}{\ding{55}}%

\newcommand*{\ditto}{-\texttt{"}-}

\usepackage{fancyhdr}
\begin{document}
\thispagestyle{empty}
\chead{F. BATOOL}
\begin{center}
\textbf{\Large Initialization methods for optimum average silhouette width clustering }
 
\textbf{Fatima Batool}
 
      \textit{  Department of Statistical Science,
       University College London,
       Gower Street, London WC1E 6BT, United Kingdom,\\
       Email: ucakfba@ucl.ac.uk. } \\~~\\
 
  \end{center}

\noindent\paragraph{Abstract}
A unified clustering approach that can estimate number of clusters and produce clustering against this number simultaneously is proposed. Average silhouette width (ASW) is a widely used standard cluster quality index.  A distance based objective function that optimizes ASW for clustering is defined. The proposed algorithm named as OSil, only, needs data observations as an input without any prior knowledge of the number of clusters. This work is about thorough investigation of the proposed methodology, its usefulness and limitations. A vast spectrum of clustering structures were generated, and several well-known clustering methods including  partitioning, hierarchical, density based, and spatial methods were consider as the competitor of the proposed methodology. Simulation reveals that OSil algorithm has shown superior performance in terms of clustering quality than all clustering methods included in the study. OSil can find well separated, compact clusters and have shown better performance for the estimation of number of clusters as compared to several methods. Apart from the proposal of the new methodology and it's investigation the paper offers a systematic analysis on the estimation of cluster indices, some of which never appeared together in comparative simulation setup before. The study offers many insightful findings useful for the selection of the clustering methods and indices for clustering quality assessment.
\noindent\paragraph{keywords} within cluster compactness; between cluster separation;  combinatorial optimization; local optimum; non-deterministic clustering; clustering quality.

\section{Introduction}
\label{intro}
 Clustering is a fundamental method for the data analysis in many disciplines, for instance, in pattern recognition, text analysis, community detection, image processing, and computer vision to name but a few. The high level description of the clustering task is to discover the inherent underlying structure in the data. The classification of the clustering methods are not straight forward, however, can be broadly classified into distance/centroid based methods, density/kernel based methods, grid based methods, graph/network clustering methods, constraint based methods and Bayesian parametric, semi/non-parametrics methods  (\cite{friedman2001elements}).

\par The distance based methods are broadly classify as partitioning and hierarchical methods.  The partitioning methods are based on optimization of a numerical function. They usually utilize the concepts of separation and homogeneity to perform clustering, i.e., objects within a group are  closely located (intra-cluster compactness) and have cohesive structure, and they are well separated from the objects in other clusters (inter-cluster separation).

\par All partitioning based clustering methods give flat clustering, meaning that they partition the data into non-overlapping clusters and treat these clusters at the same level in clustering i.e., clusters are not nested inside clusters. They return a single partitioning of a data instead of series of partitions, and no further structures are seen within clusters.

\par Clustering is often differentiated into crisp or fuzzy. Crisp, also known as hard clustering, is the one for which all objects in the data just belong to exactly one cluster.  Whereas, in soft or fuzzy clustering, each object belongs to each cluster with a certain cluster membership score based on how similar the object is to other objects in that cluster. 
 
\par  Deterministic clustering algorithms always arrive at the same clustering result for the given data. Examples include hierarchical or spectral clustering algorithms. In contrast, stochastic algorithms do not reach at same clustering solution if run more than once with the same parametric choices, for instance, the standard k-means algorithm, for a different set of initial points chosen as starting centres can reach different local minima resulting in different clustering solutions. Similarly, model-based clustering is also a stochastic clustering method. However, both k-means and model-based clustering can be run in deterministic fashion.

\par The  partitioning clustering methods requires some notion of proximity between observations. This is usually define by some distance measure between the pair of observations. The selection of suitable distance measure depends upon various factors, such as, meaning of closeness in a certain application, what types of clusters are to be discovered, the type of the data (e.g., binary, continuous or mixed), the space of the data (e.g., Euclidean or non Euclidean), nature of the analysis and the clustering method to use.
 
 \par In this study I have developed a non-overlapping crisp  partitioning method. The algorithmic implementations of the method can be classified as stochastic clustering.  The algorithms proposed in this study are applicable to any data types (continuous, binary, text, images, sequence e.t.c.) and data from any space given that the distance measure is possible to calculate.

 \section{Motivation and related work}
\noindent The three major challenges while performing cluster analysis are how many clusters are present in the data (\cite{mccullagh2008many}), which clustering algorithm is suitable to retain the clustering structure for the data application at hand  \cite{jain1988algorithms}, and finally how to evaluate the clustering results (\cite{milligan1981monte}, \cite{handl2005computational}).

\par Cluster validation techniques are essential for the evaluation of the clustering outcome. These are broadly classified as external and internal methods (\cite{halkidi2001clustering}, \cite{lei2016ground}).  External validation methods take knowledge of known class labels to validate clustering algorithms on data sets to learn about the performance of a method. They are also used to compare the clustering results coming from different methods or different parametric choices of same method. In situations where external labels are not known, internal validation measures can be used to validate the clustering. Internal validation measures explore how well the clustering fits the data set using some criterion. The task of clustering quality validation and estimation of number of clusters are closely related. Many of the clustering quality indices are used for the estimation of number of clusters. The number of clusters can be chosen by optimizing a quality index.  Once the number of clusters are know clustering solution is obtained using some algorithm. 
 
\par  The study aims at defining a coherent framework to estimate the number of clusters and a clustering solution using the average silhouette width (ASW) proposed by \cite{rousseeuw1987silhouettes}. A clustering method can be defined by optimizing the objective function based on the ASW index.  This index measures the clustering quality to estimate the number of clusters and has shown good performance for the estimation of the number of clusters. The motivation is that if an index is really good in estimating number of clusters then it should also be good in getting the final clustering solution. The advantage of this is that it will make the task of clustering somewhat simpler, and the users don't have to deal with the two tasks separately.

\par The ASW is a well-reputed and trusted clustering quality measure. There have been some comparisons in the literature that validate the good performance of the index as compared to other indices. The ASW has been extensively used to estimate the optimal number of clusters  (with a combination of various clustering methods), to compare the performance of clustering methods and for the quality assessment of clustering obtained from many clustering methods. Some empirical studies have also been designed to evaluate performance of the ASW in comparison with other famous indices. The index has been used for cluster analysis in a diverse range of data clustering problems and setups across disciplines, for instance,  clustering of time series: \cite{kalpakis2001distance}, for document clustering: \cite{recupero2007new}, for micro-array analysis: \cite{kennedy2003large}, \cite{bandyopadhyay2007improved}, \cite{cho2010integrative}, for genotype assesment \cite{lovmar2005silhouette}, for brain analysis: \cite{craddock2012whole}, and for image segmentation:   
 \cite{hruschka2003genetic} to mention a few.  For clustering quality measures, and clustering method comparisons see, 
\cite{chen2002evaluation},  \cite{liu2003algorithms},   \cite{reynolds2006clustering}, \cite{kannan2008new}, and \cite{ignaccolo2008analysis}.

\noindent The research questions for this study are defined as under:

\begin{enumerate}
\item  To learn which existing method can work well with the ASW index? In principle ASW can be used with any clustering method to estimate number of clusters. However, some clustering methods can better capture certain kind of pattern in the data as compared to others, therefore, the performance of ASW will vary with the clustering methods. Goal here is to  evaluate the performance of ASW with different clustering methods for two aims defined as (a) performance of ASW for finding clustering solution, and (b) performance of ASW for the estimation of number of clusters.
\item  To illustrate the characteristics, and types of clusters OSil algorithm can capture and identify. For this two aims are defined as: (a) performance of OSil for finding clustering solution, and (b) performance of OSil for the estimation of number of clusters.
\item  To find out the best way to initialization OSil since initialization can effect the algorithm's output greatly \citep{arthur2007k}. For this several initialization methods were compared and evaluated against both aims.
\item To extensively evaluate the performances of other indices  in comparative setting for the estimation of number of clusters in combination of various clustering methods. 
\end{enumerate}

\paragraph{Organization} The next section  defines the notations and proposed methodology. In Section \ref{simmodmain} and \ref{simmodmaintwo} the particular details about the experiments  are defined and results are presented. In Section \ref{rec}, the article is concluded by high lighting the strengths of experiments conducted here as compared to other such studies in literature. 
\section{Proposed methodology} 
Formally, suppose that the aim is to cluster a data set $\Lb = \{ x_1, \cdots, x_n\}$ from some space $\mathds{S}$ of size  $n \geq 2$ into $k \geq 2$ clusters, where $n \in \mathds{N}$ and $k \in \mathds{N}_n$. Each object in $\Lb$ is  a $p$-dimensional vector, where $ p \in \mathds{N}$.  A function  $d(x_i, x_h)$ calculates the pairwise dissimilarities between pairs of objects $(x_i, x_j) \in \Lb$ such that $d: \Lb \times \Lb \rightarrow \mathds{R}^+$, satisfying \textit{non-negativity, reflexivity and symmetry} properties. The clustering task is to produce the partition of the data $\Lb$ into disjoint subsets denoted as $\mathcal{C}_k = \{C_1, C_2, \dots, C_k\}$ be a clustering of any size $k$. Let $\mathcal{P}(\Lb)$ be the set of all non-trivial partitions on $\Lb$ such that $\mathcal{C}_k \in \mathcal{P}(\Lb)$. Formally,  a partitioning function  $f_k(d, \Lb) = \mathcal{C}_k$ provides $k$-partitions such that, (a) $ \nexists ~ C_i = \emptyset; i \in \mathds{N}_k$, (b) $C_i \cap C_j = \emptyset; i \neq j $, and (c) $\cup_{i=1}^k C_i = \Lb$. We also require $ \sum_{i=1}^k \lvert C_i \rvert = n$. 

\par  There are two trivial clustering cases which are not of interest in this work. They are defined now as: all the data points belong to one cluster only i.e., $k=1$, and each data point forms its own cluster i.e., $k=n$.

\par For a given partition $\mathcal{C}_k$, the clustering label vector defines the cluster memberships of all the observations in $\Lb$. For $x_i \in C_r$, $r \in \mathds{N}_k$, the label of $x_i$ for $i \in \mathds{N}_n$ is $l(i) = r$. Therefore, a complete label vector for a partition is $(l(1), \cdots, l(n))$, where $l(i)$ represents a label for object ‘i’ and each of $l(i) \in \mathds{N}_k$. Let $l(\Lb, k)$ be the brief notation for the clustering label vector. 

 \par The Average Silhouette Width (ASW)  proposed by \cite{kaufman1987clustering} to evaluate the clustering quality and to estimate the number of clusters is defined as follows. Given a clustering label vector $l(\Lb, k)$  for a  $\mathcal{C}_k$ a clustering identified by some clustering function $f_k$ on $\Lb$ the silhouette width (SW) of the object $i$, $ i \in \mathds{N}_n$, is calculated as follows:
 
\begin{equation} \label{eq:Sbarck}
S_i\big(l(\Lb, k), d\big) = \frac{b(i)  - a(i)}{ \textrm{max} \{ a(i), b(i)\}},
\end{equation}
where, 
\begin{equation*} 
a(i) =  \frac{1}{n_{{l(i)}}-1} \sum_{\substack{  l(i) = l(h)\\
                 i \neq h }} d(x_i, x_h), \quad \text{and} \quad b(i) =  \min_{r \neq l(i)} \frac{1}{n_r} \sum_{l(h) = r} d(x_i, x_h).
\end{equation*}
A global measure of clustering quality named as ASW can be obtained by averaging the SW for each member of the data. The ASW is formally calculated as follows:

 \begin{equation} \label{eq:OASWobjfunone}
 \bar{S}\big( l(\Lb, k), d\big)  = \frac{1}{n} \sum_{i=1}^n S_i\big(l(\Lb, k),d\big),
\end{equation} such that $-1 \leq \bar{S}(l(\Lb, k), d) \leq 1$.

 \par The \textit{Optimum Average Silhouette Width} (OASW) clustering of $\Lb$ is defined by maximizing the following function, over all $l^\ast(\Lb, k)  \in \mathcal{L}$, where $\mathcal{L}$ represents a set of all possible label vectors $l^\ast(\Lb, k)$ for all possible non-trivial clusterings $\mathcal{C}_k \in \mathcal{P}(\Lb)$. 
\begin{equation} \label{chapterthree:OASWobjfunthree}
 f\big( l(\Lb, k), d\big)  = \underset{ l^\ast(\Lb, k) \in \mathcal{L}}{\text{arg~max}} ~ \bar{S}_l\big(l^\ast(\Lb, k), d\big).
\end{equation} 
Substitution \eqref{eq:Sbarck} in \eqref{eq:OASWobjfunone} and the resulting expression  in \eqref{chapterthree:OASWobjfunthree} will give us a full definition of the objective function as follows:
\begin{equation}\label{chapterthree:OASWobjfunfour}
 f\big( l(\Lb, k), d\big)  =  \underset{ l^\ast(\Lb, k) \in \mathcal{L}}{\text{arg~max}} ~ \frac{1}{n} \sum_{i=1}^n \frac{b(i) - a(i)}{\text{max}\{a(i),b(i)\}},
\end{equation}
where $l^\ast(\Lb, k)$ is required for $b(i)$ and $a(i)$ defined above. 

\par ASW is a combinational index based on within and between cluster distances that tries to achieve compromise between the two aims of homogeneity and separation.  The $a(i)$ is not defined for singletons and to calculate $b(i)$ there should be at least two clusters. Therefore   SW is only defined for $k > 1$. The ASW provides a natural and intuitive definition for clustering. For a good clustering based on ASW criterion the ``within'' clusters dissimilarity  should  be less than the ``between'' clusters dissimilarity. Therefore, if $a(i)$ is much smaller than the smallest between clusters dissimilarity $b(i)$ we get evidence (larger $S_i$, close to 1 is better in this case),  that object $x_i$ is in the appropriate cluster. On the other hand, $S_i$ close to -1, points towards the wrong cluster assignment for object $x_i$. In this case $a(i) > b(i)$,  meaning that object $x_i$ is more close to its neighbouring cluster than to its present cluster. A neutral case occurs when  $S_i \approx 0$, i.e., object $x_i$ is approximately equally distant from both, its present cluster and neighbouring cluster.  
 
\par The optimum ASW (OASW) clustering is implemented through OSil algorithm abbreviated from \textbf{O}ptimum \textbf{A}verage \textbf{Sil}houette width).  The objective of OASW clustering is to find a clustering for which $\bar{S}(l(\Lb, k), d)$ is maximum from all the possible clusterings $\mathcal{C}_k$ of $\Lb$. The set $\mathcal{L}$ is determine by  all the possible combinations of an object with the membership of a cluster for an initial clustering label vector $l(\Lb, k)$.  
 
 \begin{tcolorbox}[float=h,
  blanker,
  width=0.99\textwidth,enlarge right by=0.36\textwidth,
  before skip=1pt,
  breakable,
  overlay unbroken and first={%
    \node[inner sep=0pt,outer sep=0pt,text width=0.33\textwidth,
      align=none,
      below right]
      at ([xshift=-0.36\textwidth]current page.south east)
  {%
    
  };}]
\noindent \rule[2pt]{\linewidth}{2pt}\vspace{-.5mm}\\
\textbf{OSil algorithm}\\
\noindent \rule[2ex]{\linewidth}{0.70pt} \\
{\fontsize{10}{10} \selectfont
\noindent\textbf{Initialize }
\begin{enumerate}
\item For all $(x_i, x_h) \in \Lb$, where $(i, h) \in \mathds{N}_n$ and $i \neq h$, calculate $d(x_i, x_j)$. 
\item Calculate a clustering using any crisp clustering criterion and  initialize the clustering label vector with $k$ clusters as $l(\Lb, k) = (l(1), \dots, l(n))$.     
\item Calculate $f^{(0)} = f\big( l(\Lb, k), d \big)$. 
\item Set $q = 1$. Let $ l^{(1)}(\Lb, k) = l(\Lb, k)$.
\end{enumerate}
\textbf{Swap}
\begin{enumerate}
\item For all pairs $(i, r)$ such that $l^{(q)}(i) \neq r$, for $i \in \mathds{N}_n$ and $r \in \mathds{N}_k$, assign $l(i) = r$ and denote the new label set as $l^\ast_{(i, r)}(\Lb, k) = (l^\ast(1), \dots, l^\ast(n))$. 
\item Compute $f_{(i,r)} = f(l^\ast_{(i, r)}(\Lb, k), d)$.
\item $(h, s) =  \underset{(i, r)}{\arg\max} ~~ f_{(i, r)} $, $~~f^{(q)} = f_{(h, s)}$, $~~ l^{(q)}(\Lb, k) = l^\ast_{(h, s)}(\Lb, k)$.
\end{enumerate}
\textbf{Stop} \par If $f^{(q)} \leq f^{(q-1)}$. Else $q = q+1$. Repeat \textit{Swapping:} Step (i)-(iii). \\
\textbf{Return} \par  $f^{(q)}$ and $l^{(q)}(\Lb, k)$. \\
}
\noindent \rule[2ex]{\linewidth}{0.70pt}
\end{tcolorbox}

\par For this work,  I did not define in advance, what is the definition of clusters, because from the definition of the ASW  this can not be fully specify. One can only roughly understand what ASW is aiming for.  It has it's own notion that looks for homogeneous clusters, and separation from the closest cluster. Based on this notion one can expect that OASW clustering might identify  spherical or perhaps elliptical clusters as well. However, it is not so clear how separated and homogeneous clusters ASW can deliver, and in what situations it fails. Also, during experiments it was discovered it can find uniform, elongated, small clusters and clusters of other shapes. 

\par  OSil performance has been extensively evaluated against competitors for the estimation of numbers of clusters and for the clustering solutions. For this two set of simulation studies were designed, one for the fixed known $k$ (Simulation I) and other for the estimation  of $k$ (Simulation II). These stimulations are presented in the following section.

\section{Empirical evaluation}
\subsection{Simulation I}\label{simmodmain}
As defined in step (ii) of the \textbf{Initialize} phase of the OSil algorithm an initial clustering to start the optimization process is required. To find the best initialization method for the algorithm a wide range of existing clustering methods were used as an initialization  for  $OSil$,  namely, \textit{$k$-means}, \textit{PAM}, \textit{model-based}, \textit{spectral},  agglomerative hierarchical linkage methods using  \textit{average}, \textit{complete}, \textit{single},\textit{ McQuitty} and \textit{Ward}'s method. A list of references to the original articles to these methods are presented in Table \ref{listofmethod} together with their software implementation used in this study. The simulations were conducted in the R language \citep{Rcore2019}. 

\begin{table}[!ht]
\renewcommand{\arraystretch}{0.90}
\fontsize{7}{7}\selectfont
\caption{A list of clustering methods used as initialization for OSil algorithm and their implementation used in this study} 
\begin{tabular}{ l l l l l}
\toprule
Method (Acronym) &   Reference & R function & Package \\
\midrule
kmeans &  \cite{forgy1965cluster} & kmeans() & stats; \\
Partitioning around medoids (PAM) &  \cite{kaufman1987clustering} & pam() & cluster; \cite{packagecluster} \\
PAMSIL &  \cite{van2003new}  & PAMSIL  & C; \cite{van2003new} \\
Single linkage & \cite{sneath1957application} & hclust() & stats; \\
Complete linkage & \cite{sorensen1948method} & \ditto\\
Average linkage & \cite{sokal1958statistical} & \ditto\\
Ward's method & \cite{ward1963hierarchical} & \ditto \\
McQuitty similarity & \cite{mcquitty1957elementary} & \ditto \\
Spectral clustering & \cite{ng2002spectral} & specc() & kernlab;  \cite{zeileis2004kernlab}\\
Model-based clustering & \cite{fraley1998many} & Mclust() & mclust; \cite{fraleymclust}\\

\midrule
  Calinski and Harabasz (CH)   & \cite{calinski1974dendrite}  & index.G1 &  clusterSim; \cite{walesiak2011clustersim} \\
  Hartigan (H)  & \cite{hartigan1975clustering}  &  index.H & \ditto\\
  Krzanowski and Lai (KL) & \cite{krzanowski1988criterion}  & index.KL & \ditto\\
  Gap  &	 \cite{tibshirani2001estimating} & index.Gap() & \ditto\\
   Gamma & \cite{baker1975measuring} &  intCriteria()& clusterCrit; \cite{desgraupes2013clustercrit} \\ 
  C &  \cite{hubert1976quadratic} & \ditto & \\
  Jump & \cite{sugardocumentation}	&  jump() & \cite{sugardocumentation}\\
  Prediction strength (PS)  &	 \cite{tibshirani2005cluster} &  prediction.strength() & fpc; \cite{hennig2010fpc}\\
  Bootstrap instability (BI) & \cite{fang2012selection} 	& nselectboot() & \ditto\\
    CVNN & \cite{liu2013understanding} & cvnn()  & \ditto\\
  Bayesian Information Criterion (BIC)  &\cite{schwarz1978estimating}	& mclustBIC()& mclust; \cite{fraleymclust}  \\
\midrule
Adjusted Rand Index (ARI) & \cite{hubert1985comparing} & adjustedRandIndex() & \ditto\\
\bottomrule 
\end{tabular}
     \label{listofmethod}
\end{table}

\par Several data generating process (DGPs) were designed to compare the performance of OSil with the existing methods. Each DGP has certain kinds of clustering problem(s) to solve. These DGPs cover a range of clustering characteristics defined as: clusters with different variations among observations, i.e., compact and widely spread clusters; equal and unequal sized clusters; clusters from different distributions assuming every individual cluster is coming from a single distribution. For instance, clusters from Gaussian, Student's $t$, Gamma or Beta distributions; clusters from skewed distributions; different types of clusters for instance, spherical, non-spherical, and elongated clusters;  close and far away clusters,  i.e., the distance between the means of clusters are varied; overlapping and well-separated clusters; relatively small clusters in presence of bigger clusters; nested clusters; clusters with correlated variables within clusters; different number of clusters; different  number of variables/dimensions; and combinations of these. The data sets were only generated from the euclidean space in this work although the algorithm is applicable to any kind of data.
 
\par
\begin{figure}[!htb]
\rule{-3ex}{.2in}
\subfloat[Model 1 ($k=2$, $d=2$)]{
\includegraphics[width=40mm,height=40mm]{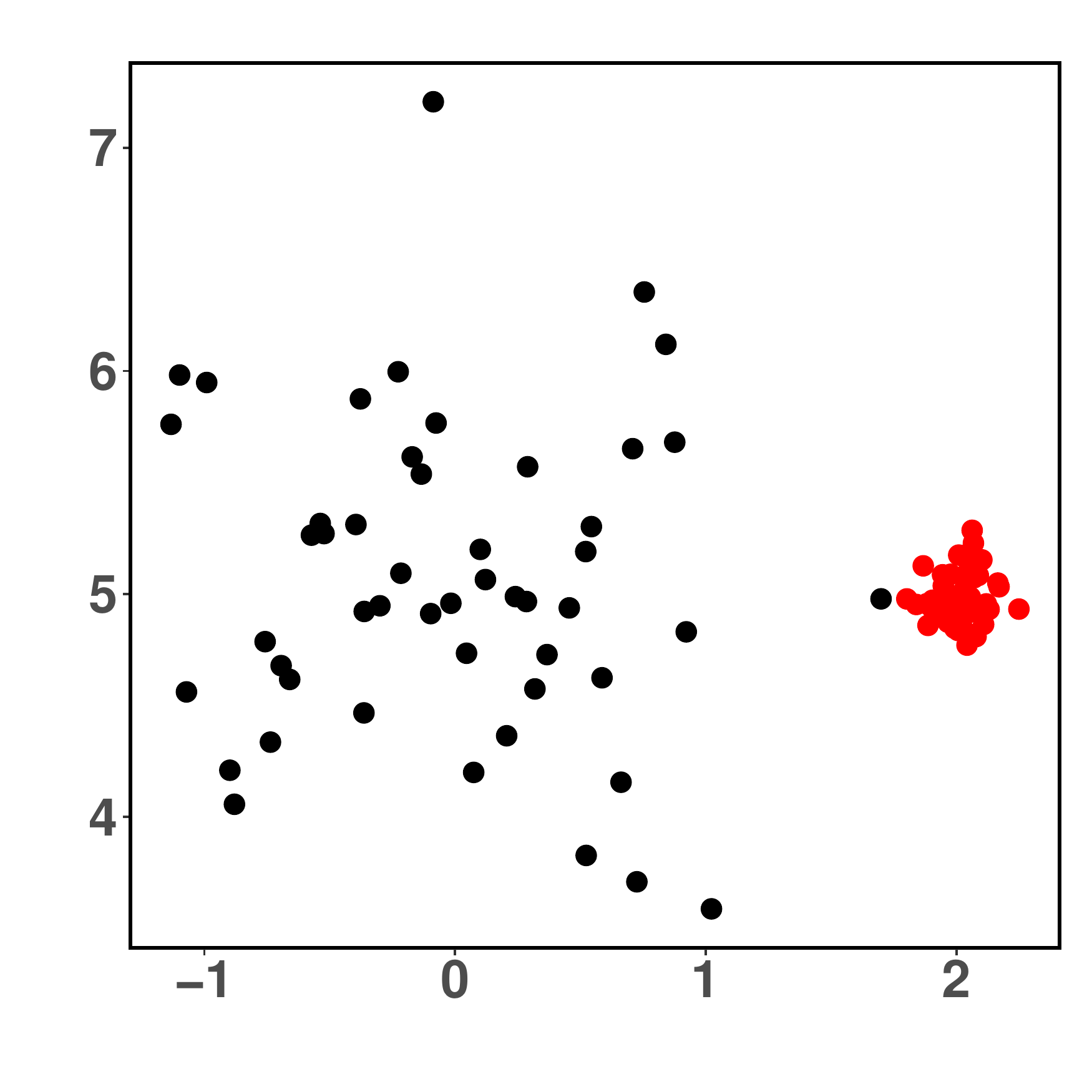} 
}
\subfloat[Model 2 ($k=3$, $d=2$)]{
  \includegraphics[width=40mm,height=40mm]{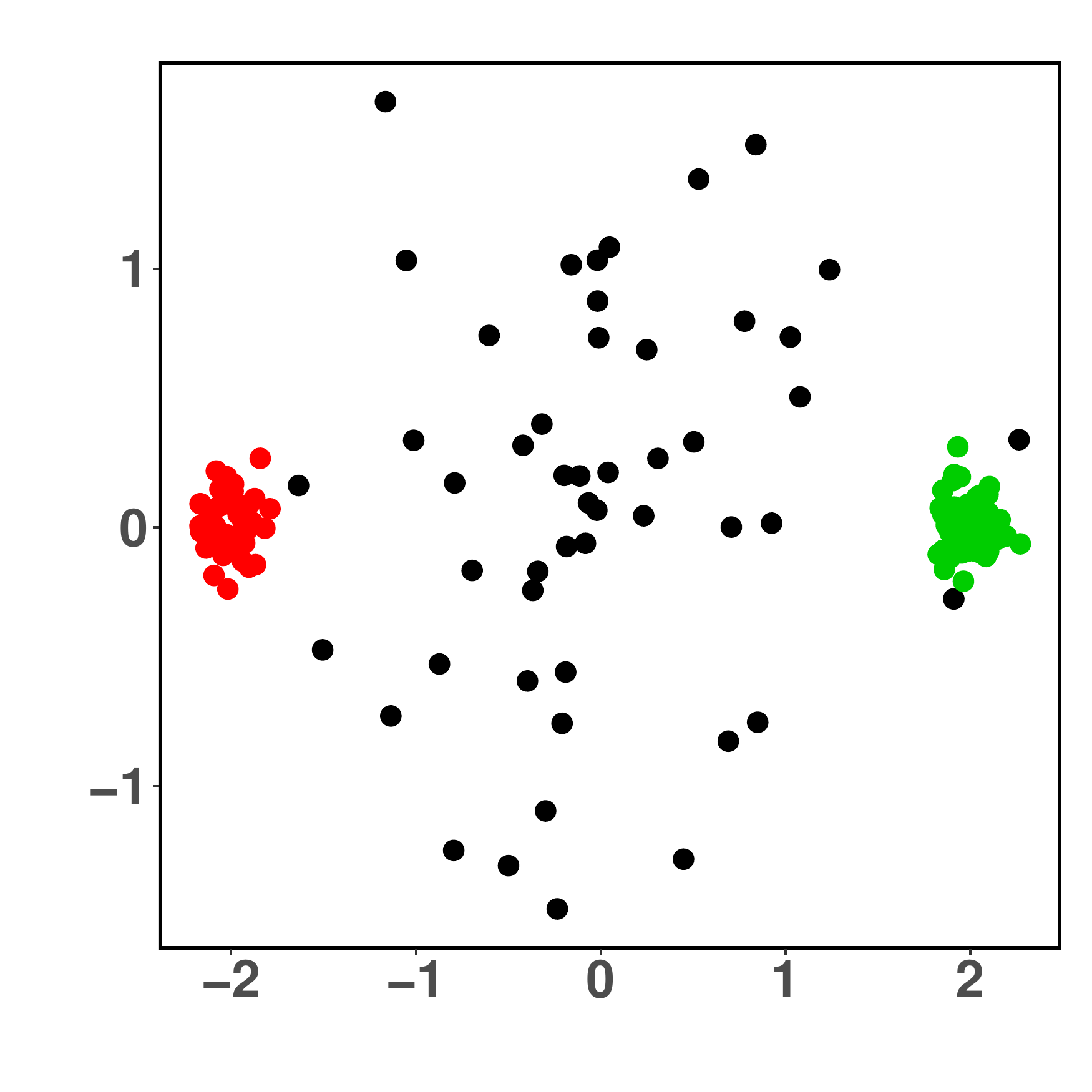}
}
\subfloat[Model 3 ($k=4$, $d=2$)]{
  \includegraphics[width=40mm,height=40mm]{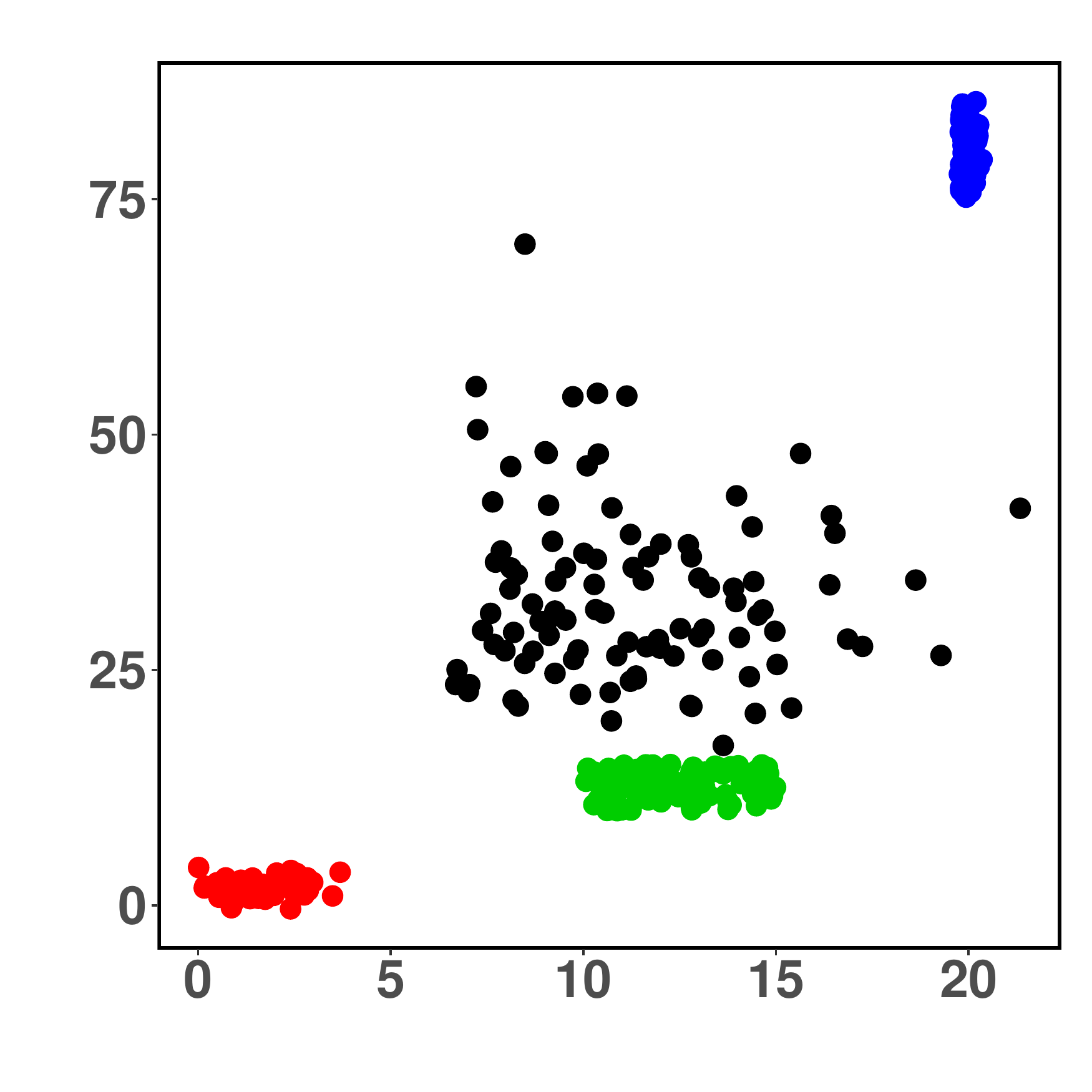}
}
\subfloat[Model 4 ($k=5$, $d=2$)]{
  \includegraphics[width=40mm,height=40mm]{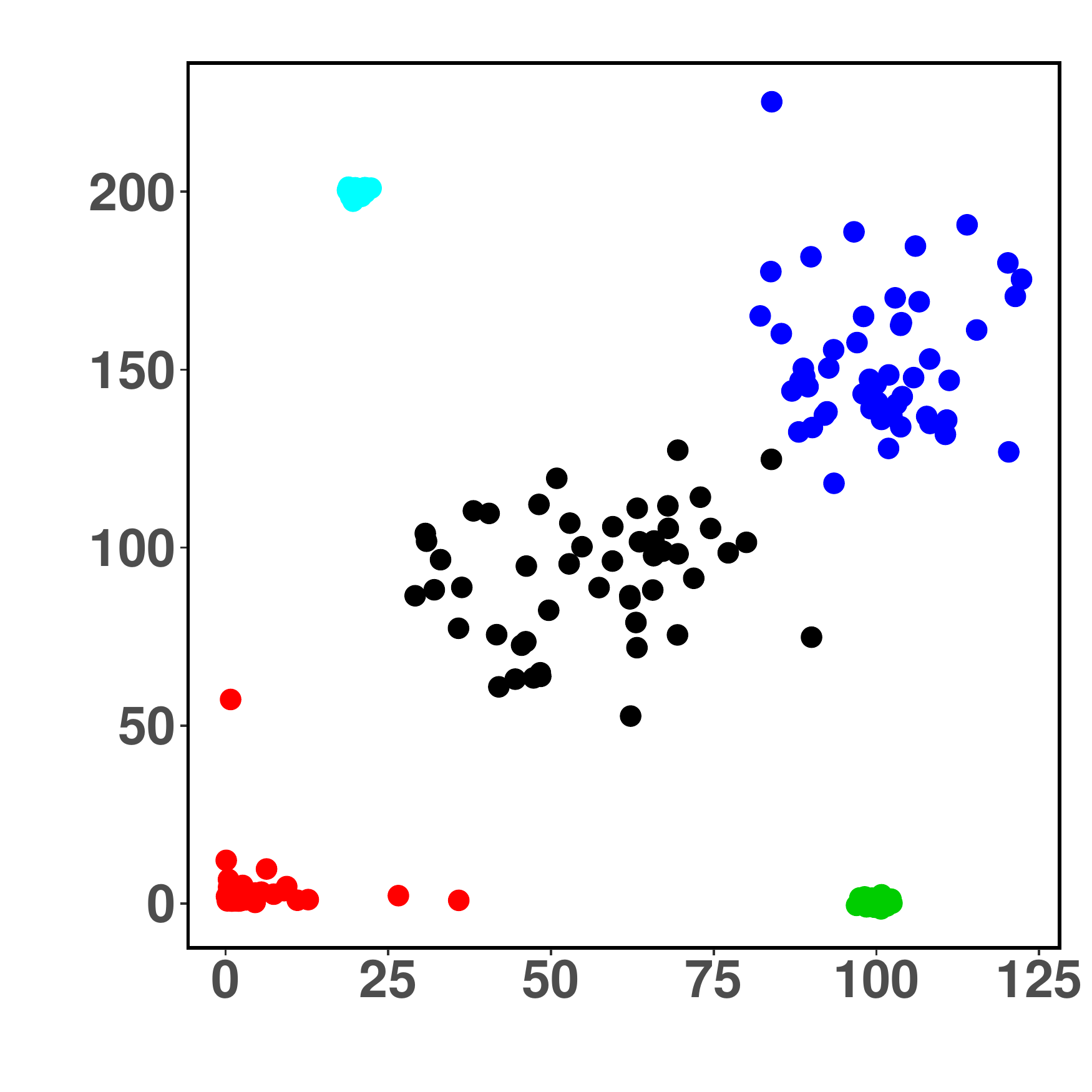}
}
\newline
\rule{-3ex}{.2in}
\subfloat[Model 5 ($k=6$, $d=2$)]{
  \includegraphics[width=40mm,height=40mm]{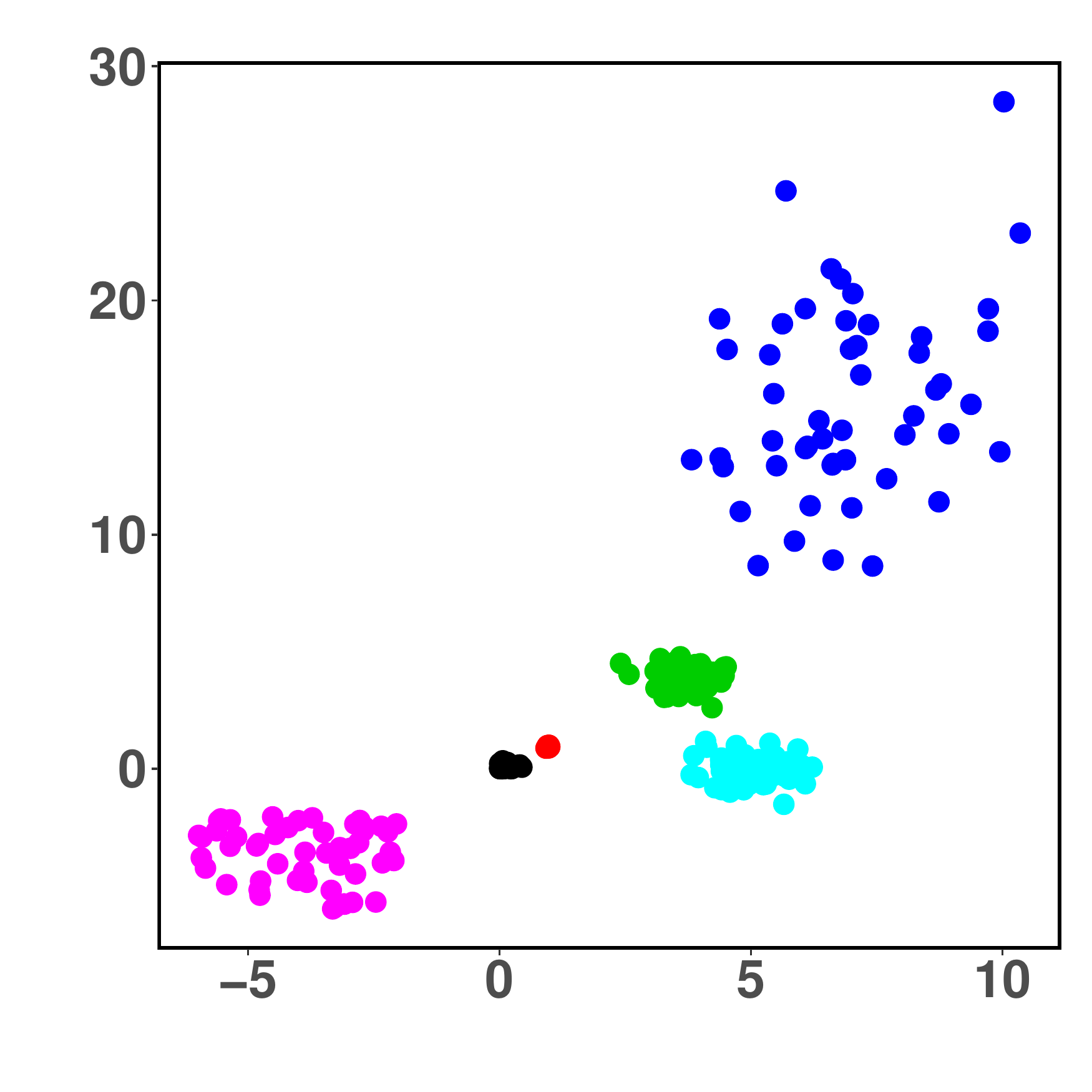}
}
\subfloat[Model 6 ($k=5$, $d=5$)]{
  \includegraphics[width=40mm,height=40mm]{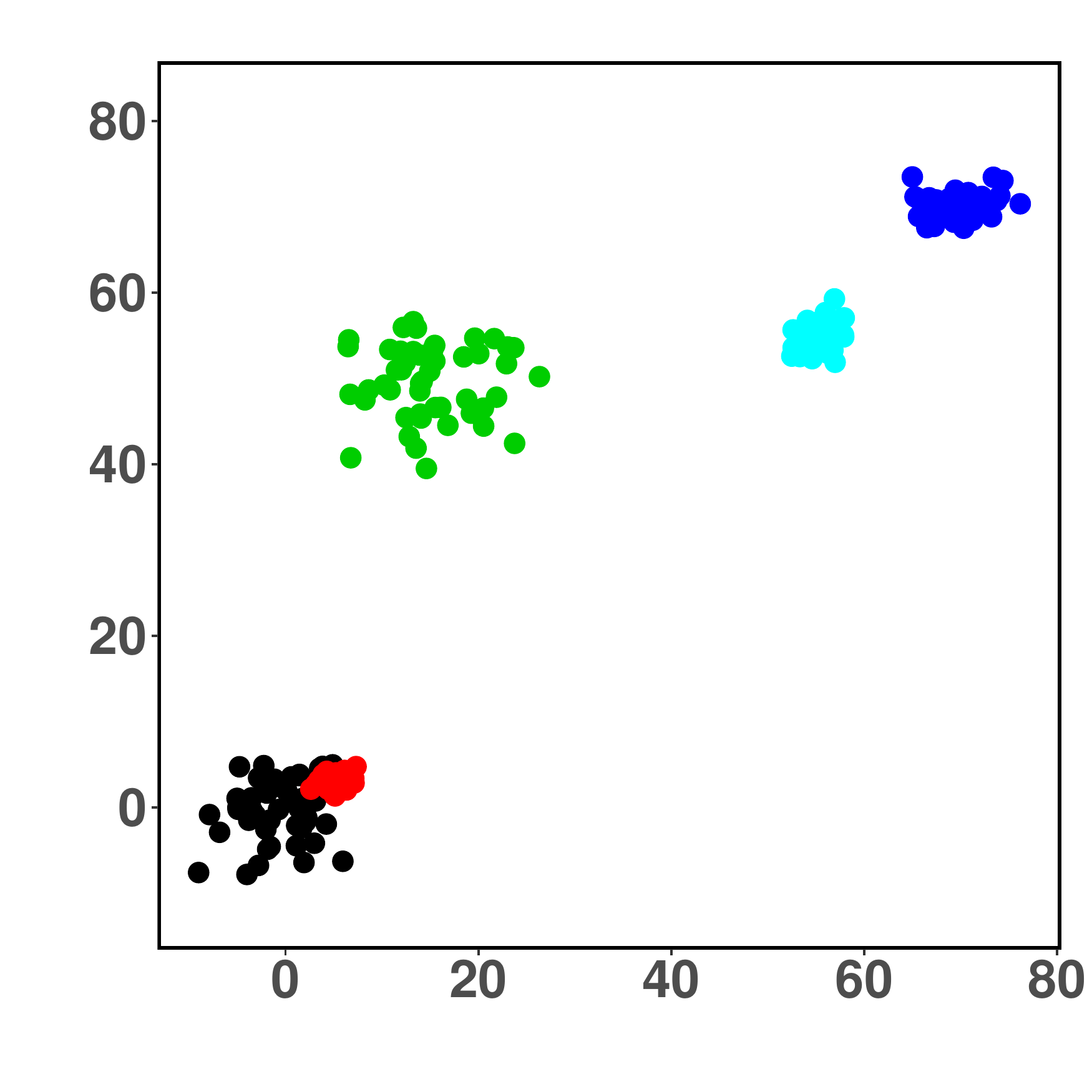}}
\subfloat[Model 7 ($k=7$, $d=10$)]{
  \includegraphics[width=40mm,height=40mm]{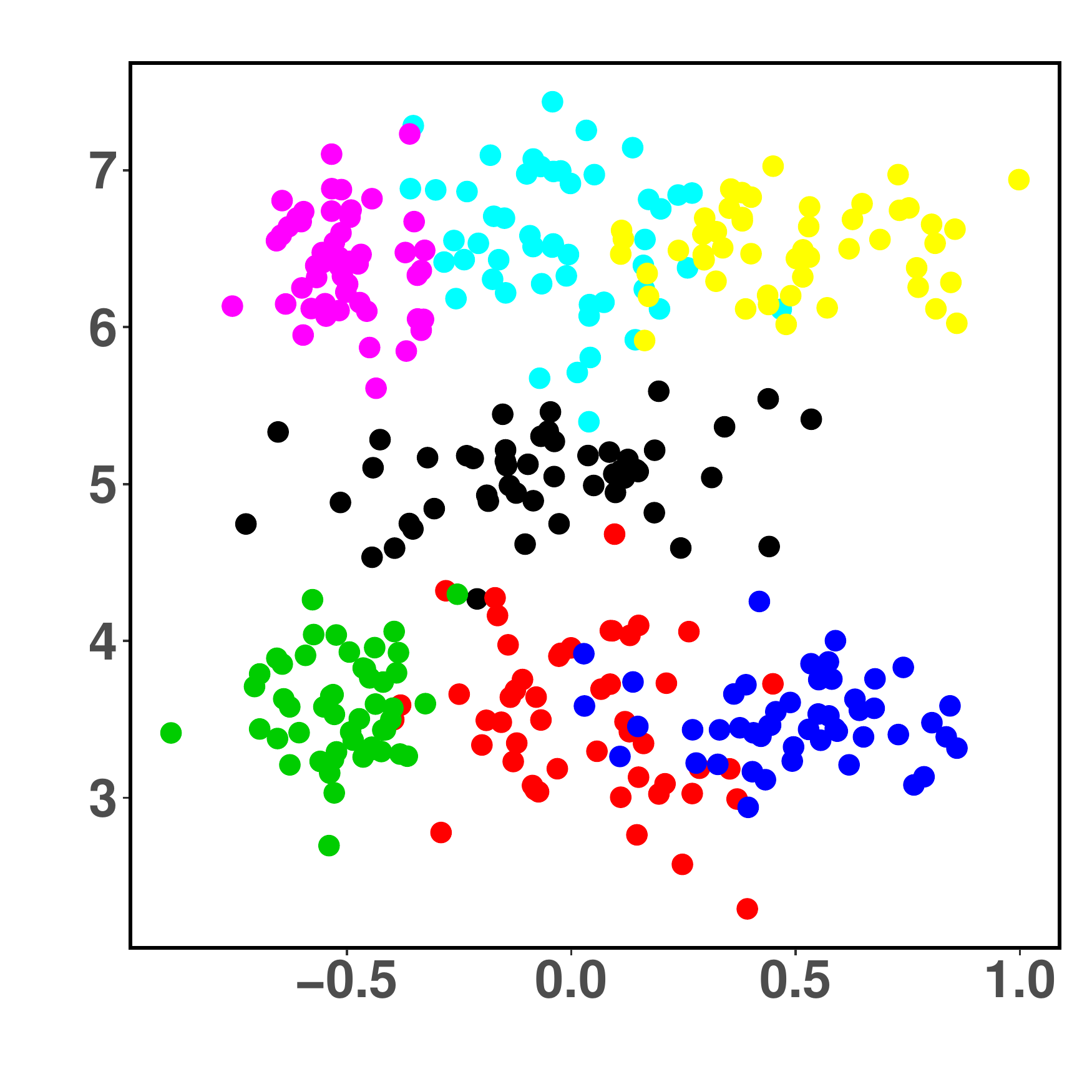}
}
\subfloat[Model 8 ($k=10$, $d=500$)]{
  \includegraphics[width=40mm,height=40mm]{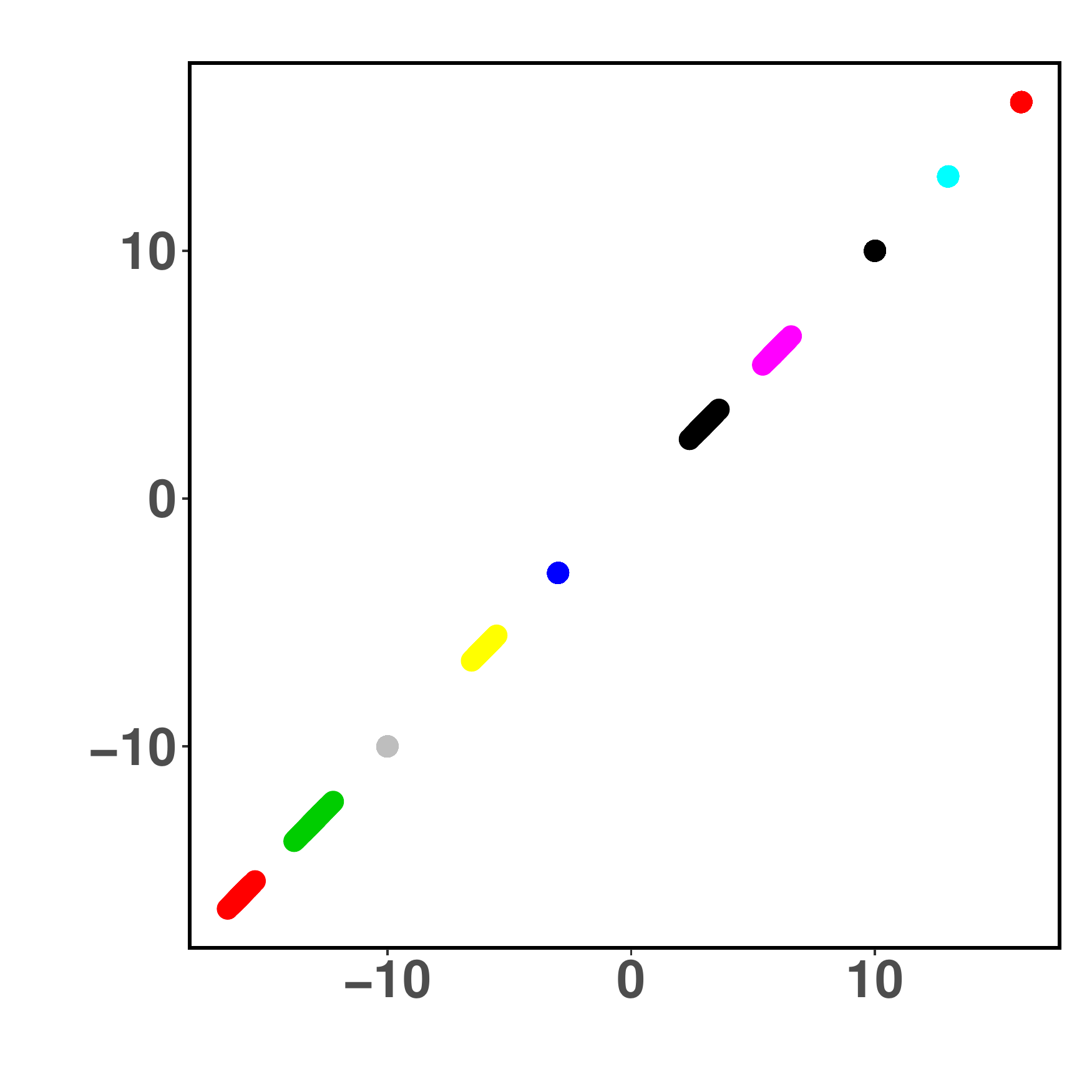}
}
\caption{Data plots of clustering structures defined for the study with an indication of true number of clusters.}
\label{plotoddatasets}
\end{figure}

 The definitions of all the DGPs used in this study are given in supplementary file of the article. The plots of a data sets generated from each of the DGPs is shown in Figure \ref{plotoddatasets}.   I expect from the algorithms to retrieve the clustering as defined by the DGPs. Through comparisons we will learn  clustering  structure OSil algorithm can identify correctly. For all the clustering methods their R functions with the default parametric choices were used except for $k$-means where the random centres was fixed at (nstart = 100). This is because, the performance of  $k$-means improves if one allows the algorithm to optimize the objective function by taking several set of cluster centres. For the ASW calculations of the clustering solutions obtained from the clustering methods other than OSil the `silhouette()' function in the R package `cluster' (\cite{packagecluster}) was used. PAMSIL proposed by  \cite{van2003new} to optimize ASW was also considered for the comparison.  Note that PAMSIL is not same as OSil initialized with PAM.

\par For all the clustering methods, the known number of clusters from the corresponding DGPs were used. For each DGP, 25 data sets were generated. Clusterings were performed using the 9 clustering methods just mentioned. The ASW and ARI values for these clusterings were calculated\footnote{Several other statistics of interest including standard errors, run time, number of iterations taken by the algorithm to converge, ARI, histograms, and boxplots were computed to compare results. All  results are not reported here due to limited space but are available upon request.}.  These clusterings were then passed  to the OSil algorithm as initializations. This will result in 9 different OSil clusterings. The ASW and ARI values together with their SE for these OSil clusterings were calculated as well.  
 
\paragraph{Performance evaluation} For comparison, the aggregated results (averages) of the ASW  value for each initialization method considered are reported.   The box-plots for the ASW  are also plotted. Since the clustering methods used to initial OSil algorithm are well developed clustering methods in their own right, therefore, I used the clustering results (referred to as the initial clustering in the following discussions) obtained from these to compare OSil clustering. Adjusted rand index is an external criterion for matching similarities between two clustering was also used to evaluate performance. Each clustering was matched with the true known DGPs labels. ARI  index ranges between  0 and 1 inclusive and a higher value is indication of higher agreement with ground truth clustering.

\subsubsection{Results}\label{subsectionsimone}

The aggregated ASW values obtained from OSil from each method is reported in Table \ref{aswoaswsumma}. Note that for the evaluation both the row comparison and the column comparison is of interest.  The comparison in the first row for each model will indicate which among the competitors has performed best with ASW (a maximum value is of interest here). The comparison in the second row for each model will indicate which method is the best contestant for the OSil initialization. Whereas, the column-wise comparison of these two rows will indicate will OSil improves the quality of the clustering as compared to the standalone clustering methods? If so how much improvement in terms of ASW value can OSil brings. The corresponding ARI values for these clusterings are reported in Table \ref{arisummary}.

 \par
\begin{figure}[!htb]
\centering
\subfloat[ ]{
  \includegraphics[width=45mm]{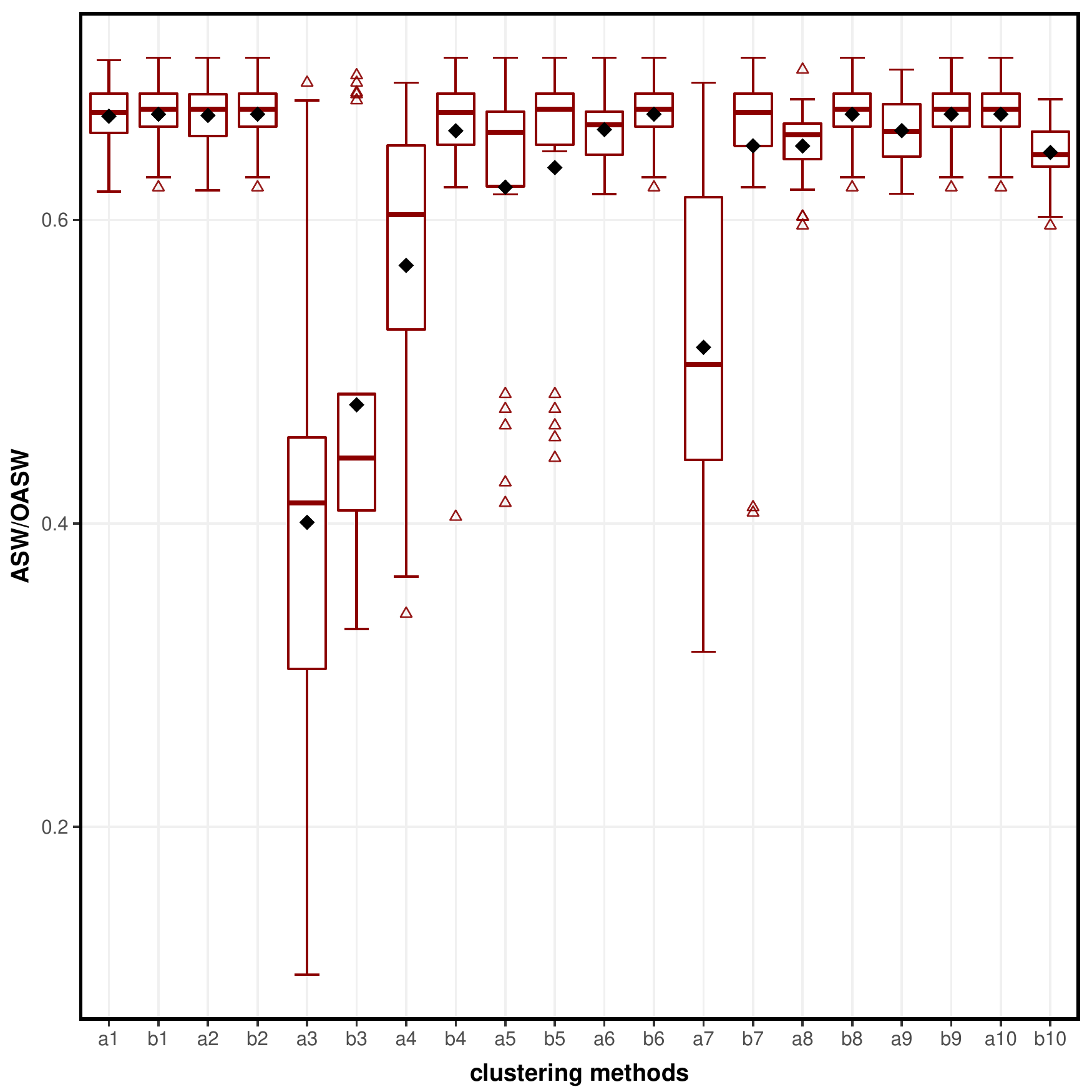}
 }
\subfloat[]{
  \includegraphics[width=45mm]{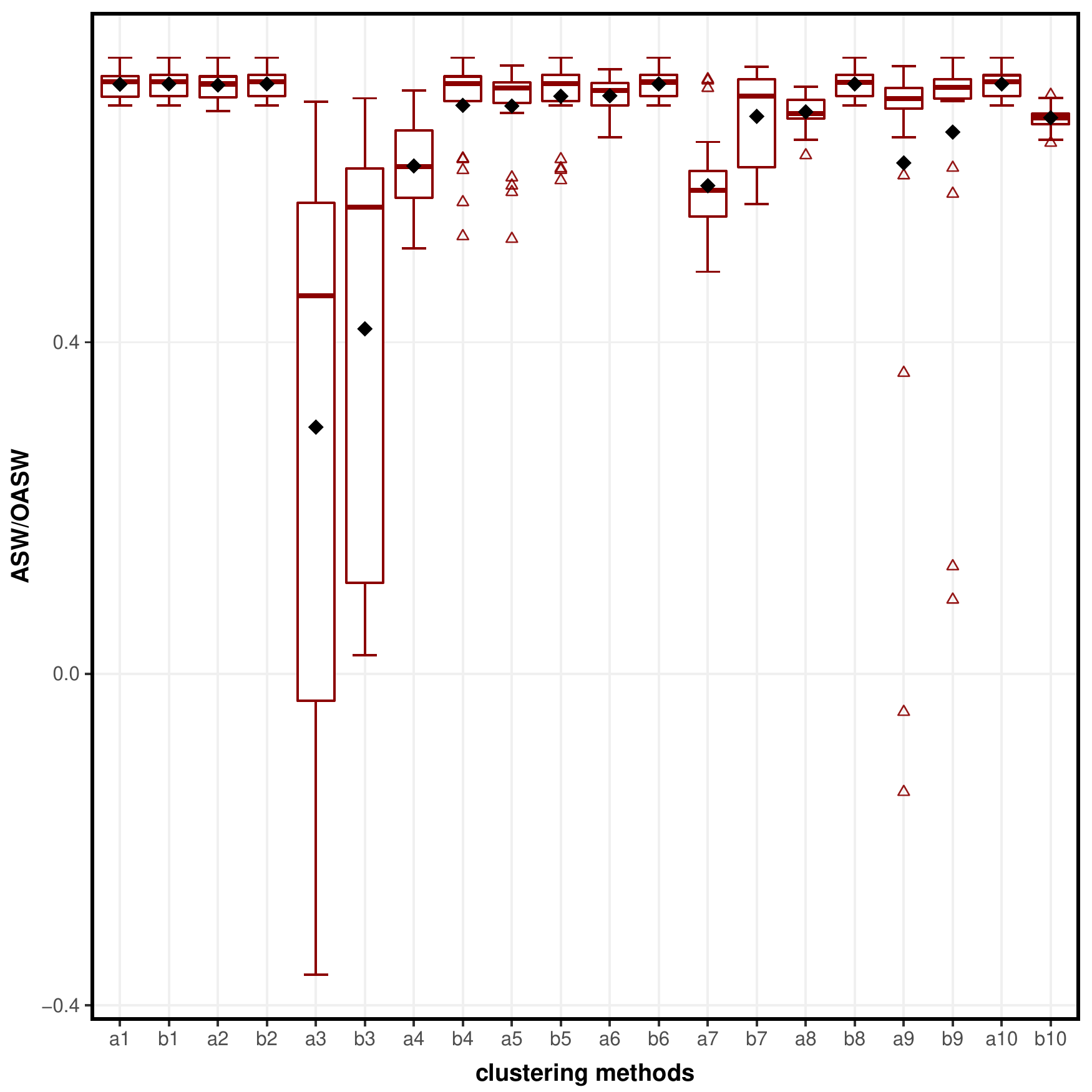}
 }
\subfloat[ ]{
  \includegraphics[width=45mm]{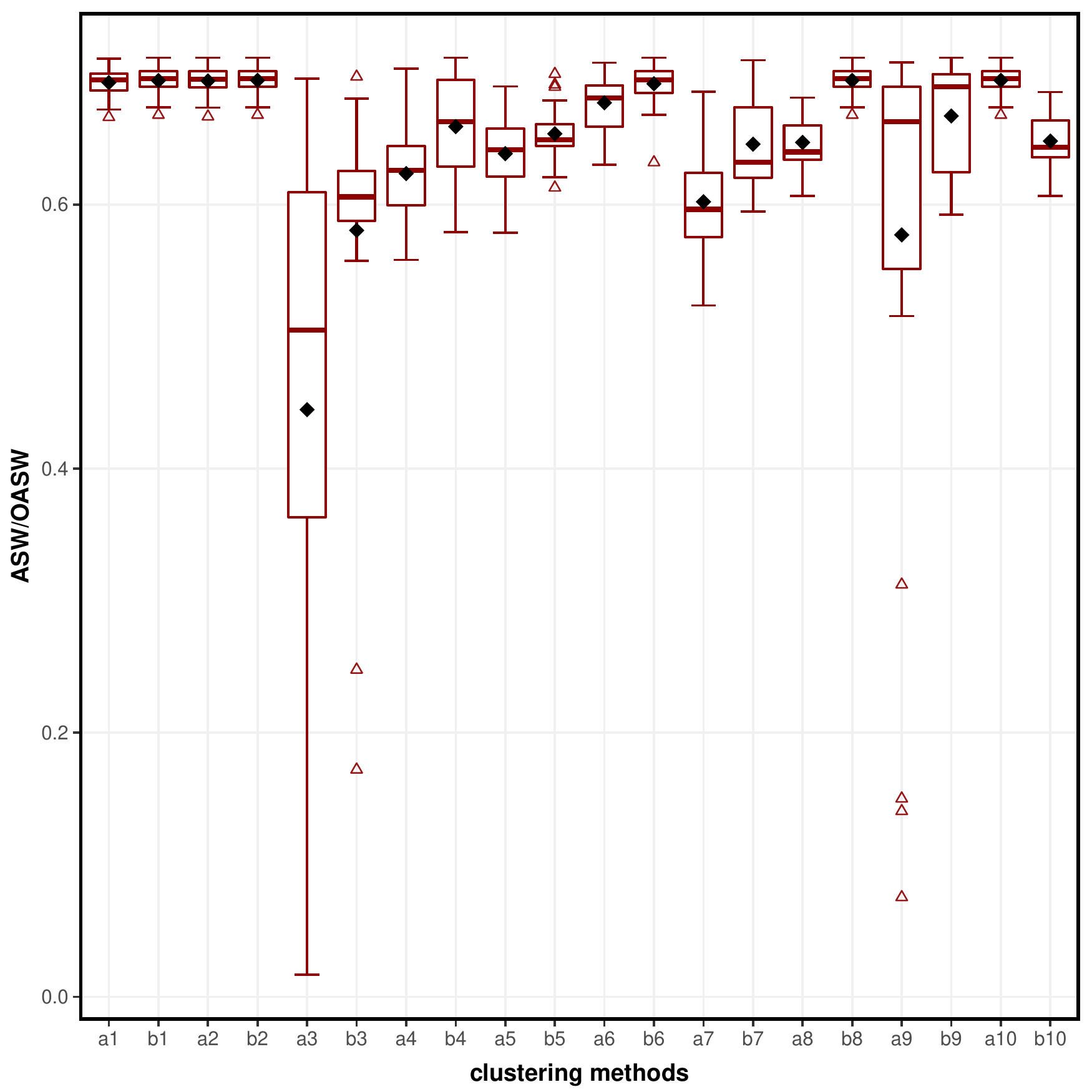}
 }
\newline
\subfloat[ ]{
  \includegraphics[width=45mm]{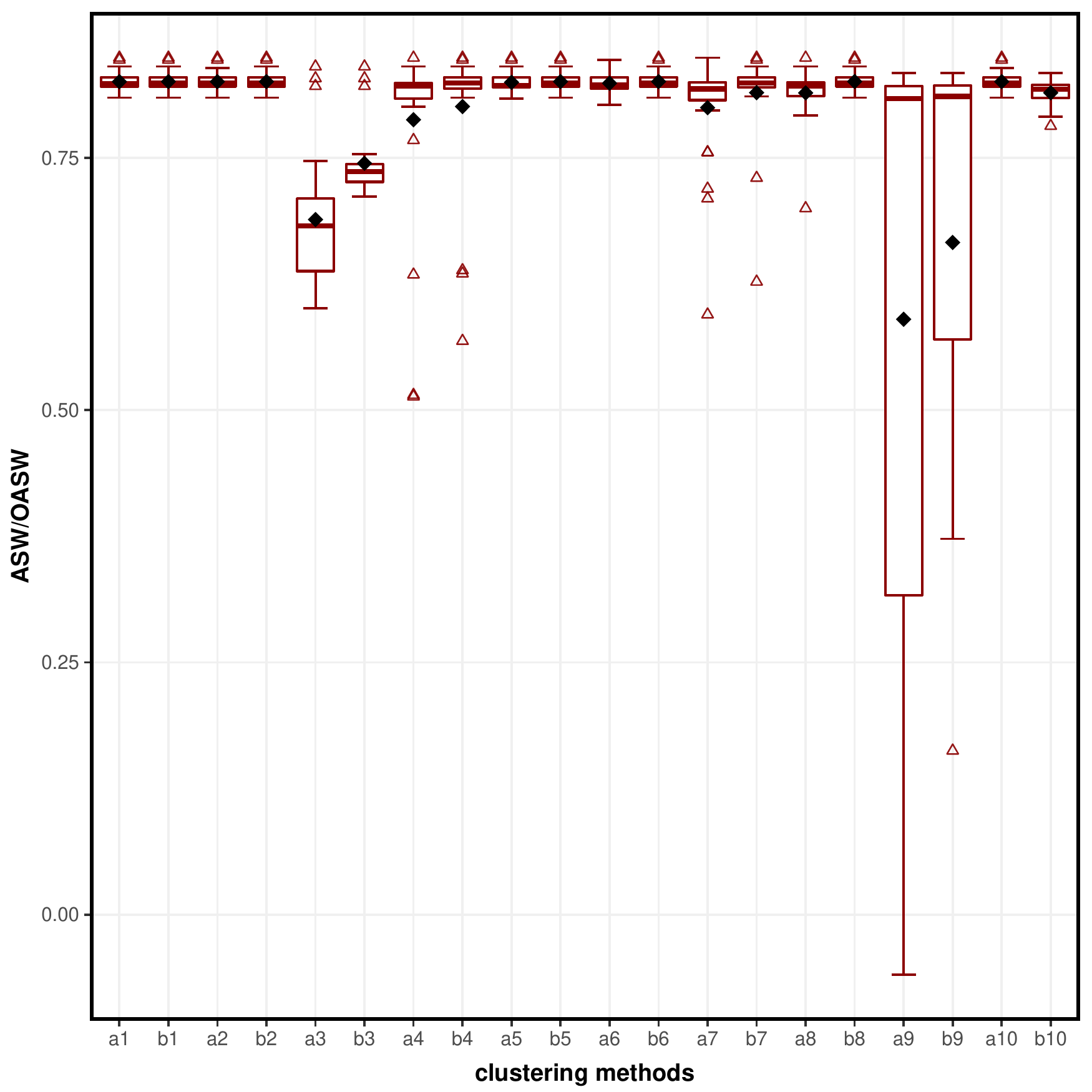}
 }
\subfloat[ ]{
  \includegraphics[width=45mm]{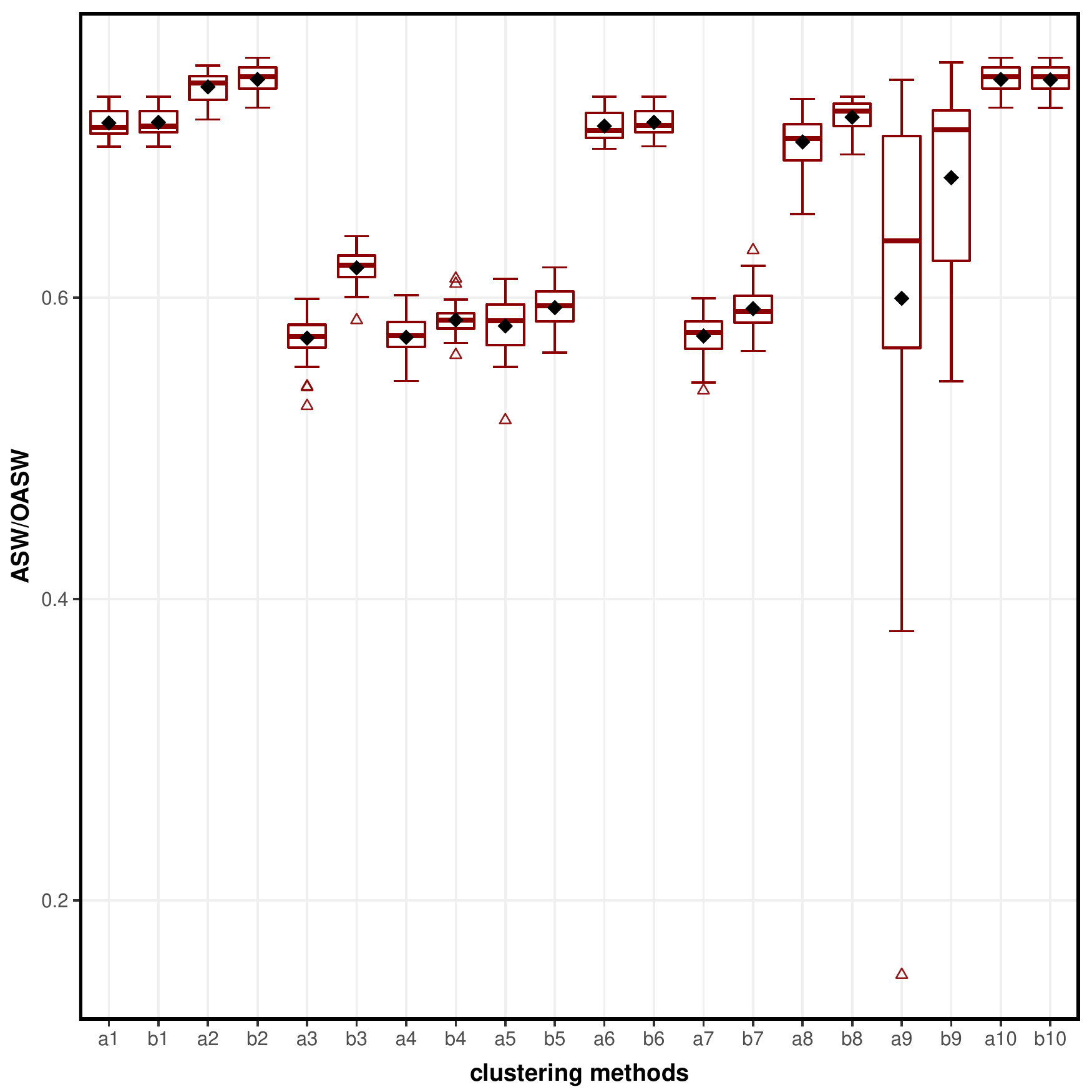}
 }
\subfloat[ ]{
  \includegraphics[width=45mm]{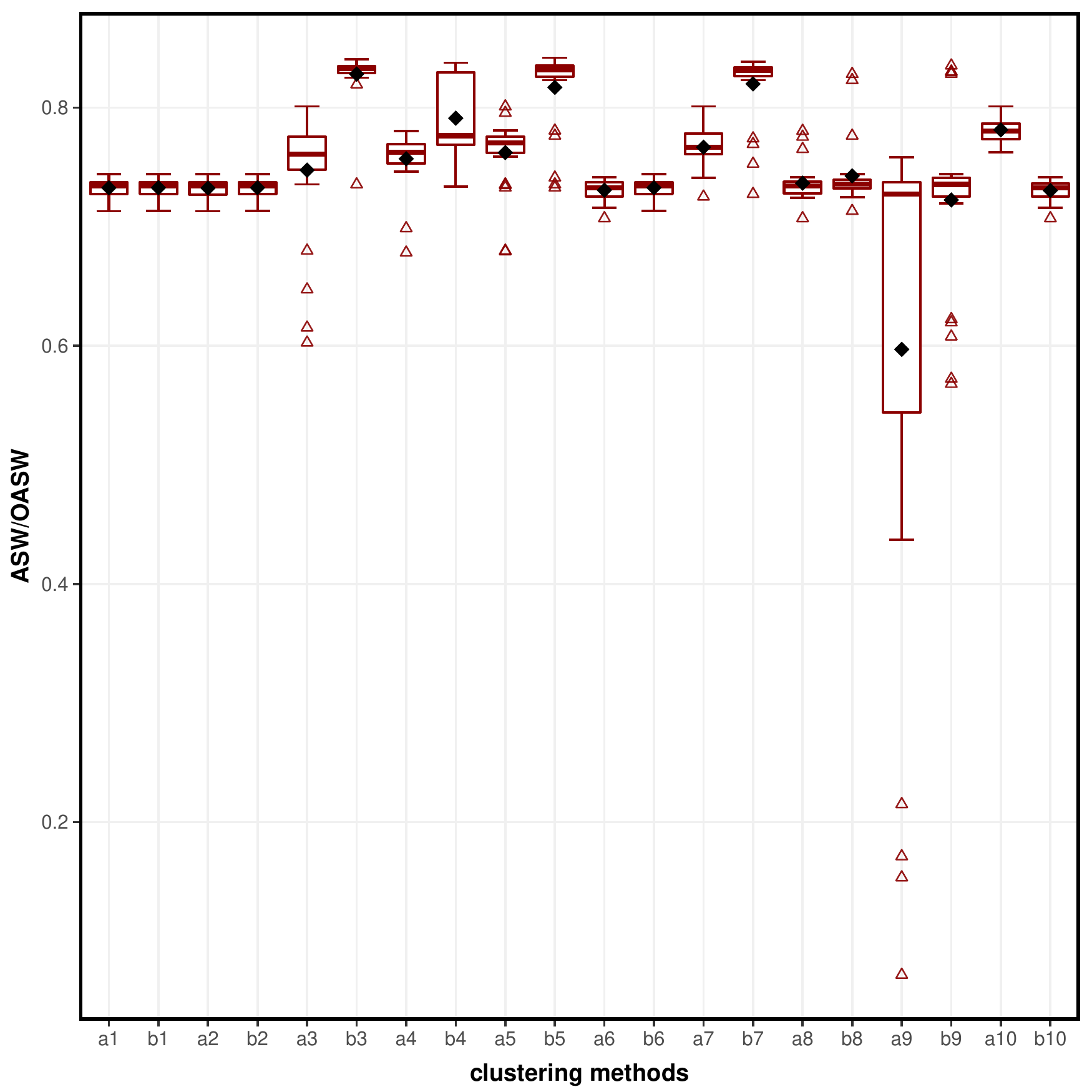}
 }
\newline
\rule{-5ex}{.2in}
\subfloat[ ]{
  \includegraphics[width=45mm]{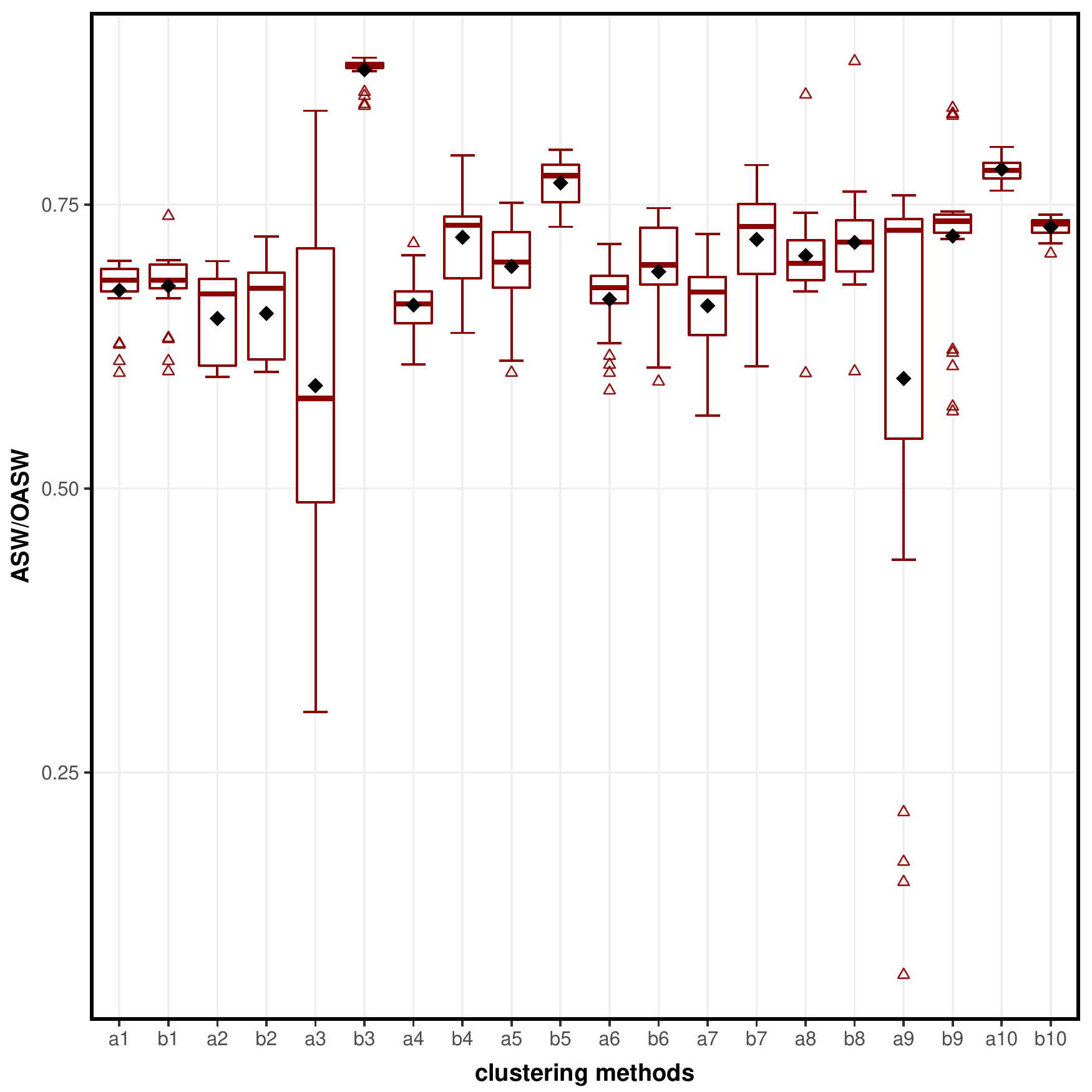}
 }
\subfloat[ ]{
  \includegraphics[width=45mm]{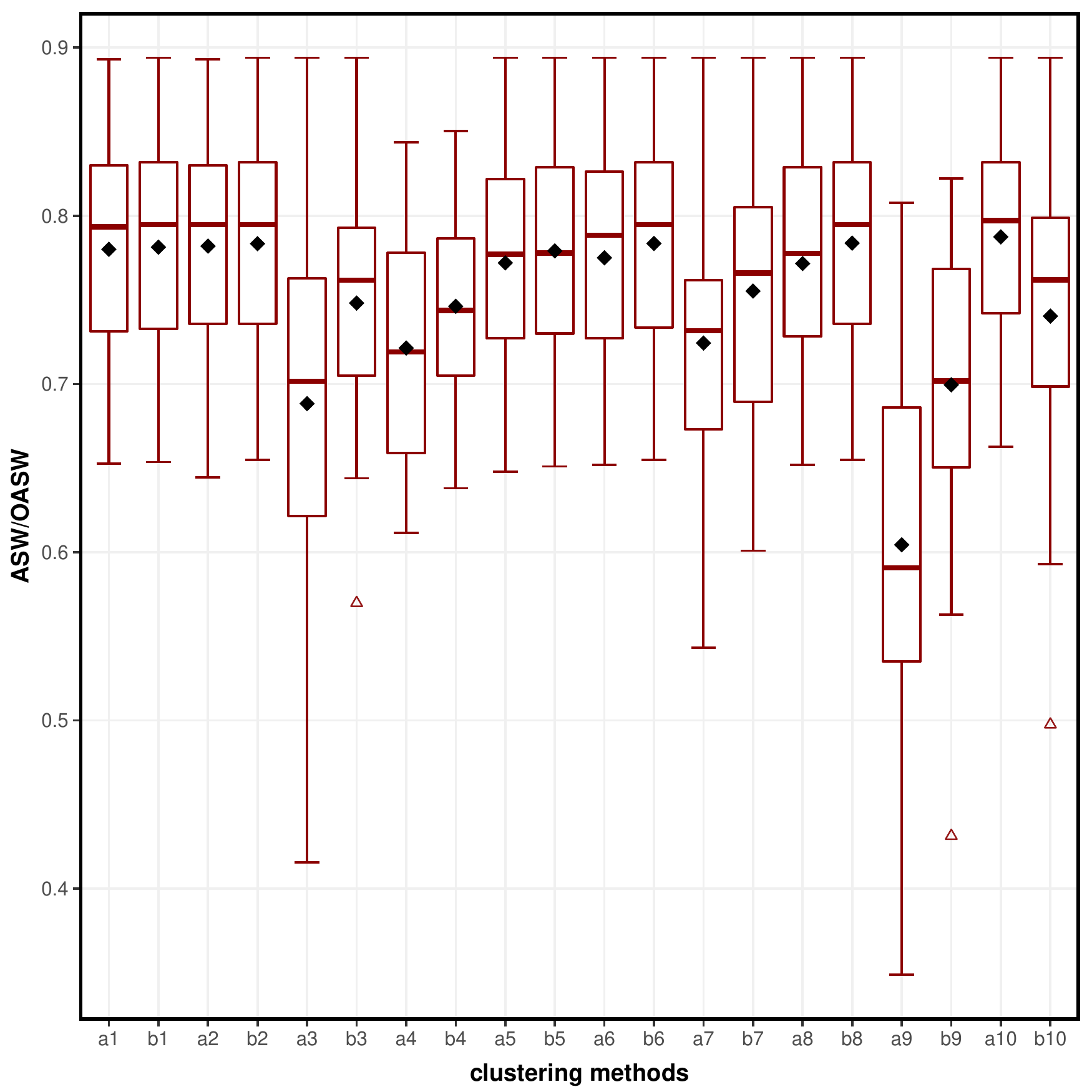}
 }
\subfloat[ ]{
  \includegraphics[width=45mm]{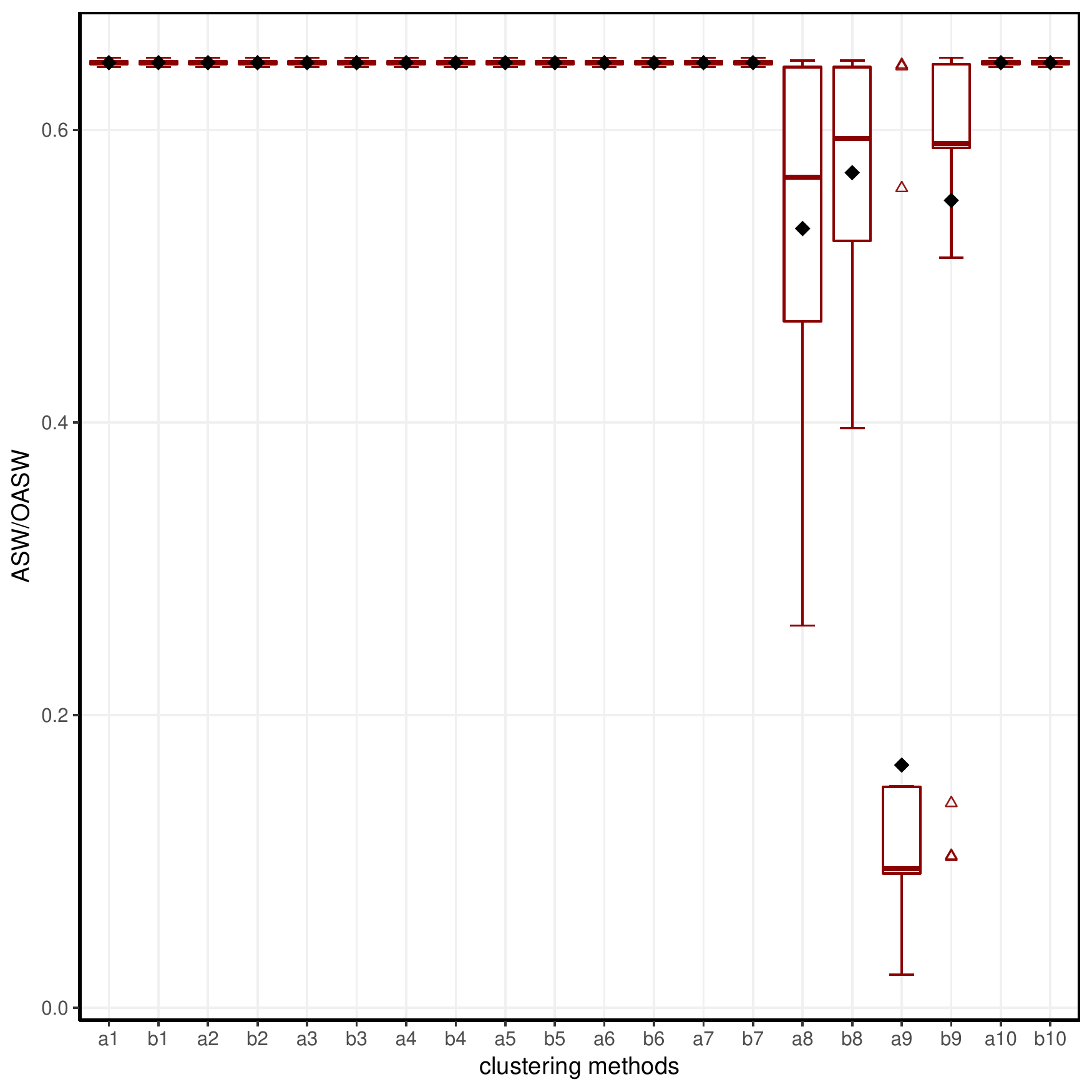}}
\caption{Boxplots for the average silhouette width values calculated against the clustering obtained from the existing clustering methods and optimum average silhouette width values (represented after /) calculated against these clustering methods when used as the initializations for $OSil$ \textemdash  a1/b1(k-means),  a2/b2(PAM), a3/b3(single), a4/b4(complete),   a5/b5(average), a6/b6(Ward), a7/b7(Mcquitty), a8/b8(model-based),  a9/b9(spectral),  a10/b10(PAMSIL/true).  The mean value for all the methods are plotted as black diamonds and the outliers are red triangles. (a) - (i) represents the results for Model 1 to Model 9.}
\label{boxplotssim1}
\end{figure}

 The boxplots for ASW values from the 20 mehtods are plotted in Figure  \ref{boxplotssim1}. The empirical distributions of the ASW vary greatly across the initialization methods. The distribution is mostly not symmetric but left or right skewed. For a few times the distribution was also observed to be left or right J shaped.

\par  Among the standalone clustering methods category the best ASW values were mostly obtained form $k$-means or PAM clustering methods whereas the best ARI values were mostly obtained for model-based clustering methods.  

\par The \textbf{single linkage} hierarchical  clustering is mathematically  well formed clustering method due to its   properties rooted in topology, for instance, see
 \cite{carlsson2013classifying}. However, it did not perform well with ASW from all  DGPs. The ASW values obtained including the OSil algorithm initialized with it were very low together with the corresponding ARI values as compared to other methods. 
 \par The \textbf{complete linkage} hierarchical  clustering method never gave the best ASW, OASW or ARI value for the data structures included in this study except for Model 9. Similar result is true for the \textbf{McQuitty similarity} method  which has only given the best ASW  for Model 6 with a very low ARI.  Other than this it has never  given maximum ASW, or ARI.  

\par The \textbf{spectral} clustering initialization has only improved ASW once (as compared to other clustering methods) namely Model 1 with the best ARI value as well. But many other methods also achieve this best value for the ARI. For all the other models spectral clustering has never achieved the best ASW value standalone or when used as an initialization for OSil, and the ARI values were also very low. 

\par \textbf{ARI discussion} The OSil clustering has decreased the  best ARI (row maximum) value  obtained  as compared to the standalone clustering methods for some DGPs. For instance for Model 1 the highest ARI value was obtained for the standalone model-based clustering method, whereas the highest value of ARI for OSil was obtained with Wards initialization  which is smaller than what is mentioned earlier.     In particular, the best ARI values from OSil clustering for the Models 1-3, 6 and 8 have decreased.      On the contrary, the OSil  has increased the ARI for Model 4 and 7. For Model 4 both PAMSIL and OSil gave same ARI value. For Model 5, PAMSIL has performed best in terms of ARI, keeping in mind OSil with PAM initialization has also performed very close to this. For Model 7, OSil is a clear winner leaving all other methods much behind in clustering quality. For Models 9 many methods gave ARI=1.

\par However, it is interesting to note that, although for Models 1-3, 6 and 8 the best ARI values obtained from OSil has decreased but OSil has improved the individual ARI values for several clustering methods. This highlights the fact that OSil has the potential to improve the clustering quality of these methods.  I now summarize these results in the Table \ref{simonesummaryARIcomp}. A \cmark, \xmark, and \textbf{=} represents that OSil has increased, decreased, and not changed the ARI values when a clustering method was used as an initialization for OSil as compared to the ARI values obtained from the same clustering methods for their standalone use.
\begin{table}[!htb]
\fontsize{8}{8}\selectfont
\renewcommand{\arraystretch}{2.0}
\caption{ASW index values obtained for all clustering  methods for the nine DGPs for the fixed $k$.  The best values are made bold in each row. } 
\begin{tabular}{ c c c c c c c c c c c c c c c c }
\toprule
DGPs  & $k$-means & PAM & single & complete & average & Ward & McQuitty & model-based & spectral & PAMSIL \\
\midrule
\multirow{2}{*}{Model 1} &   0.6684 &  \textbf{0.6689} & 0.4008 &  0.5701 & 0.6217 & 0.6596 & 0.5161  & 0.6488 & 0.6589 & -  \\
					 & \textbf{0.6697} &  \textbf{0.6697} & 0.4782 &  0.6588 & 0.6345 &  \textbf{0.6697} &  0.6489  &  \textbf{0.6697} & \textbf{ 0.6697} & \textbf{0.6697}    \\
						\cline{2-12} 
\multirow{2}{*}{Model 2}  & \textbf{0.7114} &  0.7103 & 0.2976 &  0.6128 &  0.6851 & 0.6975 & 0.5889 &  0.6780  & 0.6164  & -   \\
 			   			  &  \textbf{0.7118 } & \textbf{0.7118} &  0.4161 &  0.6859  & 0.6968 &  \textbf{0.7118} & 0.6726 & \textbf{0.7118} & 0.6536 &  0.7117  \\
						\cline{2-12} 
\multirow{2}{*}{Model 3}   & 0.6923 & \textbf{0.6936} & 0.4446 & 0.6234 & 0.6385  &  0.6770 &  0.6021 & 0.6470 &  0.5770 & -   \\
					  &   0.6939 & \textbf{0.6940} &  0.5804 & 0.6590 &  0.6535 &  0.6914 & 0.6457  & \textbf{0.6940 } & 0.6670  & \textbf{0.6940}   \\
						\cline{2-12} 
\multirow{2}{*}{Model 4}   &  \textbf{ 0.8254} & \textbf{0.8254}   & 0.6887 & 0.7876 &  0.8247 & 0.8239  & 0.7996 &  0.8143 & 0.5899 & - \\
						 &  \textbf{ 0.8255} &  \textbf{0.8255} & 0.7445 &  0.8008 & \textbf{0.8255}  &   \textbf{0.8255} & 0.8145 & \textbf{ 0.8255} & 0.6660 & \textbf{0.8255} & \\
						\cline{2-12} 
\multirow{2}{*}{Model 5}   &  0.7159 & \textbf{0.7398} & 0.5731 & 0.5737 &  0.5812 & 0.7138 & 0.5745 & 0.7033 &  0.5995 & - \\
						  &  
0.7163 & \textbf{0.7448} &  0.6197 &  0.5851 &  0.5934 & 0.7163  &  0.5927 & 0.7197 & 0.6796 & \textbf{ 0.7448}   \\
						\cline{2-12} 
\multirow{2}{*}{Model 6}   & 0.7327 & 0.7325 & 0.7476  &  0.7570 &  0.7623 & 0.7307 & \textbf{0.7668}  &  0.7366 & 0.5970  & -    \\
						  &  0.7327  & 0.7327 &  \textbf{0.8282}  &  0.7912 & 0.8170 & 0.7328  & 0.8200 & 0.7427& 0.7225 & 0.7813    \\
						\cline{2-12} 
\multirow{2}{*}{Model 7} & 0.6748 & 0.6500 & 0.5908 & 0.6618  & 0.6957  & 0.6668 & 0.6611 &  \textbf{0.7052 }  & 0.5987 & -   \\
						  & 0.6785 & 0.6543 & \textbf{0.8691 }  &  0.7215 & 0.7693 & 0.6911 & 0.7196 & 0.7167 & 0.7117 & 0.7790   \\
						\cline{2-12} 
\multirow{2}{*}{Model 8}  &  0.7801 & \textbf{ 0.7820} & 0.6884 & 0.7214 &  0.7720 & 0.7751 & 0.7244 & 0.7716 & 0.6045 & - \\
						  &   0.7814  & 0.7834 & 0.7481 & 0.7462 & 0.7792 & 0.7836 & 0.7553  & 0.7838 & 0.6995 & \textbf{0.7875}  \\
						\cline{2-12} 
\multirow{2}{*}{Model 9}  & \textbf{ 0.6461} &   \textbf{0.6461} & \textbf{0.6461} & \textbf{0.6461} & \textbf{0.6461} & \textbf{0.6461}  & \textbf{ 0.6461} &  0.1660  & 0.5327 & -  \\
						  & \textbf{0.6461} & \textbf{0.6461} & \textbf{0.6461} &  \textbf{0.6461}  & \textbf{0.6461} & \textbf{0.6461} &  \textbf{0.6461 } & 0.5520 & 0.5709 & \textbf{0.6461}    \\
\bottomrule
\end{tabular}
\begin{tablenotes}
    \item   The first row for each model represents the ASW values against the standalone methods, whereas the second row represents the optimized value of the objective function from OSil.
    \end{tablenotes}
     \label{aswoaswsumma}
\end{table}

\begin{table}[!htb]
\fontsize{8}{8}\selectfont
\renewcommand{\arraystretch}{2.0}
\caption{ARI results for the clustering results for fixed $k$. } 
\begin{tabular}{ c c c c c c c c c c c c }
\toprule
DGMs    & $k$-means & PAM & single & complete & average & Ward & McQuitty & model-based & spectral & PAMSIL \\
\midrule
\multirow{2}{*}{Model 1}						 &  0.8197 & 0.8351 & 0.1172 & 0.4831 & 0.6652  & 0.9387 & 0.3569& \textbf{0.9920} & 0.9575 & -  \\
						  & 0.8573 & 0.8573 & 0.1742 & 0.8239 & 0.6836 & 0.8603 & 0.7823 & 0.8603 & 0.8603 & 0.8448   \\
						\cline{2-11} 
\multirow{2}{*}{Model 2} 			   			  &  0.8463 & 0.8491  &  0.3844 &  0.6039 & 0.8161 &  0.9140 & 0.5912 & \textbf{0.9880} &  0.8704  &  -   \\
						  & 0.8556 &  0.8556 & 0.3535 & 0.7631 &  0.7985 & 0.8570 &  0.7313 &   0.8577 & 0.7993 & 0.7967 \\
						\cline{2-11} 
\multirow{2}{*}{Model 3}						  & 0.8711  & 0.8857 &  0.4219 &  0.7054  &  0.6681 & 0.9125 & 0.6704 & \textbf{0.9920 }& 0.8321 & - 
 \\
					  & 0.8881 & 0.8871  & 0.3999 & 0.7643 & 0.6661 & 0.8866& 0.7122 & 0.8910& 0.7941&  0.9352 \\
						\cline{2-11} 
		\multirow{2}{*}{Model 4}				   &  0.9845 &  0.9837 &  0.8029& 0.9374 &   0.9830 &  0.9846& 0.9444& 0.9744  & 0.8457  &- \\
						  &  0.9845&  \textbf{0.9853} &  0.7990 &  0.9568& 0.9845 & 0.9845  & 0.9686&  \textbf{0.9853}&  0.8478& \textbf{ 0.9853} \\
						\cline{2-11} 
\multirow{2}{*}{Model 5}						 & 0.7716  &  0.9806 &  0.5218 &   0.2845 & 0.3012&   0.7750 &0.2810 & 0.7762 & 0.7535 & -   \\
						 & 0.7733  & 0.9984 & 0.5170 & 0.2856  & 0.3006& 0.7743& 0.2830& 0.7822& 0.7841& \textbf{1} \\
						\cline{2-11} 
	\multirow{2}{*}{Model 6} 					  & 0.9795 & 0.9784 & 0.7842 & 0.7386& 0.7909& \textbf{0.9992}& 0.7511  & 0.9697 & 0.8967 & -  \\
					  & 0.9822 & 0.9822 & 0.7786 & 0.7586 & 0.7926  & 0.9834 & 0.7710& 0.9565  & 0.8930& 0.7657 \\
						\cline{2-11} 
\multirow{2}{*}{Model 7}						  & 0.6434 & 0.6535 & 0.5351 & 0.6444 & 0.6839& 0.6477 & 0.6421& 0.6473& 0.6790 & - \\
						  & 0.6469 & 0.6509& 0.5315 & 0.6701 & \textbf{0.6939} & 0.6558 & 0.6702 & 0.6537& 0.6907 & 0.4923  \\
						\cline{2-11} 
\multirow{2}{*}{Model 8}						   & 0.9019 & 0.9145 & 0.7855& 0.7963 & 0.9028& 0.9139& 0.8279&\textbf{ 0.9418}& 0.7614 & - \\
					  & 0.9101 &  0.9185  & 0.7787& 0.8335 & 0.9034 & 0.9119 & 0.8495 & 0.9173 & 0.7608 & 0.7885 \\
						\cline{2-11} 
\multirow{2}{*}{Model 9}					  & \textbf{1}  & \textbf{1} & \textbf{1} & \textbf{1} & \textbf{1} & \textbf{1} & \textbf{1}  & 0.4788 & 0.7491  & - \\
						  & \textbf{1}   & \textbf{1} & \textbf{1} & \textbf{1} & \textbf{1} & \textbf{1} & \textbf{1} & 0.9123 & 0.805 & \textbf{1} \\
\bottomrule
\end{tabular}
\begin{tablenotes}
    \item 
The first row for each model represents the ARI obtained for the standalone clustering methods  whereas the second row represents the ARI values for the clustering obtained from OSil.
\end{tablenotes}
   \label{arisummary}
\end{table}

{
\renewcommand{\arraystretch}{0.9}
\begin{table}[!htbp]
\fontsize{9}{9}\selectfont
\caption{Summary table for the comparison of the  ARI values obtained from OSil clustering and the maximum ASW clustering for Simulation I}
\begin{threeparttable}
\begin{tabular}{ c c c c c c c c c c }
\toprule
DGMs & $k$-means & PAM & single & complete & average & Ward & McQuitty & spectral & model-based  \\
\midrule 
Model 1 & \cmark & \cmark &  \cmark   & \cmark  &  \cmark & \xmark & \cmark & \xmark & \xmark   \\
Model 2 &  \cmark & \cmark &  \xmark & \cmark & \xmark & \xmark &\cmark & \xmark & \xmark    \\
Model 3 & \cmark & \cmark  &  \xmark & \cmark  & \xmark & \xmark & \cmark &  \xmark & \xmark   \\
Model 4 &  \textbf{=} & \cmark & \xmark & \xmark  & \cmark & \textbf{=} & \cmark & \cmark & \cmark  \\
Model 5 & \cmark & \cmark   & \xmark & \cmark & \xmark & \xmark  & \cmark & \cmark & \cmark  \\
Model 6 & \cmark & \cmark & \cmark & \cmark &  \cmark &  \xmark   &  \cmark & \xmark & \xmark  \\
Model 7 & \cmark & \xmark & \xmark & \cmark &   \cmark & \cmark & \cmark & \cmark &\cmark    \\
Model 8 & \cmark &  \cmark & \xmark & \cmark & \cmark & \xmark  &  \cmark & \xmark &  \xmark \\
Model 9 &  \textbf{ =}   &  \textbf{ =}  & \textbf{ =}  &  \textbf{ =}  & \textbf{ =}  & \textbf{ =}  & \textbf{ =}  & \cmark & \cmark\\
\bottomrule
\end{tabular}
     \end{threeparttable}
     \label{simonesummaryARIcomp}
\end{table}
}

\par \textbf{Which clustering method is the best in optimizing ASW?}
OSil has always increased the ASW value for all models and all initialization methods included in the study. PAMSIL has also achieved the same ASW value as obtained by OSil  for Models 1, 3-5, 9. However, it has given the higher value of ASW only for Model 8.

\par \textbf{Which method is the best in terms of ARI?}
No single method wins here. The standalone model-based clustering method has given the best ARI values for Model 1-3, and 8. For Model 4, both OSil and PAMSIL gave the same value, for Model 5 only PAMSIL, for Model 6 standalone Wards method, for Model 7 only OSil and for Model 9 many methods gave ARI=1.

\par\textbf{ OSil versus PAMSIL }  The superiority of OSil as compared to PAMSIL is evident from the following points.
\begin{enumerate}
\item  The ASW values obtained from OSil are much higher than those of PAMSIL for Models 4, 6, and 7.
\item  Out of these, the ARI for Models 6 and 9 for PAMSIL were very low as compared to OSil.
\item The models 1 and 2 for which PAMSIL gave similar values of OASW as that from OSil, its resulting ARI values are very low as compared to OSil. 
\item  For Model 8  PAMSIL gave the highest ASW value but with very low ARI as compared to OSil.
\end{enumerate}

\par \cite{van2003new} has shown that PAMSIL is good for detecting small sized clusters in presence of bigger sized clusters, where the clusters have same covariances. By small sized clusters they mean number of observations in clusters. This is not always true or partially true and is strongly connected with the distance between the means of clusters. This is confirmed from the results (Table \ref{aggregationofresults}) of Model 6, where PAMSIL and related methods (ASW, OSil) fail to estimate 5 as a number of clusters.


\subsection{Simulation II}\label{simmodmaintwo}
\par For the estimation of the number of clusters, the comparison of the proposed method was considered for a broad spectrum of existing methods. Among the  existing estimation methods for the number of clusters: H, Gamma, C, KL, CH, Gap, Jump, PS, BI, CVNN, model-based, ASW were considered. For each of these indices nine clustering methods, namely $k$-means, $PAM$, namely single, complete, average, Ward's, McQuitty, spectral clustering and model-based clustering methods were used. Moreover, OSil was also used for the estimation of $k$ with these nine clustering methods. In addition, PAMSIL and model-based clustering with BIC were considered. The nine DGPs defined in the Appendix \ref{dgpdefs} were used for generating the data.

\par  A list of references to all indices together with their R language implementation used in the study are given in Table \ref{listofmethod}. 
Let the maximum number allowed to estimate the number of clusters is $K \in \mathds{N}_n$, such that the number of clusters are estimated for 2 to K clusters.  The maximum number allowed in the simulation for $K$ was 12. For each DGP 25 data sets were generated. The model-based and spectral clustering methods are not available to estimate the number of clusters with the Gap statistics with it's current R implementation. 

\par Therefore, in a single run of a simulation I have estimated the number of clusters, from range 2 to K, with 105 methods (10 indices $\times$ 9 clustering methods + Jump method with 6 transformation powers + PAMSIL + model-based clustering with BIC - 2 (exclude two clustering methods for Gap method) + 9 (OSil initialized with 9 clustering methods)) for a single data set. In total 105 $\times$ 25 (runs) $\times$ 9 (data models) = 23,625 times the numbers of clusters were estimated.  All the simulations were done on a 2.8 GHz Intel core i7 processor.

\paragraph{Performance evaluation} The estimation of number of clusters from OSil was decided based on the best OASW value obtained for the  number of clusters in the range 2 to $K$.    Nine tables were generated one for each DGP reporting the frequency counts each method gave as an estimate for number of clusters from 2 to $K$ (11 columns) for an in depth analysis. An overall table consist of frequency counts only for the true number of clusters for all DGPs for all the clustering methods across the indices form these nice tables were prepared. These results are presented in Table \ref{aggregationofresults}. 

\subsection{Results}\label{somesummary}

\noindent \textbf{CH} has not performed well except Model 5 (only with PAM, Wards, model-based clustering). For the rest of the models,  CH has either failed to estimate $k$ with many clustering methods or has performed below 40$\%$. \\
\noindent\textbf{H} index has performed very poorly for all models except for Model 8. PAM has never estimated the correct number of clusters with the H index except for Model 8. \textbf{Gamma} and the \textbf{C} indices have consistently performed poorly except for Model 4 $\&$ 8. These two indices have failed with many clustering methods in the estimation of the number of clusters. \textbf{KL} has also performed poorly with all clustering methods. \\

\noindent\textbf{Gap} method has performed above 60$\%$ except with single linkage for Model 1. Gap in combinations with all clustering methods has performed  poorly for Models 2, 3, 5, 6, 7 and well for Model 4 (except with single linkage). \\
\noindent \textbf{Jump} has estimated the correct number of clusters  with p/3 (87.5\%) for Model 1, p/3 (97\%) for Model 2, p/5 (58\%) for Model 3, p/2 (100\%) for Model 4 and 6 and never for Model 5, 7, and 8.\\
\noindent  \textbf{PS} has performed poorly with complete linkage clustering. It has performed 100\% with model-based clustering for Model 1, with $k$-means and model-based clustering for Model 2, with PAM and model-based clustering for Model 4, with PAM for Model 5, with PAM, Ward and model-based clustering for Model 6, with PAM, single, complete, average, Ward, McQuitty and model-based clustering for Model 8. It has estimated the desired number of clusters for Model 3 with $k$-mean, PAM and Ward about 3\%, single linkage 37\%, average linkage about 12\%, McQuitty about 10\%, model-based clustering about 82\% and for Model 7 only with single linkage (about 30\%).  \\ 
\noindent \textbf{BI} has never been  able to estimate the correct number of clusters for Model 6. Only a few clustering methods performed well in combination with this index for Model 1 and 2. Model-based clustering with BI has never been able to estimate the correct number of clusters. BI  has performed well only for Model 4 with all clustering methods except $k$-means and single linkage clustering. \\
\noindent \textbf{CVNN} has performed well only for Models 1, 2, 3, 4 (except single linkage). It has performed poorly for Models 5, 7, and 8 in combinations with all clustering methods.  It has performed well with Model 6 only with Ward's and model-based clustering. \\

\noindent \textbf{BIC} in combination with model-based clustering method has estimated the correct number of clusters 100$\%$ of the times for Models 1, 2, 6, 35$\%$ for Model 3, less than 30$\%$ for Model 4, very poorly for Model 5 and never for Models 7 and 8. \\
 \noindent \textbf{ASW} shows an overall good performance with Models 1 and 2, a very good performance for Models 4, 8, and 10, and a poor performance for model 3. It also  performed well for Model 5, but only with a few clustering methods, and it was never able to estimate the correct number of clusters for Models 6 and 7.  ASW mostly showed better performance than PAMSIL in combination with $k$-means and spectral clustering.\\
\noindent \textbf{ PAMSIL} has estimated the correct number of clusters for 100\% of the simulations for Models 4, 5, 8, and 9. The performance rate is 80\% for Model 1, 28\%  for Model 2, 12\% for Model 3 and 0\% for Models 6 and 7.\\
\noindent \textbf{OSil} has 88\% performance rate for Model 1 for the estimation of number of clusters. It has shown good performance (100\%) for the estimation of number of clusters for Models 4, 5, 8, 9 with various initialization methods. It performed poorly for Models 2, Model 3(estimated number of clusters as 2 instead of 3 majority of the times), 6 (always estimated 4 as a number of clusters instead of 5), 7 (always estimated number of clusters as 3 instead of 7).
 
{
\setlength{\tabcolsep}{1.5pt}
\renewcommand{\arraystretch}{0.70}%
\fontsize{9}{9}\selectfont
\begin{longtable}{@{\extracolsep{\fill}}l c c c c c c c c c c  @{}}
 \caption{Frequency table of indication of cluster estimation at correct level for all the indices in combination with all the methods for Model 1-9} \label{aggregationofresults}\\
\toprule
Models & M1 & M2 & M3 & M4 & M5 & M6 & M7 & M8  & M9\\
No.of dims. &  2 & 2 & 2 &  2 & 2 & 5 & 10 & 500  & 60 \\
No. of clusters & 2 & 3 & 4 & 5 & 6 & 5 & 7 & 10  & 7 & Overall\\
\midrule
& \multicolumn{10}{c}{CH} \\
\cline{2-11}
kmeans      & 18 &2 & 0 & 4  & 4 & 7 & 5 & 2    & 12 & 57\\
PAM         & 14 &1 & 0 & 0  & 0 & 25 & 1 & 2   & 25 & 68\\
Single      & 3 & 6 & 0 & 2  & 3 & 0 & 1 & 18   &  25 & 65\\
Complete    & 3 & 0 & 0 & 1  & 0 & 0 & 1 & 0    &  25 & 30\\
Average     & 6 & 1 & 0 & 11 & 0 & 3 & 0 & 10   & 25 & 106\\
Ward        & 6 & 0 & 0 & 0  & 2 & 25 & 1 & 0   & 25 & 59\\
McQuitty    & 2 & 0 & 0 & 3  & 0 & 0 & 1 & 0    & 25 & 31\\
Model-based & 6 & 4 & 1 & 7  & 7 & 25 & 10 & 0  & 14 & 74 \\
Spectral    & 19 &8 & 1 & 9  & 8 & 7 & 3 & 6    & 2 & 53\\
\hline
& \multicolumn{10}{c}{H} \\
\cline{2-10}
kmeans       & 5 & 4 & 7 & 2 & 2 & 0 & 1 & 0  & 0 & 21\\
PAM          & 0 & 0 & 0 & 0 & 0 & 0 & 0 & 25 & 0  & 25\\
Single       & 0 & 0 & 1 & 3 & 2 & 0 & 0 & 24 & 0  & 30\\
Complete     & 4 & 0 & 2 & 0 & 2 & 3 & 0 & 25 & 0  & 36\\
Average      & 3 & 2 & 1 & 3 & 2 & 7 & 1 & 23 & 0  & 42\\
Ward         & 0 & 0 & 0 & 0 & 0 & 0 & 0 & 25 & 0  & 25\\
McQuitty     & 2 & 2 & 2 & 1 & 3 & 4 & 1 & 25 & 0  & 40\\
Model-based  & 8 & 5 & 5 & 3 & 1 & 3 & 3 & 25 & 0  & 53\\
Spectral     & 4 & 5 & 4 & 1 & 3 & 4 & 1 & 0  & 0  & 22\\
\hline
& \multicolumn{10}{c}{Gamma} \\
\cline{2-10}
kmeans      & 1 & 0 & 1 & 16  & 4 & 0 & 0 & 0   & 13 & 35\\
PAM         & 0 & 0 & 0 & 20  & 0 & 0 & 0 & 25  & 25 & 70\\
Single      & 0 & 0 & 1 & 1   & 2 & 0 & 0 & 25  & 25 & 54\\
Complete    & 0 & 0 & 0 & 14  & 0 & 0 & 0 & 25  & 25 & 64\\
Average     & 0 & 0 & 0 & 7   & 1 & 0 & 0 & 25  & 25 & 58\\
Ward        & 0 & 0 & 0 & 16  & 0 & 0 & 0 & 25  & 25 & 66\\
McQuitty    & 0 & 0 & 0 & 8   & 0 & 0 & 0 & 25  & 25 & 58\\
Model-based & 0 & 0 & 0 & 13  & 0 & 0 & 0 & 25  & 14 & 52 \\
Spectral    & 0 & 1 & 0 & 10  & 4 & 0 & 2 & 8 & 0 & 25\\
\hline
& \multicolumn{10}{c}{C} \\
\cline{2-10}
kmeans        & 1 & 0 & 1 & 17 & 5 & 0 & 0 & 0   & 13 & 37\\
PAM           & 0 & 0 & 0 & 18 & 0 & 0 & 0 & 25  & 25 & 68 \\
Single        & 0 & 0 & 1 & 2  & 2 & 0 & 0 & 25  & 25 & 55\\
Complete      & 0 & 0 & 0 & 10 & 1 & 0 & 0 & 25  & 25 & 61\\
Average       & 0 & 0 & 0 & 6  & 3 & 0 & 0 & 25  & 25 & 59\\
Ward          & 0 & 0 & 0 & 14 & 0 & 0 & 0 & 25  & 25 & 39\\
McQuitty      & 0 & 0 & 0 & 7  & 3 & 0 & 0 & 25  & 25 & 63\\
Model-based   & 0 & 0 & 0 & 14 & 0 & 0 & 1 & 25  & 14 & 54\\
Spectral      & 0 & 0 & 0 & 11 & 5 & 0 & 2 & 8   & 0 & 26\\
\hline
& \multicolumn{10}{c}{KL} \\
\cline{2-10}
kmeans        & 8 & 3 & 0 & 2 & 5 & 0 & 0 & 0  & 7 & 25\\
PAM           & 7 & 4 & 4 & 4 & 7 & 0 & 0 & 0  & 0 & 26\\
Single        & 7 & 5 & 4 & 6 & 6 & 0 & 1 & 0  & 0 & 39\\
Complete      & 3 & 7 & 6 & 5 & 0 & 0 & 0 & 0  & 0 & 21\\
Average       & 5 & 2 & 3 & 1 & 7 & 0 & 0 & 0  & 0 & 18\\
Ward          & 6 & 2 & 5 & 1 & 5 & 0 & 0 & 0  & 0 & 19\\
McQuitty      & 0 & 6 & 6 & 0 & 2 & 0 & 1 & 0  & 0 & 15\\
Model-based   & 2 & 6 & 1 & 2 & 1 & 0 & 0 & 0  & 4 & 12\\
Spectral      & 2 & 2 & 3 & 6 & 5 & 0 & 1 & 0  & 5 & 11\\ 
\hline
& \multicolumn{10}{c}{Gap} \\
\cline{2-10}
kmeans    & 20 & 9  & 4 & 10 & 1  & 3 & 1 & 0 & 4 & 52\\
PAM       & 15 & 8  & 0 & 10 & 0  & 4 & 0 & 0 & 21 & 58\\
Single    & 3  & 3  & 0 & 2  & 0  & 4 & 0 & 0 & 24 & 36\\
Complete  & 14 & 5  & 3 & 16 & 0  & 3 & 0 & 0 & 25 & 67\\
Average   & 18 & 11 & 1 & 22 & 0  & 0 & 0 & 0 & 25 & 77\\
Ward      & 13 & 9  & 0 & 21 & 11 & 0 & 0 & 0 & 23 & 82 \\ 
McQuitty  & 13 & 8  & 5 & 14 & 0  & 0 & 5 & 0 & 25 & 70\\
\hline
& \multicolumn{10}{c}{Jump} \\
\cline{2-10}
$p$/2 & 6  & 8  & 1  & 25 & 0 & 25 & 0 & 0 & 0 & 65\\
$p$/3 & 22 & 24 & 3  & 25 & 0 & 25 & 0 & 0 & 0 & 99\\
$p$/4 & 2  & 20 & 8  & 25 & 0 & 8 & 0 & 0  & 0 & 63\\
$p$/5 & 0  & 1  & 11 & 25 & 0 & 0 & 0 & 0  & 0 & 37\\
$p$/6 & 0  & 0  & 2  & 25 & 0 & 0 & 0 & 0  & 0 & 27\\
$p$/7 & 0  & 0  & 0  & 4  & 0 & 0 & 0 & 0  & 0 & 4\\
\hline
& \multicolumn{10}{c}{PS} \\
\cline{2-11}
kmeans     & 24 & 25 & 1  & 1  & 0  & 0 & 0 & 0    & 0 & 51\\
PAM        & 21 & 0  & 1  & 25 & 25 & 25 & 0 & 25  & 25 & 147\\
Single     & 12 & 2  & 9  & 0  & 0  & 14 & 7 & 25 & 0 & 69\\
Complete   & 0  & 0  & 0  & 20 & 0  & 0 & 0 & 25  & 25 & 70\\
Average    & 4  & 0  & 3  & 22 & 0  & 10 & 0 & 25  & 14 & 78\\
Ward       & 5  & 0  & 1  & 25 & 0  & 25 & 0 & 25  & 25 & 106\\
McQuitty   & 2  & 0  & 2  & 21 & 0  & 7 & 0 & 25   & 15 & 72\\
Model-based & 25 & 25 & 20 & 24 & 0  & 25 & 0 & 25  & 2 & 146\\
Spectral    & 18 & 1  & 0  & 0  & 0  & 0 & 0 & 0    & 0 & 19\\
\hline
& \multicolumn{10}{c}{BI} \\
\cline{2-10}
kmeans        & 25&22 & 0  & 5    & 1 & 0 & 1 & 0  & 1 & 46\\
PAM           & 6 & 0 & 4  & 24   & 6 & 0 & 0 & 0  & 25 & 65\\
Single        & 11 & 2 & 3 & 1    & 3 & 0 & 0 & 0  & 25 & 45\\
Complete      & 0 & 0 & 0  & 19   & 0 & 0 & 0 & 5  & 25 & 49\\
Average       & 0 & 0 & 0  & 21   & 0 & 0 & 0 & 0 & 25 & 46\\
Ward          & 0 & 0 & 0  & 23   & 0 & 0 & 0 & 6  & 25 & 54\\
McQuitty      & 0 & 0 & 1  & 17   & 0 & 0 & 0 & 7  & 25 & 50\\
Model-based   & 25& 25& 7  & 17   & 0 & 0 & 0 & 0  & 0 & 74\\
Spectral      & 5 & 1 & 0  & 1    & 0 & 0 & 3 & 0  & 0 & 20\\
\hline
& \multicolumn{10}{c}{CVNN} \\
\cline{2-10}
kmeans        & 9  & 21 & 7  & 17 & 3 & 3 & 0 & 0  & 25 & 85\\
PAM           & 11 & 21 & 16 & 25 & 0 & 8 & 0 & 0  & 25 & 106\\
Single        & 11 & 4  & 0  & 2  & 0 & 0 & 0 & 0  & 25 & 42\\
Complete      & 12 & 8  & 5  & 21 & 1 & 0 & 0 & 0  & 25 & 72\\
Average       & 3  & 20 & 1  & 23 & 0 & 3 & 0 & 0  & 25 & 75\\
Ward          & 10 & 21 & 16 & 23 & 5 & 15 & 0 & 0 & 12 & 102\\
McQuitty      & 17 & 8  & 5  & 20 & 0 & 0 & 0 & 0  & 25 & 75\\
Model-based   & 8  & 24 & 19 & 25 & 0 & 15 & 0 & 0  & 14 & 105\\
Spectral     & 17 & 3  & 4  & 7  & 3 & 0 & 1 & 0 & 0 & 36\\
\hline
& \multicolumn{10}{c}{BIC} \\
\cline{2-10}
Model-based & 25 & 25 & 10 & 11 & 1 & 25 & 0 & 0   & 15 & 126\\
\hline
PAMSIL & 20 & 7 & 2 & 25 & 25 & 0 & 0 & 25   & 25 & 129  \\
\hline
& \multicolumn{10}{c}{ASW} \\
\cline{2-10}
kmeans      & 21 & 13 & 1 & 17 & 9  & 0 & 0 & 2  & 13 & 76\\
PAM         & 20 & 12 & 2 & 25 & 24 & 0 & 0 & 25  & 25 & 133\\
Single      & 4  & 5  & 0 & 2  & 0  & 0 & 0 & 25  & 25 & 61\\
Complete    & 7  & 1  & 0 & 20 & 0  & 0 & 0 & 25 & 25 & 78\\
Average     & 13 & 11 & 1 & 24 & 0  & 0 & 0 & 25 & 25 & 99\\
Ward        & 19 & 8  & 2 & 25 & 11 & 0 & 0 & 25  & 25 & 115\\
McQuitty    & 7  & 2  & 1 & 20 & 0  & 0 & 0 & 25 & 25 & 80\\
Model-based & 20 & 13 & 3 & 24 & 0 & 0 & 0 & 25  & 14 & 110 \\
Spectral    & 24 & 12 & 1 & 14 & 10  & 0 & 0 & 7 & 0 & 68\\
\hline
& \multicolumn{10}{c}{OSil} \\
\cline{2-10}
kmeans         & 21 & 12 & 2 & 17 & 9  & 3 & 0 & 2  & 20 & 86\\
PAM            & 20 & 9  & 2 & 25 & 25 & 0 & 0 & 25   &  25& 131\\
Single         & 7 & 3   & 0 & 2  & 0  & 0 & 0 & 25 & 25 & 62\\
Complete       & 19 & 6  & 0 & 21 & 0  & 0 & 0 & 25   & 25& 96\\
Average        & 17 & 9  & 1 & 24 & 0  & 0 & 0 & 25   &25 & 101 \\
Ward           & 20 & 10 & 2 & 25 & 12 & 0 & 0 & 25   &25 & 114\\
McQuitty       & 15 & 7  & 1 & 22 & 0  & 0 & 0 & 25  &25 & 95\\
Model-based    & 21 & 9  & 3 & 25 & 0  & 0 & 0 & 25  & 14 & 97\\
Spectral       & 22 & 7  & 1 & 14 & 6  & 0 & 0 & 12 &   0 & 62\\
\bottomrule
\end{longtable}
}

\par The comprehensive findings from the Simulation II are given below.
\begin{enumerate}
\item The distance between clusters turned out to be a very significant characteristic for many clustering methods and indices to estimate the correct number of clusters. Many clustering methods, especially H, Gamma, C and  KL performed badly for the models with unequal difference between clusters' locations. 
Also, varying spread among the observations between the clusters is hard for many combinations to determined correctly. This includes different shaped clusters in the data like compact as well as wide clusters or other shapes like as generated from Uniform distributions. The  Gaussian clusters with different shapes and orientations across the different dimensions were not identified by the methods correctly. Even the methods that are designed especially for Gaussian data failed quite often, for instance, see the model-based clustering results for Models 6, 7, and 8. For ASW and OASW (including OSil and PAMSIL), the difference between the clusters mean locations turns out to be a challenging situation. 

\item \par A ranking of the indices included in the study is plotted in Figure \ref{barplotcounts}. The bars in the plot is made from the block sums of each index in Table \ref{aggregationofresults}. For instance, for CH sum of overall column is 543 and success rate is calculated by (543/25*9*9)*100. The model-based clustering with BIC, PAMSIL and Jump methods were not added in the figure as they do not corresponding to the same total.

\begin{figure}[H]
\centering
\includegraphics[width=50mm,height=40mm]{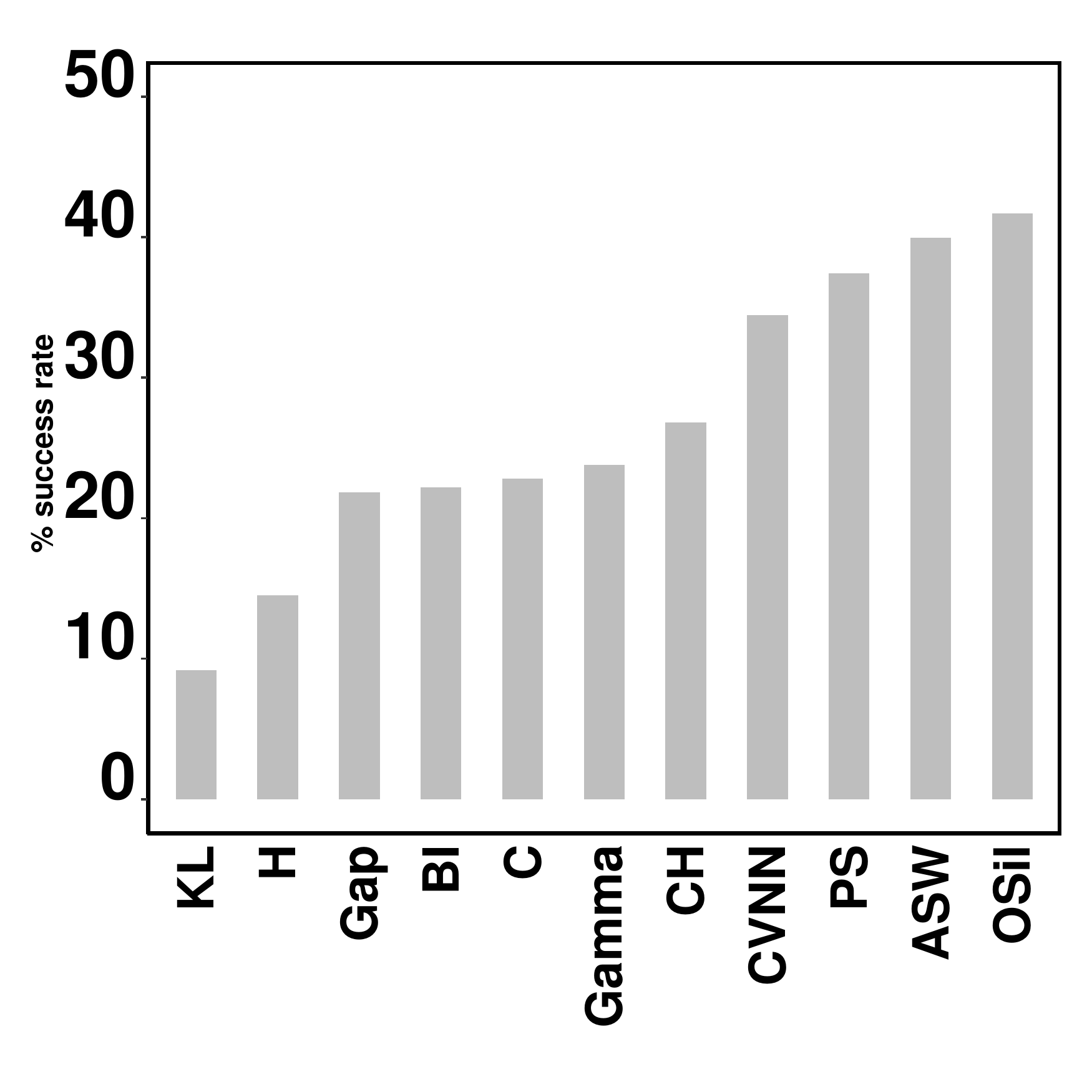} 
\caption{Overall results for the number of clusters estimation methods aggregated for  Simulation II across all DGPs.} 
\label{barplotcounts}
\end{figure}

\item The choice of clustering method with the indices matters and has an effect on the estimation of the number of clusters. For instance, for Model 2, PS and BI were never able to estimate the desired number of clusters using PAM, complete, single, Ward's, McQuitty and spectral. On the other hand they were able to estimate clusters at the desired level 100\% of the times with $k$-means and model-based clustering. There is plenty of other such evidence. The performance of each index with one model across these clustering methods differs significantly.   

\par As defined in above paragraphs, various indices has shown good performance with a particular clustering method only.   Figure \ref{bartwopaper} shows the \% success rate of each clustering index with the top performing clustering method. One best row for each index in Table \ref{aggregationofresults} was used to construct the plot. 

\begin{figure}[H]
\centering
\includegraphics[width=50mm,height=50mm]{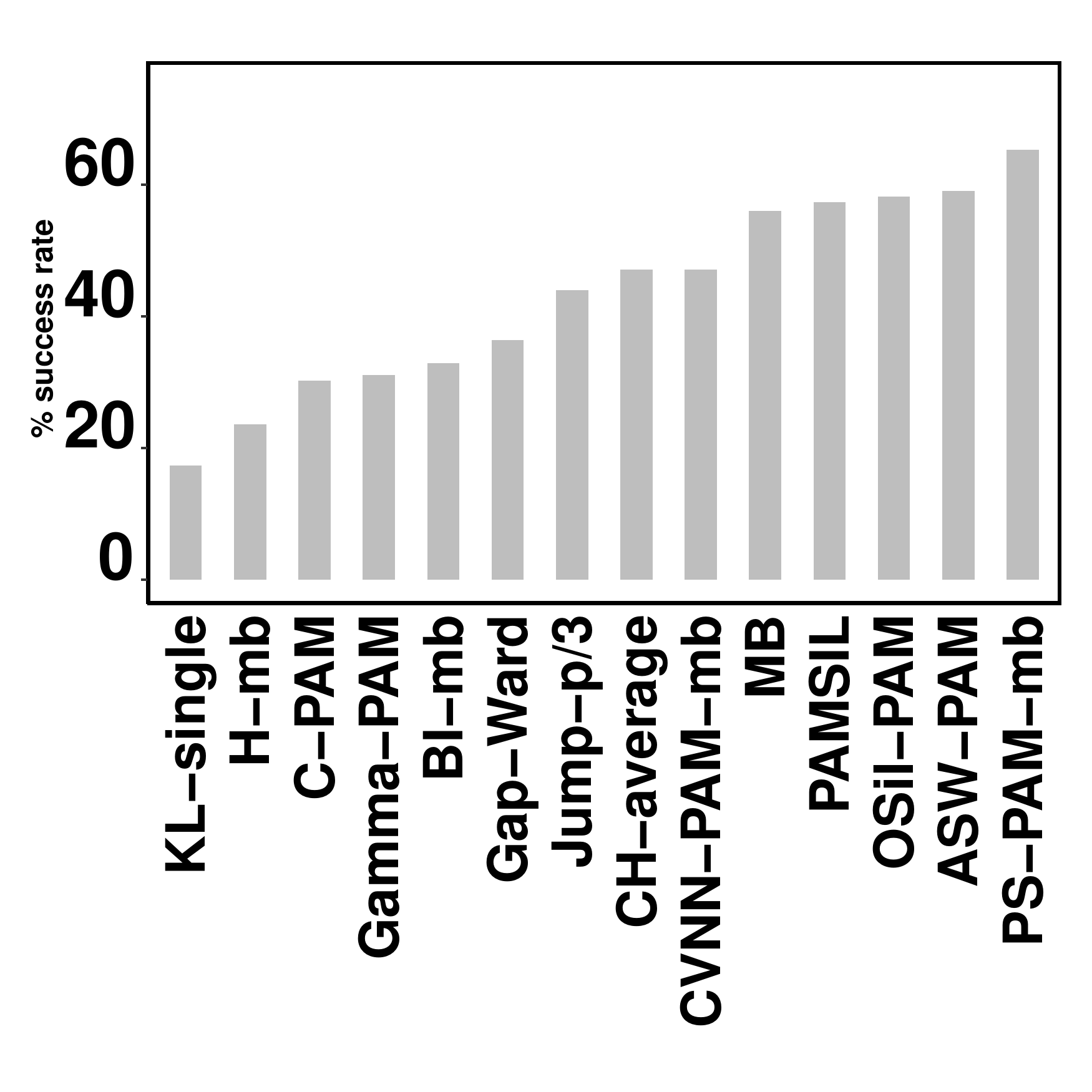} 
\caption{Overall results for  Simulation II aggregated over for all DGPs for all the indices in combination with clustering methods. } 
\label{bartwopaper}
\end{figure}

\item Table \ref{aswoaswsummarycomp}  shows the performance of ASW using OSil as compared to ASW for all the clustering methods and DGPs for the estimation of number of clusters. The comparison is done using the percentage performances rate (PPR) of estimating the correct numbers of clusters. The OASW has either performed the same or better than ASW for all the clustering methods except for a new cases. 
\begin{table}[H]
\renewcommand{\arraystretch}{0.9}
\fontsize{8}{8}\selectfont
\centering
\caption{Performance comparison of OSil as compared to ASW for all DGPs for Simulation II. }
\begin{threeparttable}
\begin{tabular}{ c c c c c c c c c c c c  }
\toprule
DGMs & $k$-means & PAM & sing. & comp. & avg. & Ward & McQ & MB & spec. &\\
\midrule
Model 1 & $=$ & $=$ & $\checkmark$ & $\checkmark$ & $\checkmark$ & $\checkmark$  & $\checkmark$ & $\checkmark$ & $\times$  \\
Model 2 & $\times$ & $\checkmark$ & $\times$ & $\checkmark$ & $\times$ & $\checkmark$ &  $\checkmark$ & $\checkmark$ & $\times$   \\
Model 3 & $\checkmark$ & $=$ & & & $\checkmark$ & $\checkmark$ &$\checkmark$ & $\checkmark$ & $\times$  \\
Model 4 & $=$ & $=$ & $=$ & $\times$ & $=$ & $=$  & $\checkmark$ & $\checkmark$ & $\times$ \\
Model 5 & $=$ & $\checkmark$ &  & & & $\checkmark$ & & & $\times$\\
Model 6 & $\checkmark$ & & & & &  & &  \\
Model 7 & & & & & & &  & \\
Model 8 & $=$ & $=$ & $=$ & $=$ & $=$ & $=$ & $=$  & $=$ & $\checkmark$ \\
Model 9 &  $\checkmark$ &  $=$ & $=$ & $=$ & $=$ & $=$ & $=$ & $=$ & \\
\bottomrule
\end{tabular}
\begin{tablenotes}
    \item   $=$, $\checkmark$, $\times$ represent the same, increase, and decrease in percentage performance of OASW as compared to ASW, respectively.  $\star$ means only OASW was able to estimate the correct number of clusters, whereas an empty box represent that neither OASW nor ASW were able to estimate the number of clusters at the desired value. 
    \end{tablenotes}
     \end{threeparttable}
     \label{aswoaswsummarycomp}
\end{table}

\item  OSil initialized with spectral clustering method has slightly decreased the performance rate  as compared to the ASW value obtained from the standalone spectral clustering method for Models 1 to 5. ASW has shown a poor performance with the spectral clustering method to estimate the correct number of clusters for Model 3 (8\%), has shown good performance for Model 2 (48\%), Model 4 (56\%), Model 5 (40\%) and Model 8 (28\%). This combination has a performance rate of  96\% for Model 1. However, this combination never worked for Models 6 and 7. OSil has further reduced the performance rate for the estimation of k with spectral clustering for Models 1 to 5. This combination also never worked for Models 6 and 7. However, OSil only improved the performance rate for Model 8 from 28\% up to 48\%. Overall, ASW approach does not look a good fit for spectral clustering to estimate number of clusters. 

\item OSil has always increased the frequency count for all the methods to estimate the true number of clusters except a slight decay  for model-based clustering.  The OASW clustering approach is indeed better than the ASW approach  for the estimation of the number of clusters, as the overall counts for the two methods over all the clustering methods included are 844 and 809, respectively.

\item For the Jump method, $p/3$ is the best transformation power choice for the majority of DGPs. 

\item The performance for the number of clusters estimated from model-based clustering using the ASW index is much higher as compared to BIC method for Model 4 and 8. For Model 4, model-based with BIC has 44\% PPR whereas both ASW and OSil initialized with model-based clustering have 100\% PPR. For Model 8, model-based with BIC has 0\% PPR whereas ASW indea and OSil initialized with model-based clustering have 100\% PPR.
 
\end{enumerate}

\begin{figure}[!htb]
\centering
\subfloat[ ]{
   \includegraphics[width=48mm]{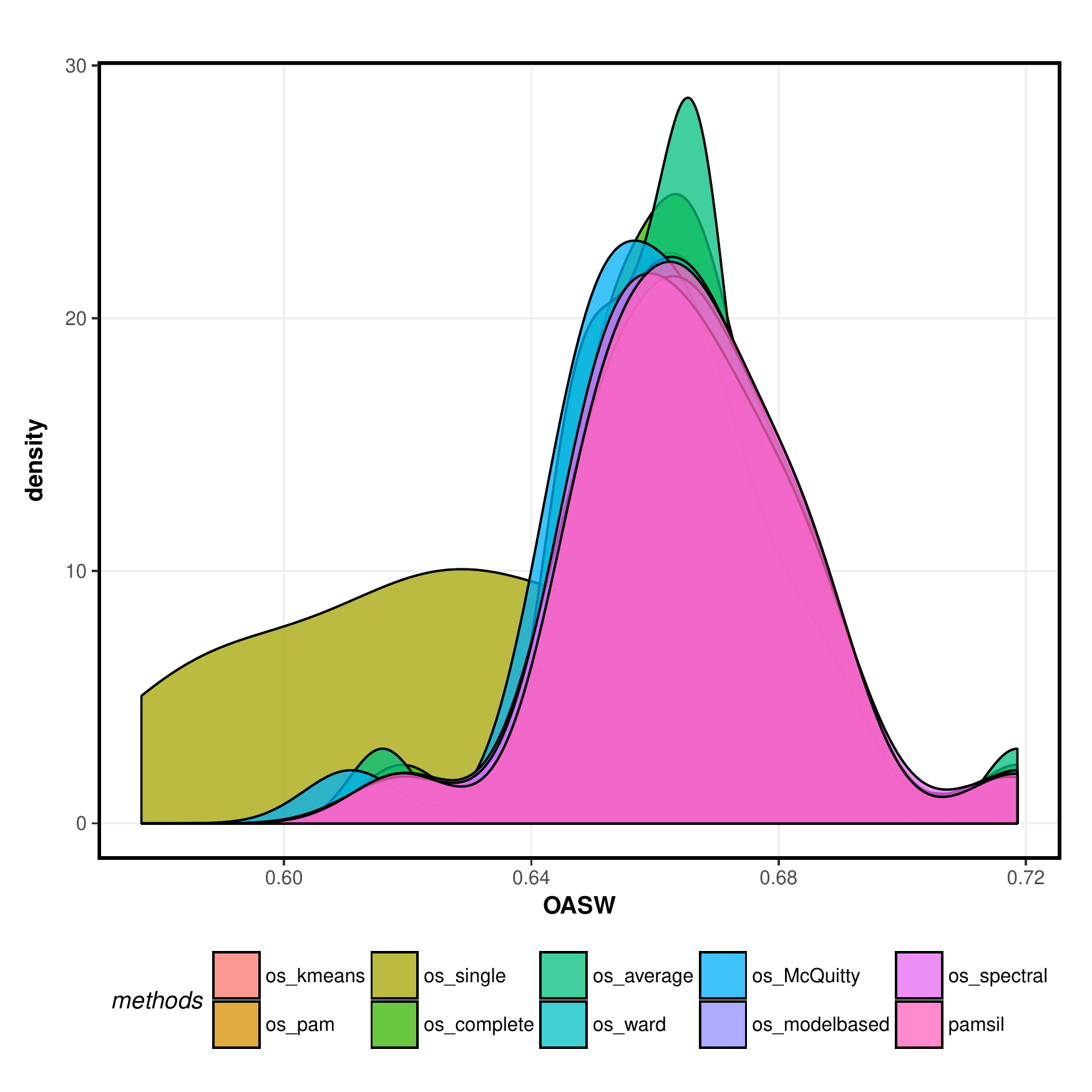}
\label{appendix:densitymodelone}}
\subfloat[ ]{
  \includegraphics[width=48mm]{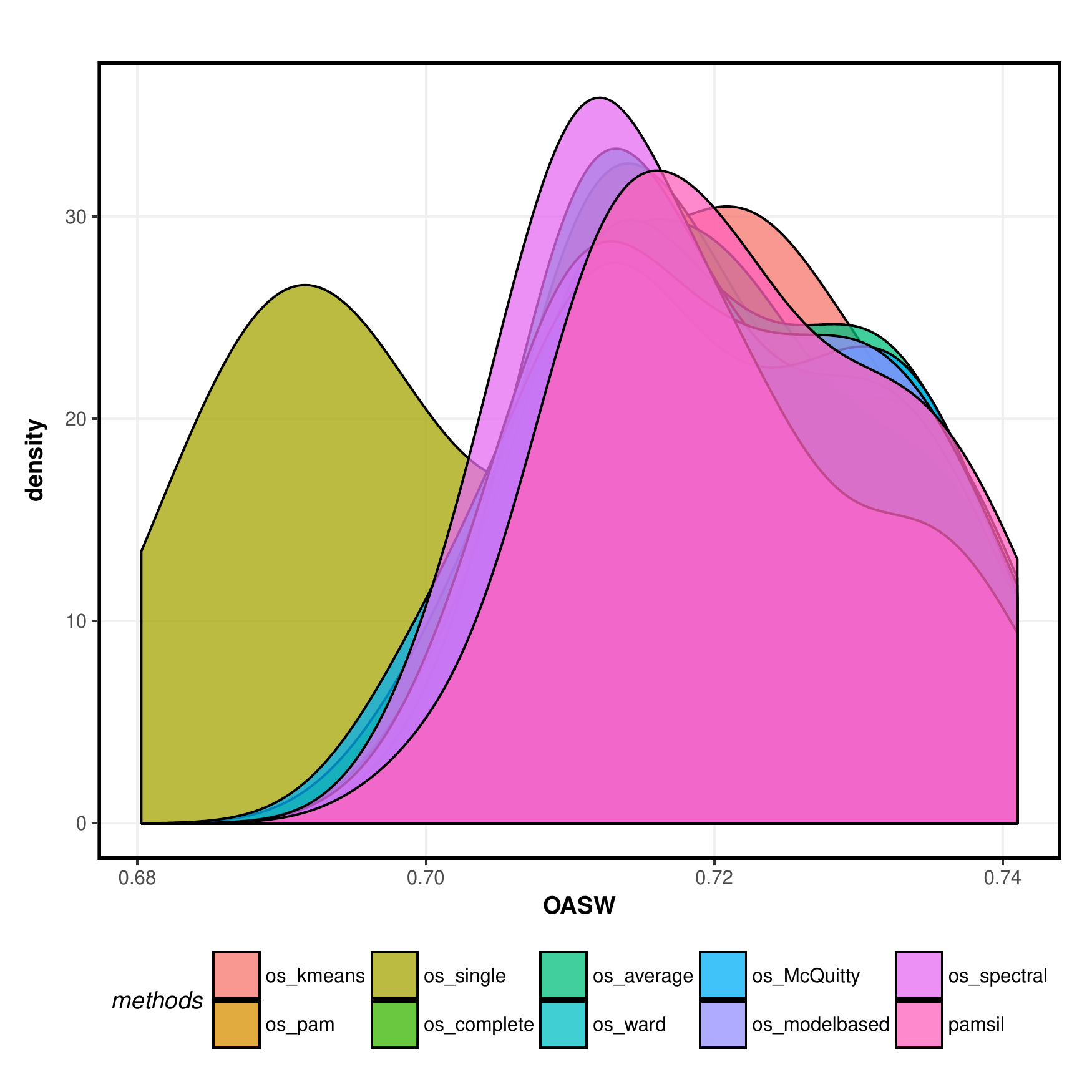}
\label{appendix:densitymodeltwo}}
\subfloat[ ]{
  \includegraphics[width=48mm]{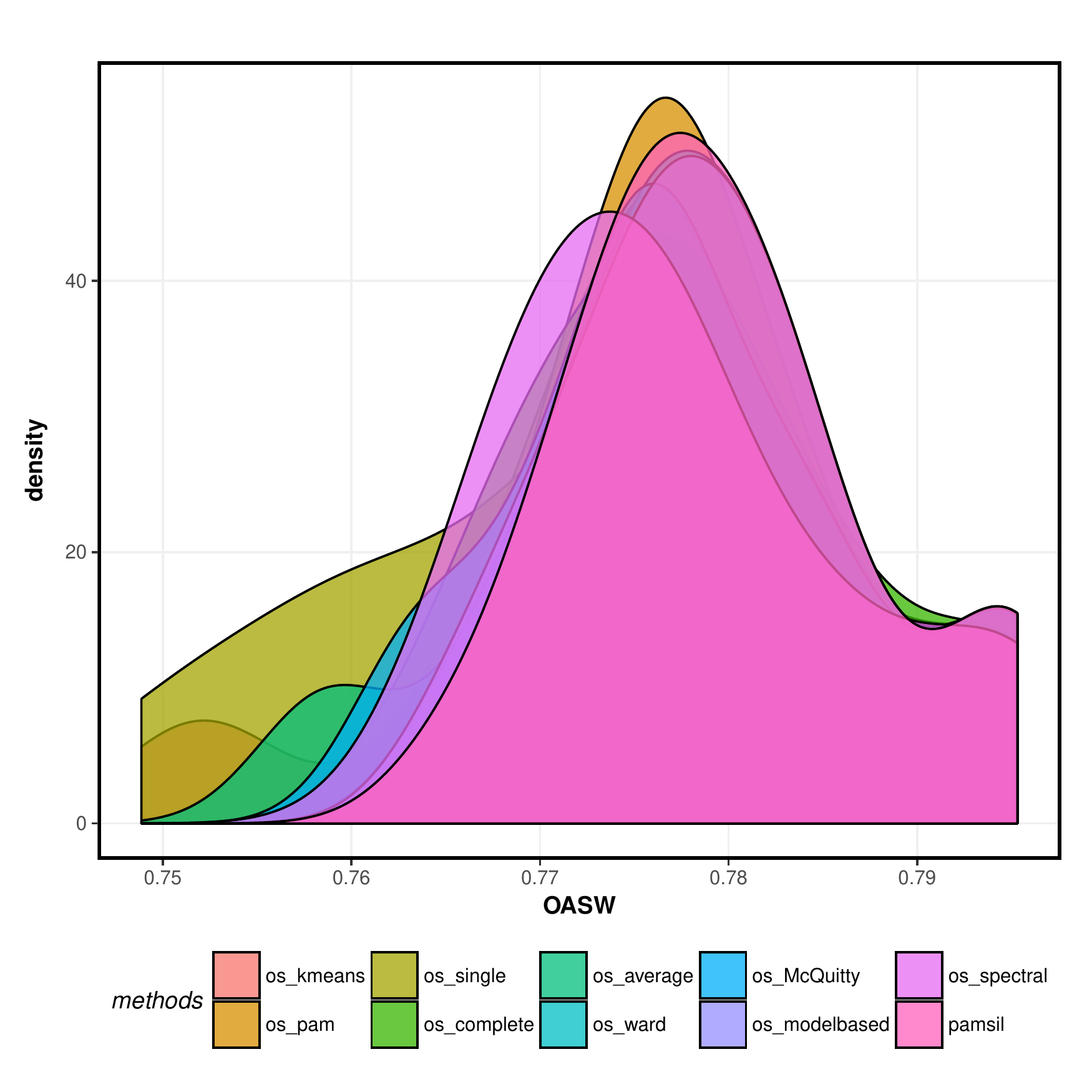}
\label{appendix:densitymodelthree}}
\newline
\subfloat[ ]{
  \includegraphics[width=48mm]{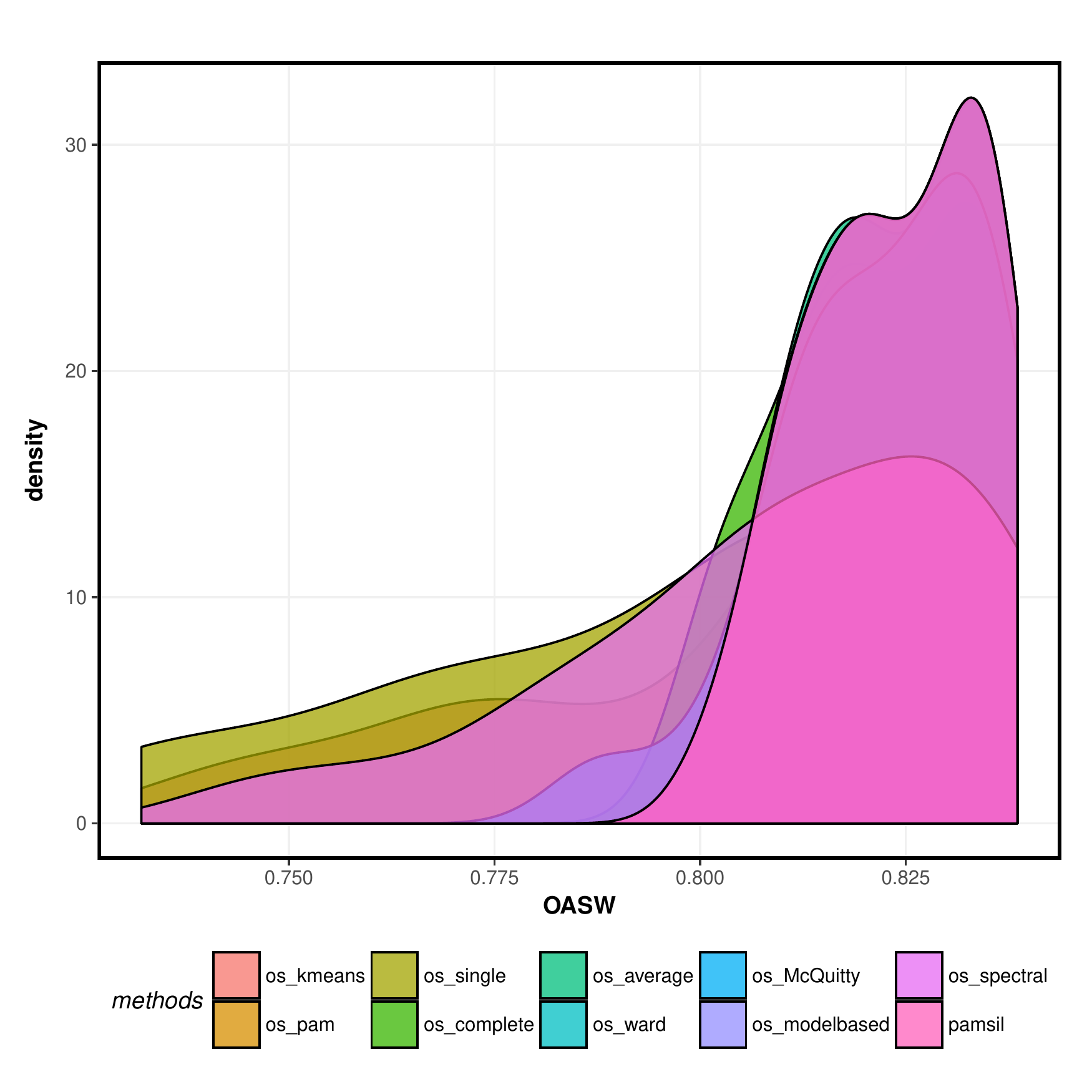}
\label{appendix:densitymodelfour}}
\subfloat[ ]{
  \includegraphics[width=48mm]{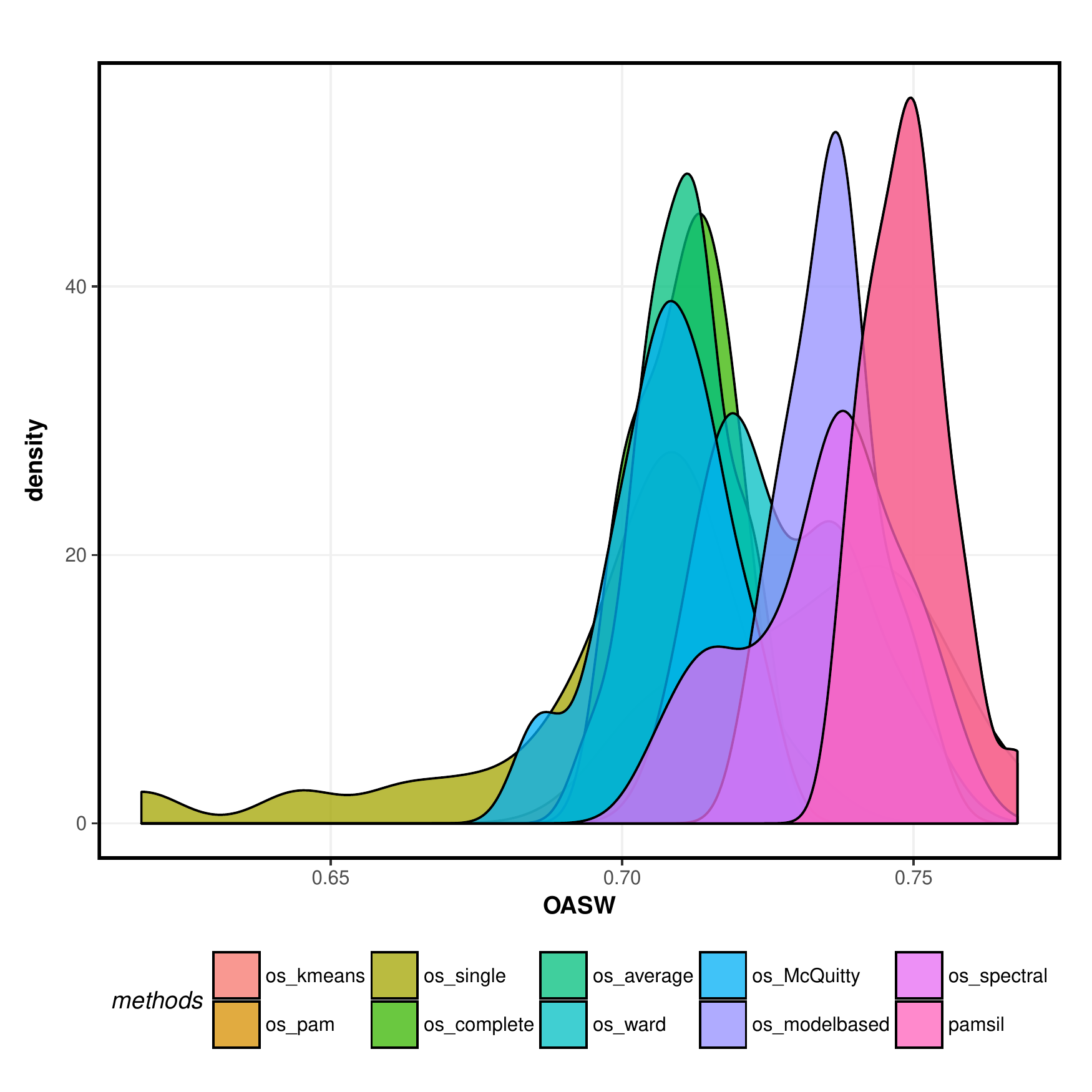}
\label{appendix:densitymodelfour}}
\subfloat[ ]{
 \includegraphics[width=48mm]{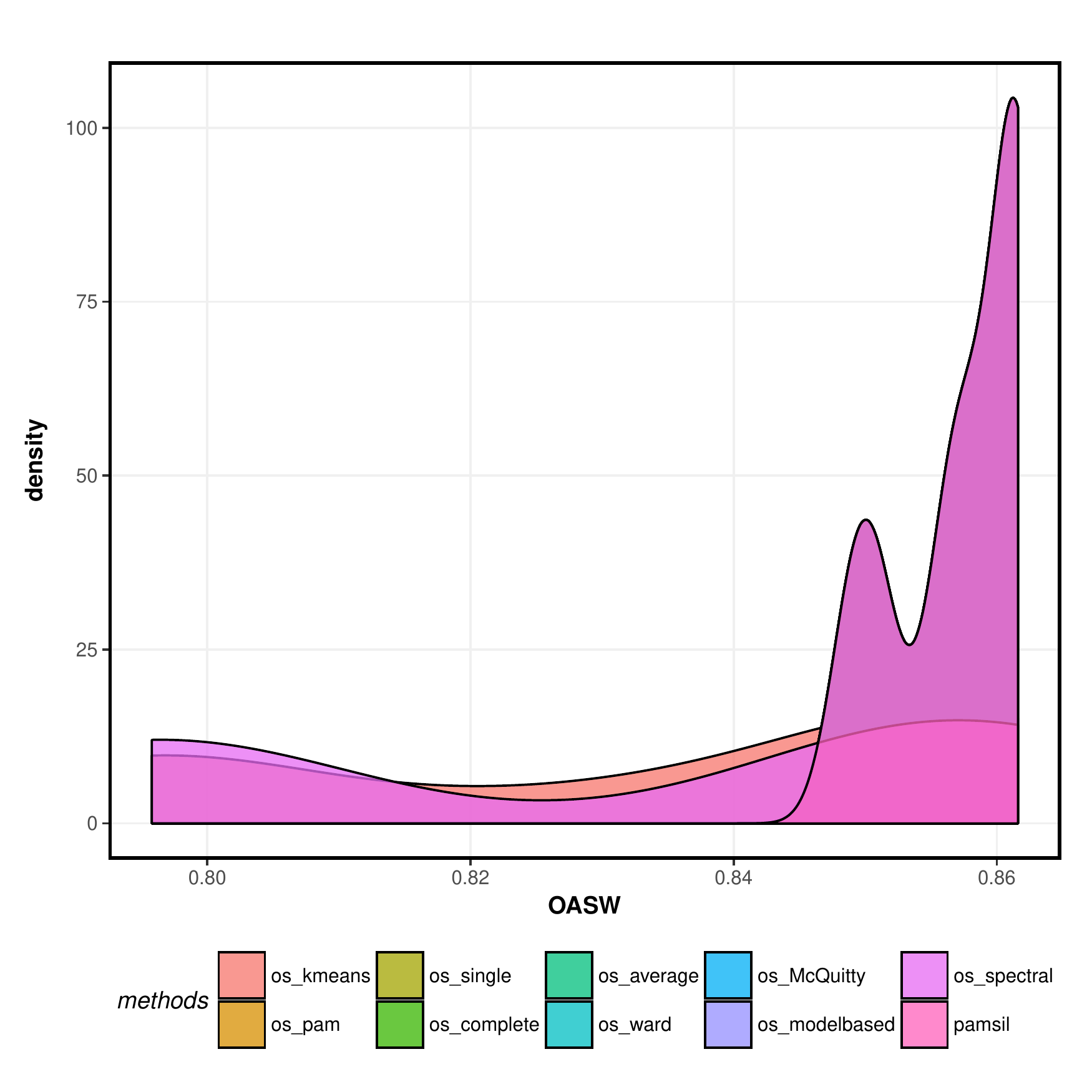}
\label{appendix:densitymodelsix}}
\newline
\subfloat[ ]{
  \includegraphics[width=48mm]{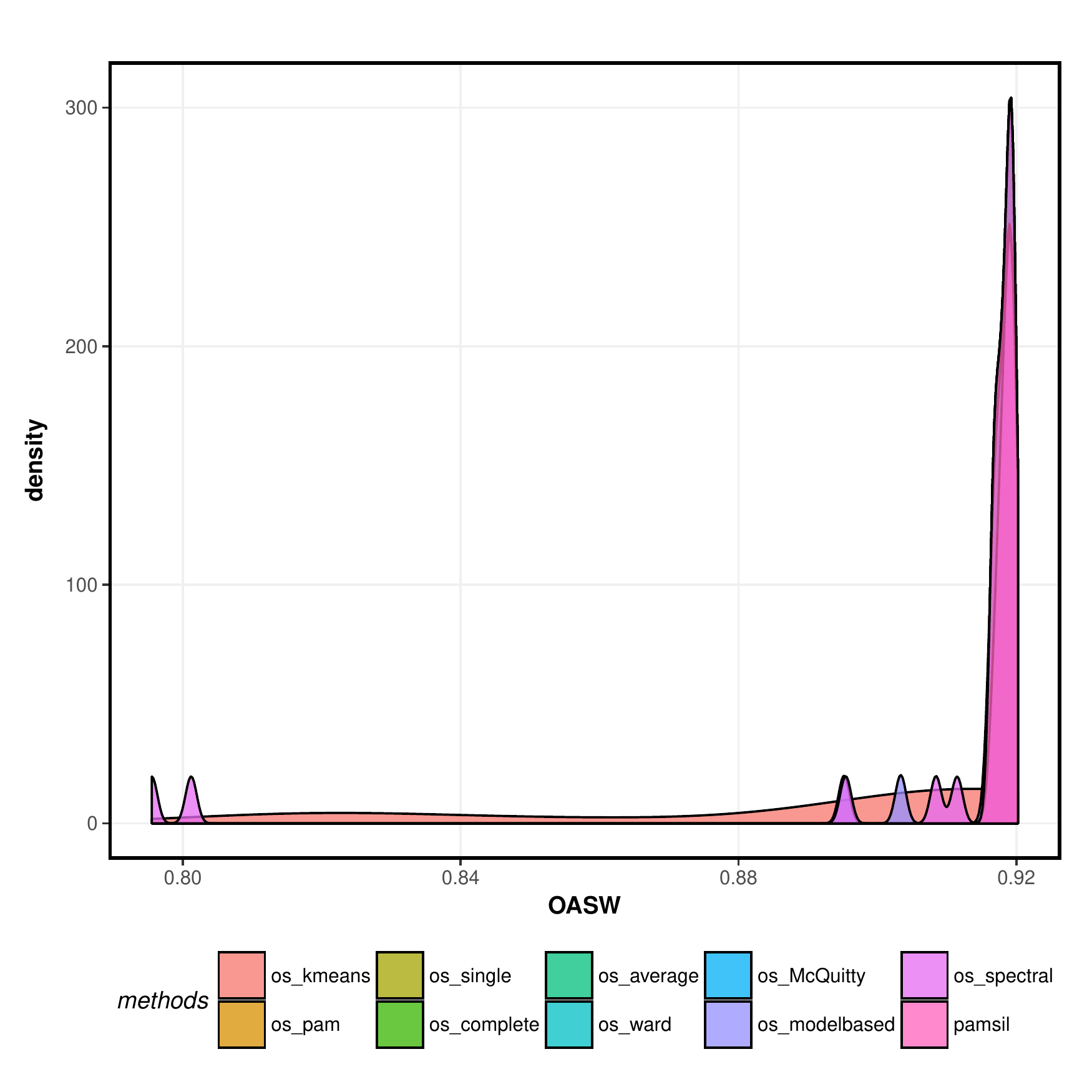}
\label{appendix:densitymodelseven}}
\subfloat[ ]{
  \includegraphics[width=48mm]{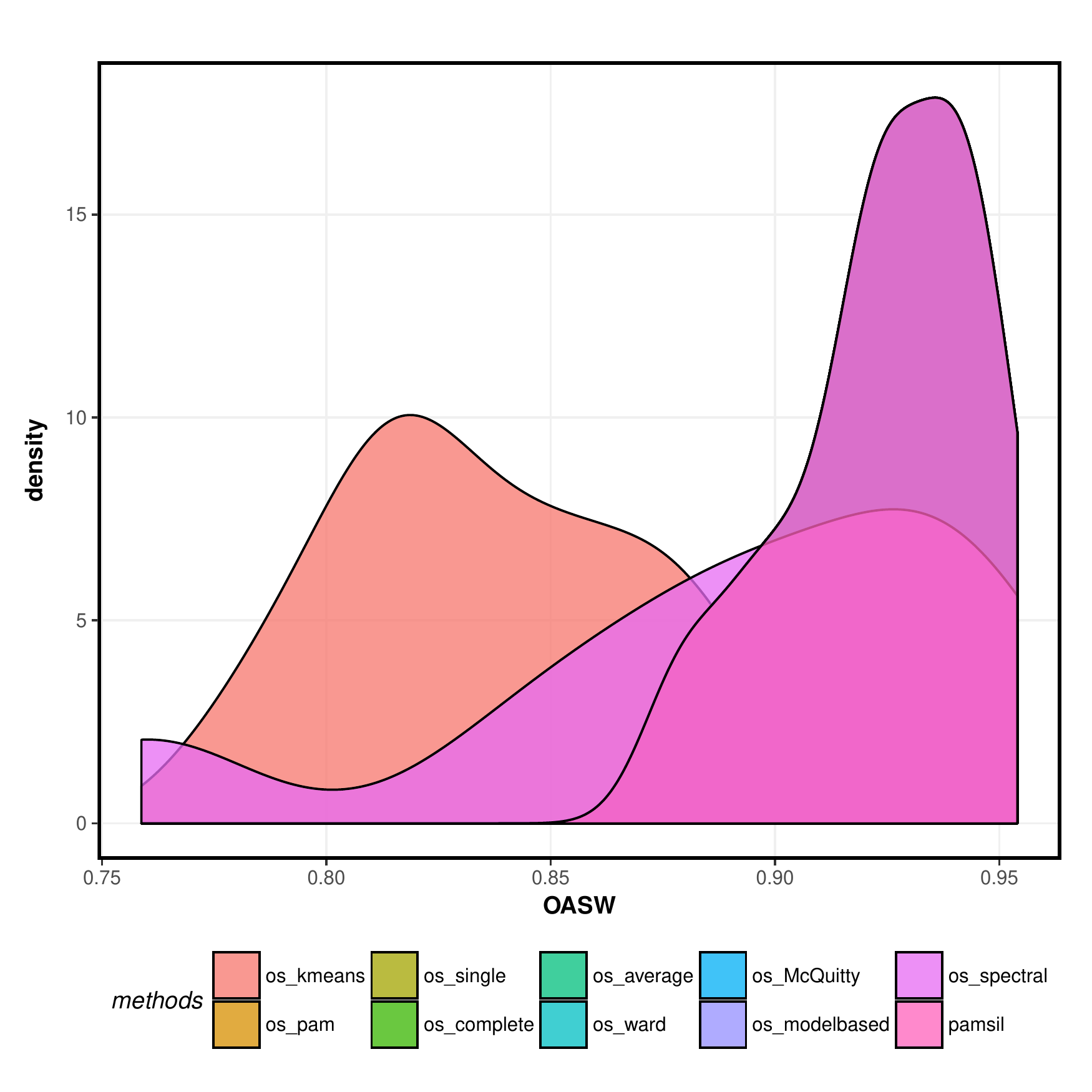}
\label{appendix:densitymodeleight}}
\subfloat[ ]{
  \includegraphics[width=48mm]{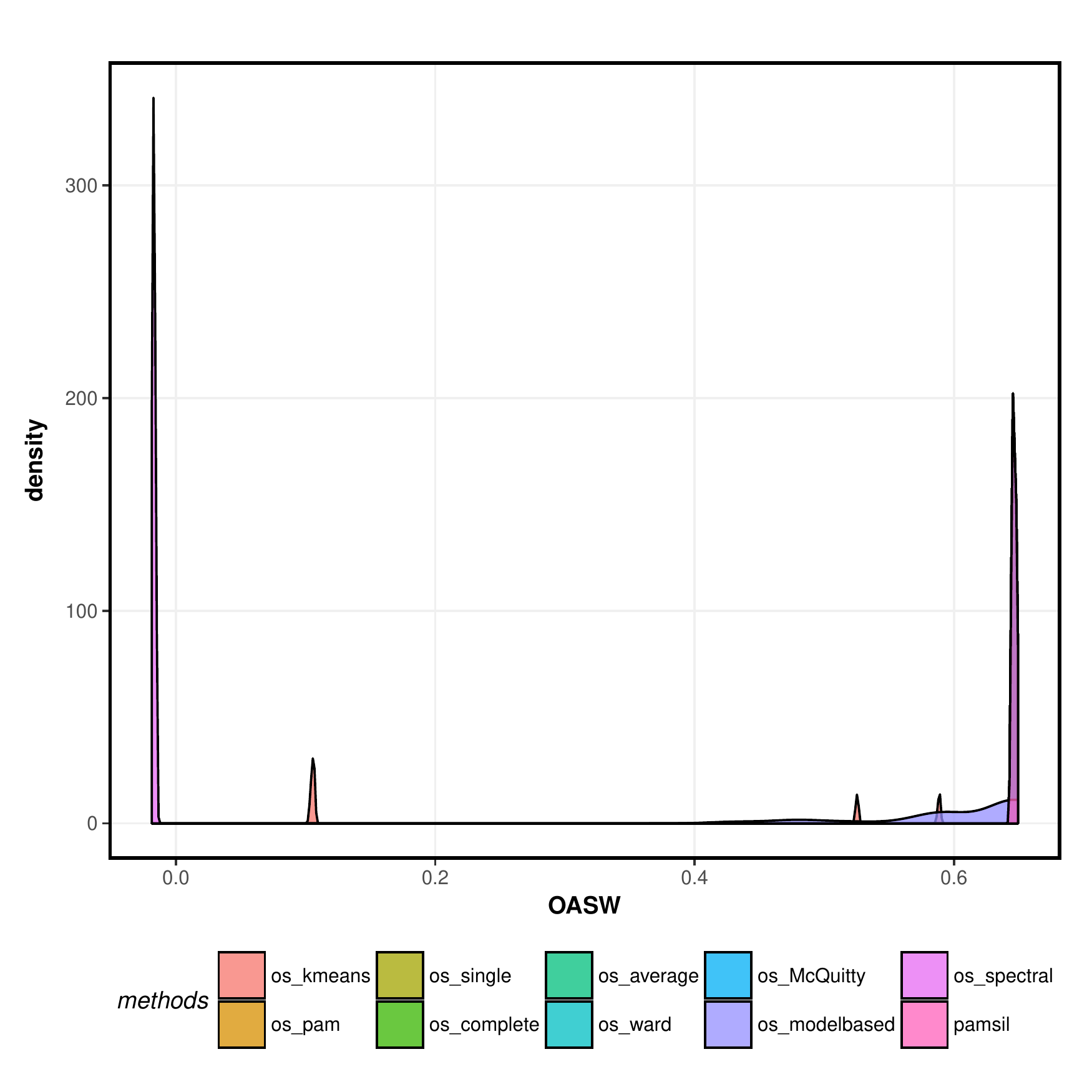}
\label{appendix:densitymodelnine}}
\caption{Estimation of k case. Density curve  plots for ASW values obtained against each initialization methods for model 1 to 9. }
\label{densityplotssim2}
\end{figure}

The plots represents the density of ASW values obtained by OSil across the  nine initialization methods and  for PAMSIL for all DGPs included in the Simulation II are presented in Figure \ref{densityplotssim2}.

\section{Final comments }\label{rec}  
In this study, a distance based clustering algorithm is proposed computed on pairwise proximity between the observations, namely OSil. While there are a large amount of clustering algorithms proposed in the literature, the clustering results of existing algorithms usually depend on user-specified parameters heavily, and it is usually difficult to determine the optimal parameters. With the pairwise data dim-similarity matrix as the input, OSil clustering has  shown to be an effective data clustering, due to its ability to find out the underlying data structure and determine the number of clusters automatically.
 
 \par The most promising and widely used indices were chosen for the comparative study here. 
These widely used indices are paired with various fundamentally different clustering methodologies that are in use across disciplines. Gamma and C were the top performing indices in \cite{milligan1981monte} and \cite{milligan1985examination}. CH was first best performing index in \cite{milligan1985examination} and third in \cite{arbelaitz2013extensive}. ASW was top performing index in \cite{arbelaitz2013extensive}.

\par The purpose of the experiments was not only to compare the performance of the proposed method with the existing  competitors, but also to find out whether the idea of OASW clustering is worthwhile at all for the estimation of number of clusters in comparison to the ASW. Another motivation for setting up this  experiment is to systematically investigate the behaviour of the existing estimation indices as well. \cite{milligan1981monte} conducted a study to evaluate the clusterings obtained from 4 hierarchical clustering methods using 30 internal criterion.  The clustering methods used were single link, complete link, group average, and Ward's minimum variance. The data model considered had a strong clustering structure with 4 clusters. The clusters were compact and well separated.  The study covered the indices  proposed the period of 1967 and 1980.

In another study, \cite{milligan1985examination} has conducted an experiment with 4 artificially generated data sets having 2 to 5 number of clusters.  30 methods to estimate number of clusters were considered with four hierarchical clustering methods mentioned previously. The indices covered in the study were proposed  during the period of 1965 to 1983.
Recently, \cite{arbelaitz2013extensive} have conducted the cluster validation study using the 30 indices proposed during the period of 1973 to 2011. For many indices, their several versions were included to compare their performance. The experiment was conducted to estimate the numbers of clusters from each index. They considered three clustering methods, namely average linkage, Ward's method and $k$-means. The artificial data sets were generated considering five factors, that were number of clusters (allowed values were 2, 4, and 8), dimensions,  cluster overlap, cluster density, and noise level. They also considered 20 real data sets ranging from 2 to 15 clusters. 

\par The strength of the experiments done here is that they include a larger number of clustering methods for evaluation that were never considered together in any previous study with indices covering wide spectrum of statistical methods. Another potential of the present study is its  artificial data sets having various clustering structures than those used in previous such studies. In addition, the Gap, PS, BI, and CVNN indices never appeared in a comparative study together with other indices in such extensive systematic simulations.

\bibliographystyle{spbasic}
\bibliography{osil}
  \newpage
 \textbf{Supplementary to``Clustering by optimizing the Average Silhouette Width"}
\section{Definition of DGPs} \label{dgpdefs}
 
\subsection*{Model 1: } 2 clusters in 2 dimensions: 50 observations each were generated from  two independent Gaussian random variables, to form two spherical clusters in two dimensions, of unequal variations. One cluster has mean (0, 5) with covariance matrix as 0.1$I_2$ and the other cluster has mean (2, 5), where $t$ represents the transpose, with covariance matrix  as 0.7$I_2$.  The result is one bigger spherical cluster with wider spread lying  next to a compact spherical cluster.
\subsection*{Model 2: }  3 clusters in 2 dimensions: The observations in each of the three clusters were generated from independent Gaussian random variables centred at (-2, 0) and covariance matrix 0.1$I_2$  for cluster 1, mean  (0, 0) and covariance matrix 0.7$I_2$  for cluster 2, and  mean (2,  0) and covariance matrix 0.1$I_2$  for cluster 3. The  cluster contains 50 observations each. The clusters are of such nature that the  cluster with greater observational variation is located between the two clusters having less variations among observations. 
\subsection*{Model 3: }  4 clusters in 2 dimensions: Cluster one was generated from two independently distributed non-central $t$ variables with parameters $t_7(10)$ and $t_7(30)$. Cluster two was generated from $\mathds{U}(10, 15)$ along both dimensions independently. Cluster 3 was generated from  independent Gaussian distribution having mean  (2, 2) with covariance matrix $\Sigma = \begin{bmatrix}
 2 &  0\\
  0 & 4 
\end{bmatrix}$. Cluster four was also generated from independent Gaussian distributions  with mean  (20, 80) and  covariance matrix   $\Sigma = \begin{bmatrix}
 1 &  0\\
  0 & 2 
\end{bmatrix}$ respectively. Each cluster contains 50 observations.  
\subsection*{Model 4: }  5 clusters in 2 dimensions:  the clusters are parametrized from $\mathds{F}$, Chi-squared, Gaussian,  skewed Gaussian  and $t$ distributions respectively as: $\mathds{F}_{(2, 6)}(4)$ along first dimension and  $\mathds{F}_{(5, 5)}(4)$ along second dimension,  $\chi^2_7(35)$ and $\chi^2_{10}(60)$,  $N(100, 2)$ and $N(0, 2)$, $SN(20, 2, 2, 4)$ and $SN(200, 2, 3, 6)$,  $t_{40}(100)$ and $t_{35}(150)$ respectively. The clusters contains 50 observations each and were generated independently along both dimensions.

\subsection*{Model 5: } 6 clusters in 2 dimensions: the clusters 1 and 2 are generated from Uniform and exponential distributions as $\mathds{U}(-6, -2)$, $\textrm{Exp}(10)$ in both dimensions. The cluster 3 is $\textrm{NBeta}(2, 3, 220)$ along one dimension and $NBeta(2, 3, 120)$ across the other dimension. Cluster 4 is from $SN(5, 0.6, 4, 5)$ and $SN(0, 0.6, 4, 5)$. Cluster 5 is  $\mathds{W}(10, 4)$ across both dimensions. Cluster 6 is  $\textrm{Gam}(15, 2)$ and $\textrm{Gam}(15, 0)$ along first and second dimension respectively. The clusters contains 50 observations each and were generated independently along both dimensions.

\subsection*{Model 6: } 
5 correlated dimensions within 5 clusters are generated from multi-variate Gaussian distributions each containing 50 observations. Cluster 1 to 5 are centred at (0, 0, 0, 0, 0), (40,  80, 15, 30, 22),  (15, 40,  40,  55, 80), (70, 80, 70, 70, 70), and (100, 100, 100, 100, 100), respectively with the covariance matrices as follows: \\
{
\renewcommand{\arraystretch}{0.7}
\[
 \Sigma_1=
\begin{bmatrix}
  9 &  &   &   &   \\
  1  & 17 & &   &  \\
 1  & -1.4 & 12 & &  \\
  0.4 &  0.6 & 0.5 & 2 & \\
-1.2 & -1.6 & -1.4 & -0.6 & 16\\
\end{bmatrix}, 
\quad
\Sigma_2=
\begin{bmatrix} 
  25 & \\
  0.2 & 9 &   \\
  0.2 & -0.2 & 16  &  \\
  -0.2&  -0.2 & 0.2 & 1  & \\
 -0.2& -0.2 & -0.2 & -0.2 & 49 \\
\end{bmatrix},
\]

\[
\Sigma_3 =
\begin{bmatrix}
25  & \\
0.3 &  9  &   \\
0.3 & -0.3& 16  &  \\
-0.3& 0.3 & 0.3 & 1 & \\
-0.3& -0.3&-0.3 & -0.3 & 49\\
\end{bmatrix},
\quad
\Sigma_4 =
\begin{bmatrix}
  5 & \\
  0.1  & 0.9 &   \\
  0.1& -0.2 & 1.6 &  \\
 -0.7 &  0.2 & 0.2 &1  & \\
 -0.2& -0.9 & -0.2 & -0.2 &4.9 \\
\end{bmatrix},
\]

$\Sigma_5 =
\begin{bmatrix}
  2 & \\
  0.2 & 9 &   \\
  0.2 & -0.1 & 3  &  \\
  -0.3 & 0.2  & 0.1 & 1 & \\
 -0.1 & -0.1 & -0.2 & -0.9 & 4\\
\end{bmatrix}$.

 }
\subsection*{Model 7:} 7 clusters in 10  dimensions having 50 observations each. All the clusters are from the Gaussian distributions. The clusters are present in the first two dimensions only. Cluster 1 has mean (0, 5) with covariance matrix  $\begin{bmatrix}
  0.5 & 0\\
  0 & 0.2 \\
\end{bmatrix}$. 
Cluster 2 has mean (-0.5, 3.5) and covariance matrix  
 $\begin{bmatrix}
  0.2 & 0\\
  0 & 0.1 \\
\end{bmatrix}$. Cluster 3 has mean  (0, 3.5) with covariance matrix  
$\begin{bmatrix}
  0.4 & 0\\
  0 & 0.3\\ 
\end{bmatrix}$. 
Cluster 4 has mean (0.5, 3.5) and covariance matrix  
$\begin{bmatrix}
  0.2 & 0\\
  0 & 0.1 \\
\end{bmatrix}$.  Cluster 5 has mean (-0.5, 6.5) and covariance matrix  
$\begin{bmatrix}
  0.2 & 0\\
  0 & 0.1\\
\end{bmatrix}$.
Cluster 6 has mean   (0, 6.5) and covariance matrix  
$\begin{bmatrix}
  0.3 & 0\\
  0 & 0.2\\ 
\end{bmatrix}$.   Cluster 7 has mean (0.5, 6.5) and covariance matrix  
$\begin{bmatrix}
  0.3 & 0\\
  0 & 0.2\\
\end{bmatrix}$. Further  dimensions 3 to 6 of cluster 1 were generated by subtracting the values 3, 6, 9, 12 from its second dimension. Dimensions 7 to 10 of cluster 4  were generated by adding the values 3, 6, 9, 12 from its second dimension. Dimensions 3 to 10, of clusters 2 to 4 were generated by adding the values 3, 6, 9, 12, 15, 18, 21, 24 to the second dimensions of these clusters. For dimensions 3 to 10, of cluster 5 to 7, the values 3, 6, 9, 12, 15, 18, 21, 24 are subtracted from the second dimensions of the respected clusters.  


\subsection*{Model 8:}
10 cluster in 500 dimensions.  The clusters are centred at -21, -18, -15, -9, -6, 6, 9, 15, 18, 21. The clusters are in 500 dimensions such that the 500 dimensional mean vectors of these values were generated for all clusters.  The number of observations  for these ten clusters are 20, 40, 60, 70, and 50 each for six of the remaining clusters. The number of observations for the means of clusters were not fix. Any cluster can take  any number of observations from these such that any six clusters have equal number of observations i.e., 50 and the remaining four has different observations each, which is one out of 20, 40, 60, 70 values. The total size of the data is always 490 observations. The covariance matrix for each of these clusters is one out of 0.05$I_{500}$, 0.1$I_{500}$, 0.15$I_{500}$, 0.175$I_{500}$, 0.2$I_{500}$ matrices. The covariance matrix for each cluster was chosen randomly with replacement out of these, such that as a result, all the clusters can have same covariance matrix, two or more clusters can have same covariance matrix  or all of the 10 clusters can have different same covariance matrices.

\subsection*{Model 9: } 7 clusters in 60 dimensions with 500 observations: 
This is a data structure designed by \cite{van2003new} to simulate gene expression profiles like structure for three distinct types of cancer patients' populations. Suppose that in reality there are 3 distinct groups 20 patients each corresponding to a cancer type.  Three multivariate normal distributions were used to generate 20 samples each having different mean vectors.  For the first multivariate distribution (first cancer type) the first 25 dimensions(genes) are centred at $\log_{10}(3)$, dimensions 26-50 are centred at $(-\log_{10}(3))$  the remaining 450 dimensions are centred at 0.  For the second multivariate distribution (second cancer type) the first 50 dimensions(genes) are centred at $0$, the next 25 dimensions (51-75) are centred at $\log_{10}(3)$,  dimensions 76-100 are centred at $(-\log_{10}(3))$  and  the remaining 400 dimensions are also centred at 0. For the third multivariate distribution (third cancer type) the first 100 dimensions(genes) are centred at $0$, dimensions 101-125 are centred at $\log_{10}(3)$,  dimensions 126-150 are centred at $(-\log_{10}(3))$  and dimensions 151-500 are also centred at 0.  The three multivariate distributions has diagonal covariance matrix with diagonal elements as $(\log(1.6))^2$. Note that the described data has 20 samples each of 3 types of cancer patients each containing 500 genes. The purpose here is to cluster genes not patients. Therefore, the transpose of the data is required to transfer it to the standard format and the number of clusters to seek are 7 in 60 dimensions of 500 observations.

\section{Estimation of $k$ box and density plots}
The plots represents the ASW values obtained by OSil in Simulation II across the  nine initialization methods and for PAMSIL for all DGPs.
\newpage
\begin{figure}[H]
\centering
\subfloat[ ]{
   \includegraphics[width=44mm]{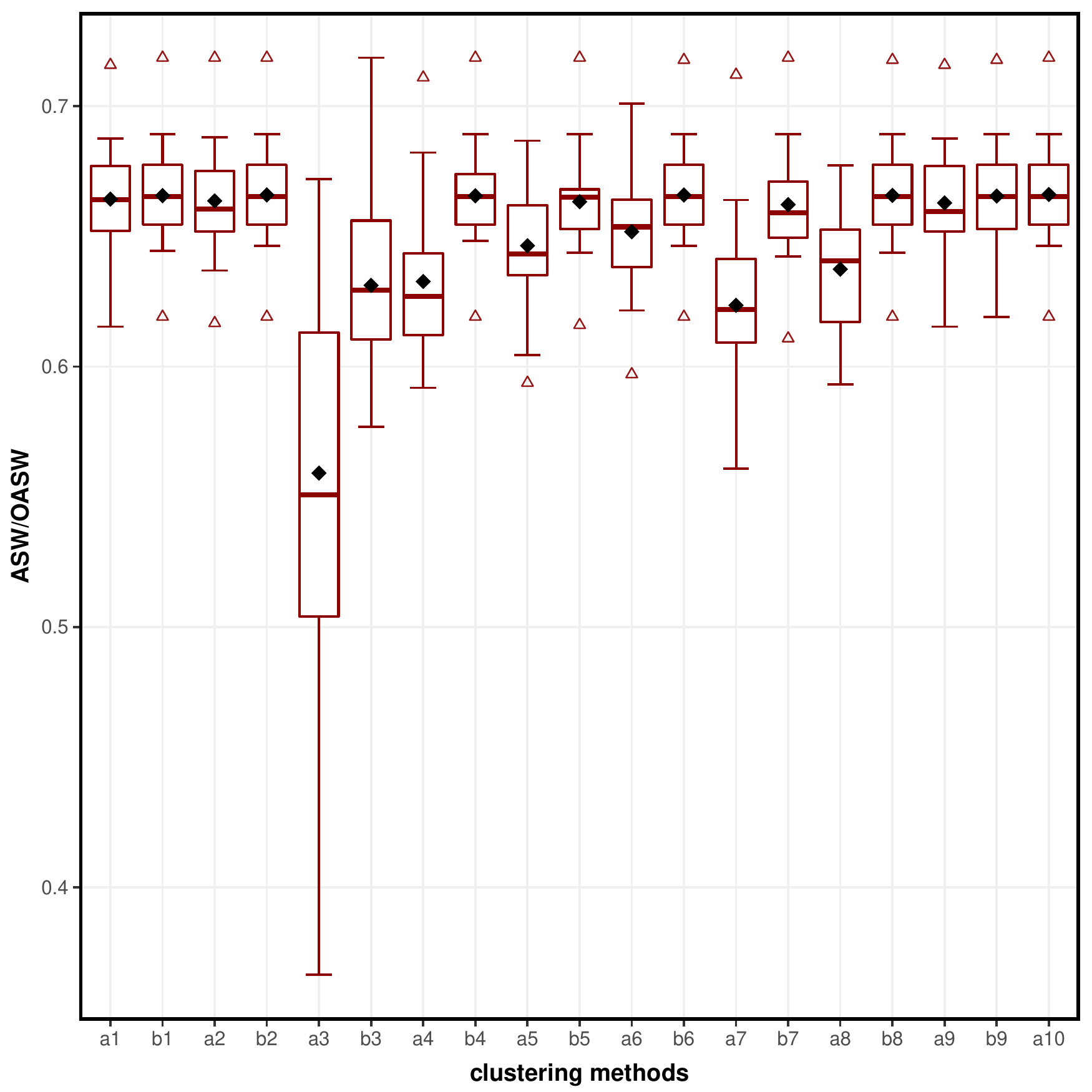}
\label{appendix:boxmodeloneone}}
\subfloat[ ]{
  \includegraphics[width=44mm]{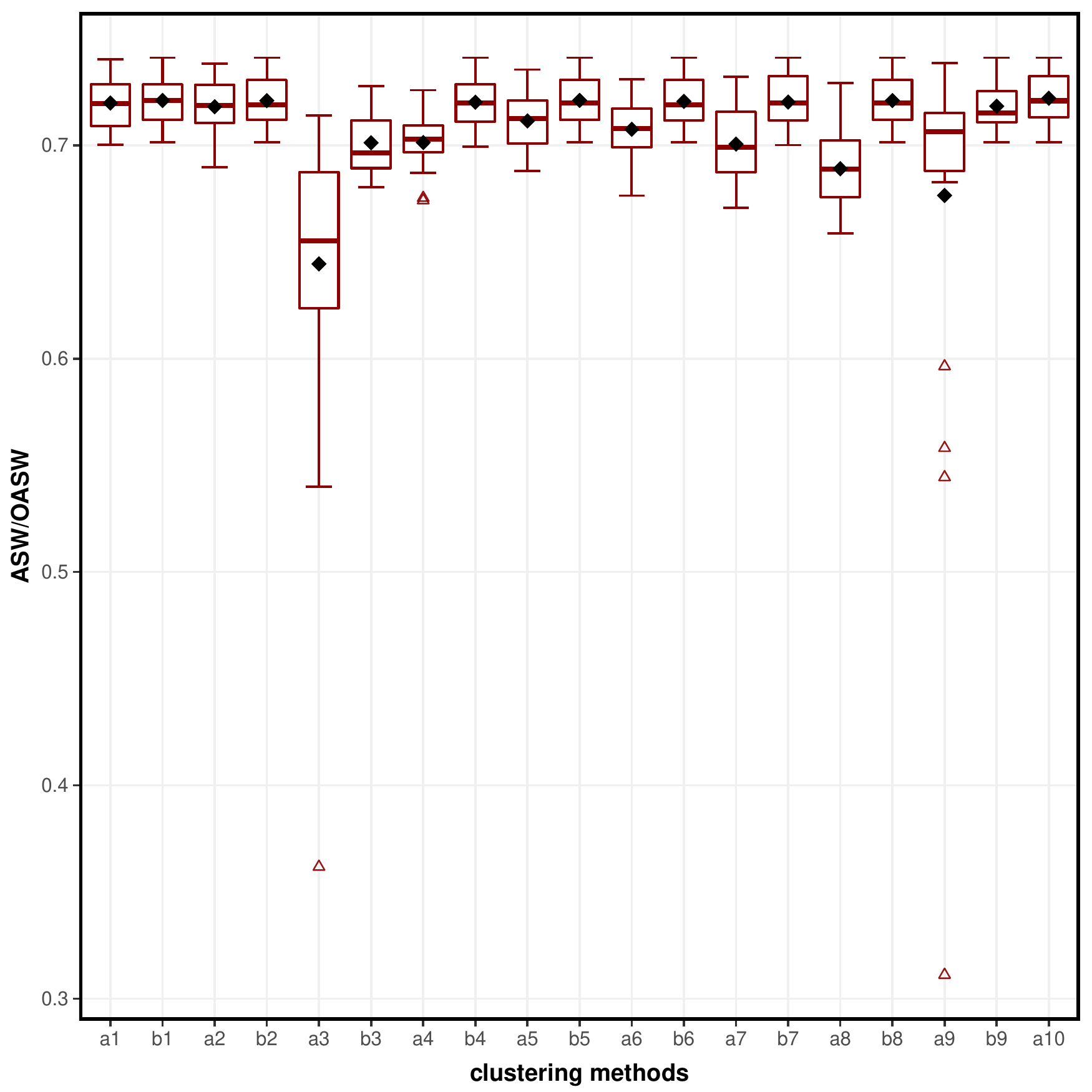}
\label{appendix:boxmodeltwotwo}}
\subfloat[ ]{
 \includegraphics[width=44mm]{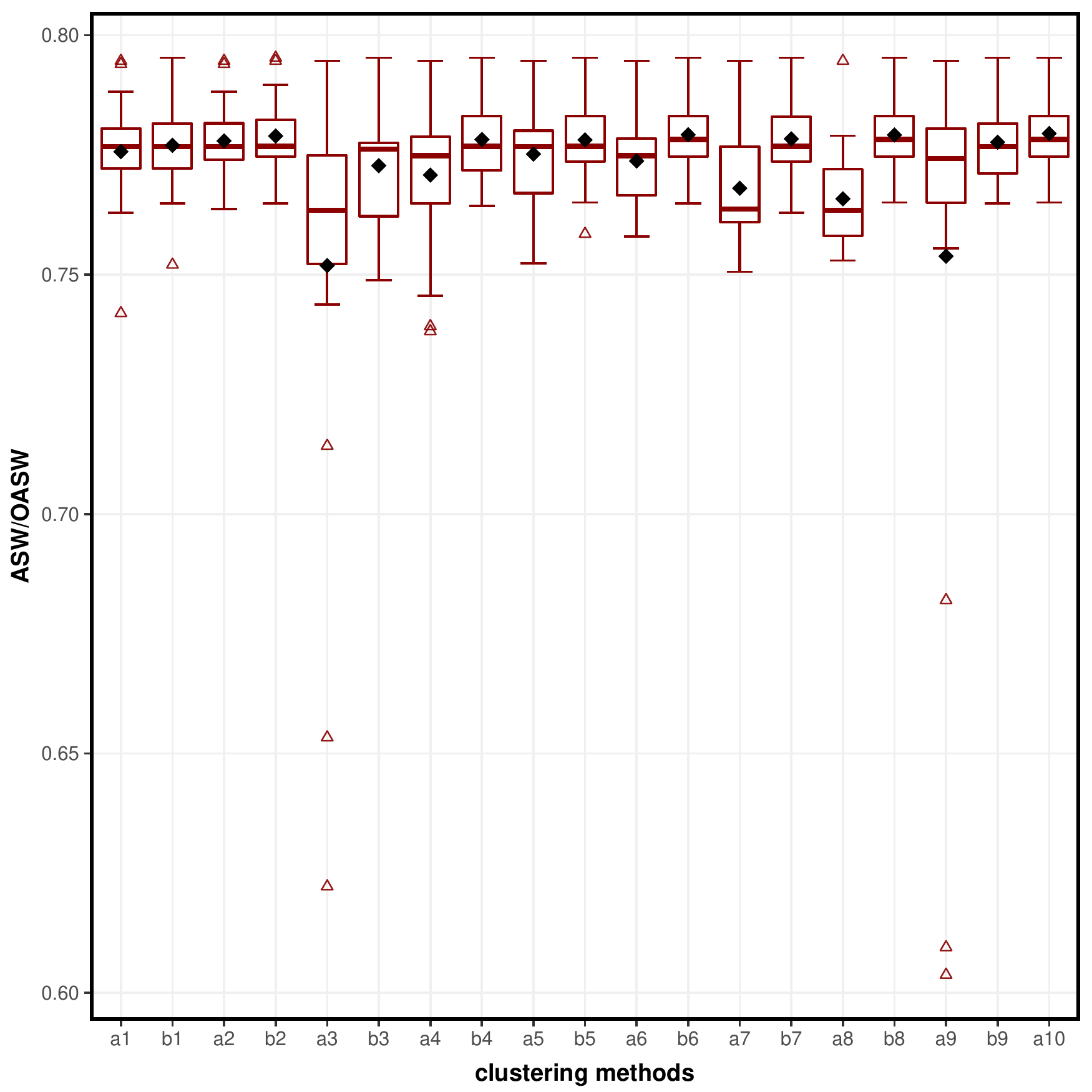}
\label{appendix:boxmodelthreethree}
 }
\newline
\subfloat[ ]{
  \includegraphics[width=44mm]{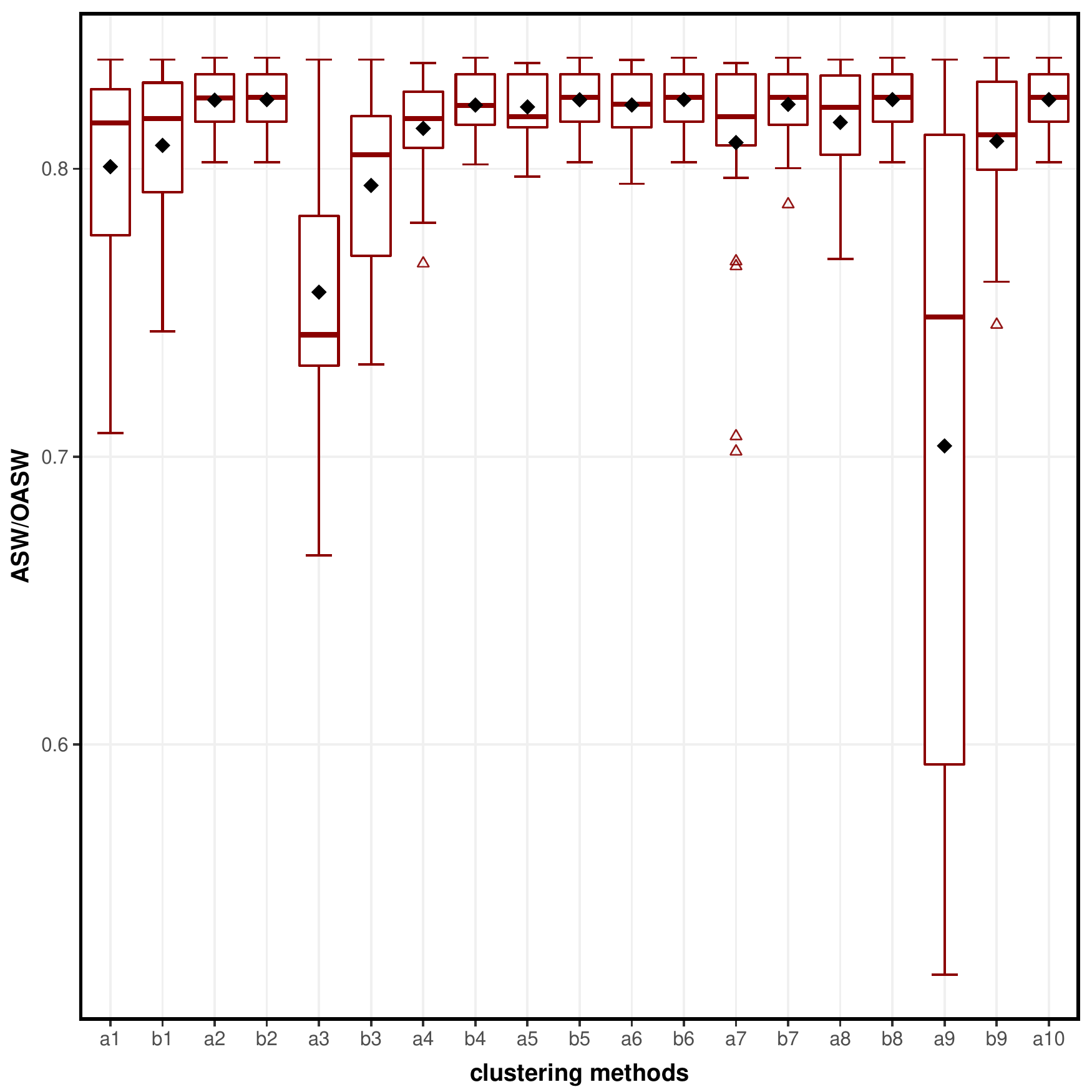}
\label{appendix:boxmodelfourfour}
 }
\subfloat[ ]{
   \includegraphics[width=44mm]{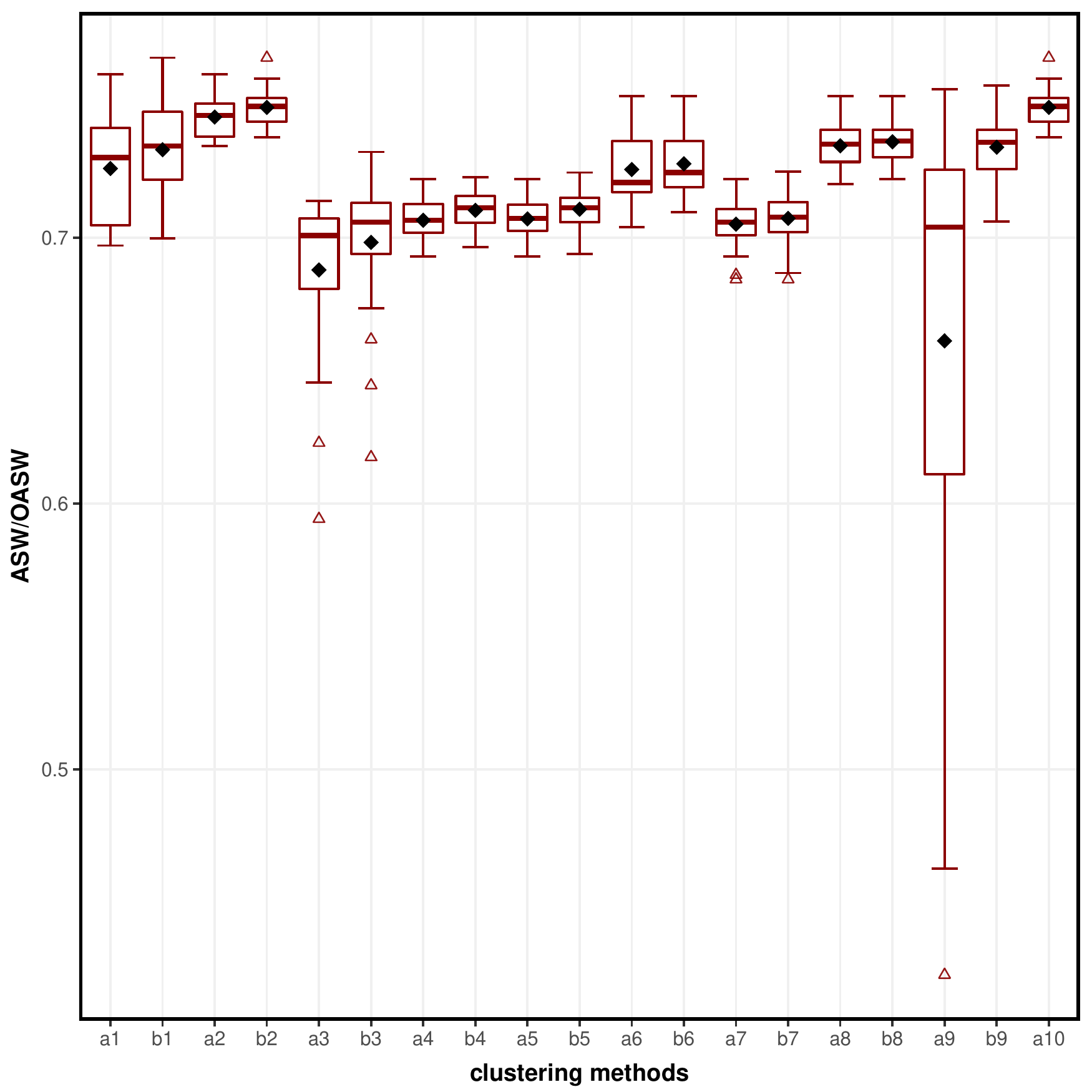}
\label{appendix:boxmodelfivefive}
 }
\subfloat[ ]{
  \includegraphics[width=44mm]{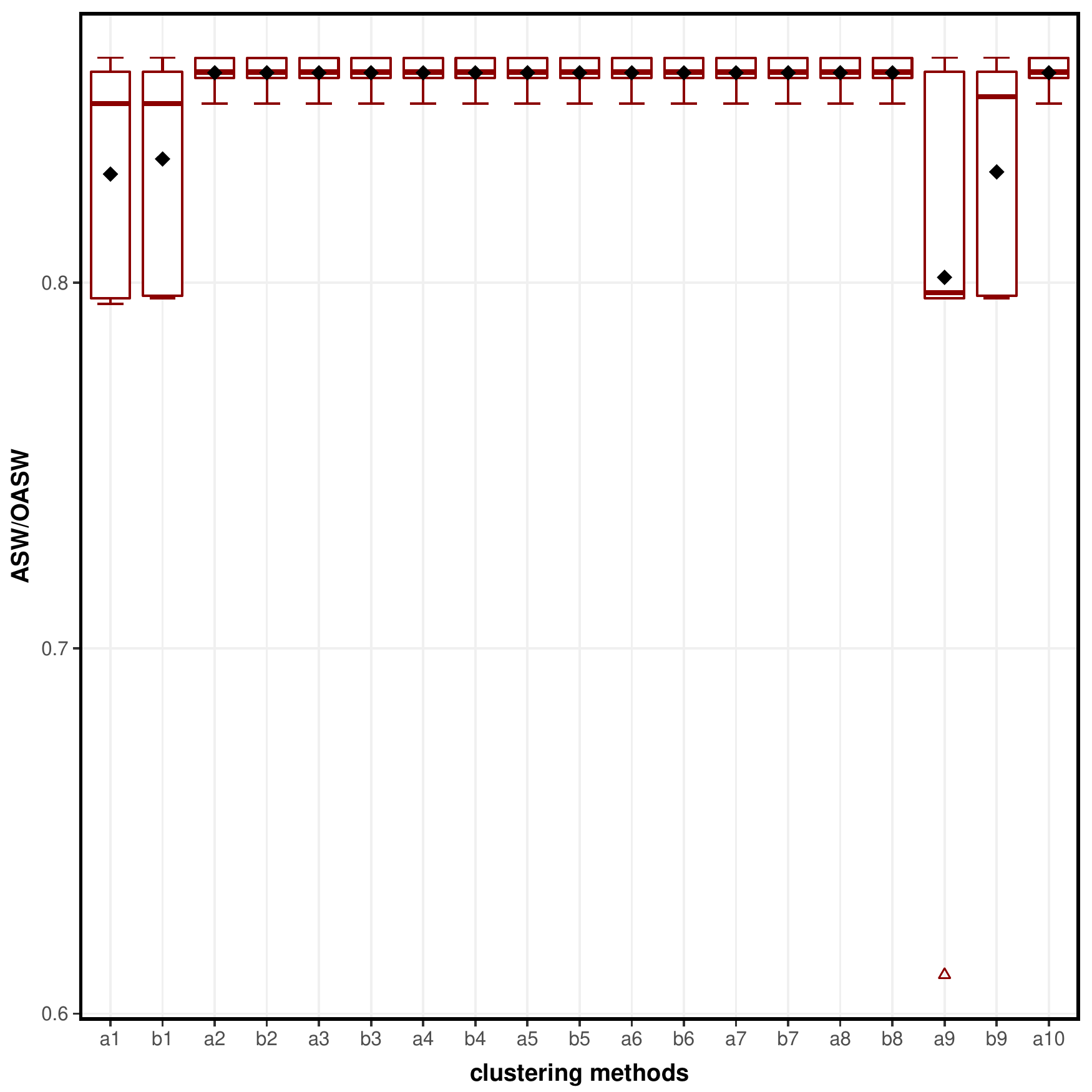}
\label{appendix:boxmodelsixsix}
 }
\newline
 \rule{-5ex}{.2in}
\subfloat[ ]{
   \includegraphics[width=44mm]{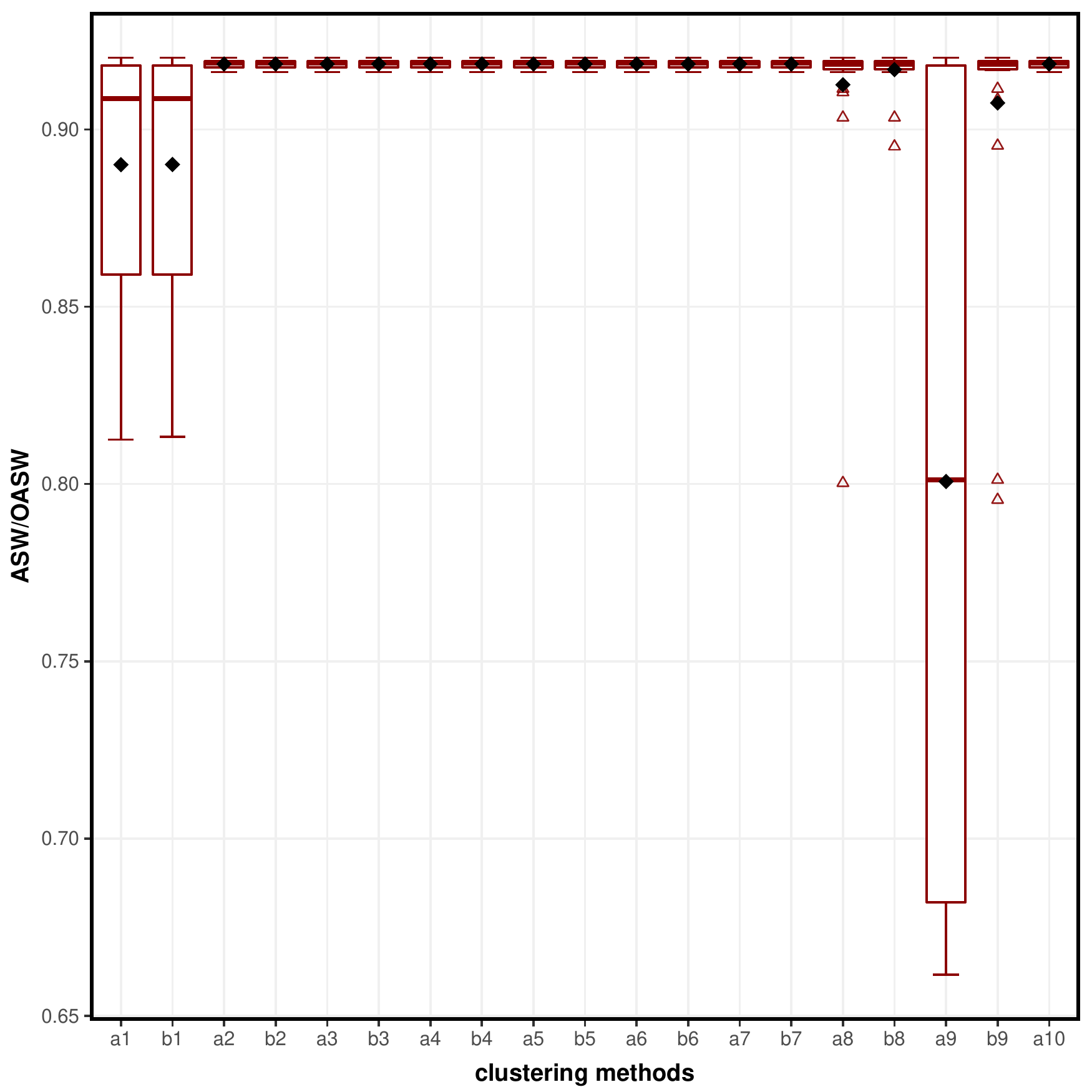}
\label{appendix:boxmodelsevenseven}
 }
\subfloat[ ]{
  \includegraphics[width=44mm]{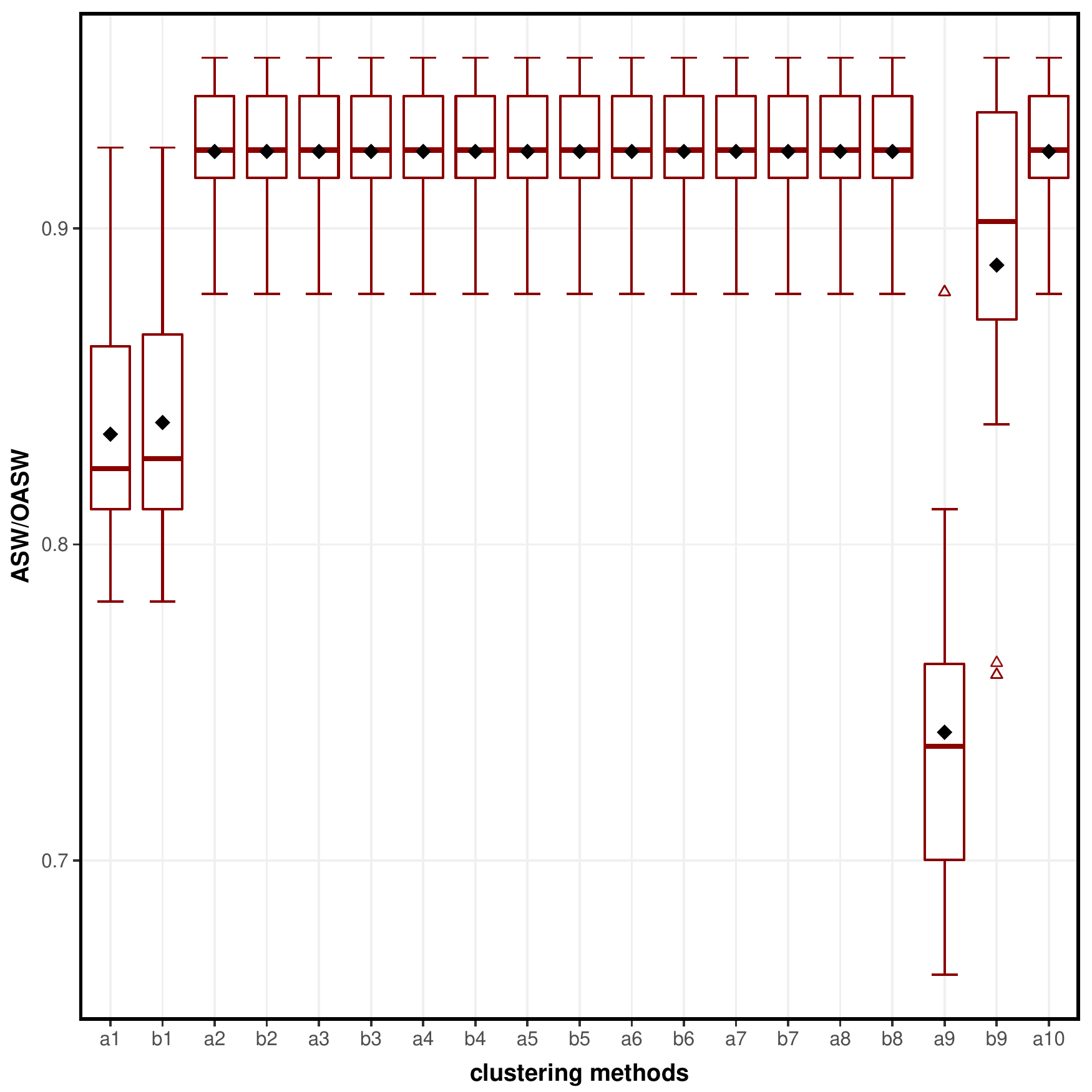}
\label{appendix:boxmodeleighteight} }
\subfloat[ ]{
  \includegraphics[width=44mm]{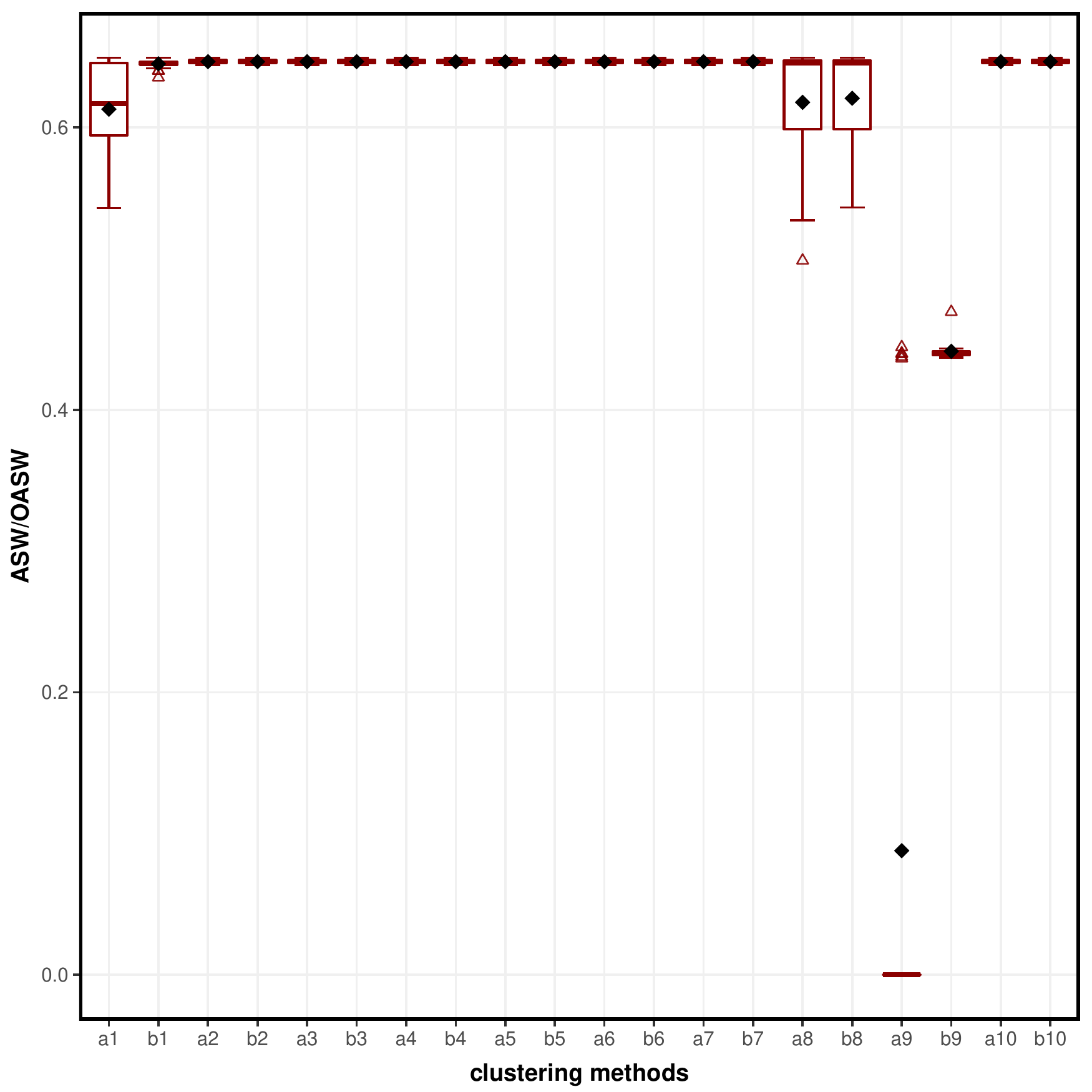}
\label{appendix:boxmodelninenine}}
\caption{Estimation of k case. (a) Boxplots of ASW  values for Model 1-9. a1/b1(k-means),  a2/b2(PAM), a3/b3(single), a4/b4(complete),   a5/b5(average), a6/b6(Ward), a7/b7(Mcquitty), a8/b8(model-based),  a9/b9(spectral),  a10(PAMSIL).  The mean value for all the methods are plotted as black diamonds and the outliers are red triangles.  }
\label{boxplotssim2}
\end{figure}
\newpage

\section{Estimation of $k$ PPR plots}
\par Nine tables were generated one for each DGP reporting the counts each method gave as an estimate for number of clusters from 2 to $K$ (11 columns) for an in depth analysis in Simulation II. From these tables the percentage performance rate (PPR) of indices were calculated. For instance, if an index estimates the known number of clusters correctly for 15 out of 25 runs, the index performance rate is  60$\%$.  The bars in the charts also represent this percentage.

\begin{figure}[!hbtp]
\centering
\subfloat[CH]{
  \includegraphics[width=35mm]{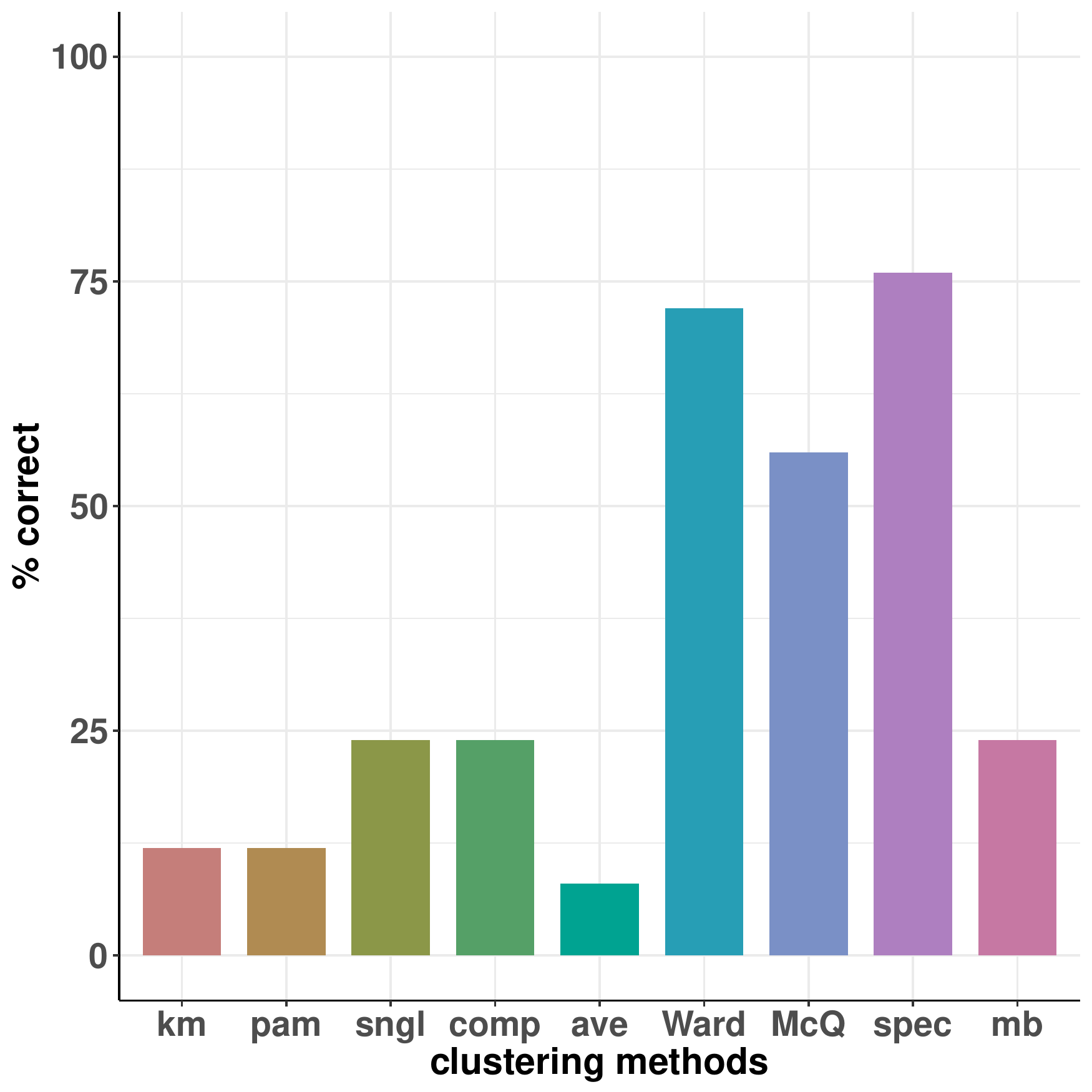}
}
\subfloat[H]{
  \includegraphics[width=35mm]{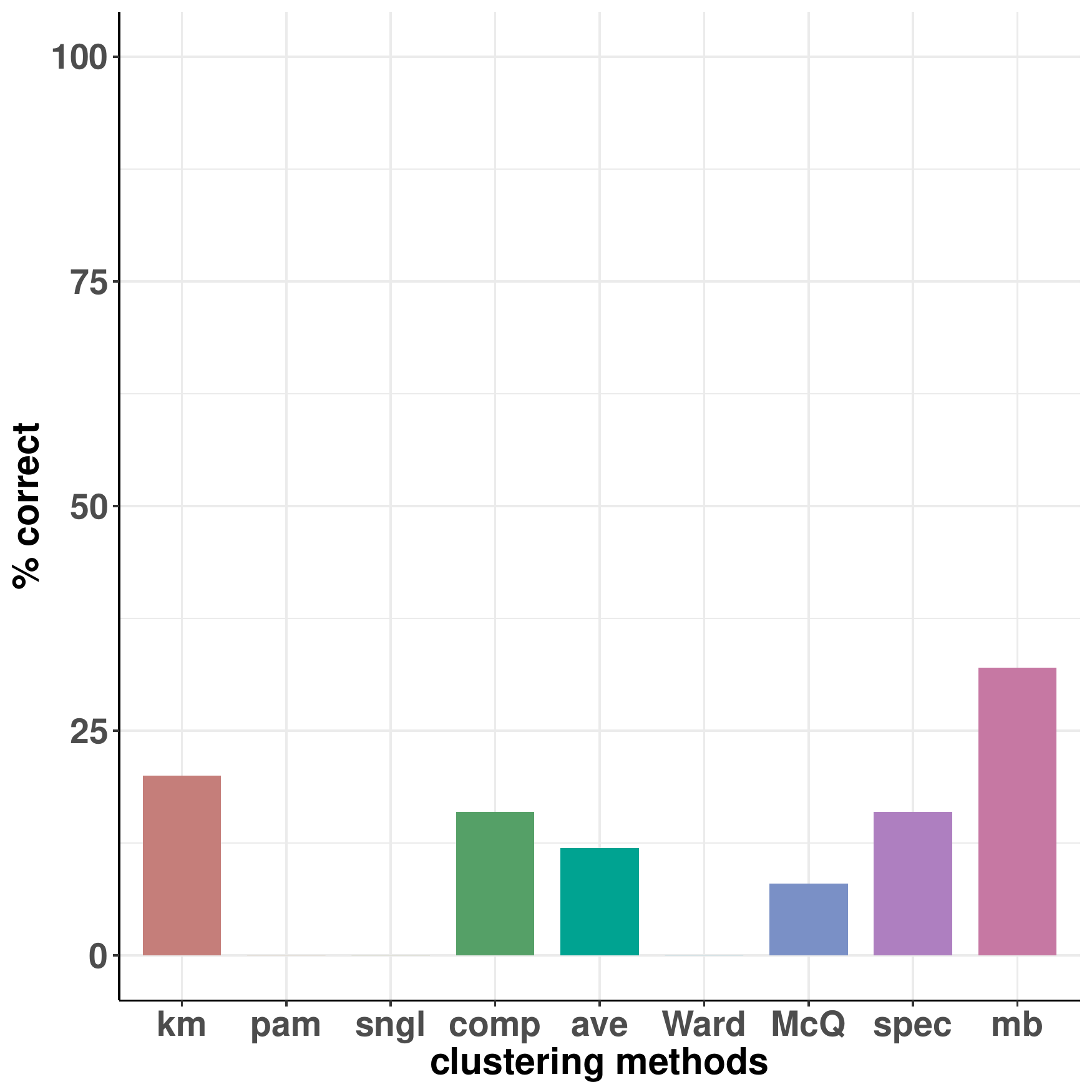}
}
  \subfloat[Gamma]{
  \includegraphics[width=35mm]{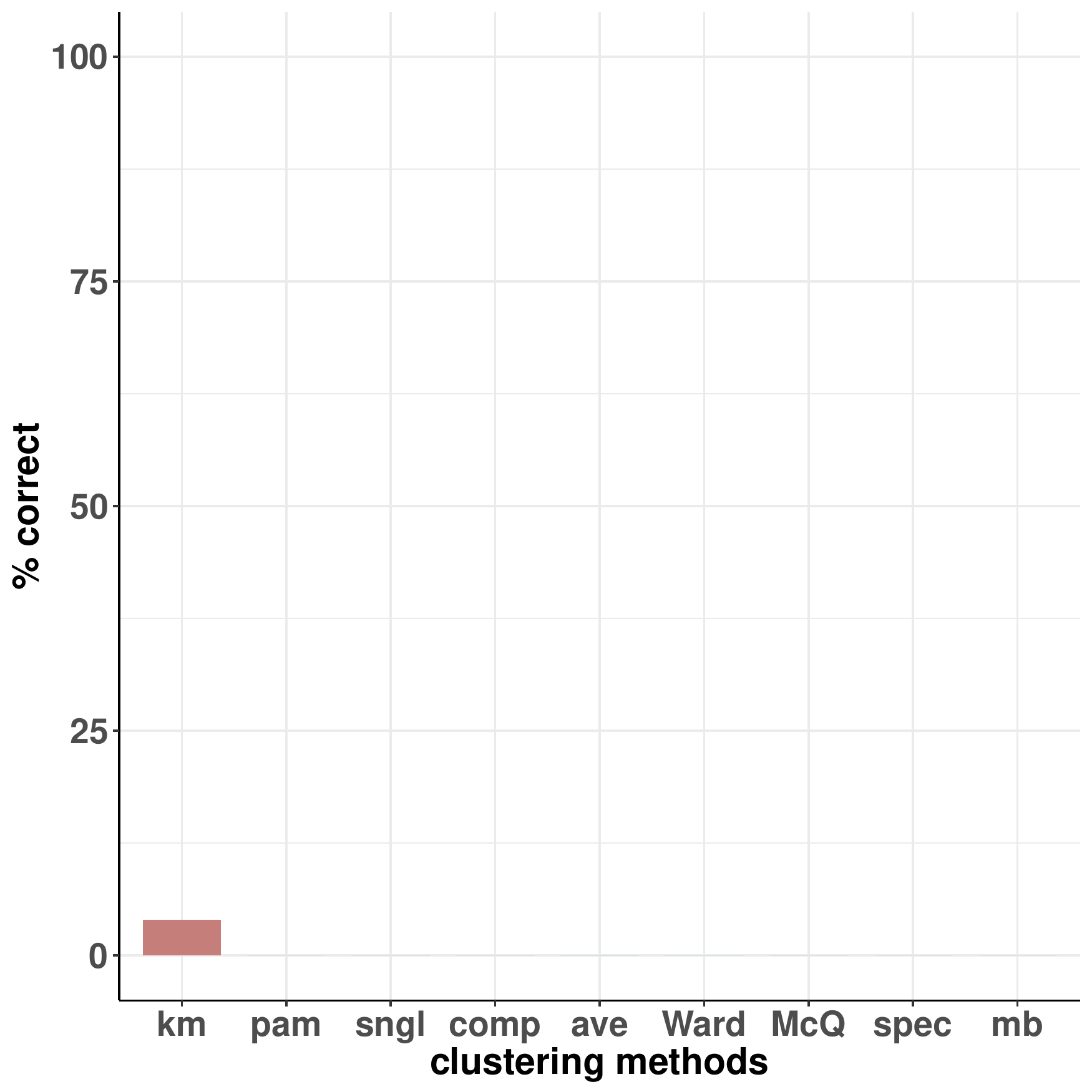}}
  \subfloat[C]{
  \includegraphics[width=35mm]{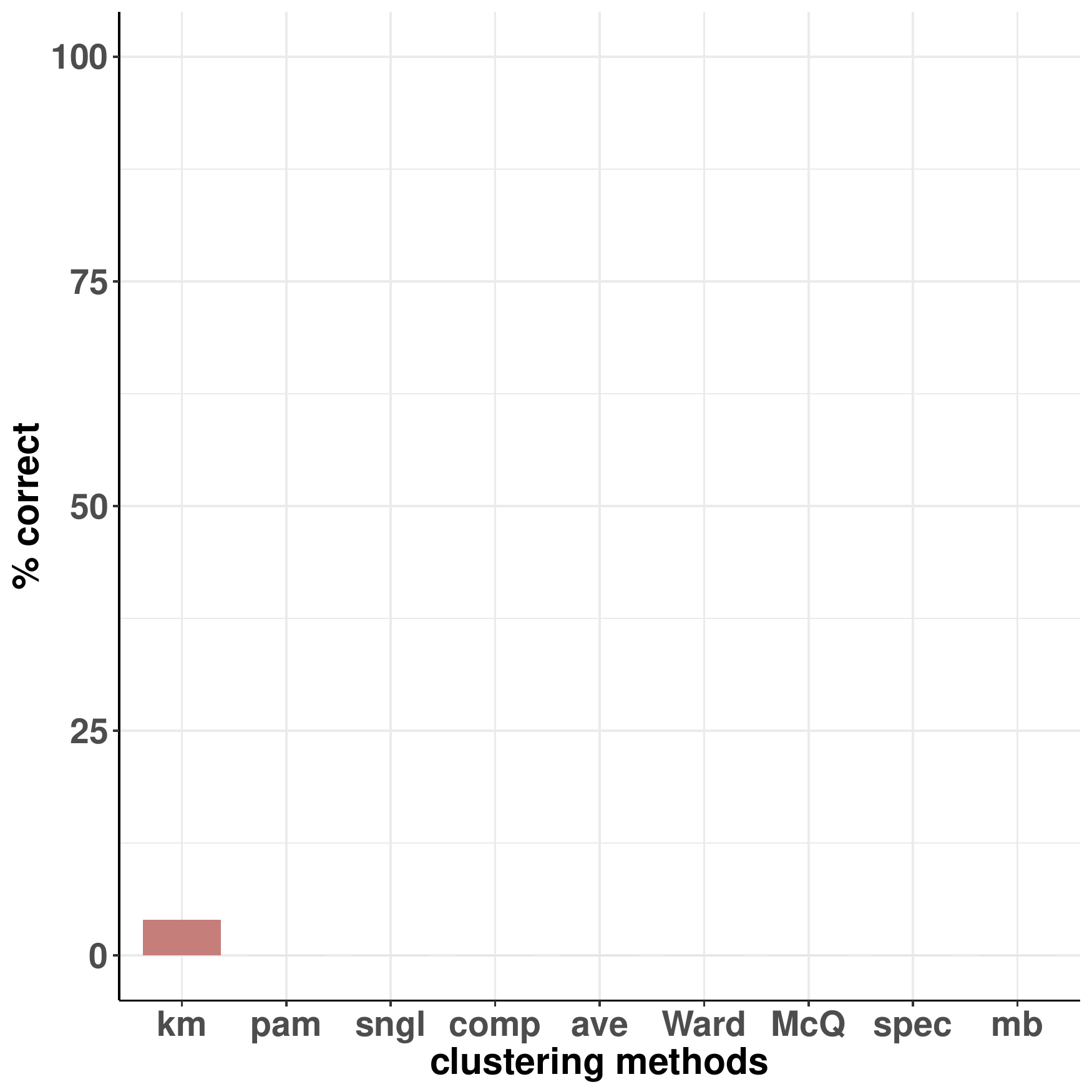}
}
\newline
\subfloat[KL]{
  \includegraphics[width=35mm]{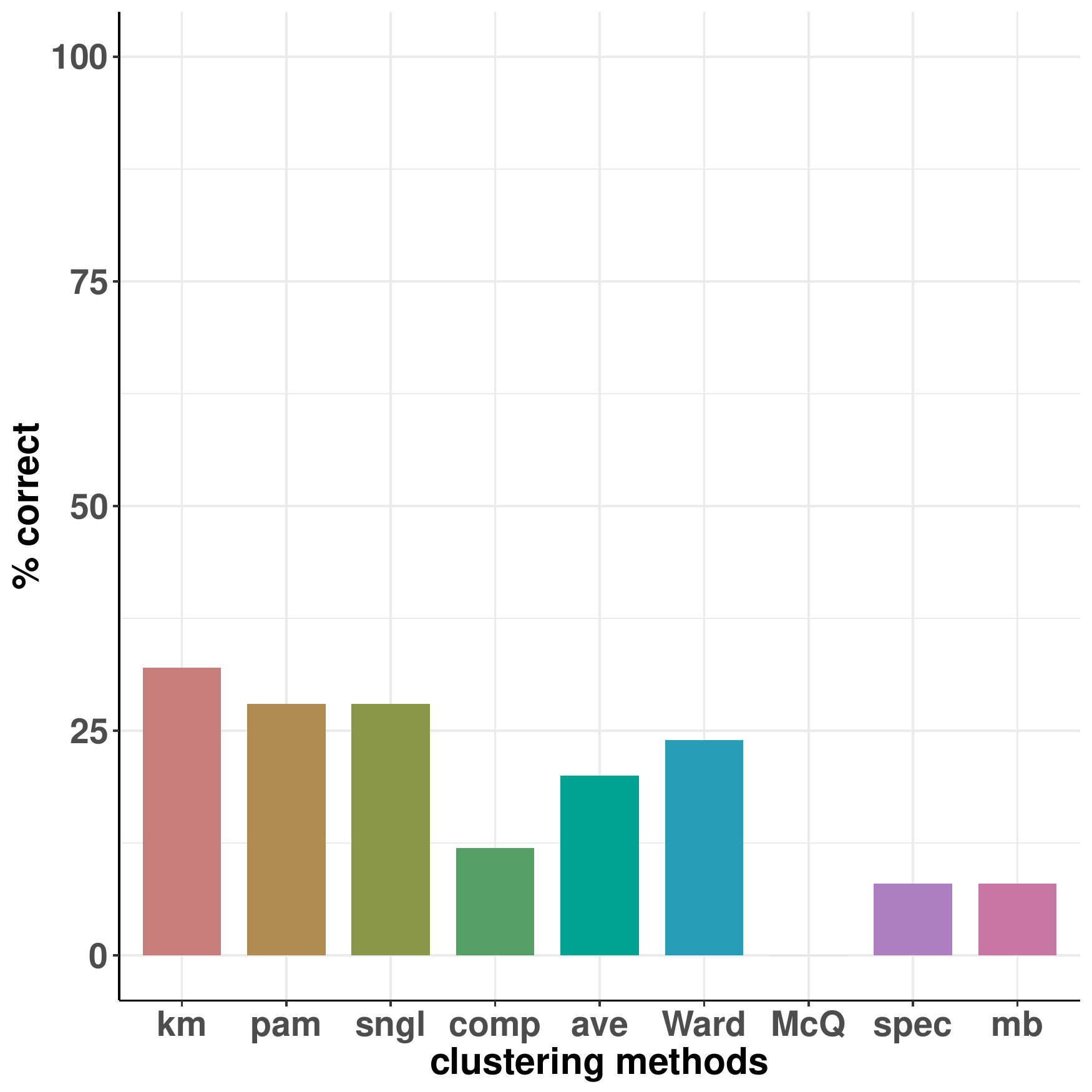}
}
\subfloat[gap]{
  \includegraphics[width=35mm]{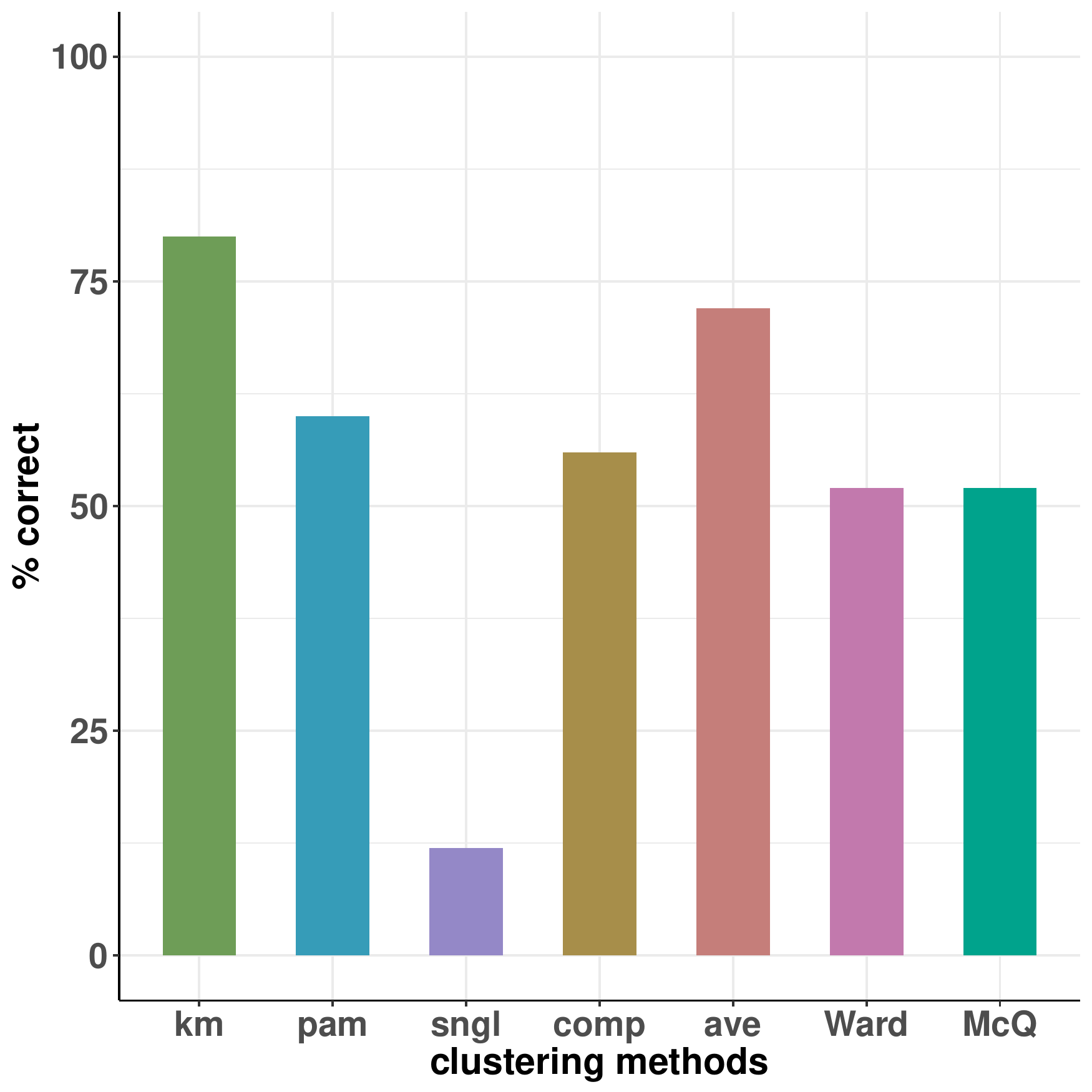}
}
\subfloat[jump]{
  \includegraphics[width=35mm]{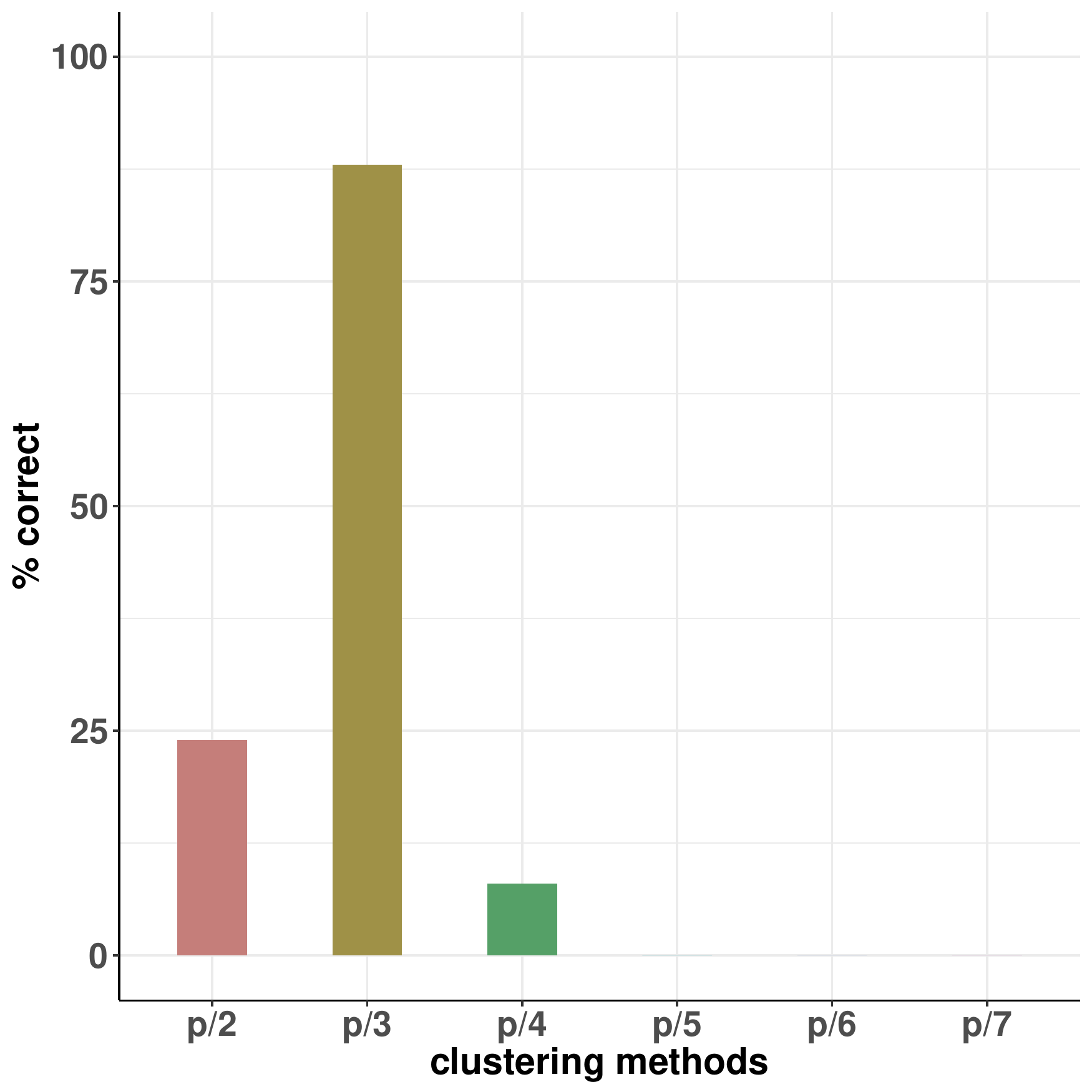}
}
\subfloat[PS]{
  \includegraphics[width=35mm]{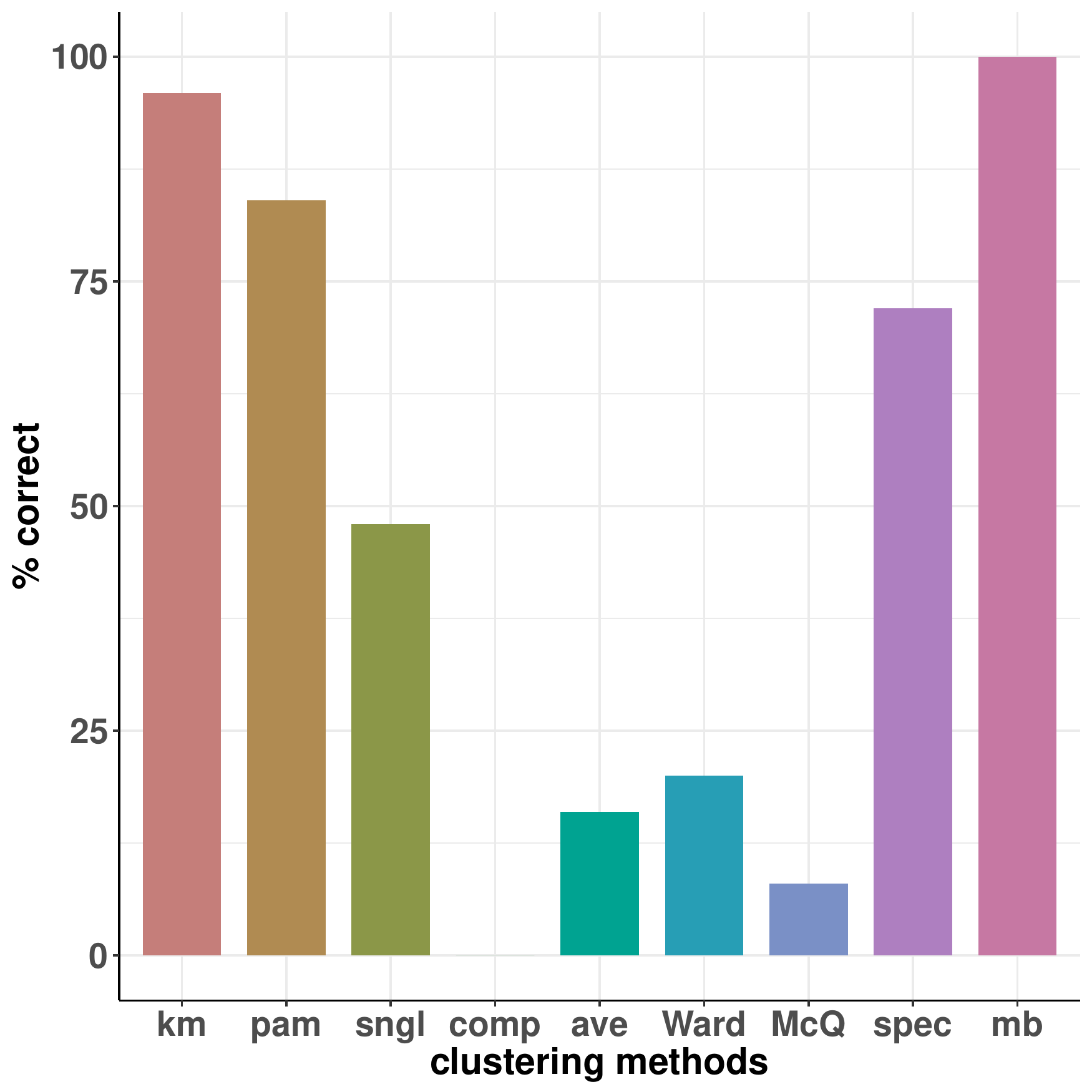}
}
\newline
\subfloat[BI]{
  \includegraphics[width=35mm]{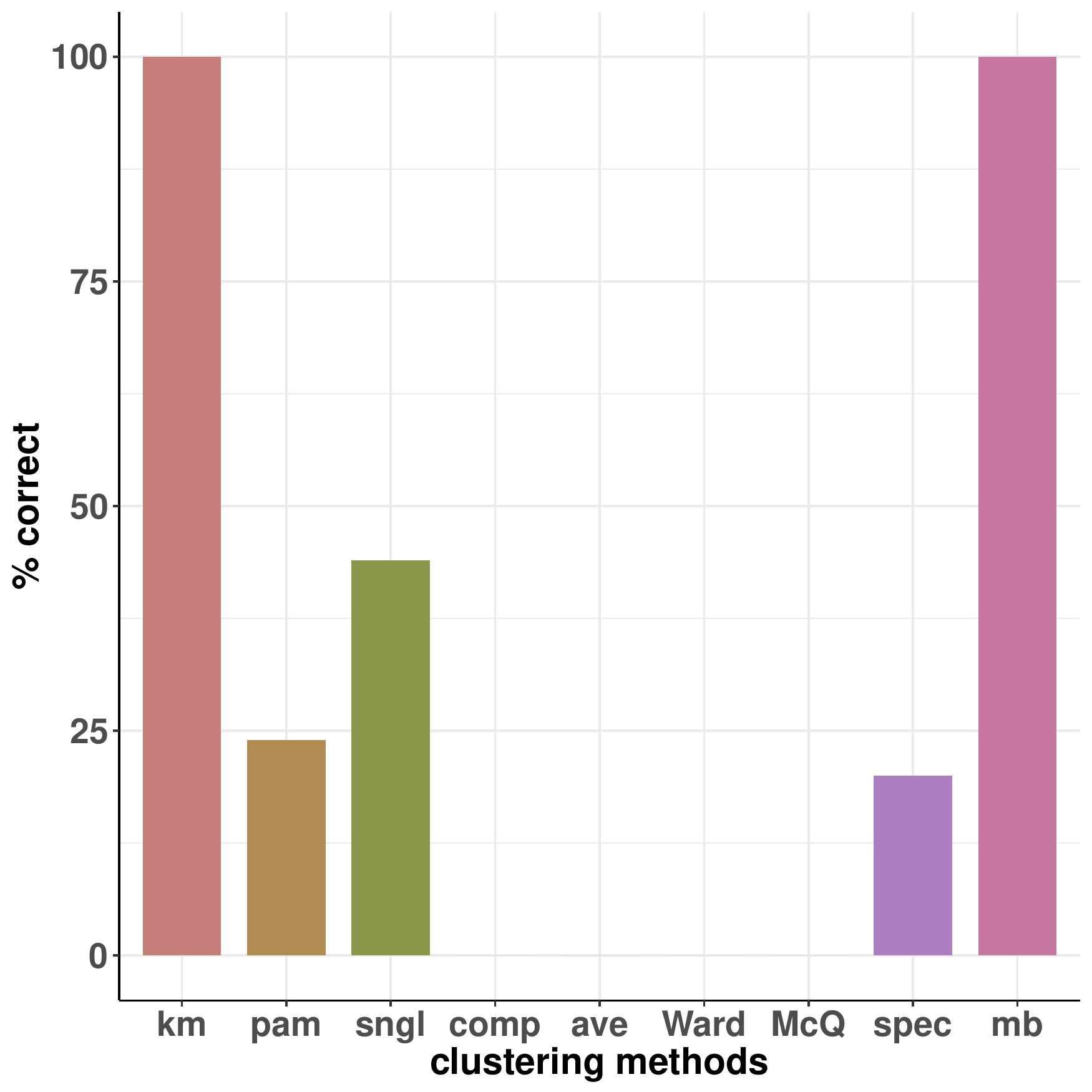}
}
\subfloat[CVNN]{
  \includegraphics[width=35mm]{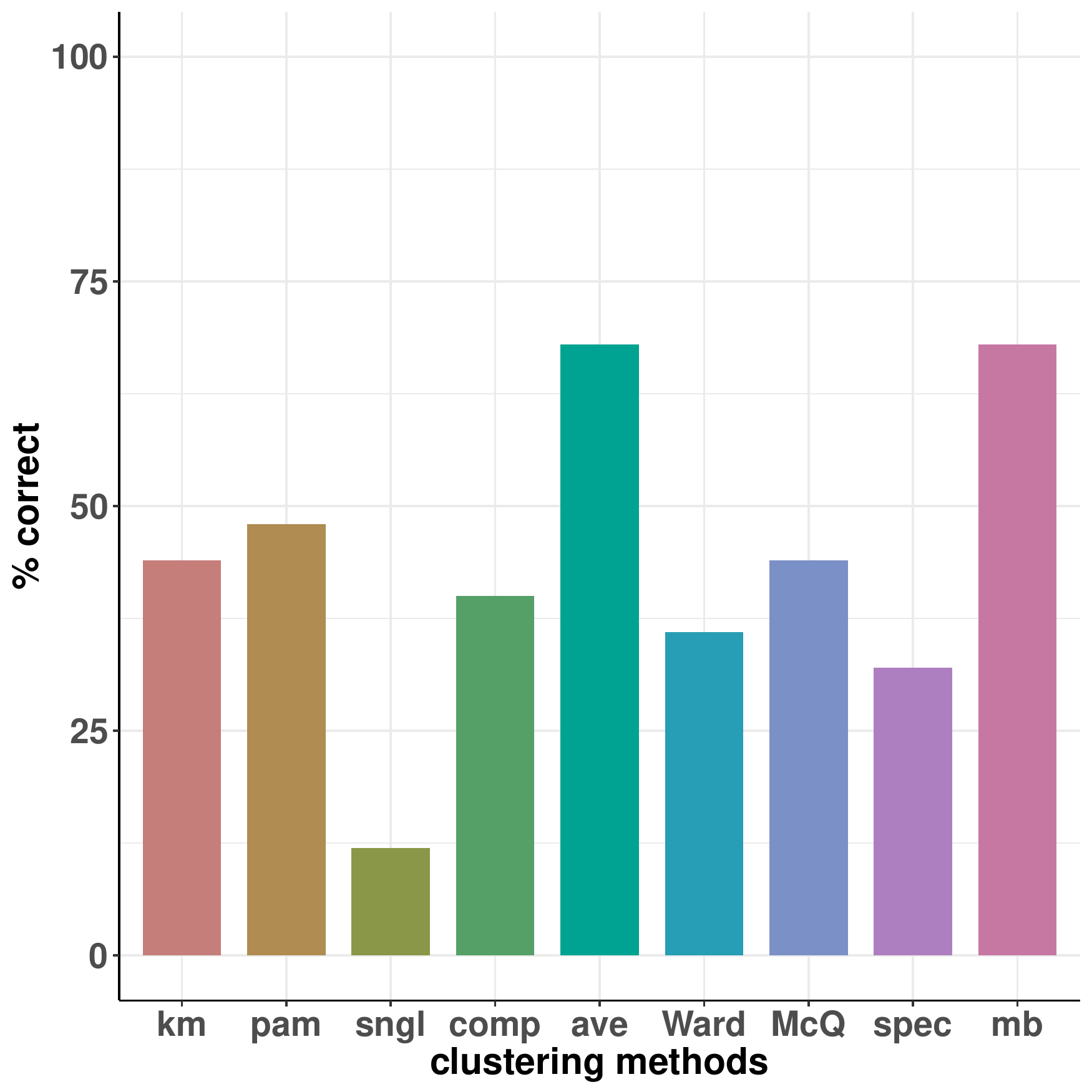}
}
\subfloat[BIC/PAMSIL]{
  \includegraphics[width=35mm]{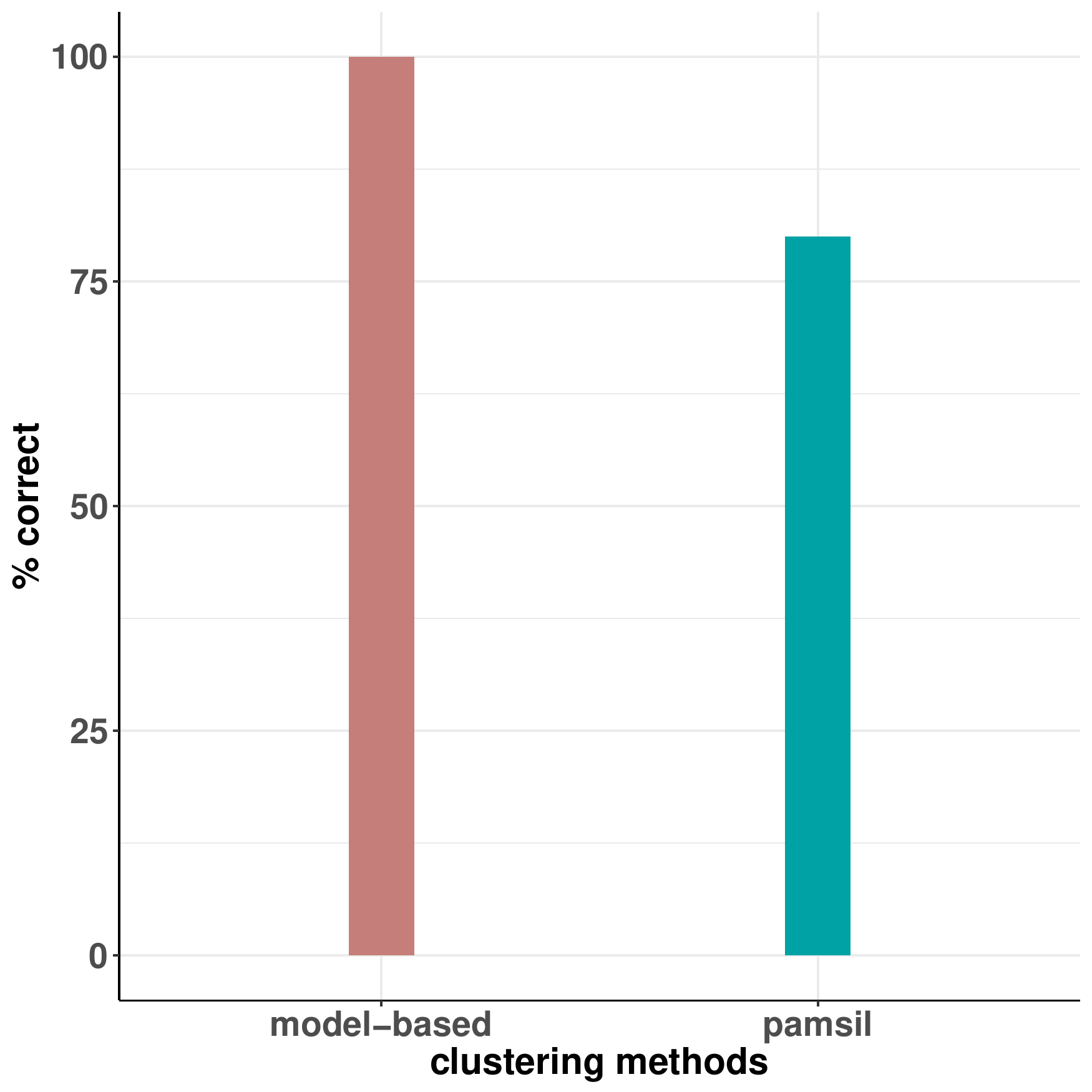}
}
\subfloat[ASW]{
  \includegraphics[width=35mm]{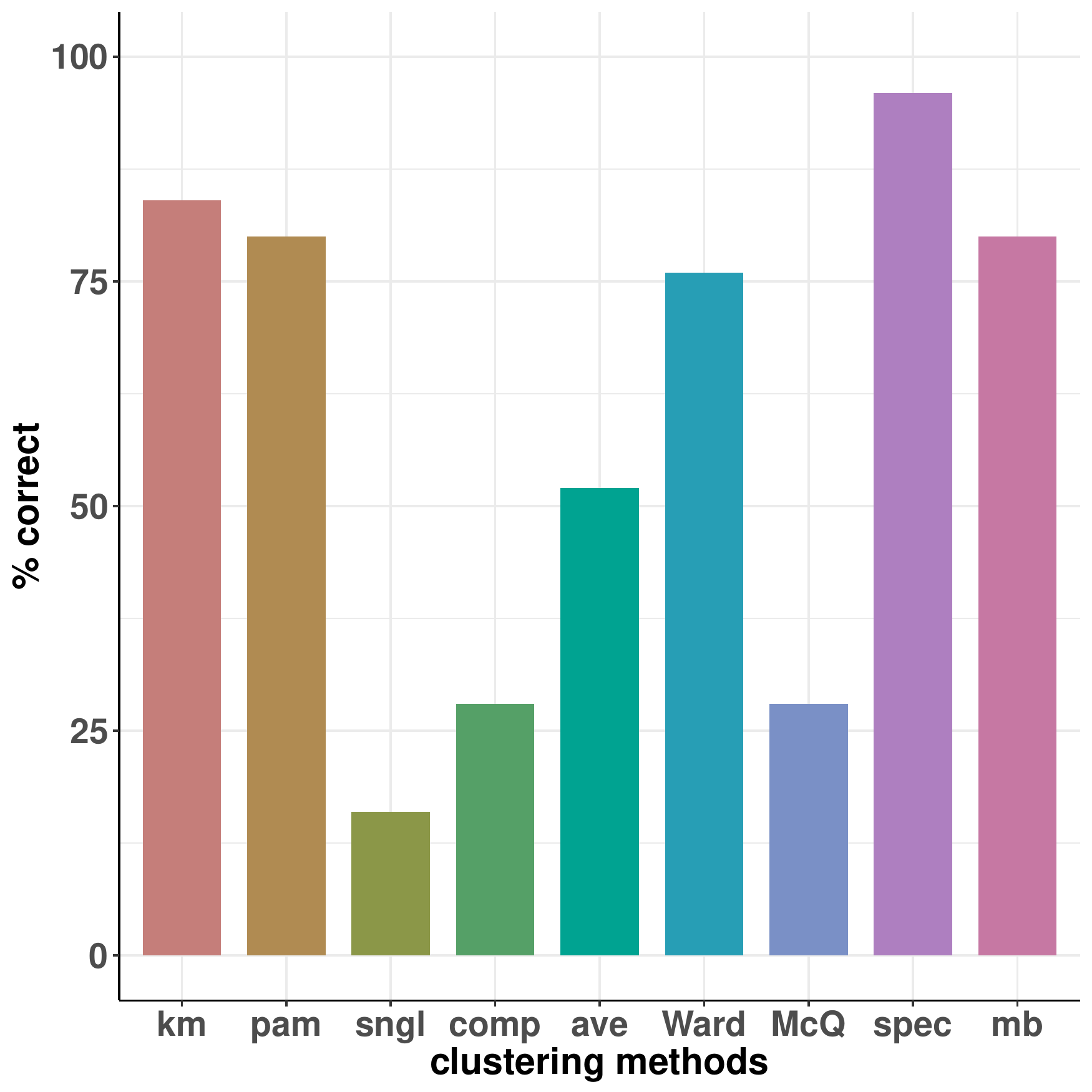}
}
\newline
\rule{-33ex}{.2in}
\subfloat[OASW]{
  \includegraphics[width=30mm]{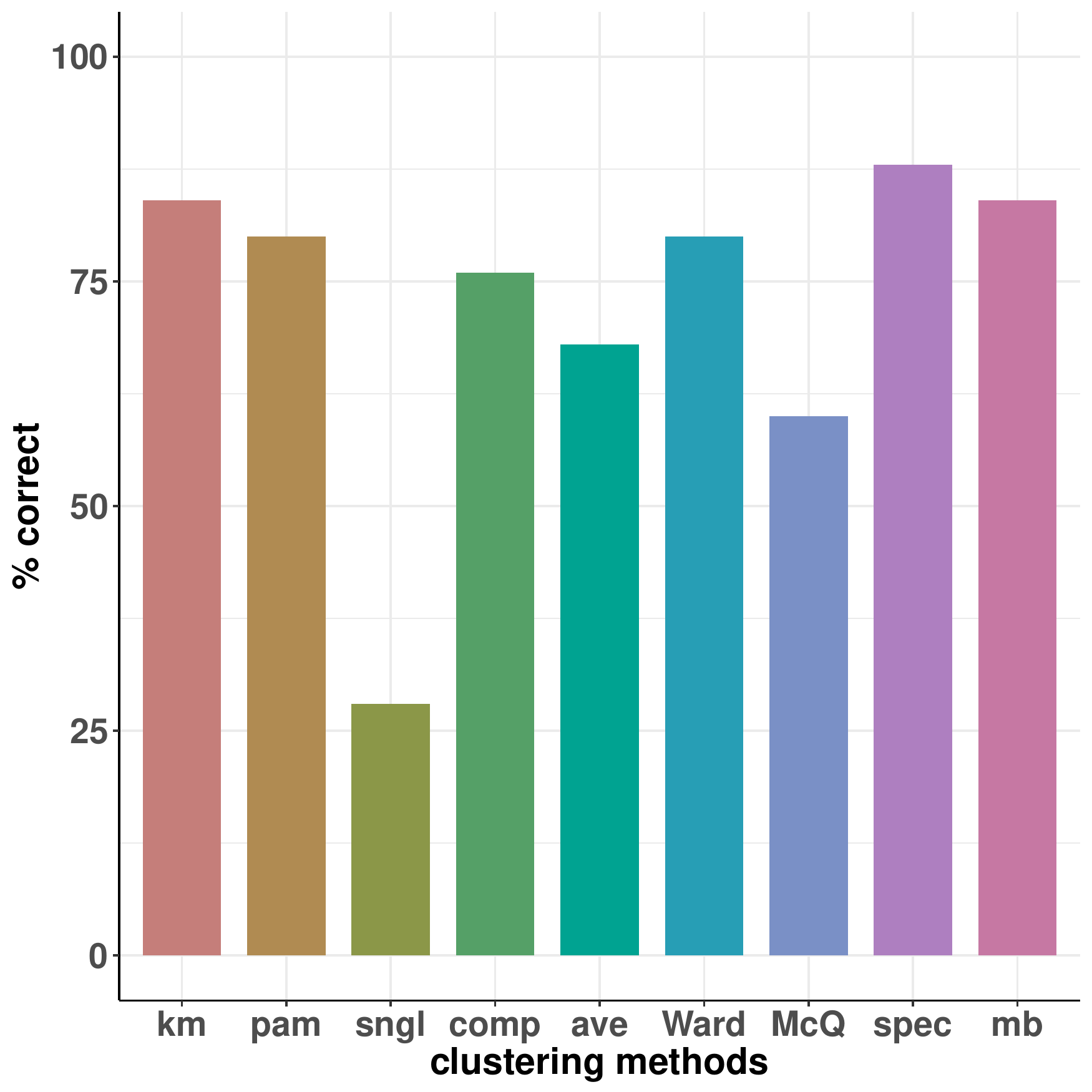}
}
\subfloat[key]{
  \hspace*{1cm}\includegraphics[width=45mm, trim=0cm 20 0 -1cm]{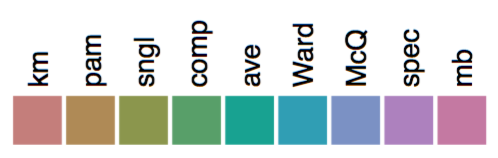}
}
\caption{Bar plots for the estimation of k for Model 1. Each bar represents the percentage count of correct estimate of k. In each panel the bars shows the results for $k$-means, PAM, single, complete, average, Ward, McQuitty, spectral and model-based clustering methods. }
\label{appendix:estkmodelone}
\end{figure}

\begin{figure}[!hbtp]
\centering
\subfloat[CH]{
  \includegraphics[width=35mm]{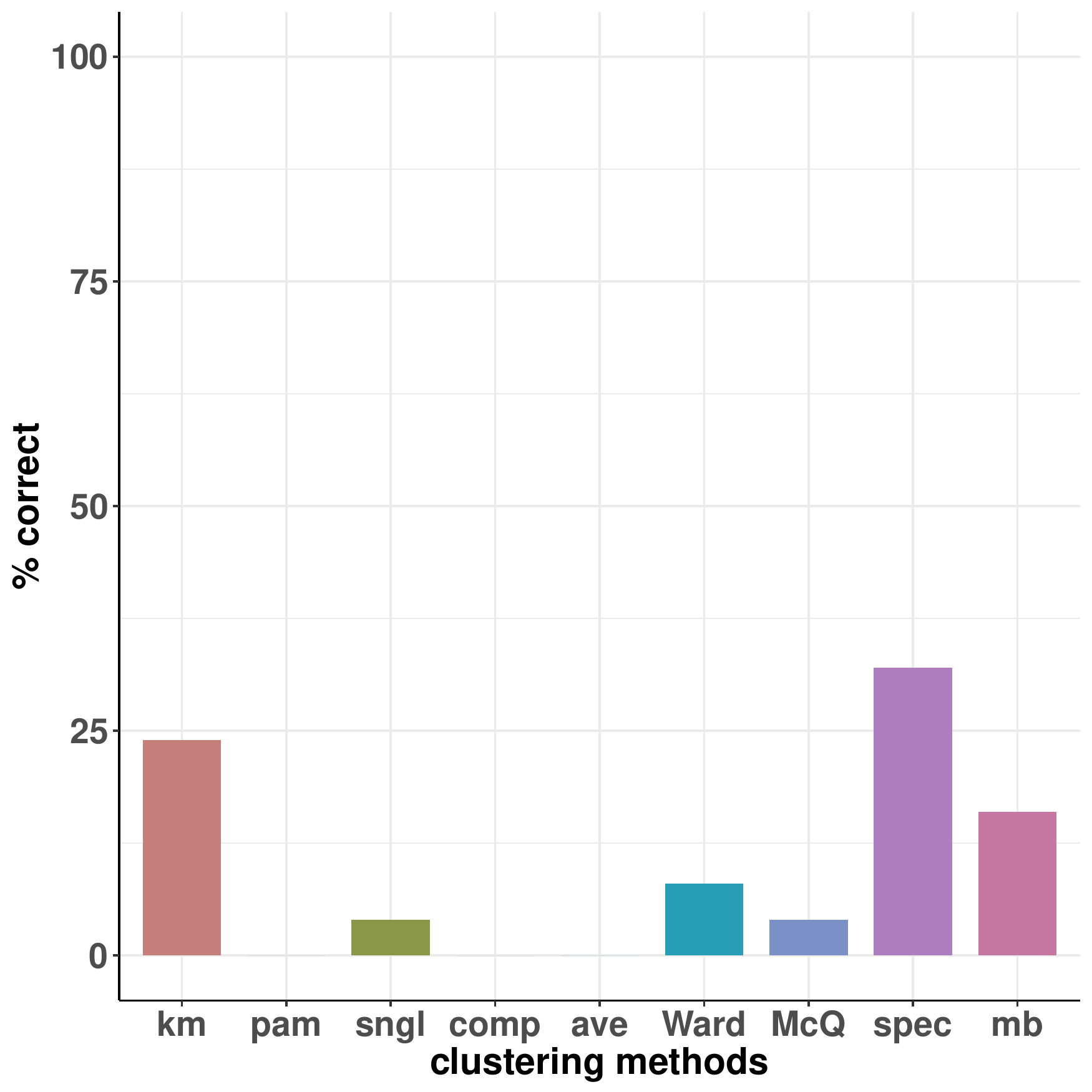}
}
\subfloat[H]{
  \includegraphics[width=35mm]{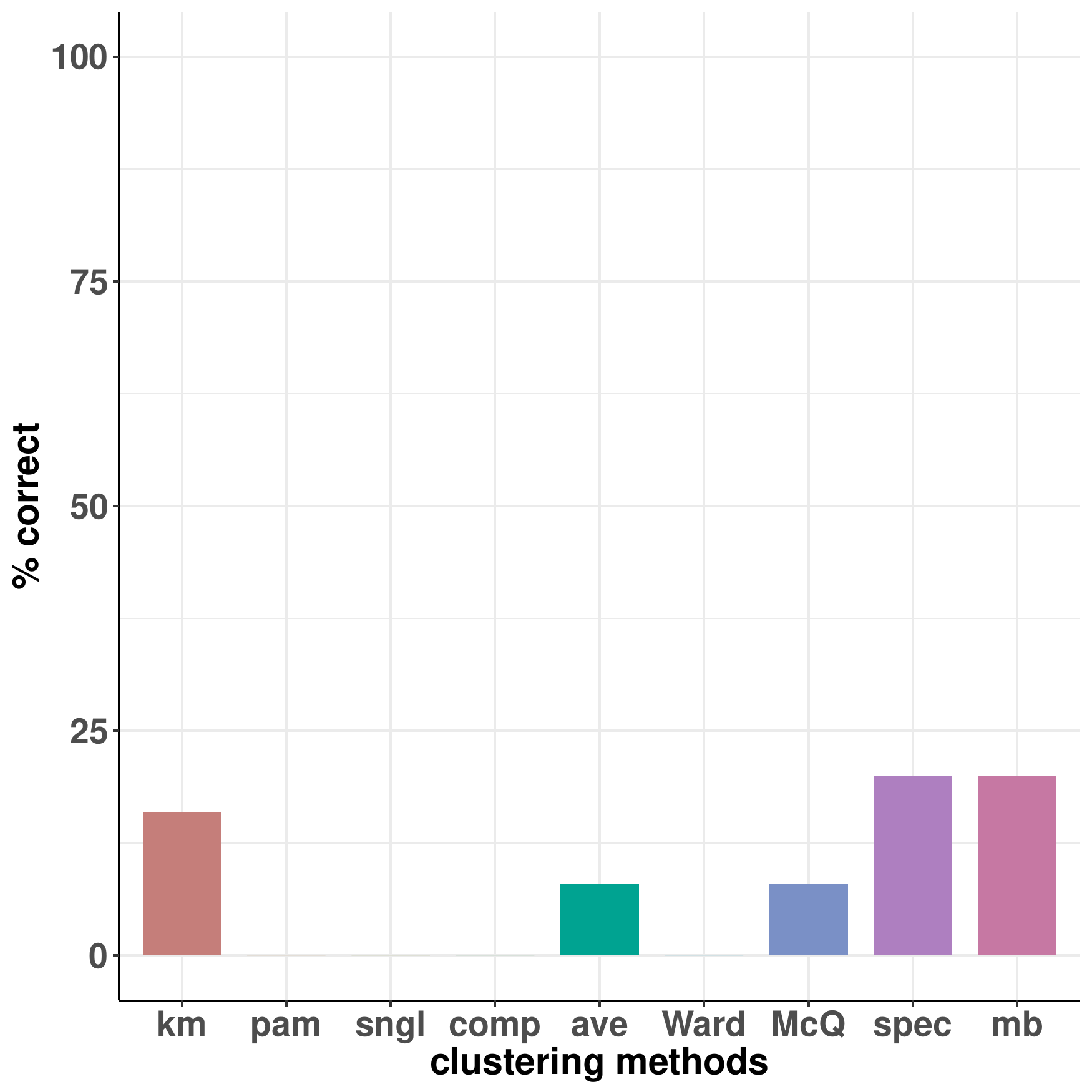}
}
  \subfloat[Gamma]{
  \includegraphics[width=35mm]{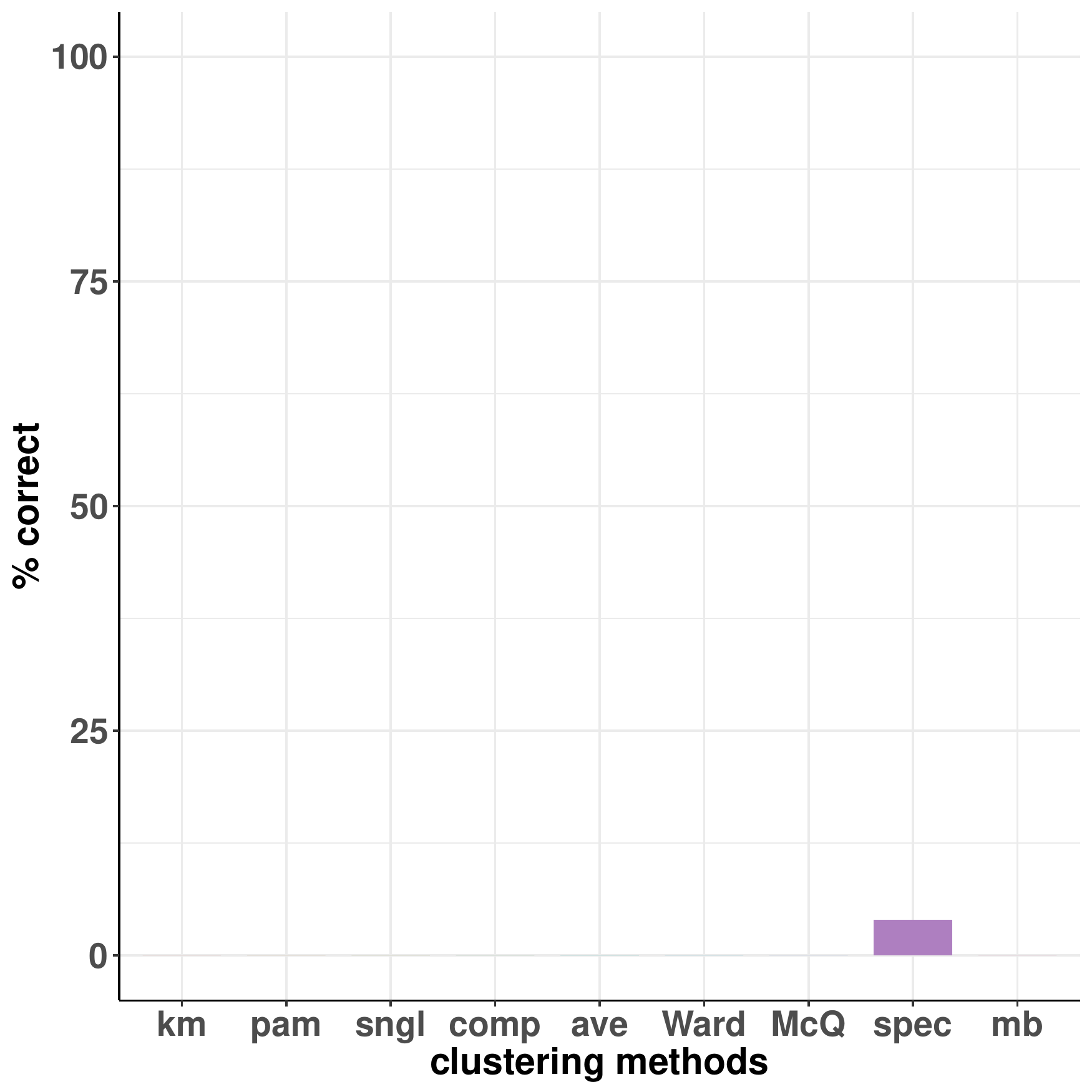}}
\subfloat[KL]{
  \includegraphics[width=35mm]{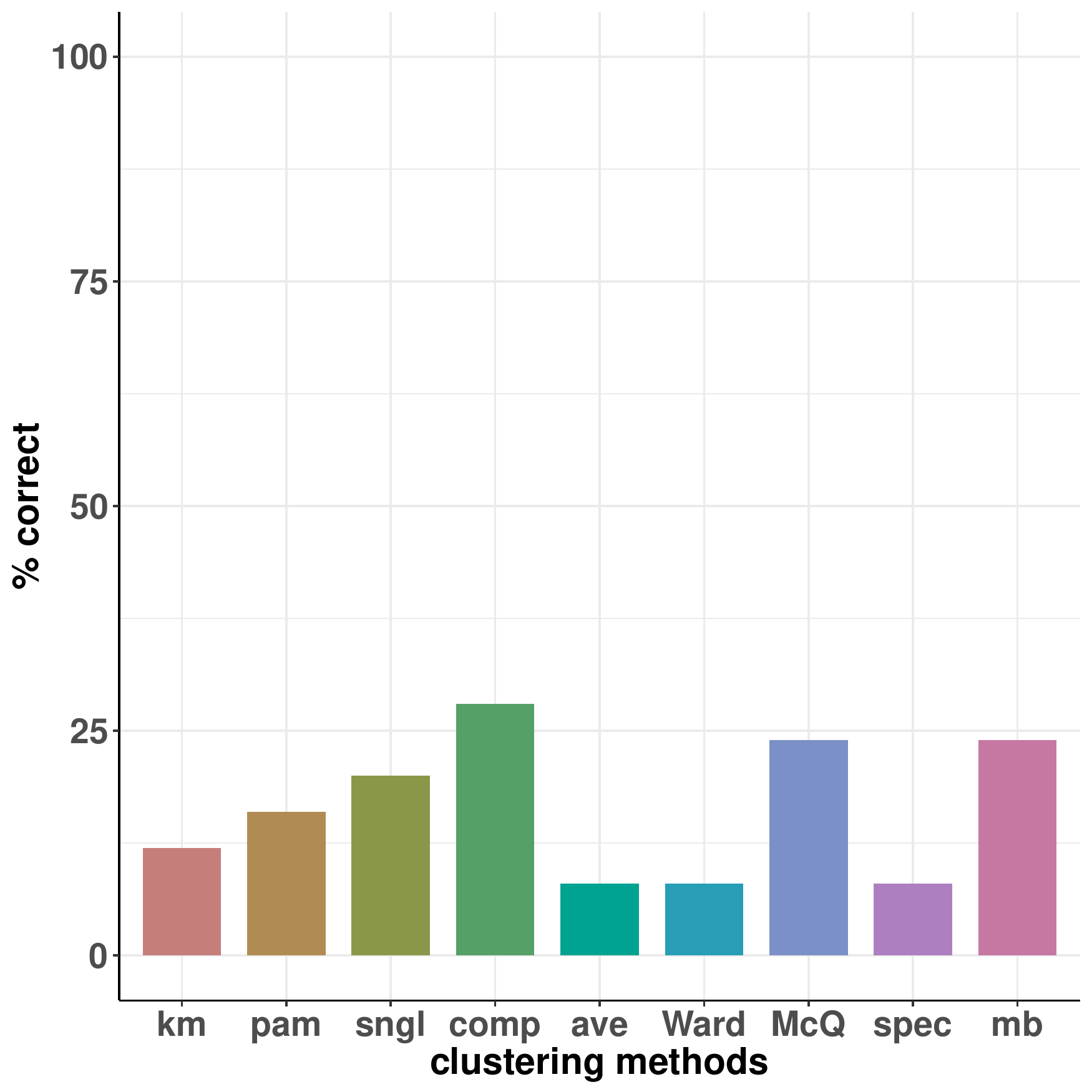}
}
\newline
\subfloat[gap]{
  \includegraphics[width=35mm]{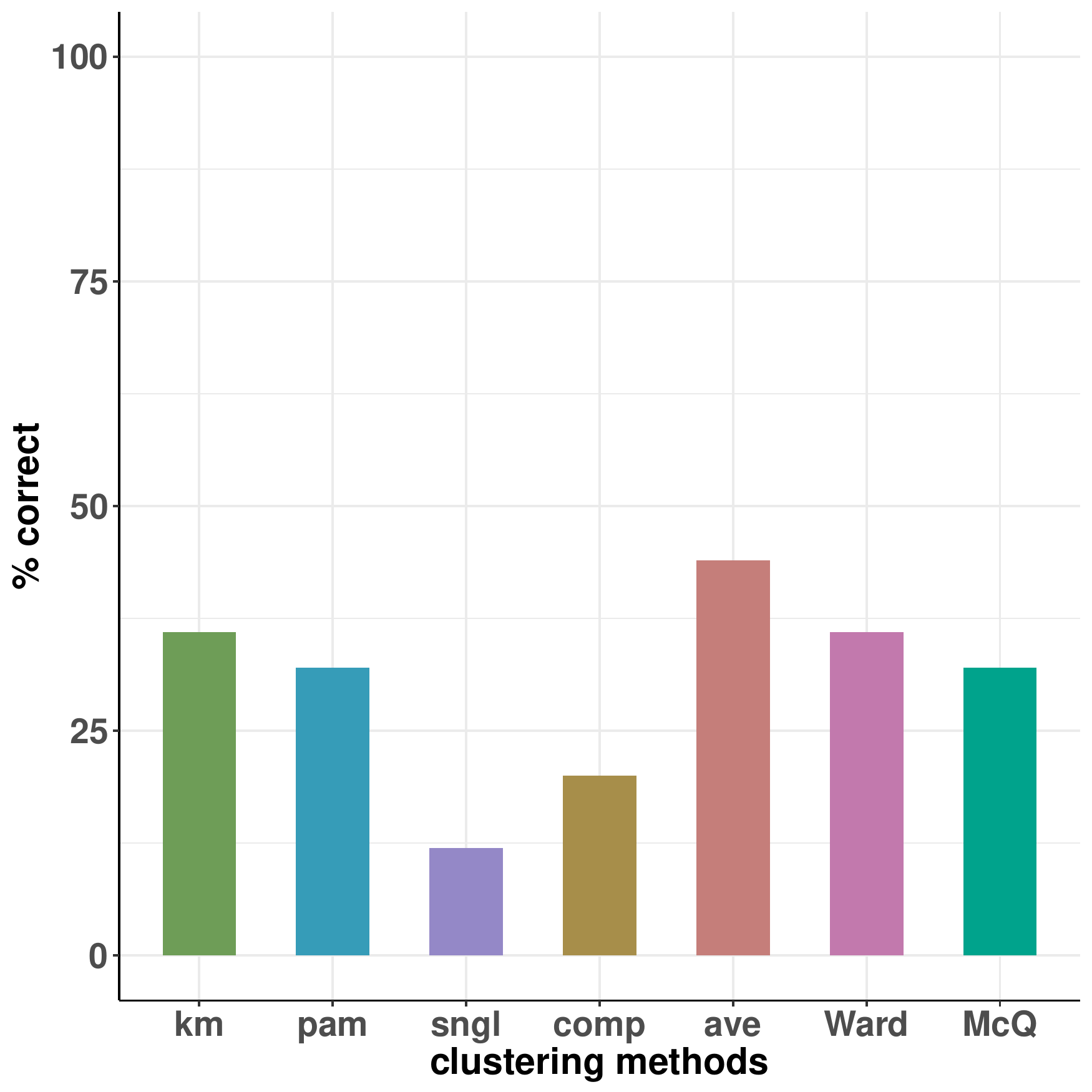}
}
\subfloat[jump]{
  \includegraphics[width=35mm]{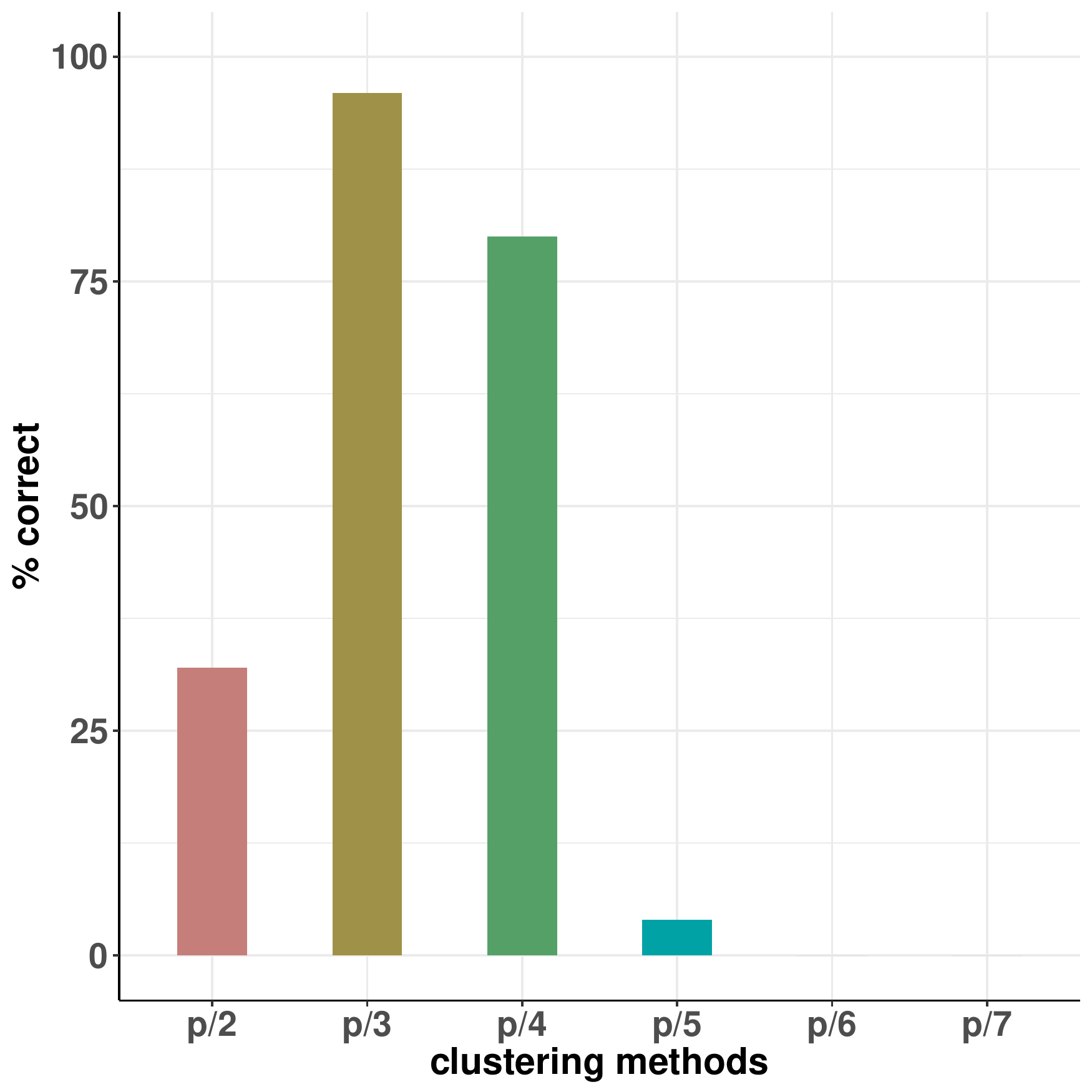}
}
\subfloat[PS]{
  \includegraphics[width=35mm]{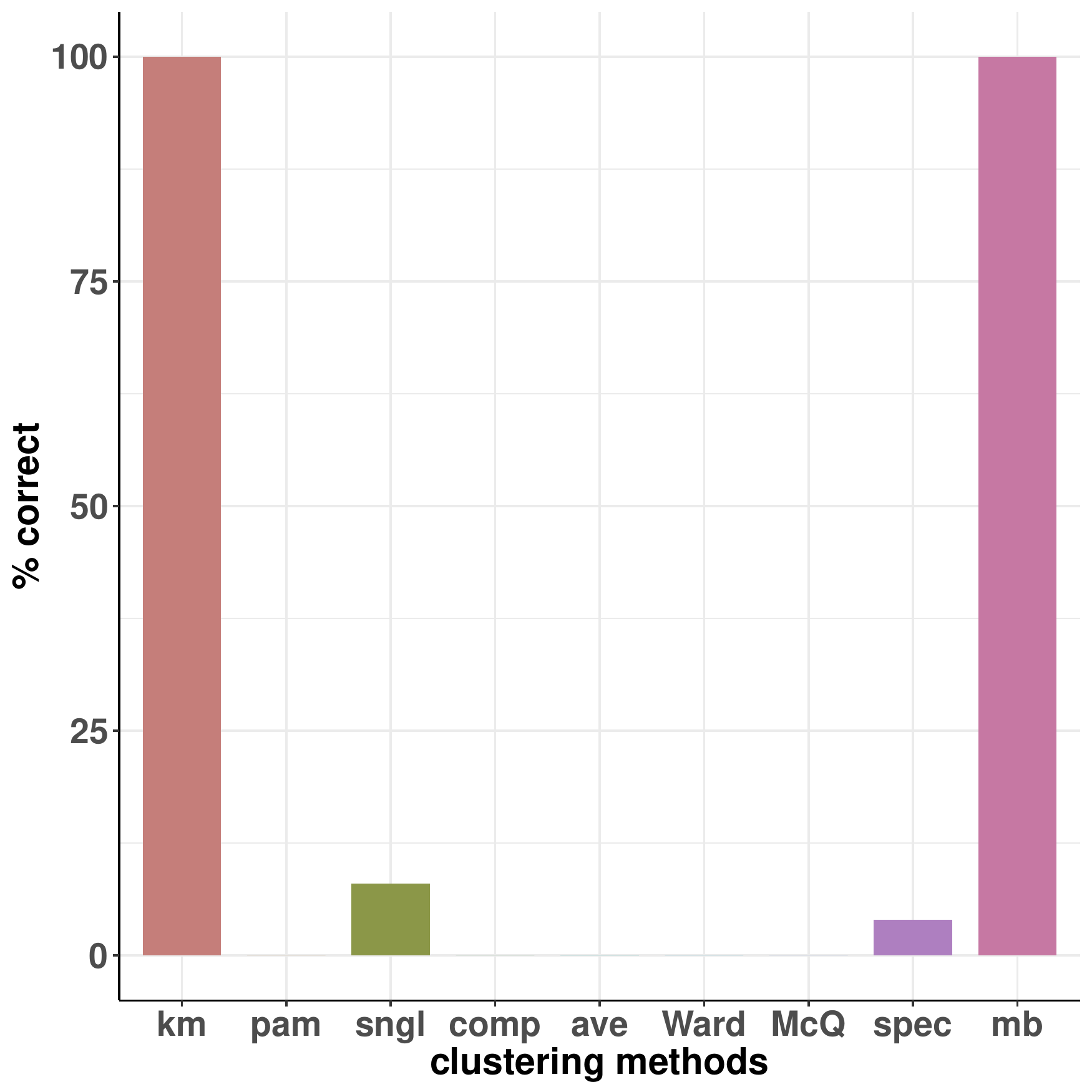}
}
\subfloat[BI]{
  \includegraphics[width=35mm]{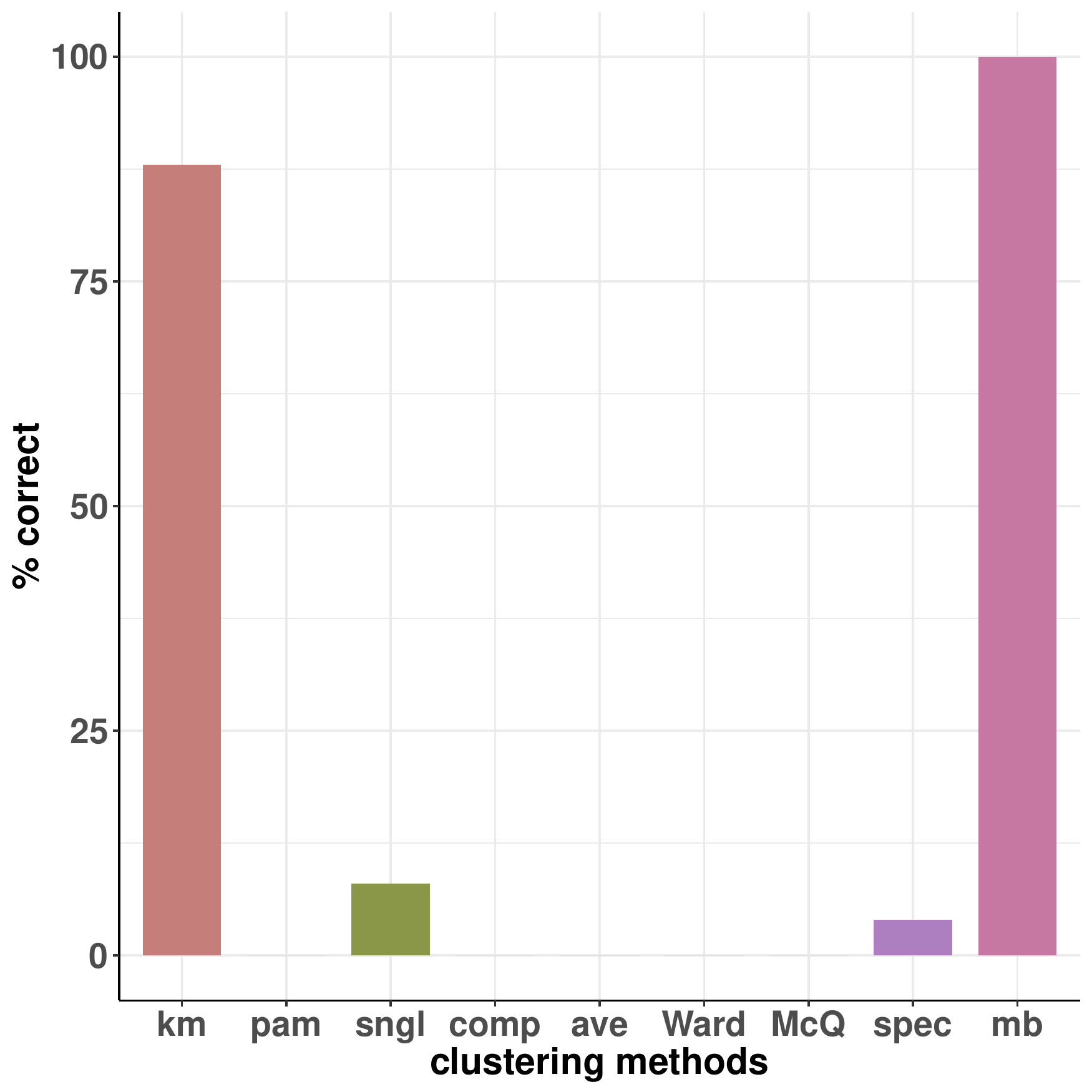}
}
\newline
\subfloat[CVNN]{
  \includegraphics[width=35mm]{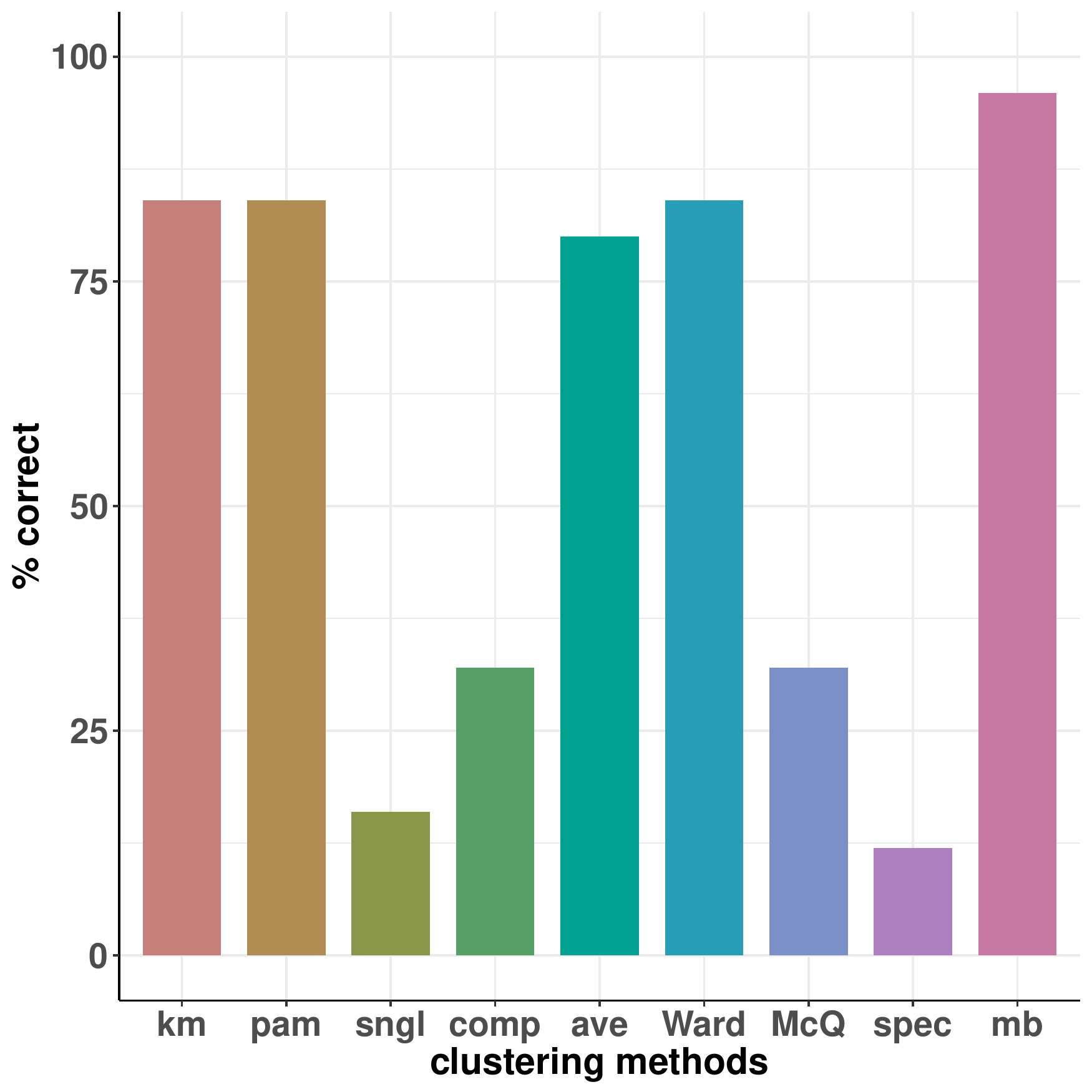}
}
\subfloat[BIC/PAMSIL]{
  \includegraphics[width=35mm]{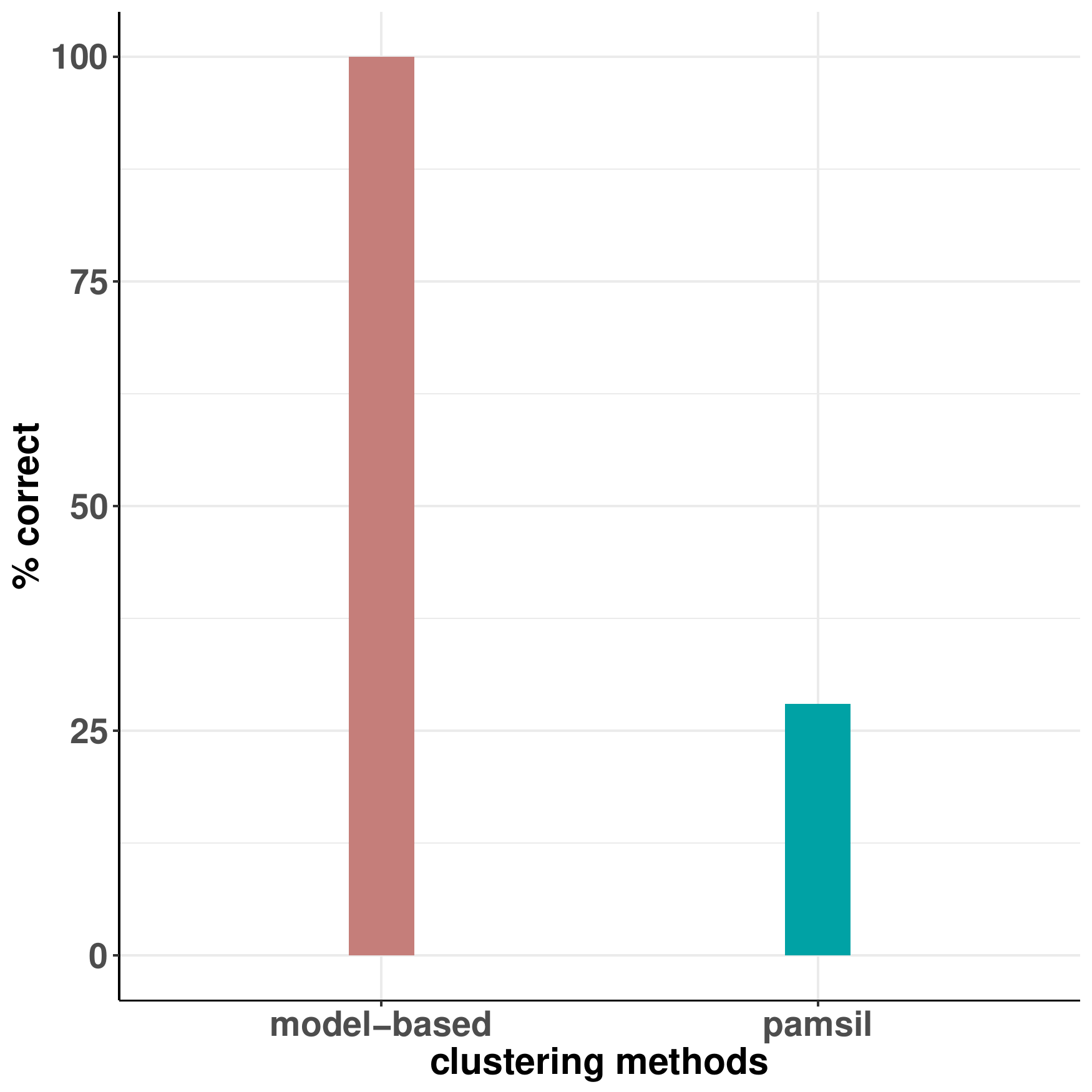}
}
\subfloat[ASW]{
  \includegraphics[width=35mm]{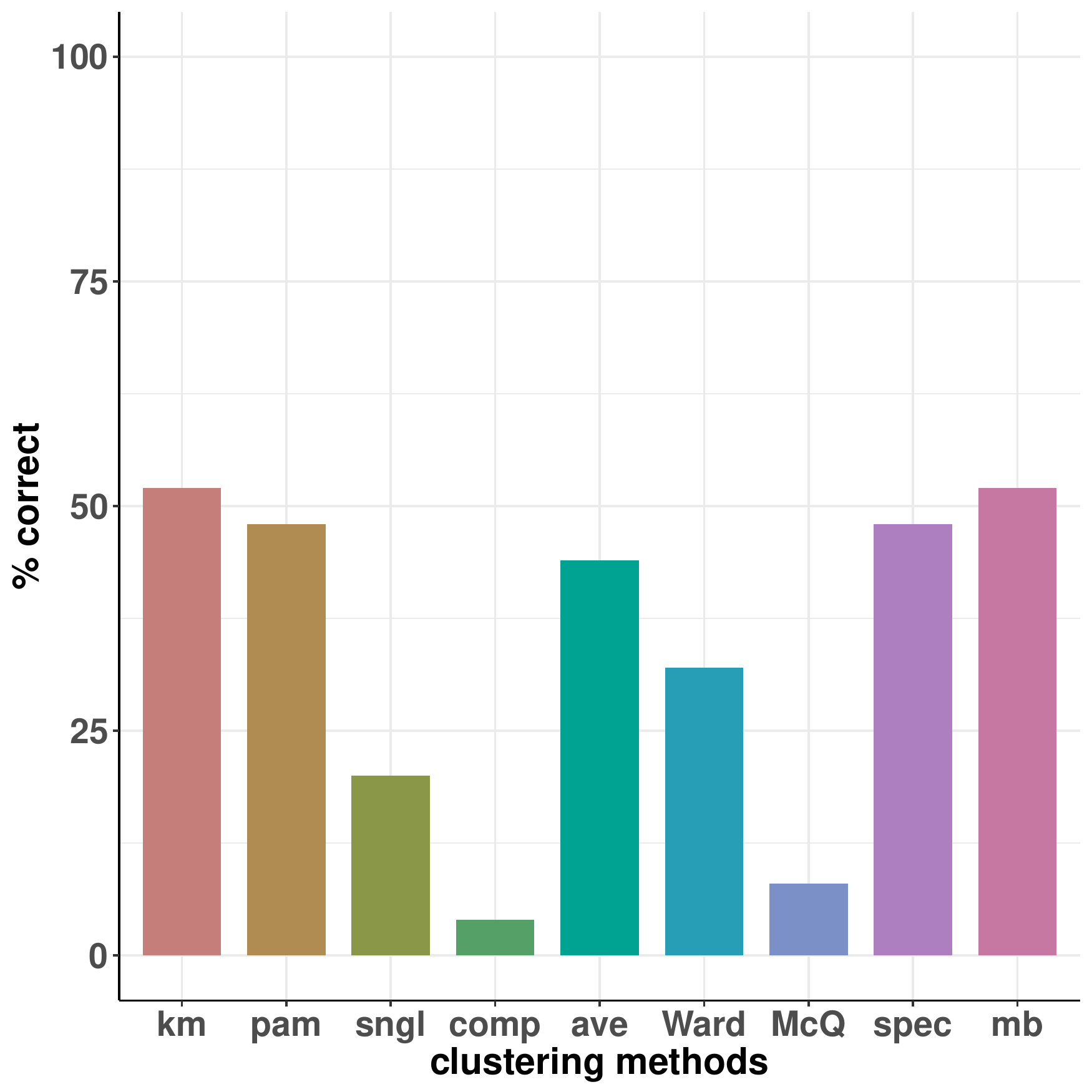}
}
\subfloat[OASW]{
  \includegraphics[width=35mm]{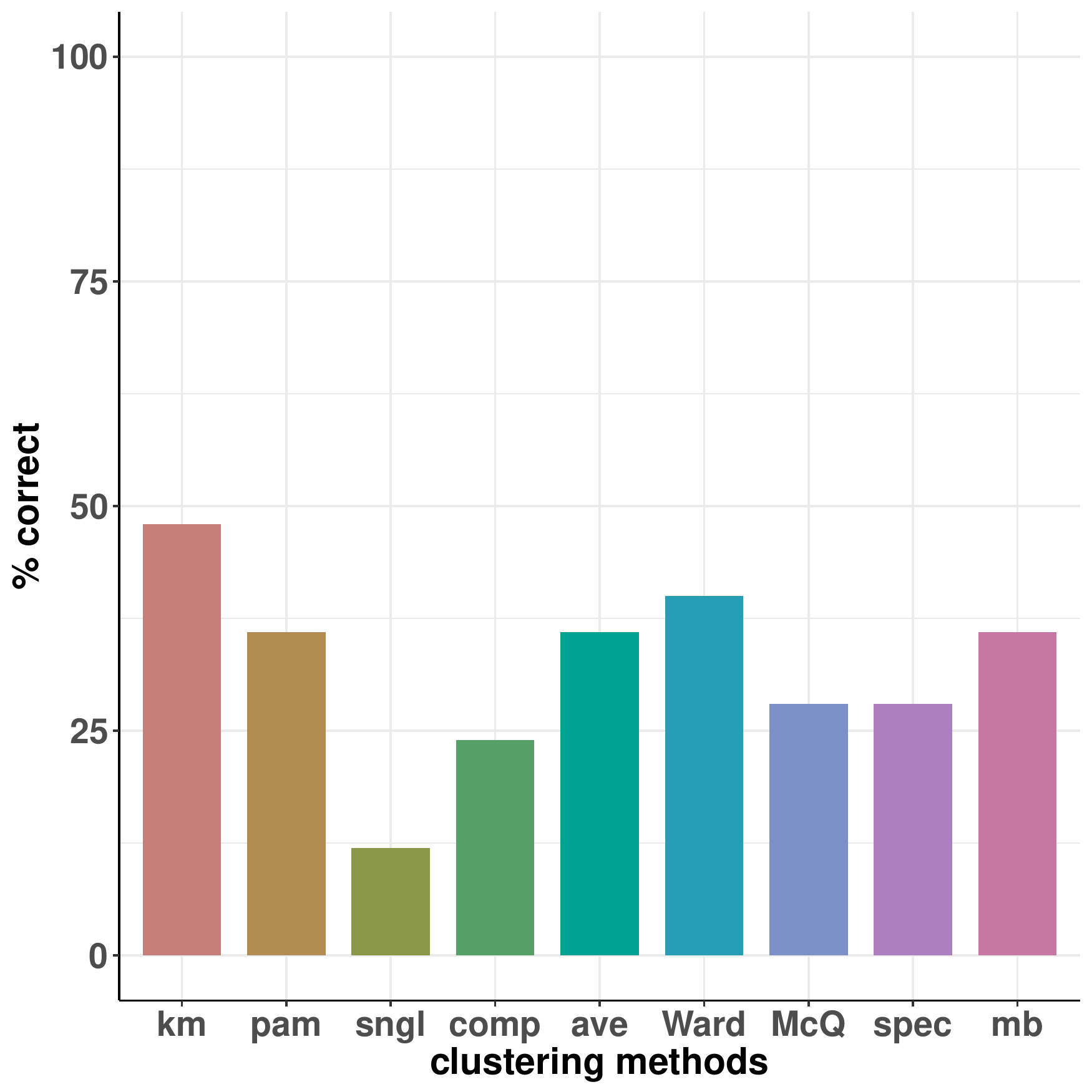}
}
\caption{Bar plots for the estimation of k for Model 2. The C index was never able to estimate correct number of clusters for Model 2. }
\label{appendix:estkmodeltwo}
\end{figure}

\begin{figure}[!hbtp]
\centering
\subfloat[H]{
  \includegraphics[width=35mm]{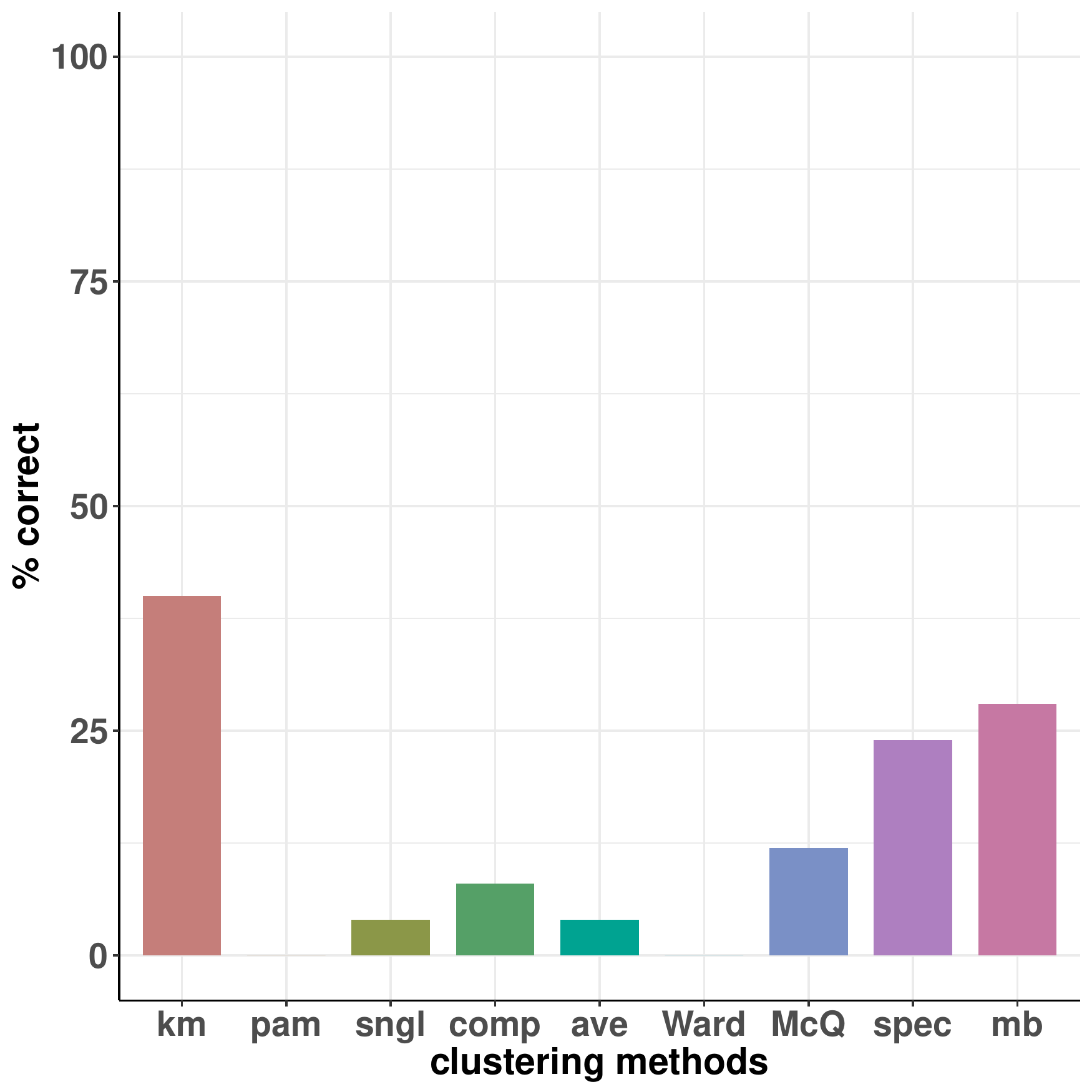}
}
  \subfloat[Gamma]{
  \includegraphics[width=35mm]{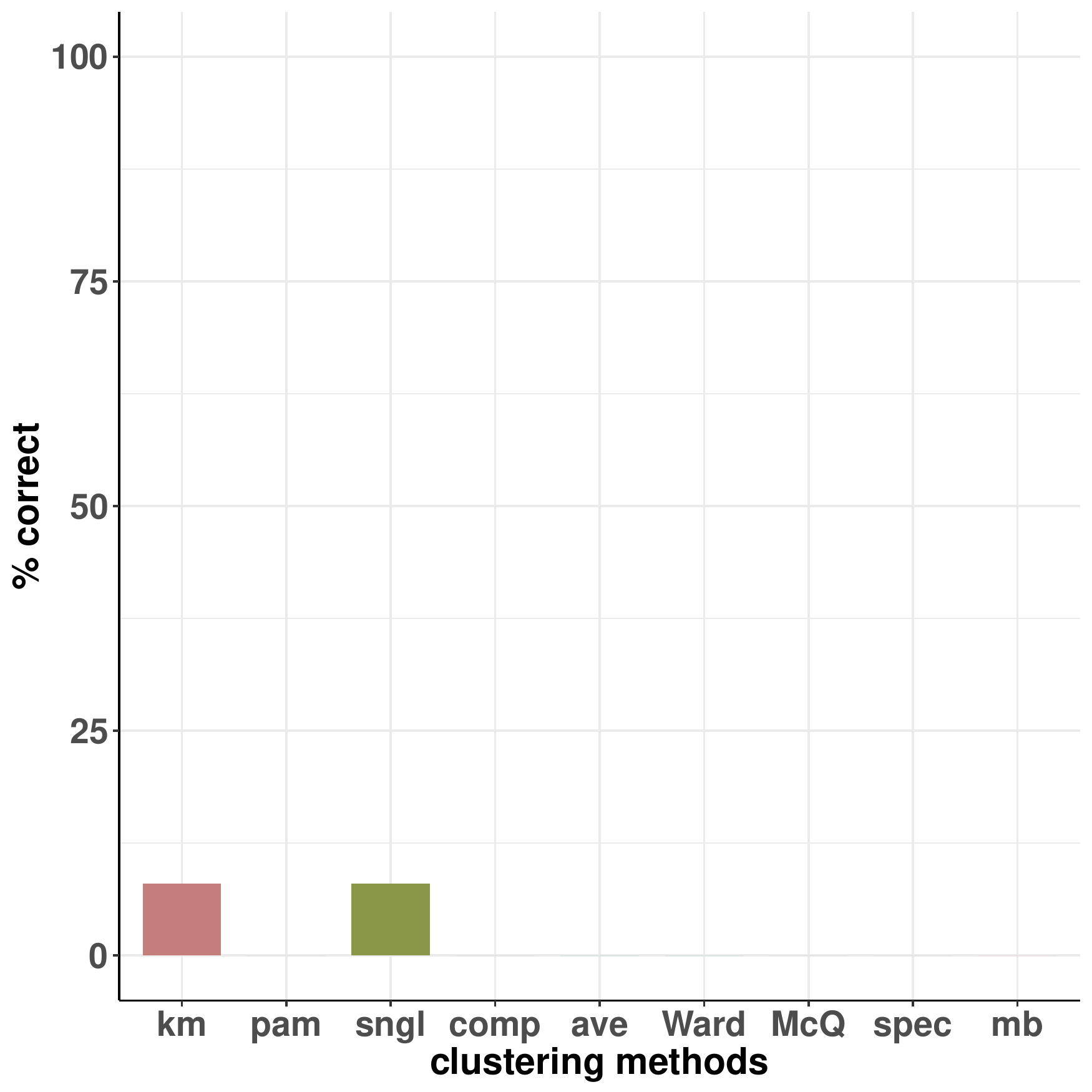}}
  \subfloat[C]{
  \includegraphics[width=35mm]{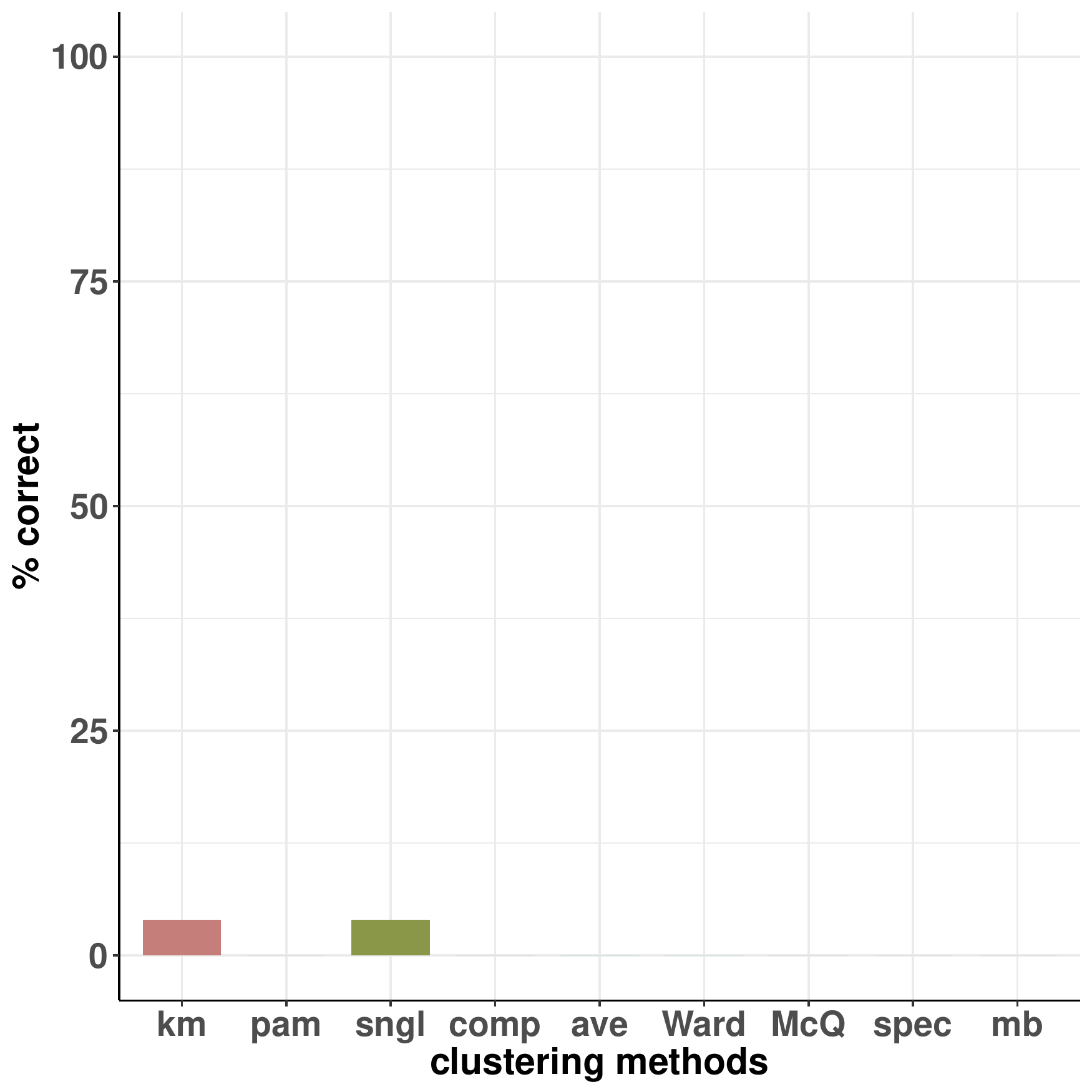}
}
\subfloat[KL]{
  \includegraphics[width=35mm]{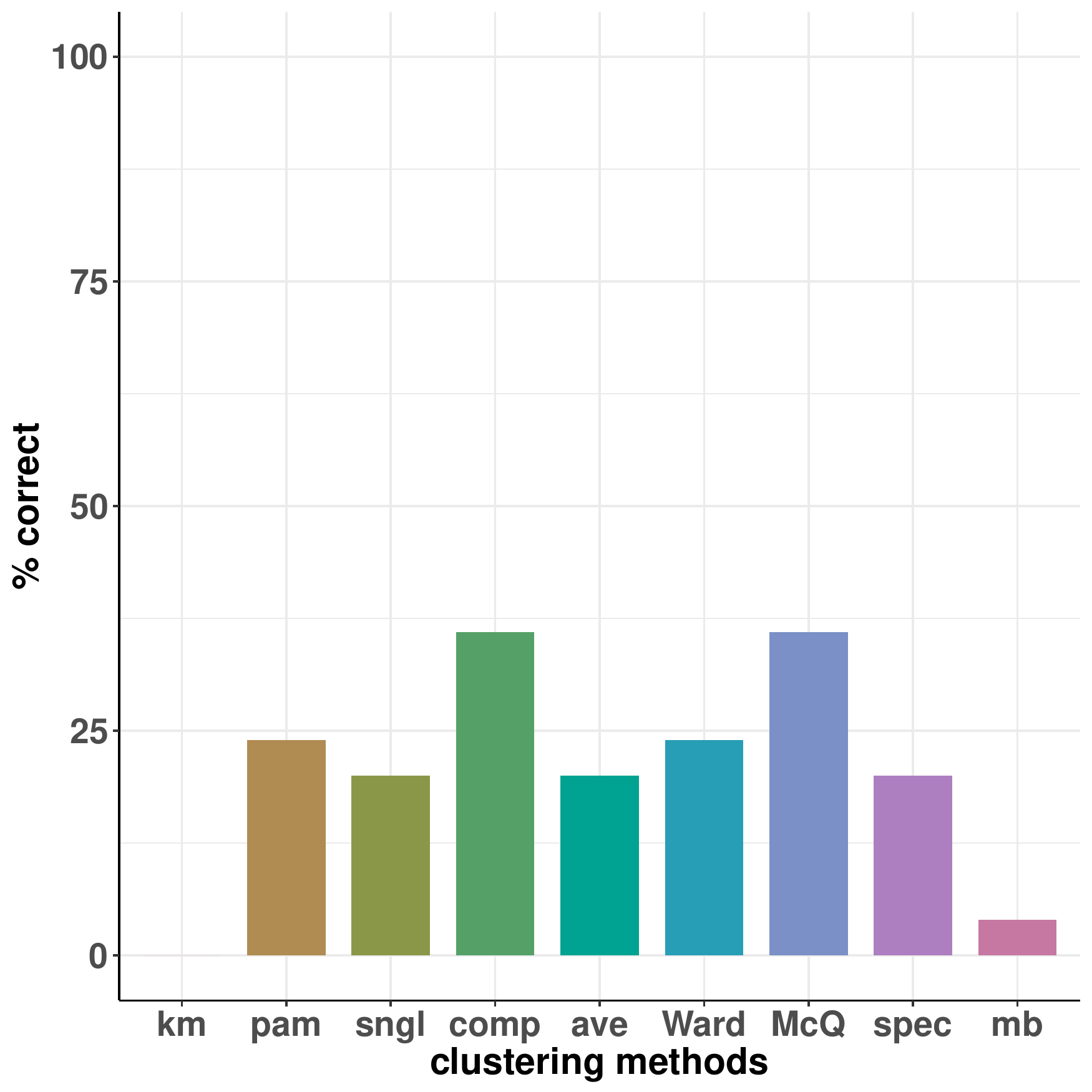}
}
\newline
\subfloat[CH]{
  \includegraphics[width=35mm]{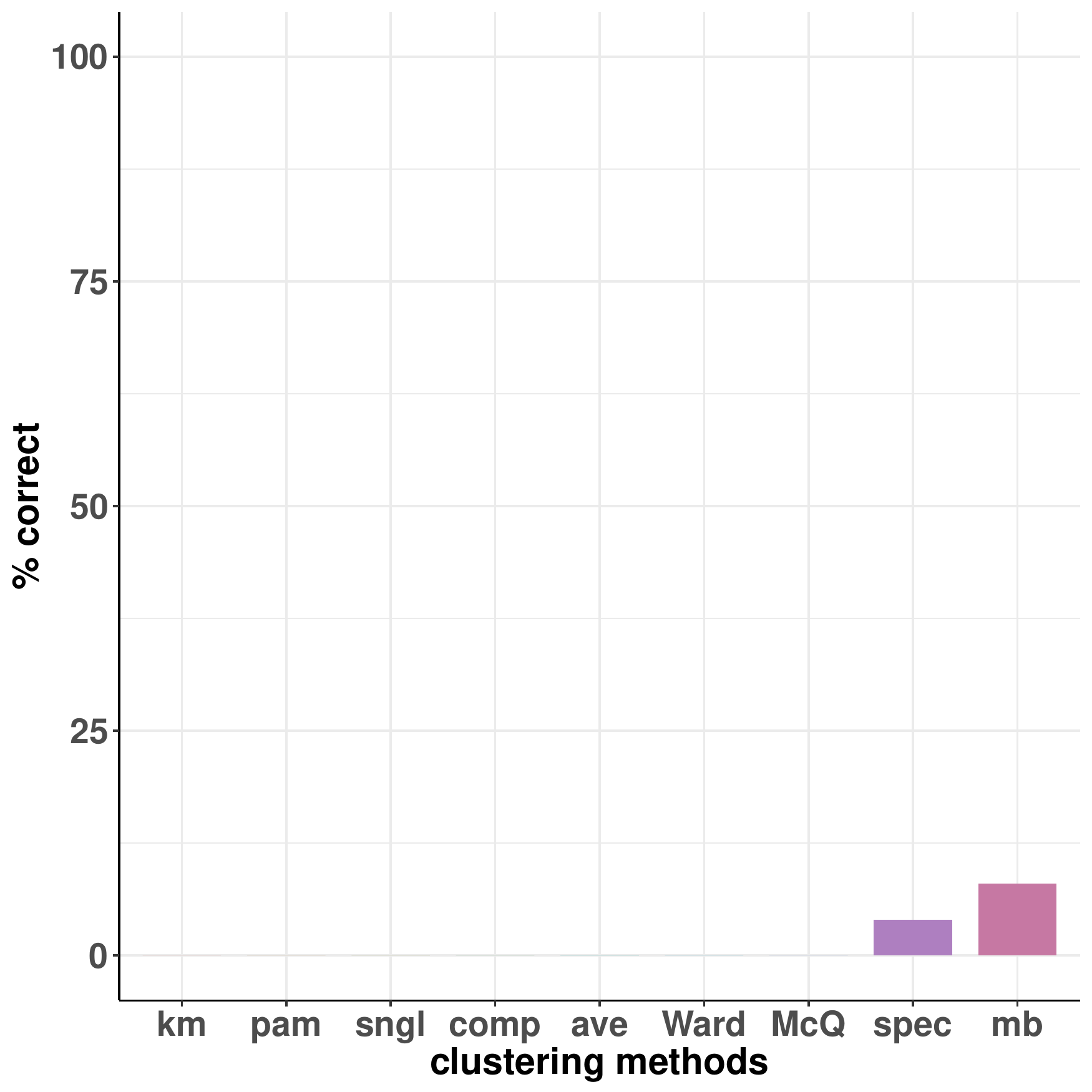}
}
\subfloat[gap]{
  \includegraphics[width=35mm]{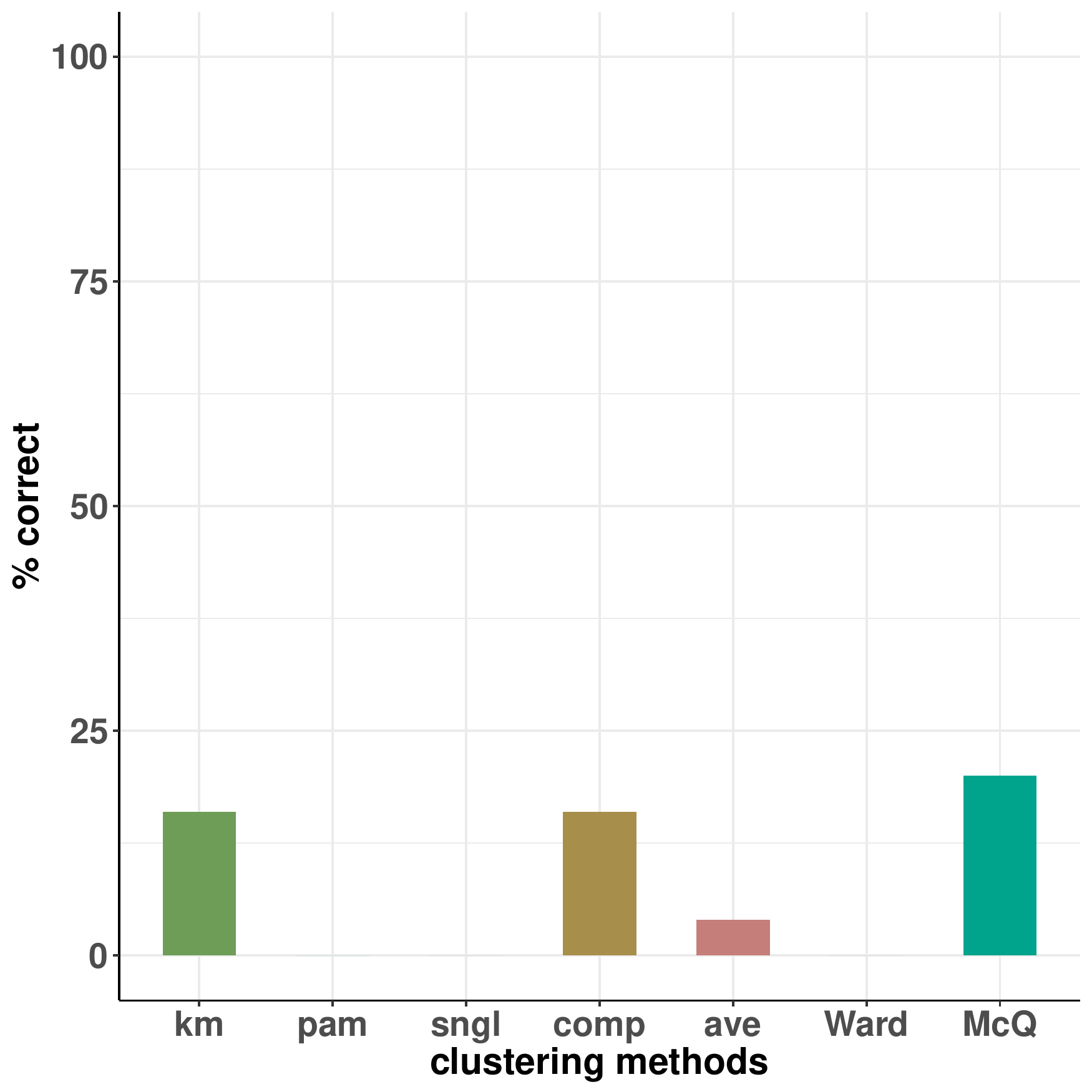}
}
\subfloat[jump]{
  \includegraphics[width=35mm]{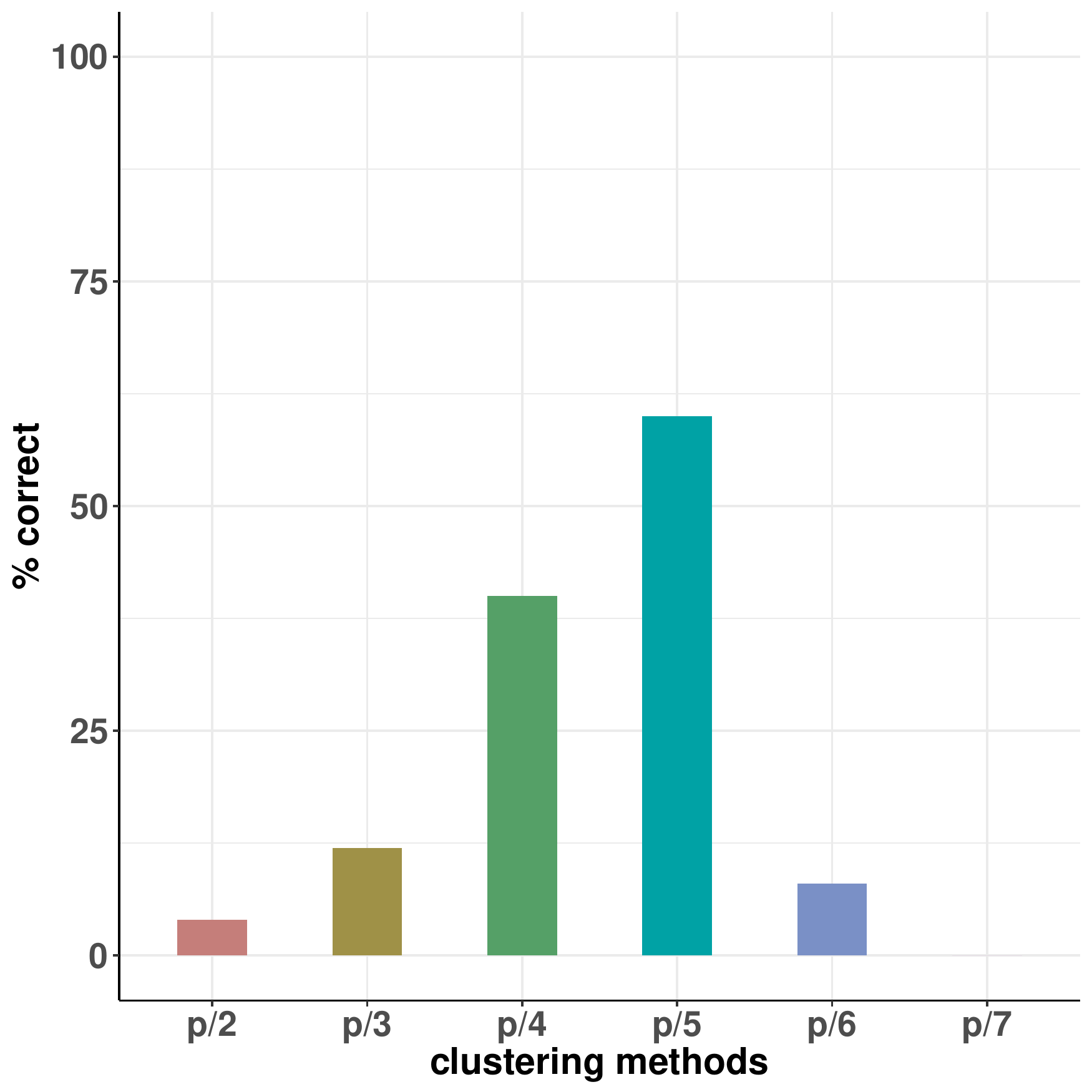}
}
\subfloat[PS]{
  \includegraphics[width=35mm]{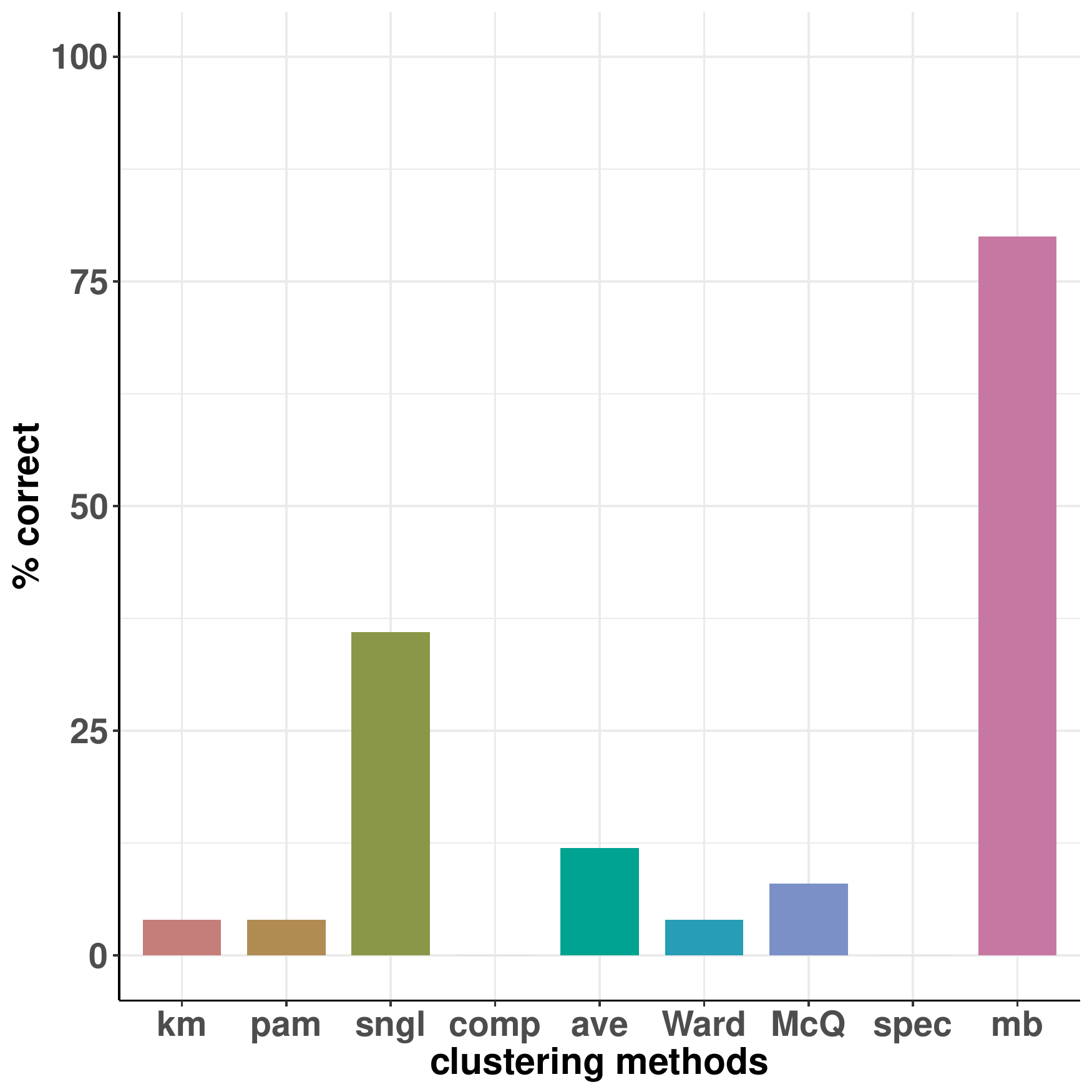}
}
\newline
\subfloat[BI]{
  \includegraphics[width=35mm]{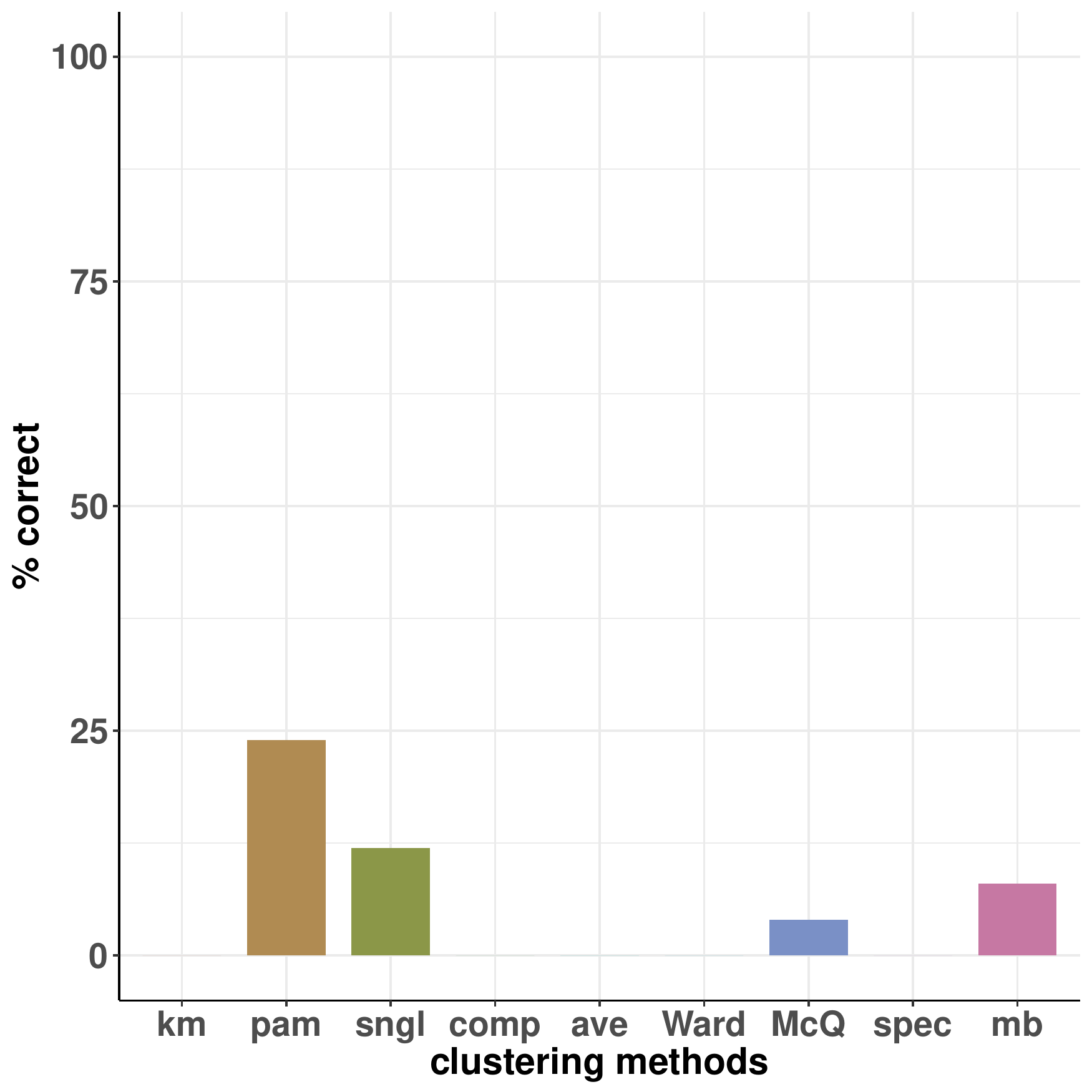}
}
\subfloat[CVNN]{
  \includegraphics[width=35mm]{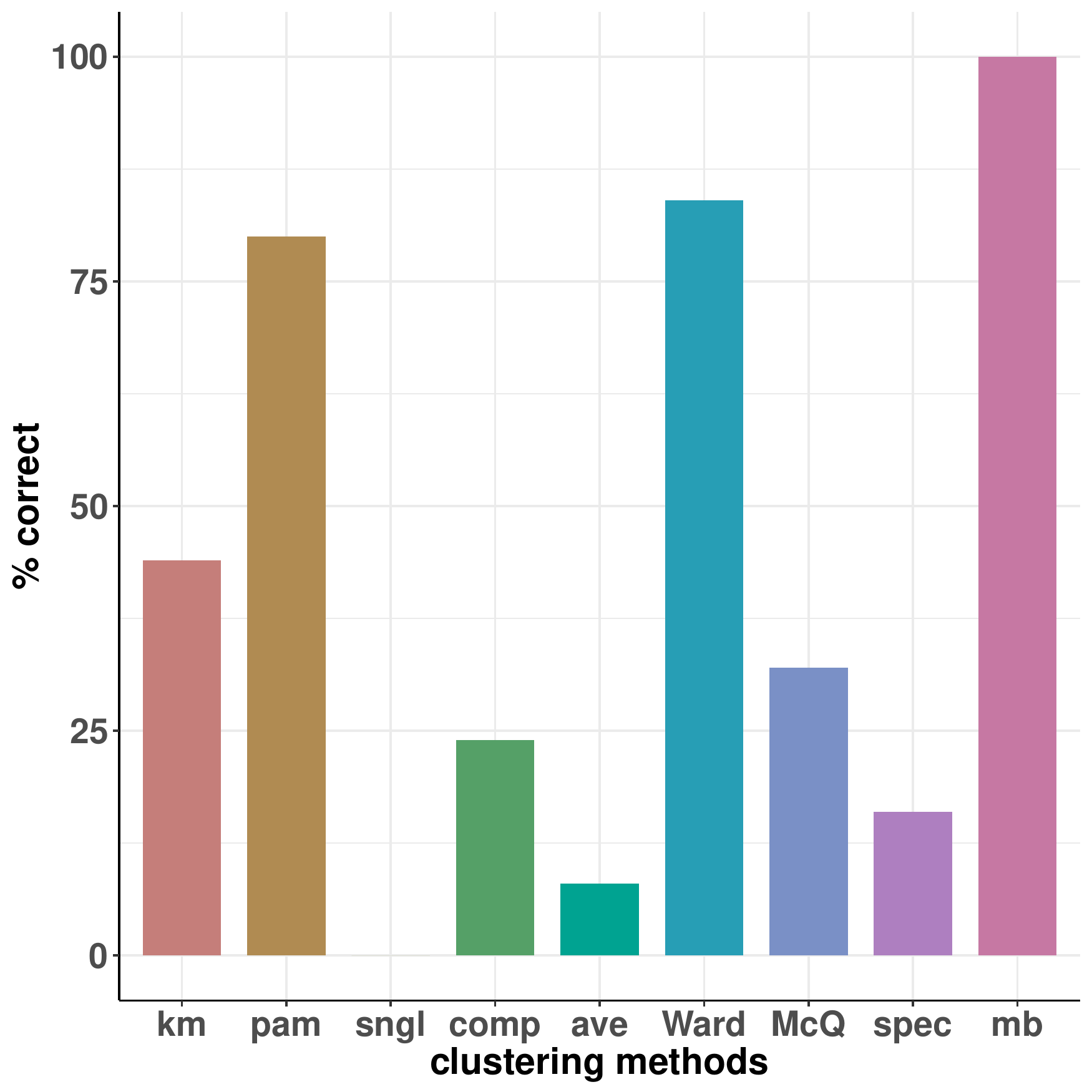}
}
\subfloat[BIC/PAMSIL]{
  \includegraphics[width=35mm]{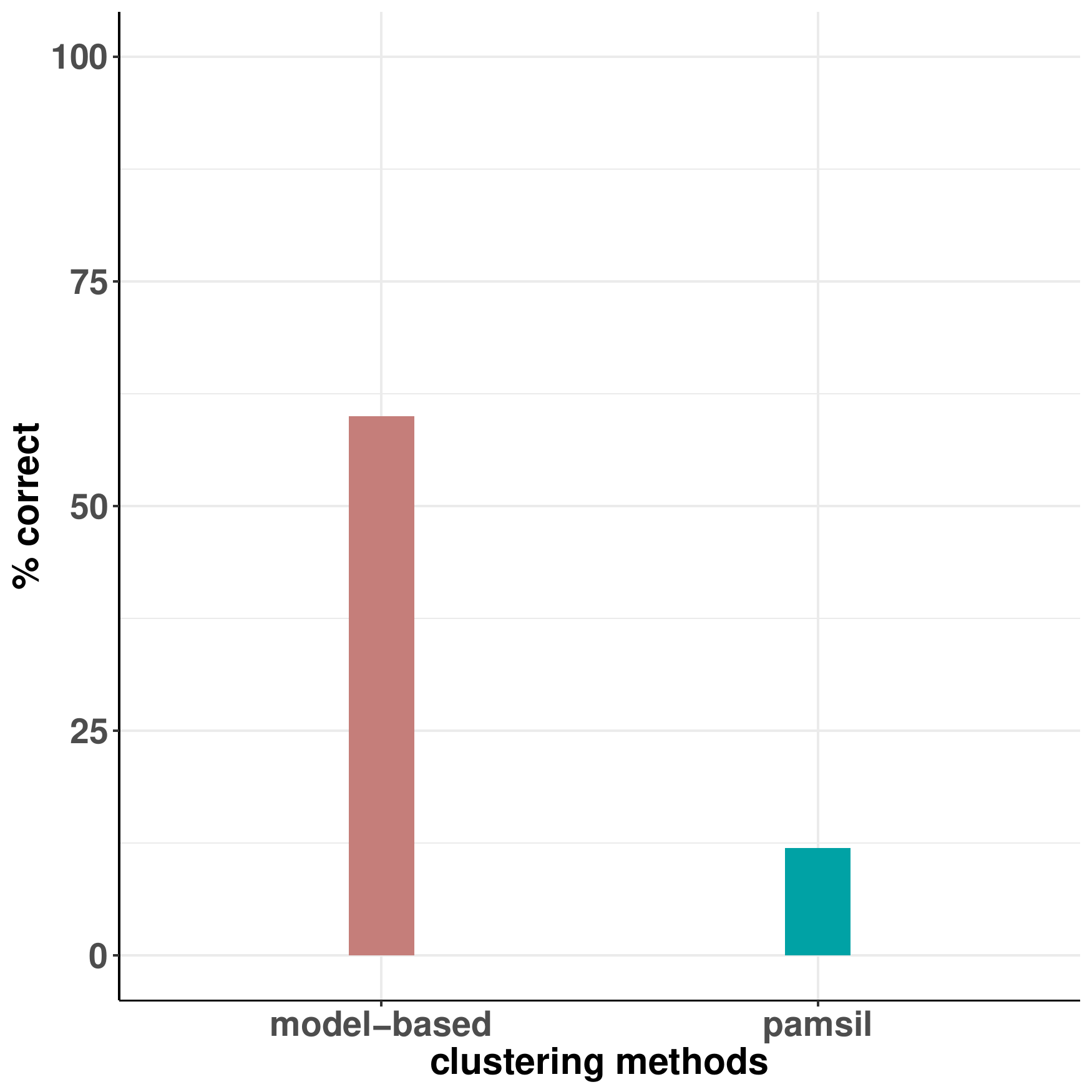}
}
\subfloat[ASW]{
  \includegraphics[width=35mm]{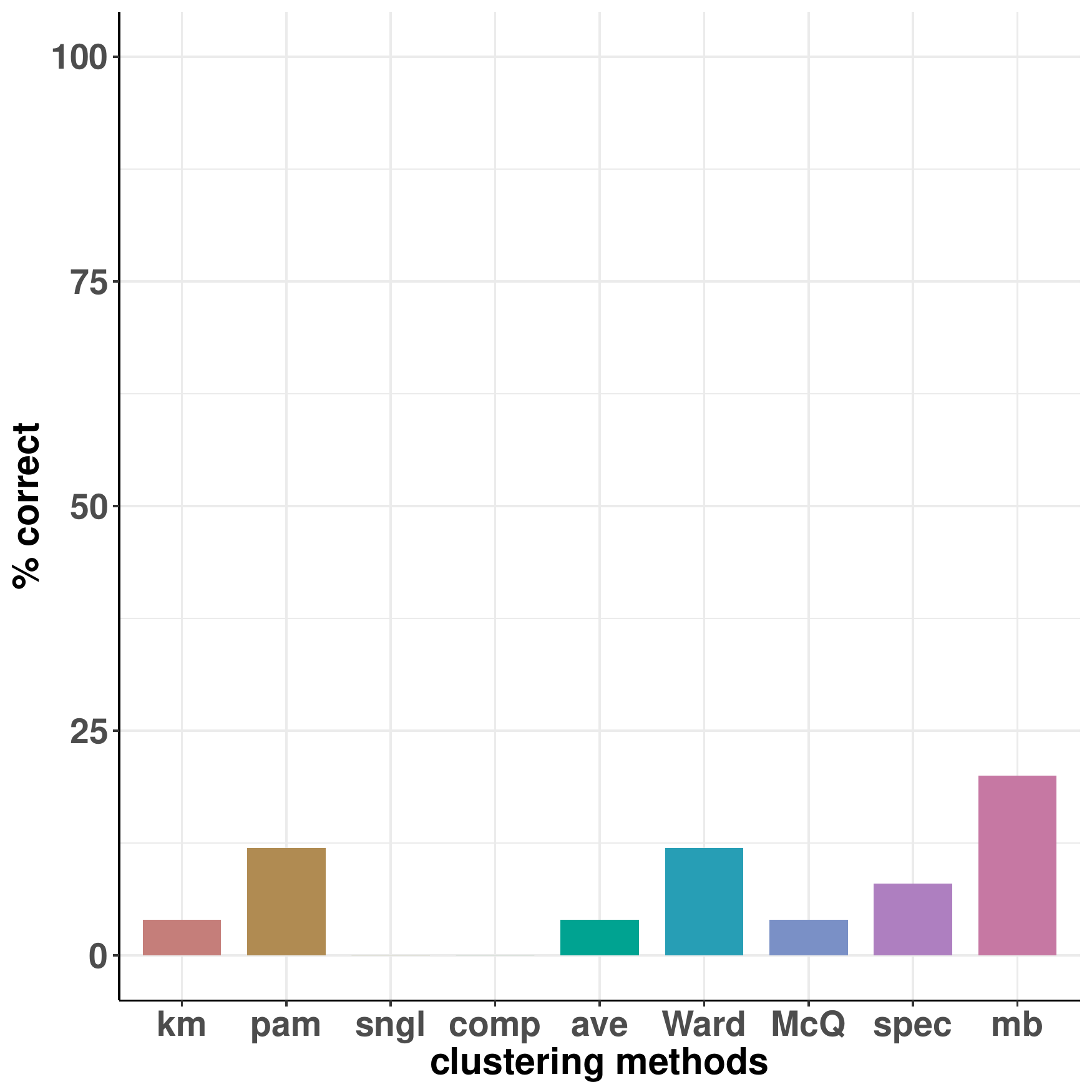}
}
\newline
\rule{-60ex}{.2in}
\subfloat[OASW]{
  \includegraphics[width=35mm]{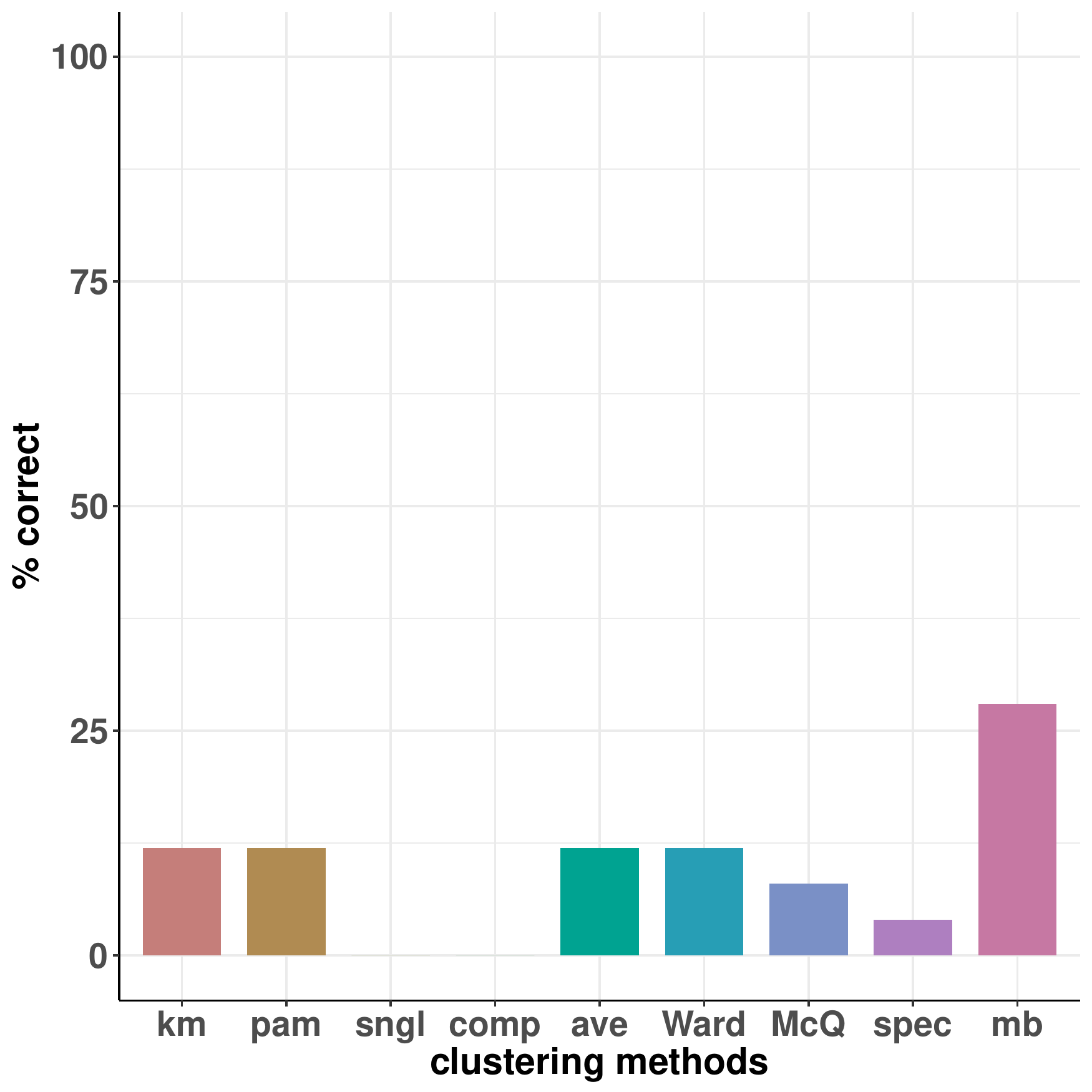}
}
\caption{Bar plots for the estimation of k for Model 3.}
\label{appendix:estkmodelthree}
\end{figure}

\begin{figure}[!hbtp]
\centering
\subfloat[CH]{
  \includegraphics[width=35mm]{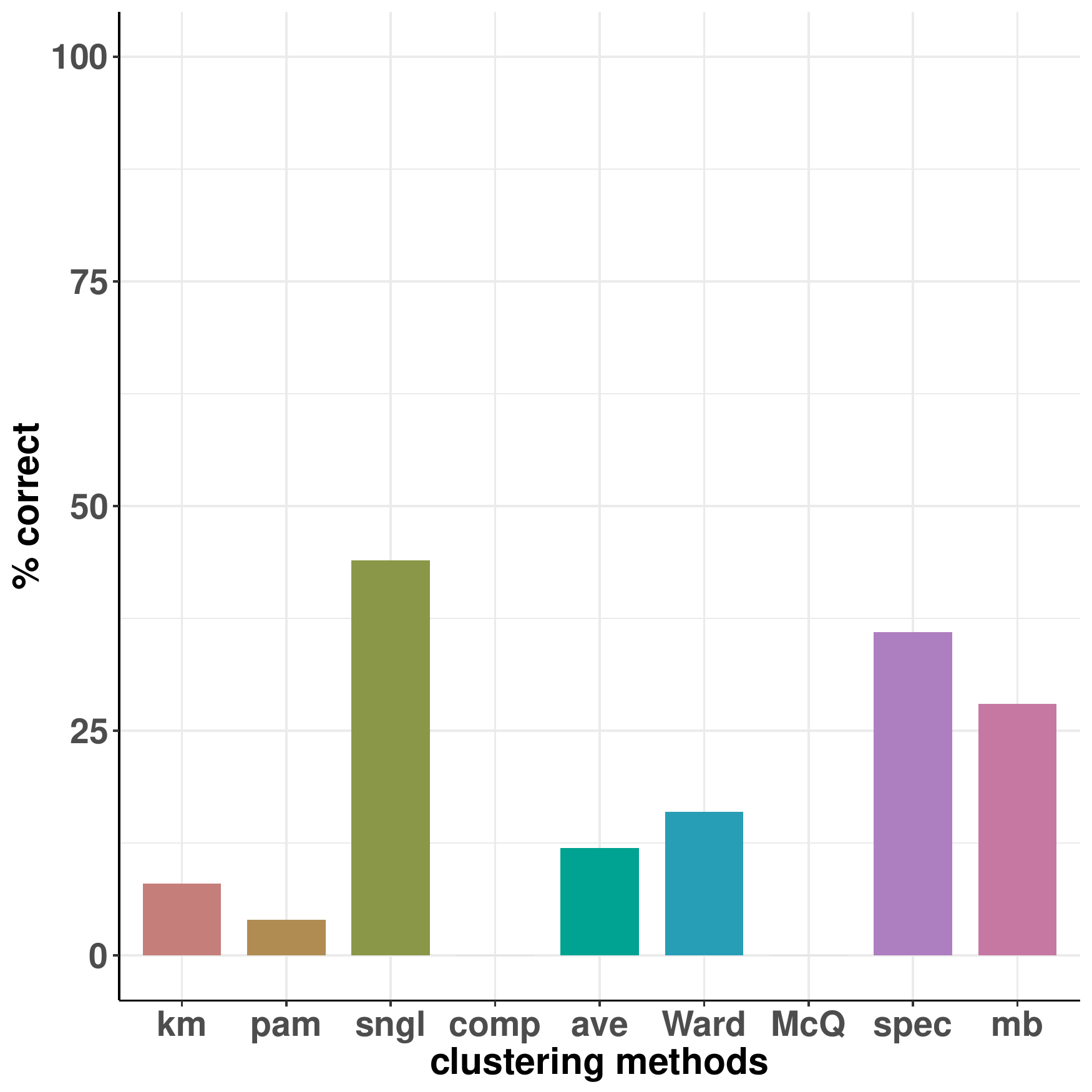}
}
\subfloat[H]{
  \includegraphics[width=35mm]{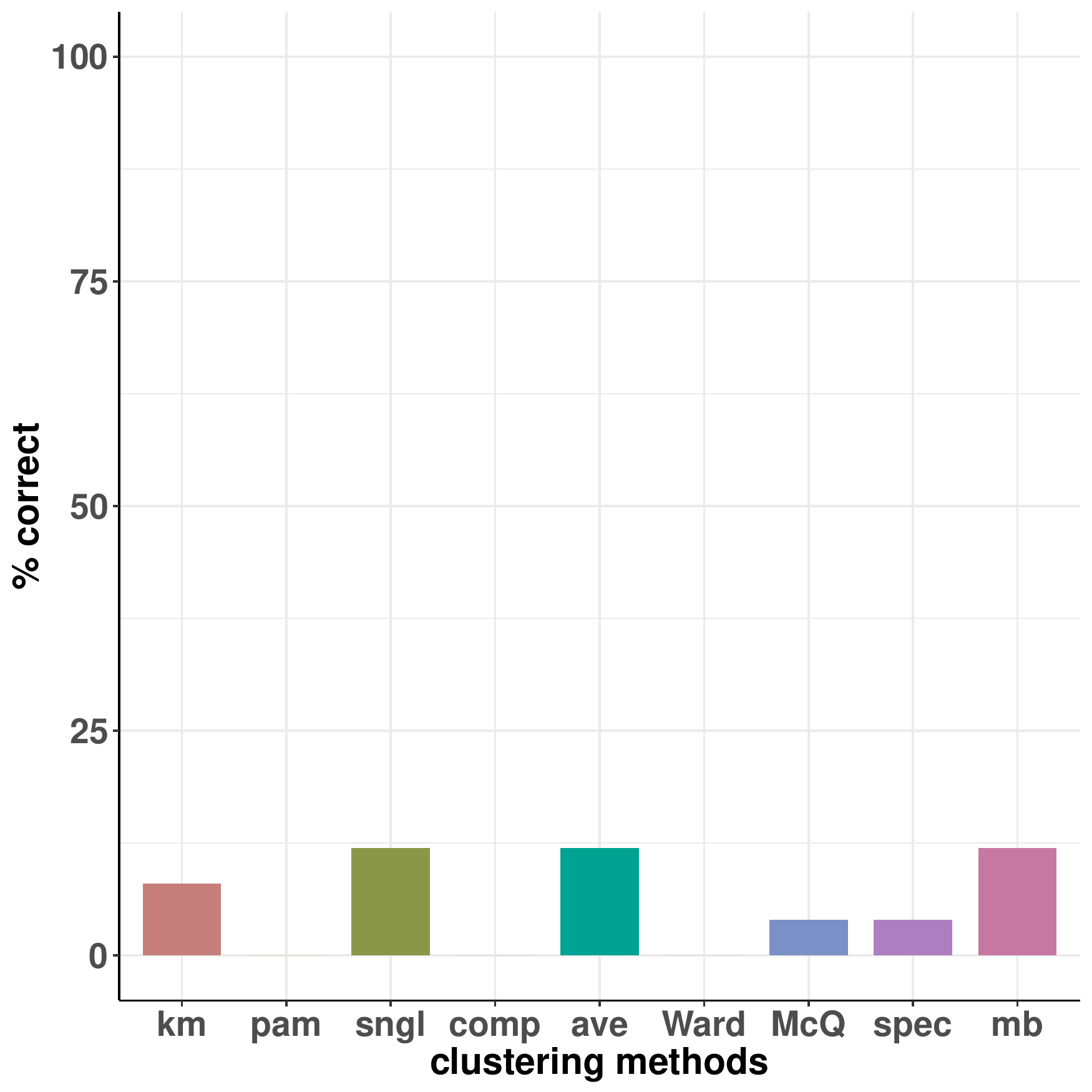}
}
  \subfloat[Gamma]{
  \includegraphics[width=35mm]{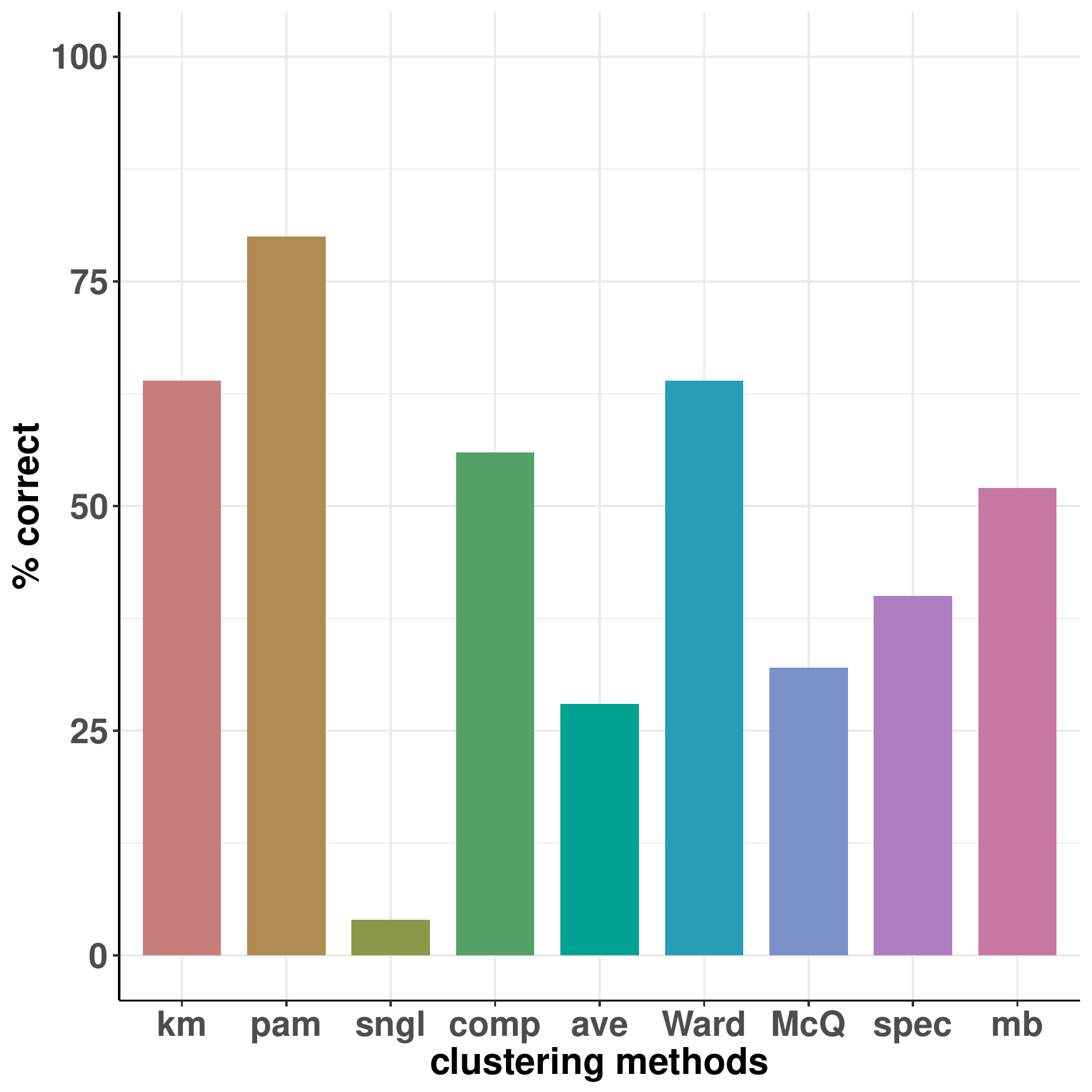}}
  \subfloat[C]{
  \includegraphics[width=35mm]{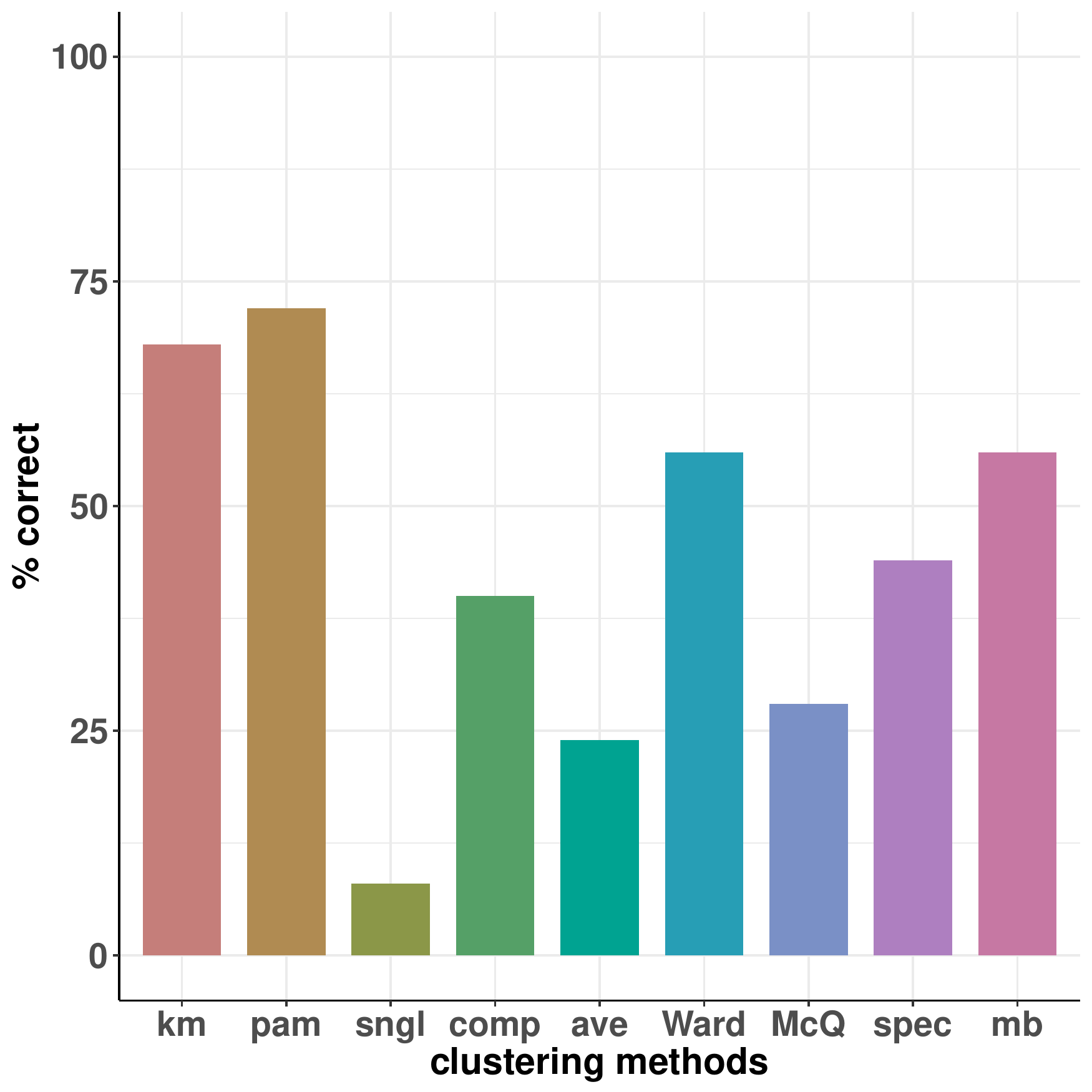}
}
\newline
\subfloat[KL]{
  \includegraphics[width=35mm]{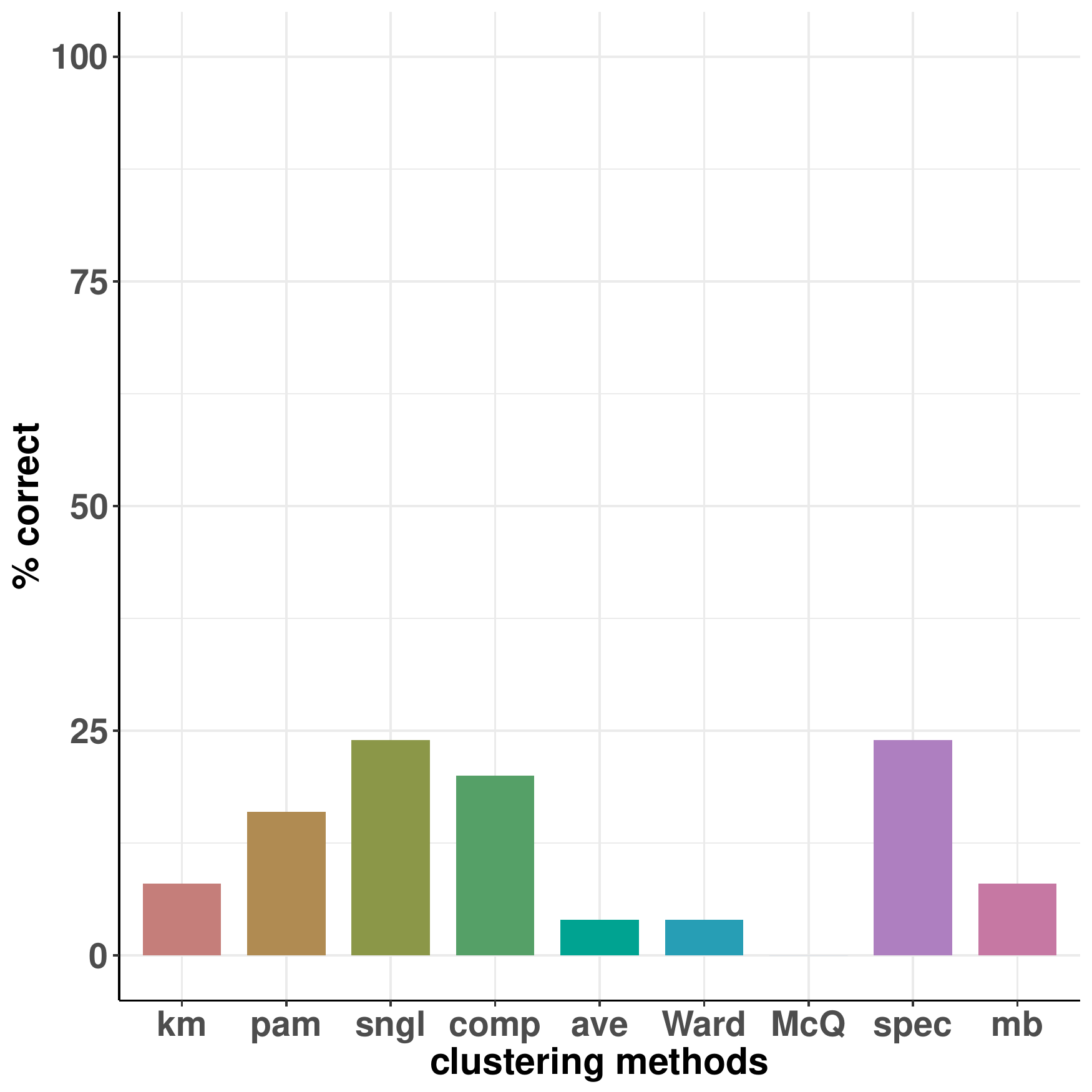}
}
\subfloat[gap]{
  \includegraphics[width=35mm]{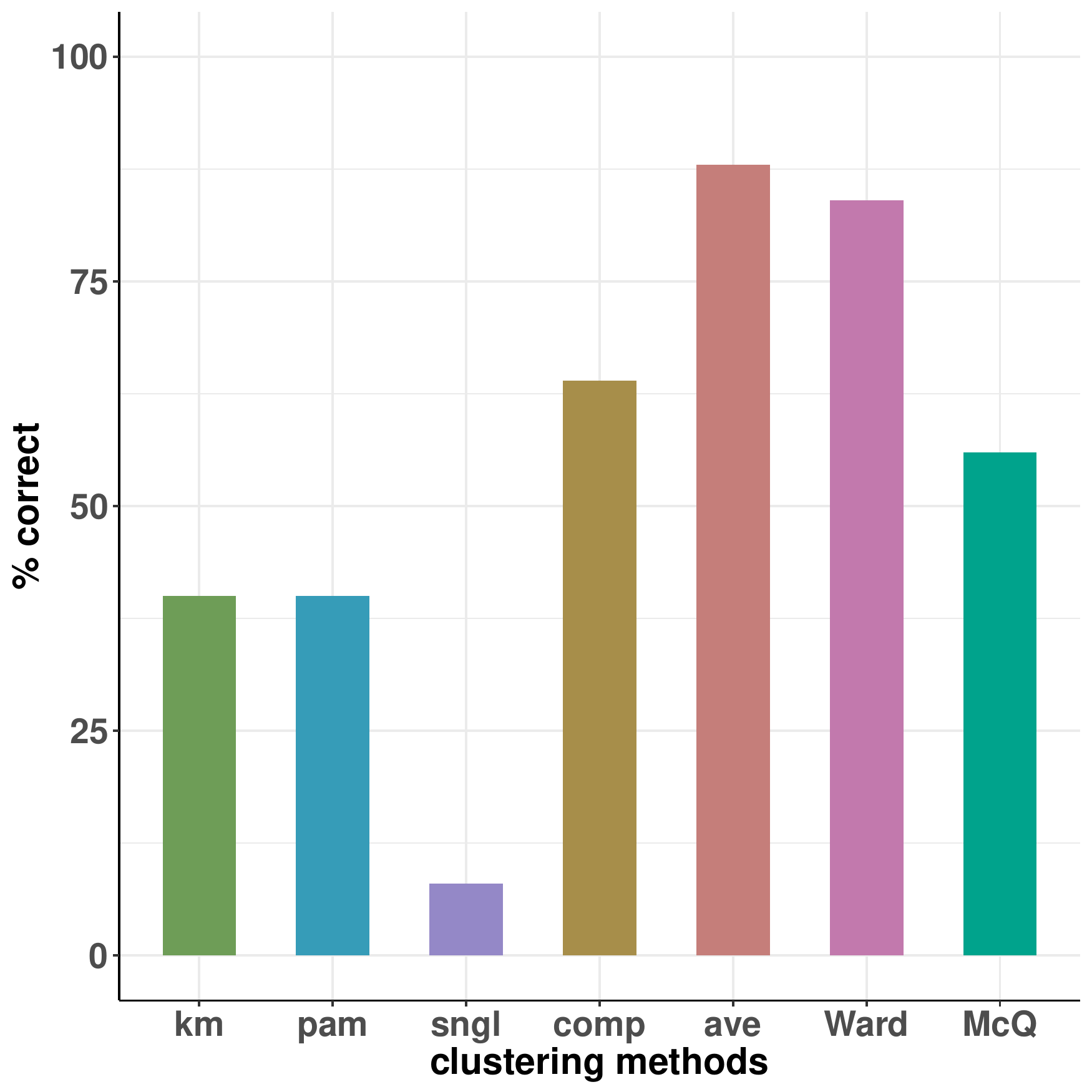}
}
\subfloat[jump]{
  \includegraphics[width=35mm]{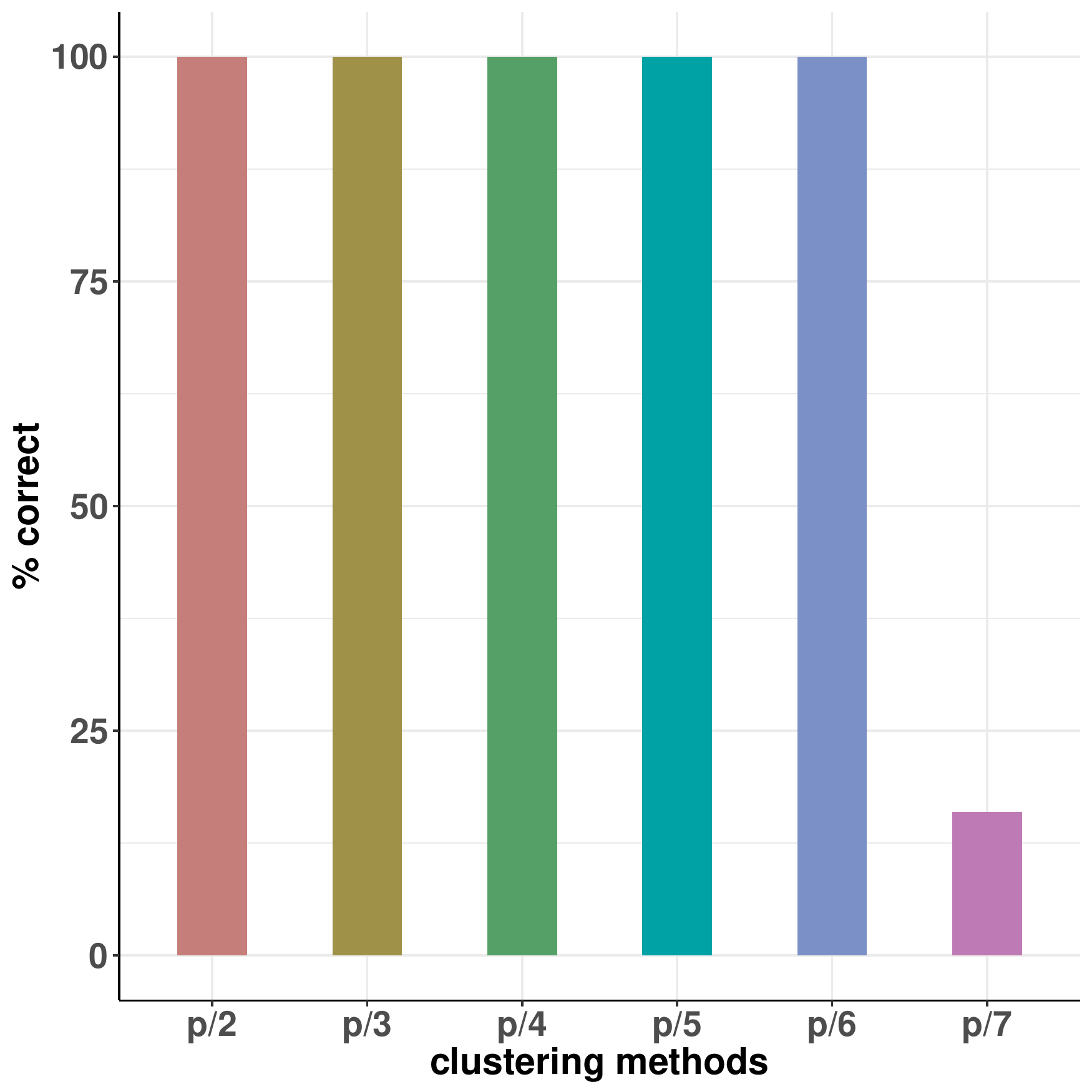}
}
\subfloat[PS]{
  \includegraphics[width=35mm]{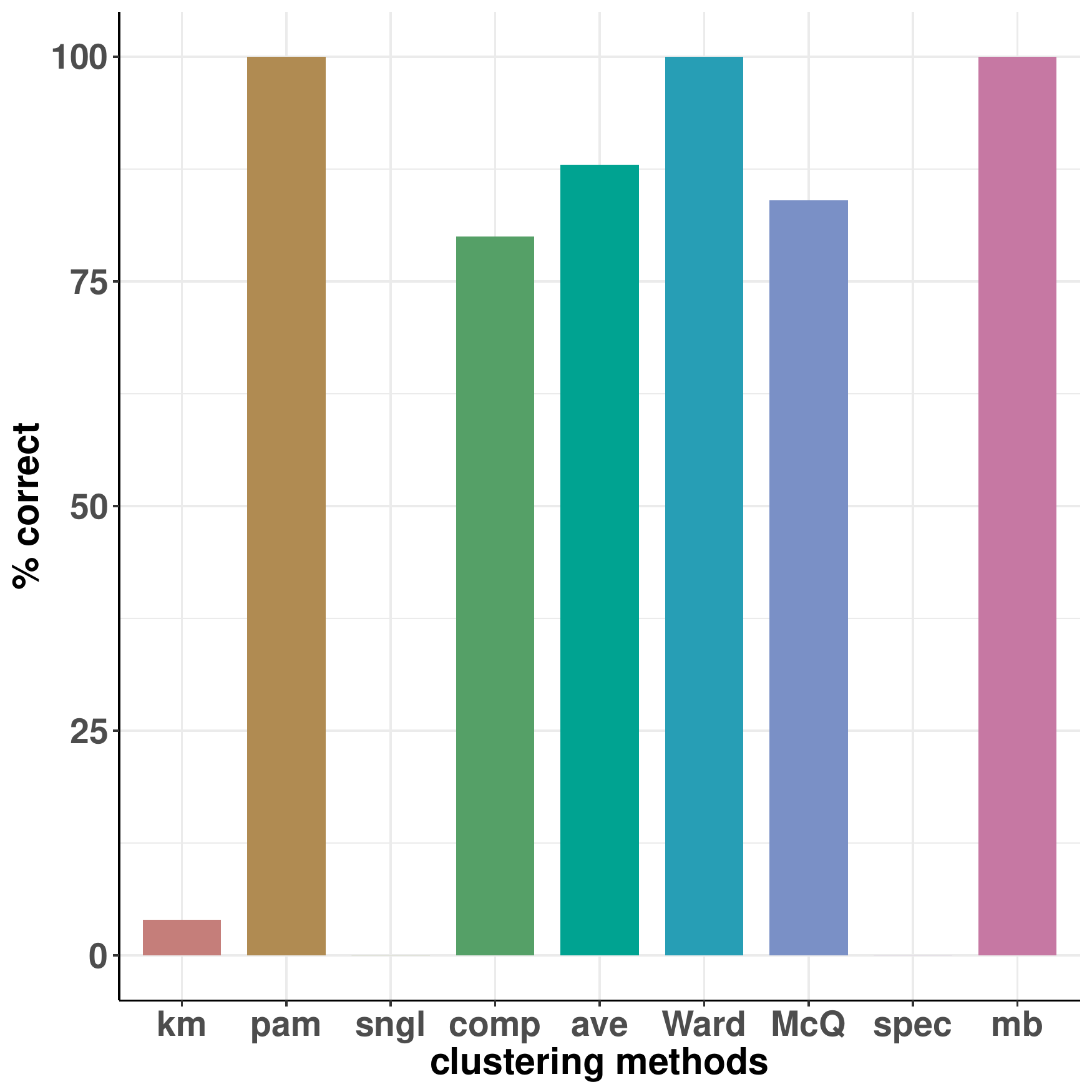}
}
\newline
\subfloat[BI]{
  \includegraphics[width=35mm]{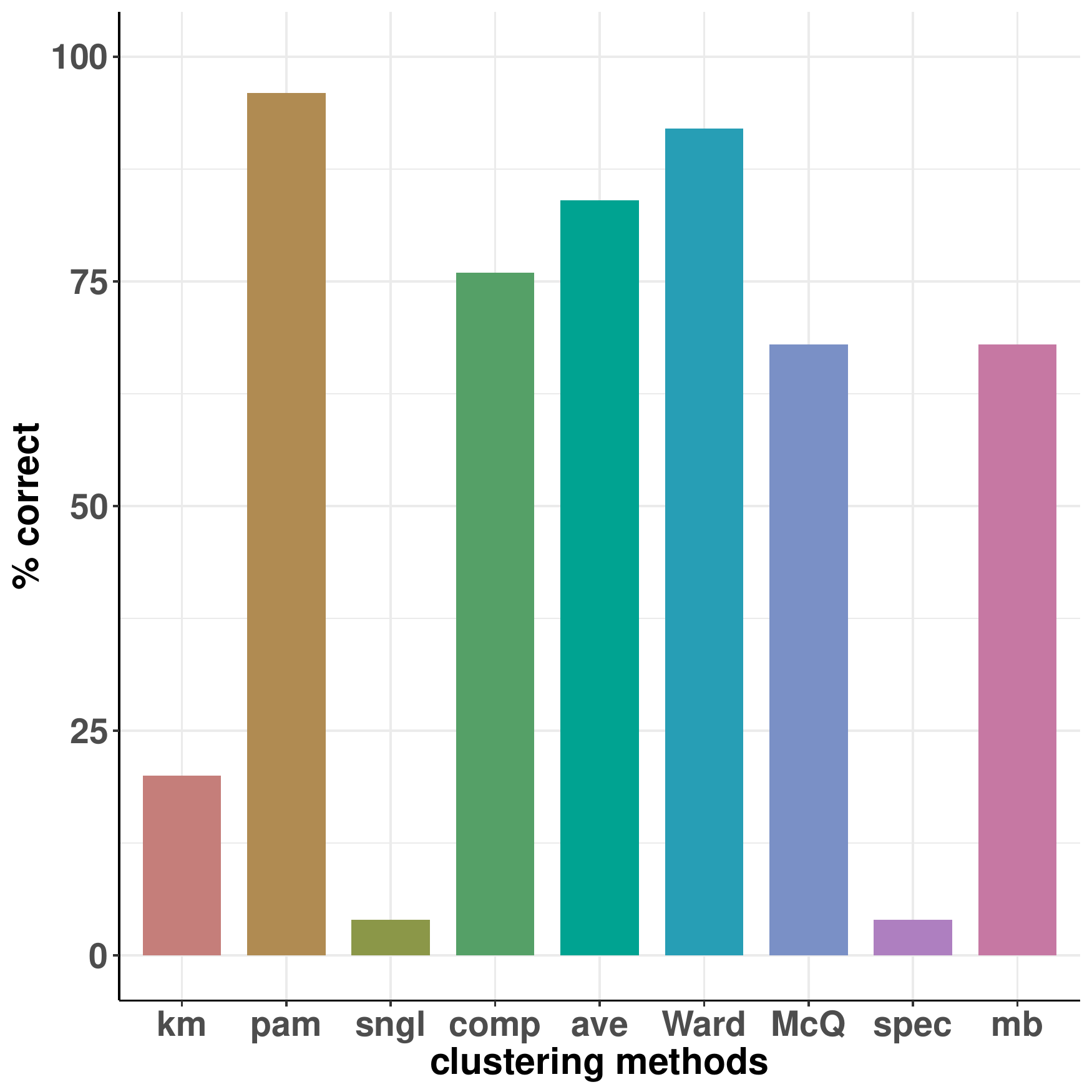}
}
\subfloat[CVNN]{
  \includegraphics[width=35mm]{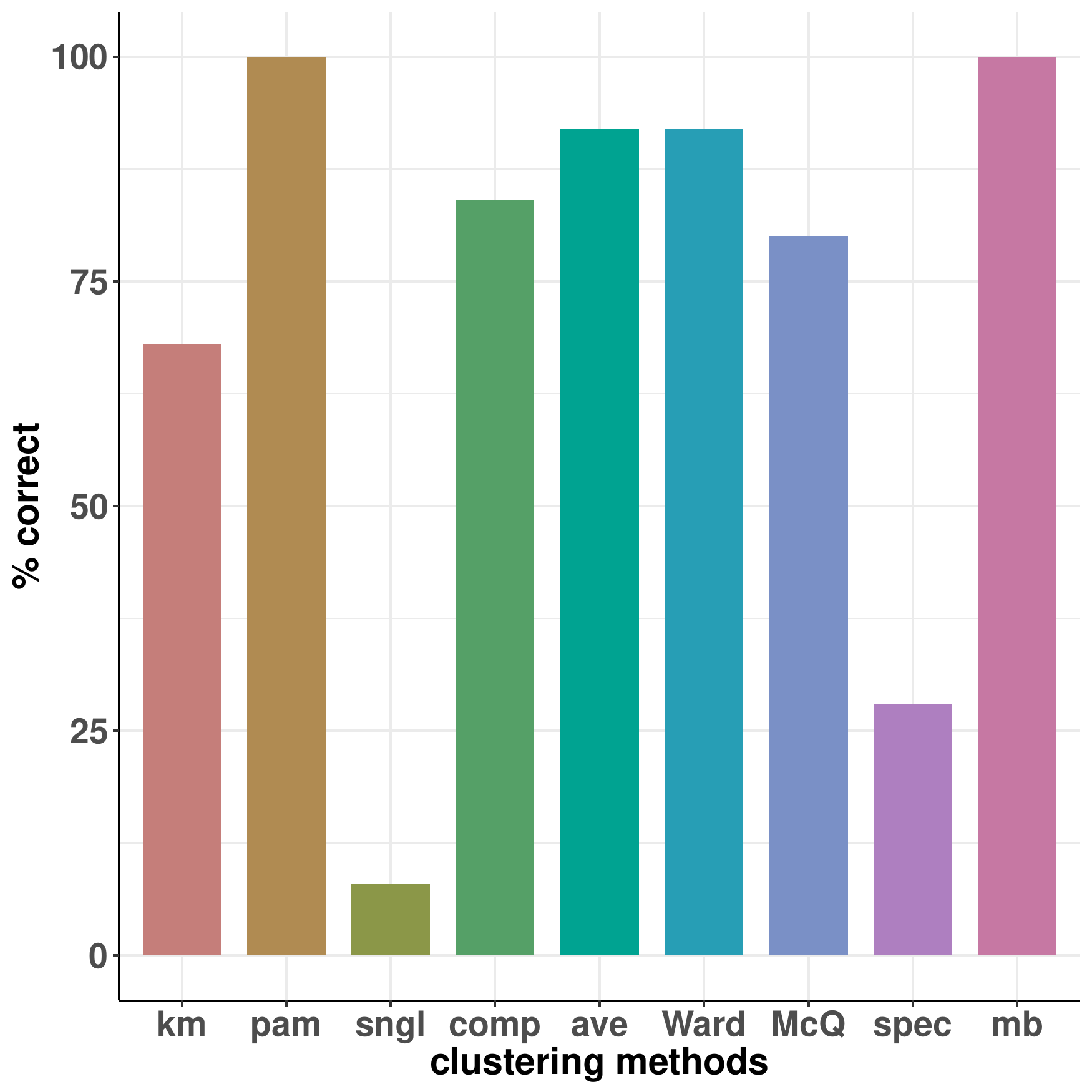}
}
\subfloat[BIC/PAMSIL]{
 \includegraphics[width=35mm]{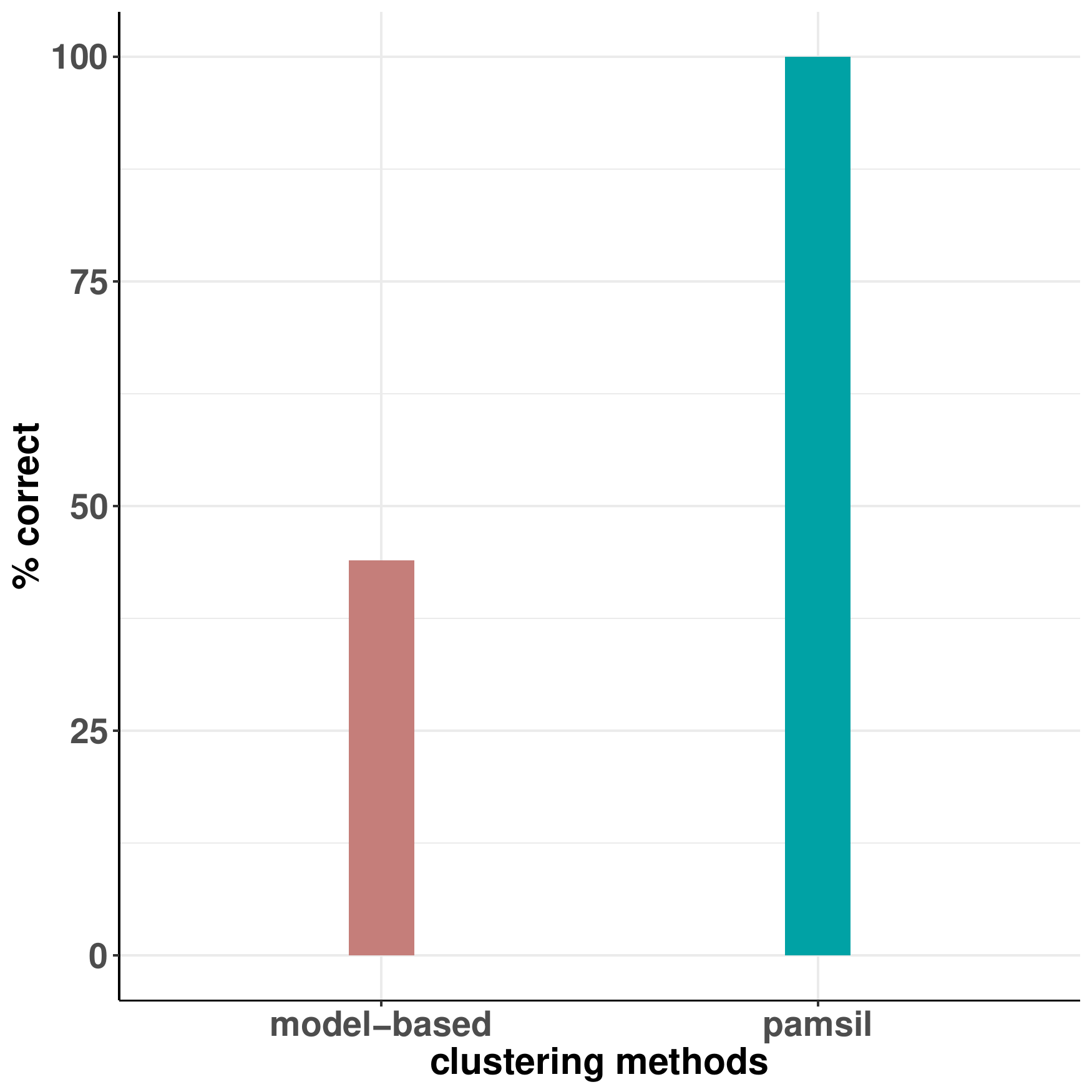}
}
\subfloat[ASW]{
  \includegraphics[width=35mm]{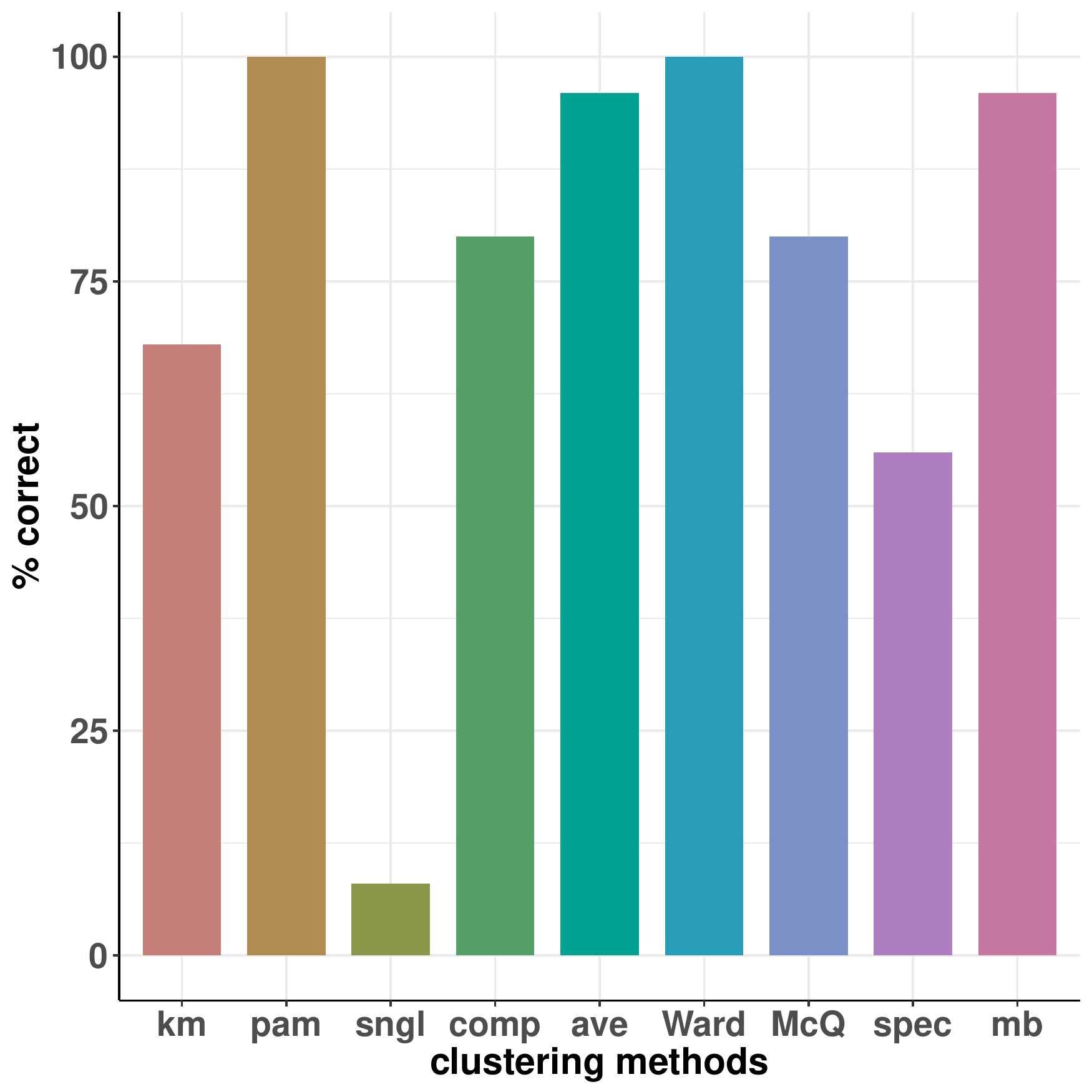}
}
\newline
\rule{-60ex}{.2in}
\subfloat[OASW]{
  \includegraphics[width=35mm]{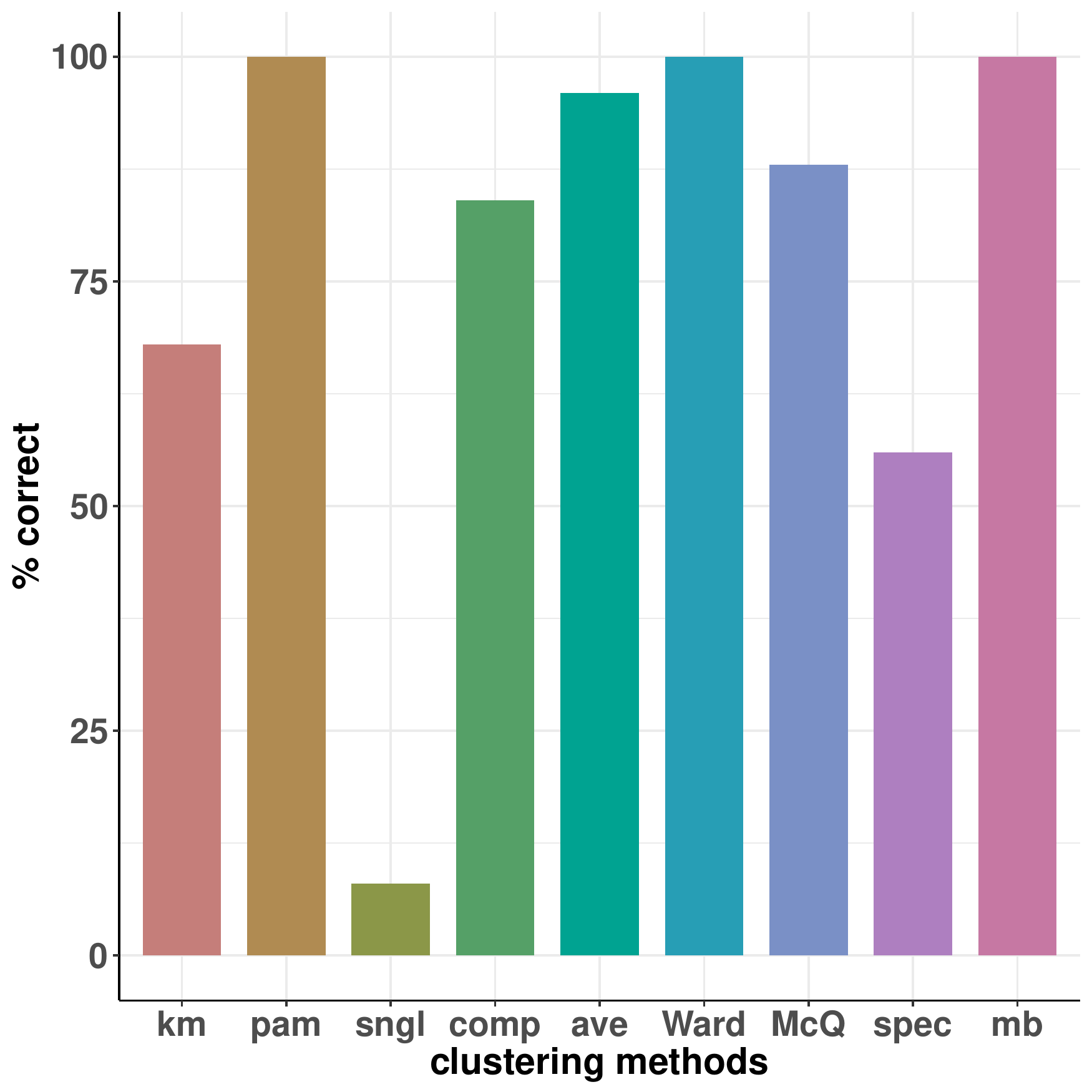}
}
\caption{Bar plots for the estimation of k for Model 4.}
\label{appendix:estkmodelfour}
\end{figure}
\newpage
\begin{figure}[hbt!]
\centering
\subfloat[CH]{
  \includegraphics[width=35mm]{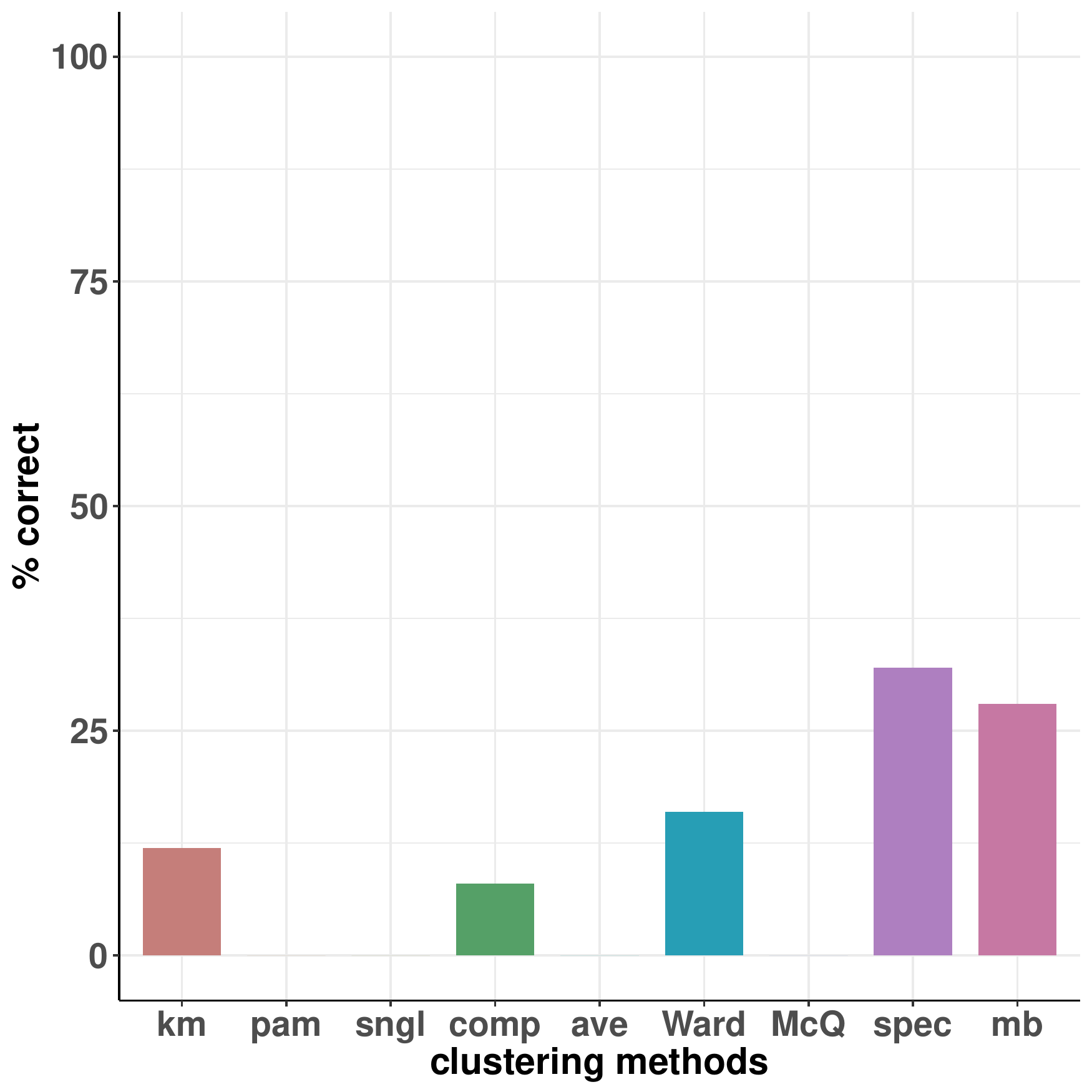}
}
\subfloat[H]{
  \includegraphics[width=35mm]{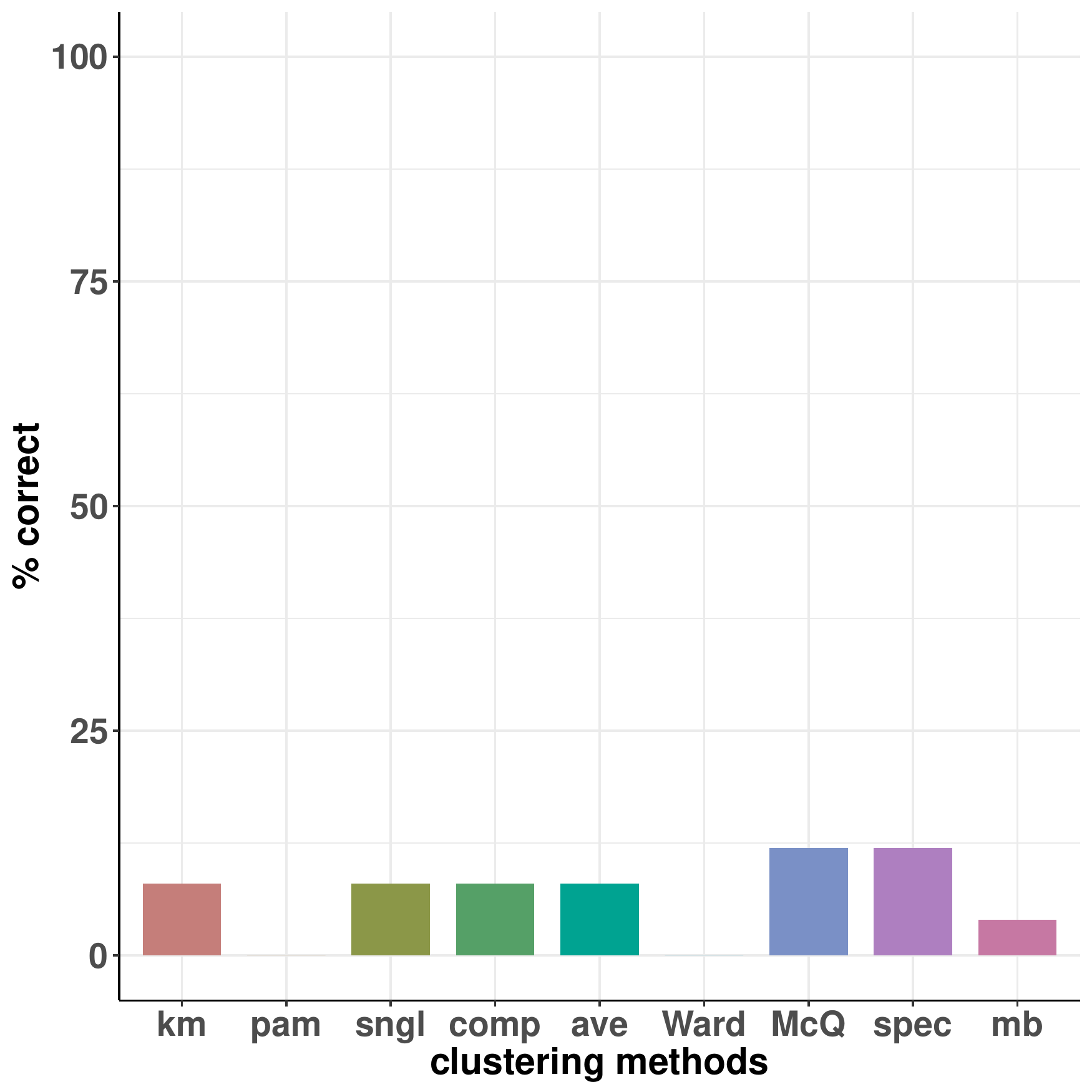}
}
  \subfloat[Gamma]{
  \includegraphics[width=35mm]{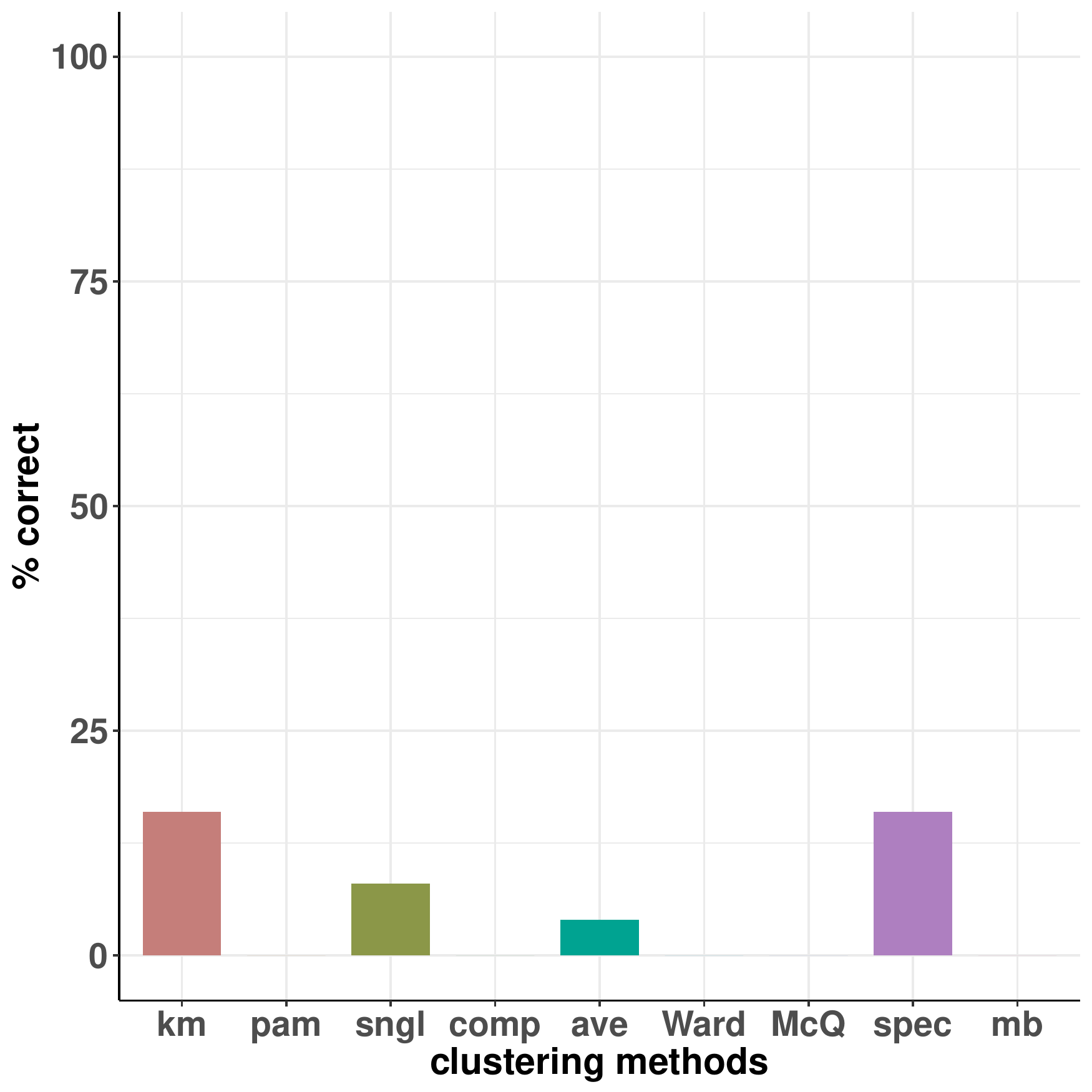}}
  \subfloat[C]{
  \includegraphics[width=35mm]{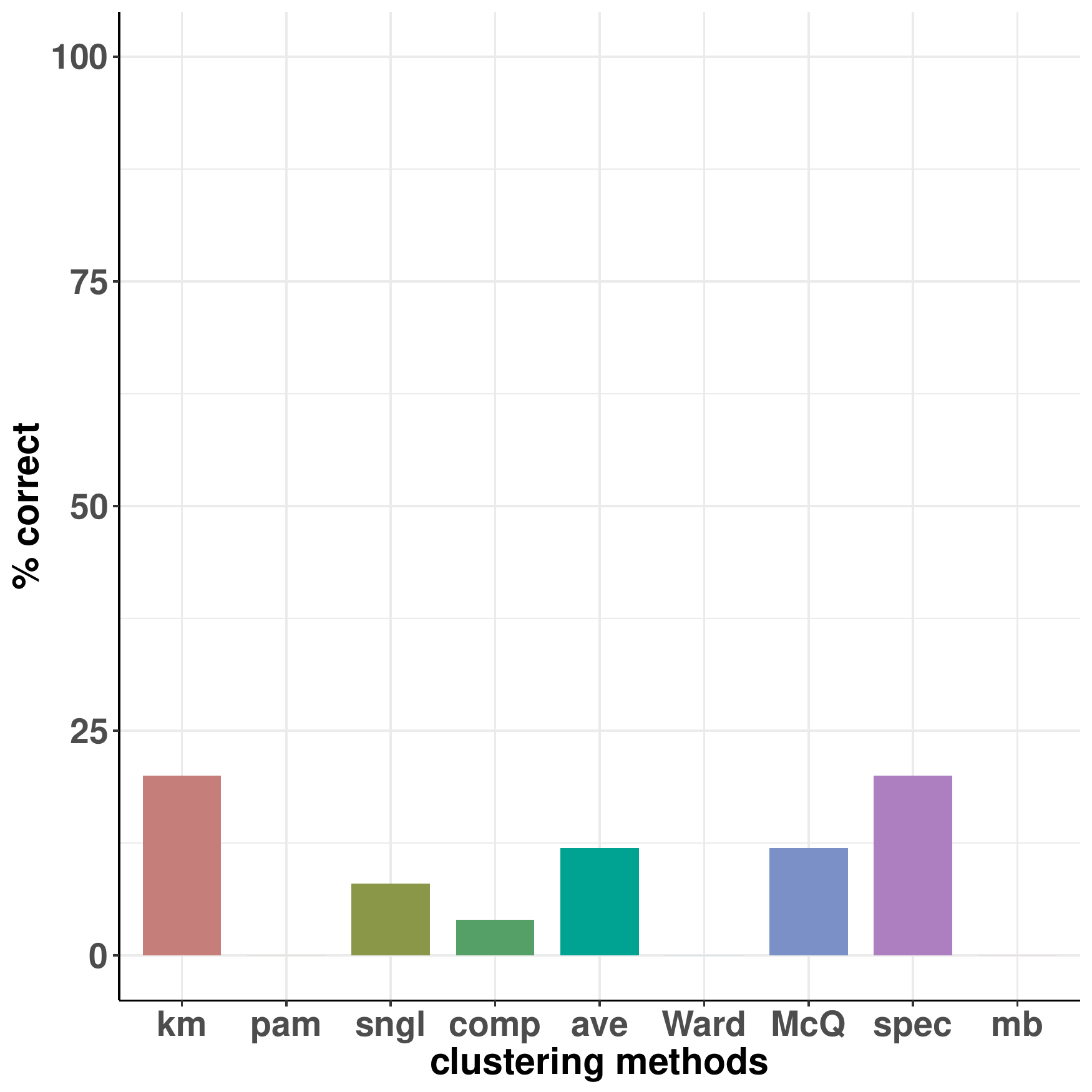}
}
\newline
\subfloat[KL]{
  \includegraphics[width=35mm]{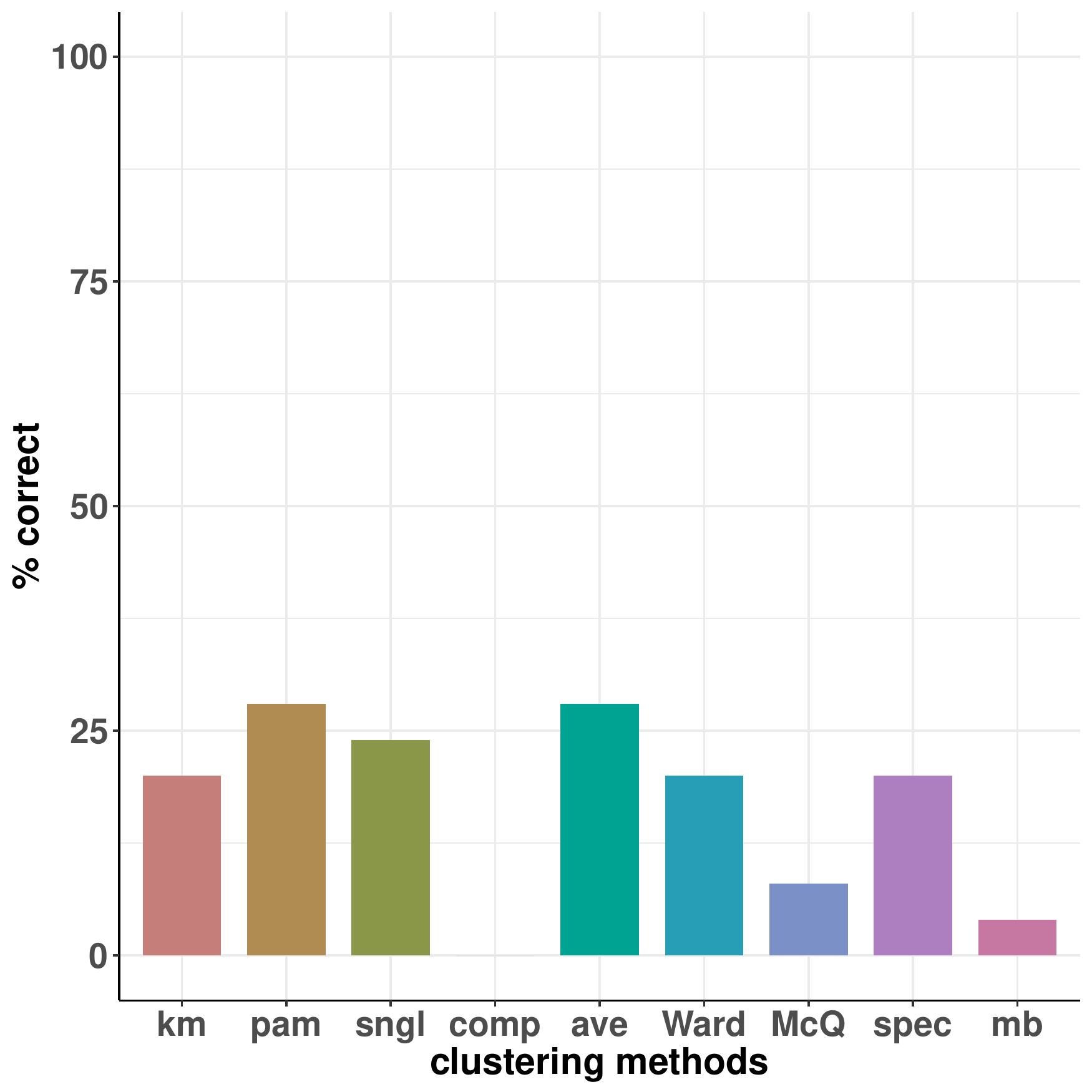}
}
\subfloat[gap]{
  \includegraphics[width=35mm]{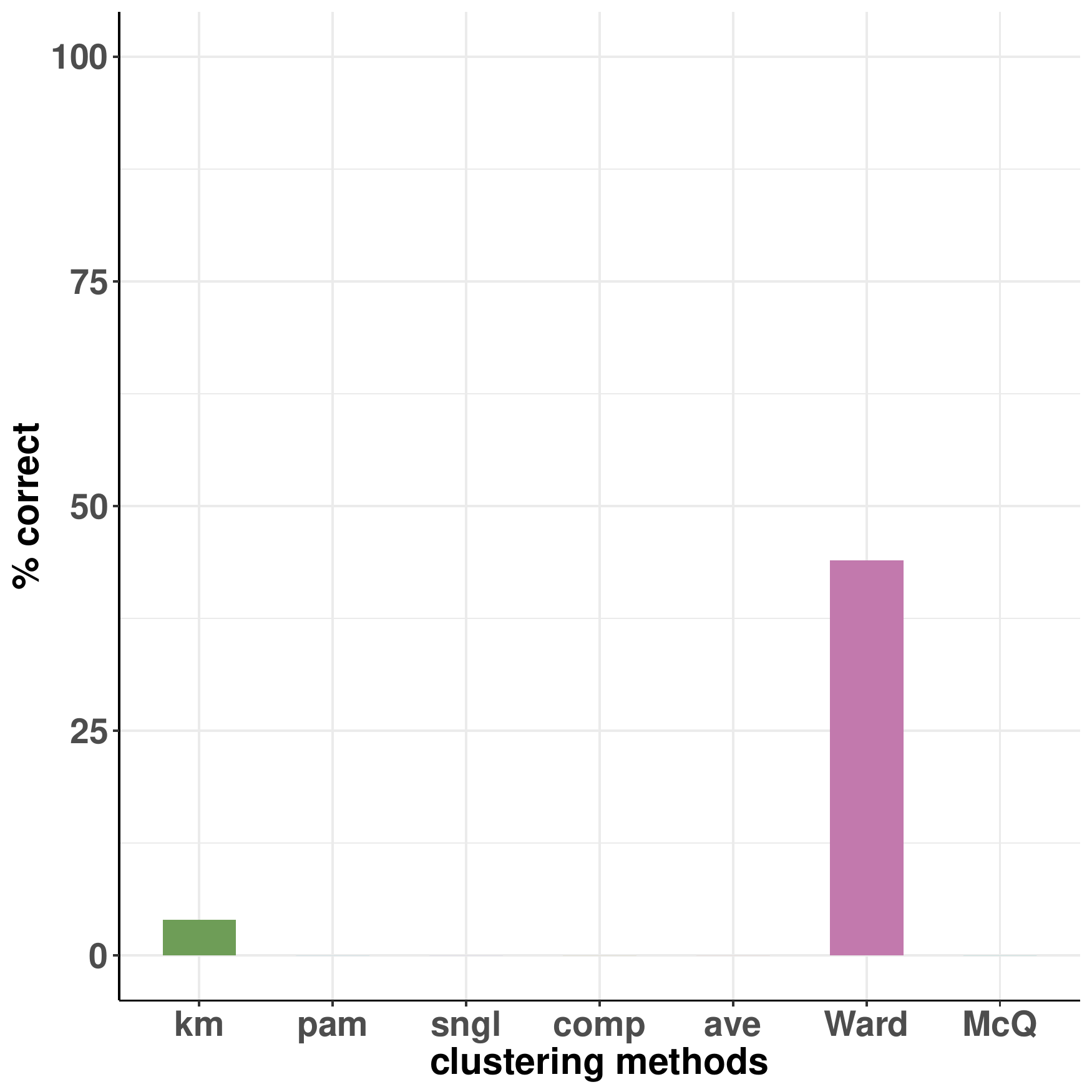}
}
\subfloat[PS]{
  \includegraphics[width=35mm]{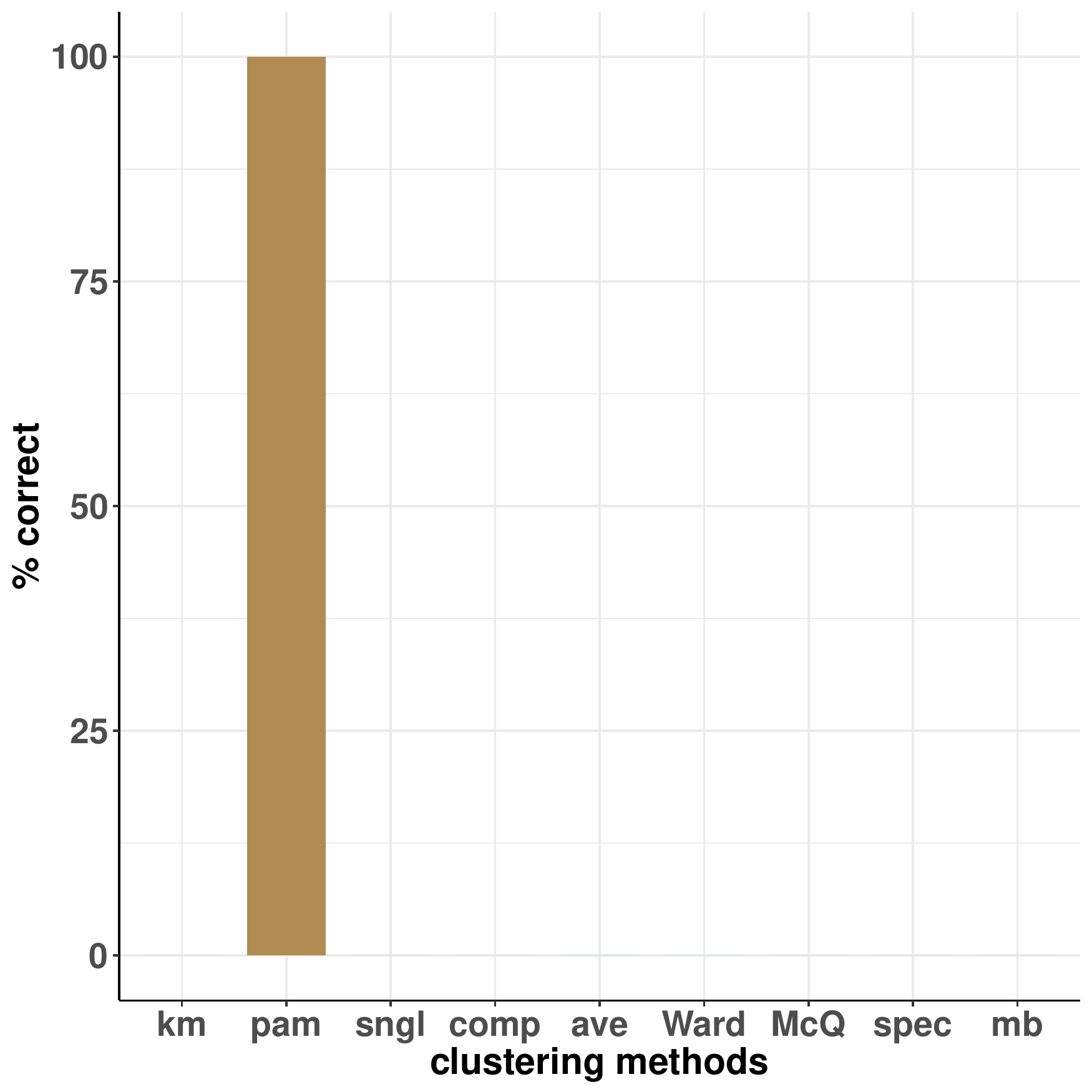}
}
\subfloat[BI]{
  \includegraphics[width=35mm]{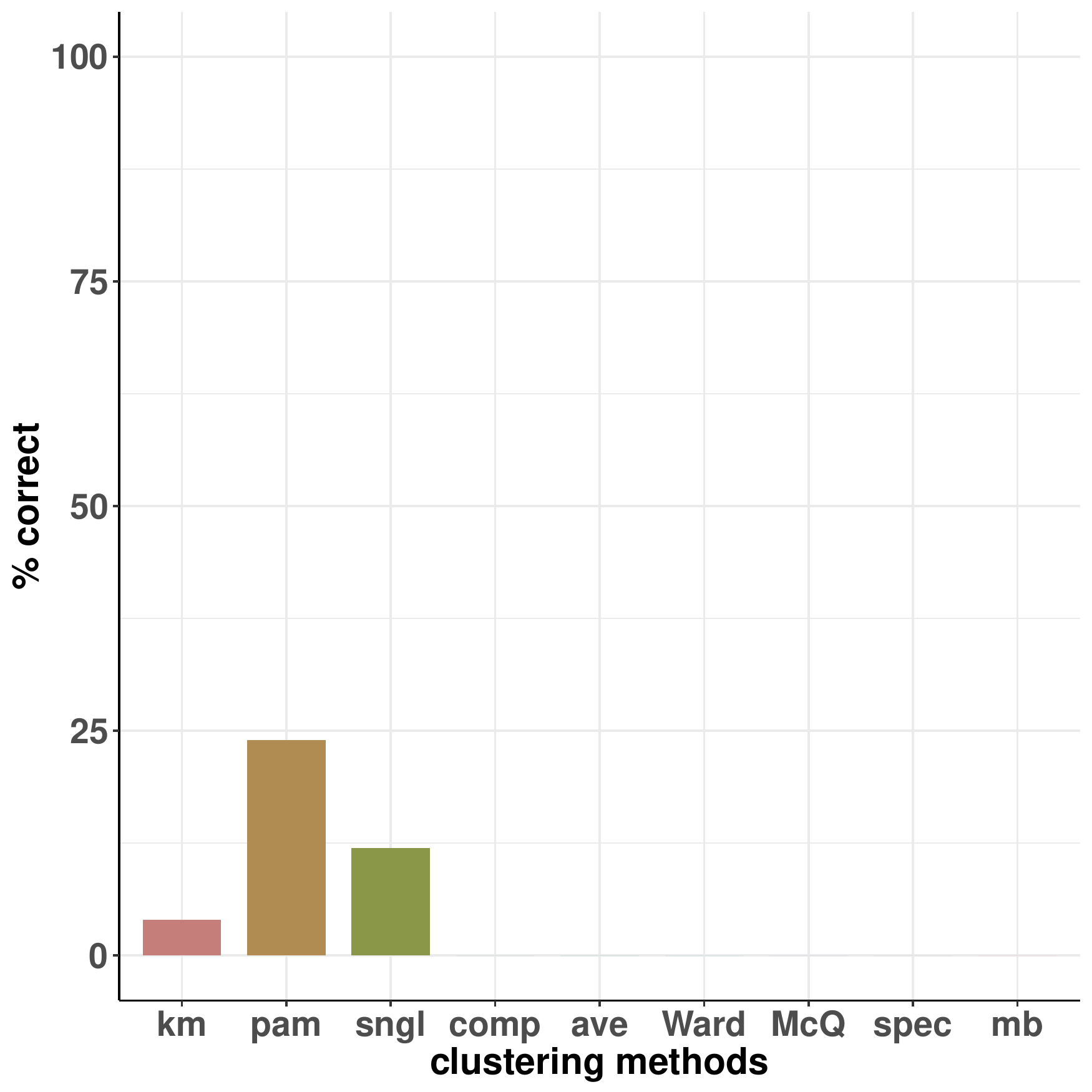}
}
\newline
\subfloat[CVNN]{
  \includegraphics[width=35mm]{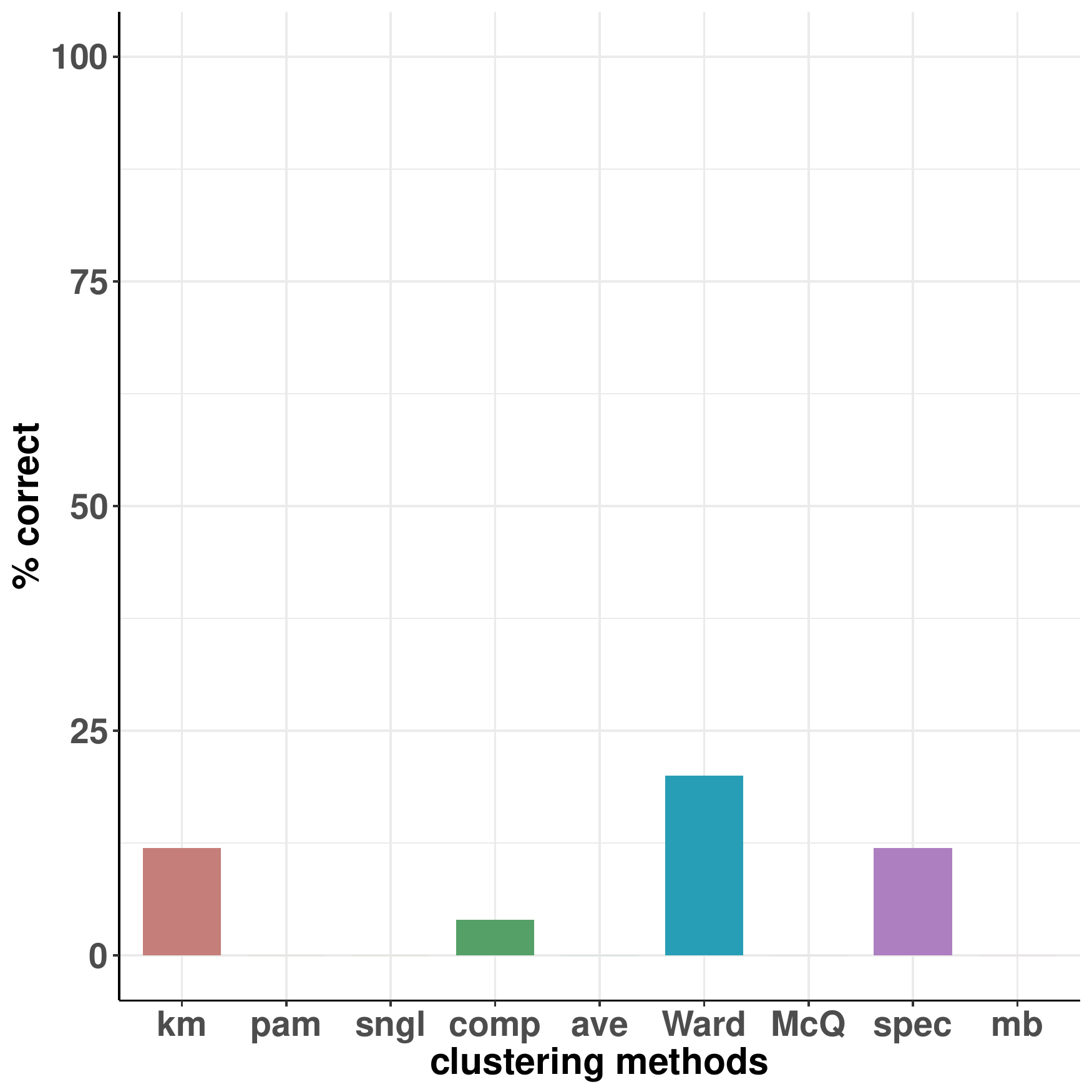}
}
\subfloat[BIC/PAMSIL]{
  \includegraphics[width=35mm]{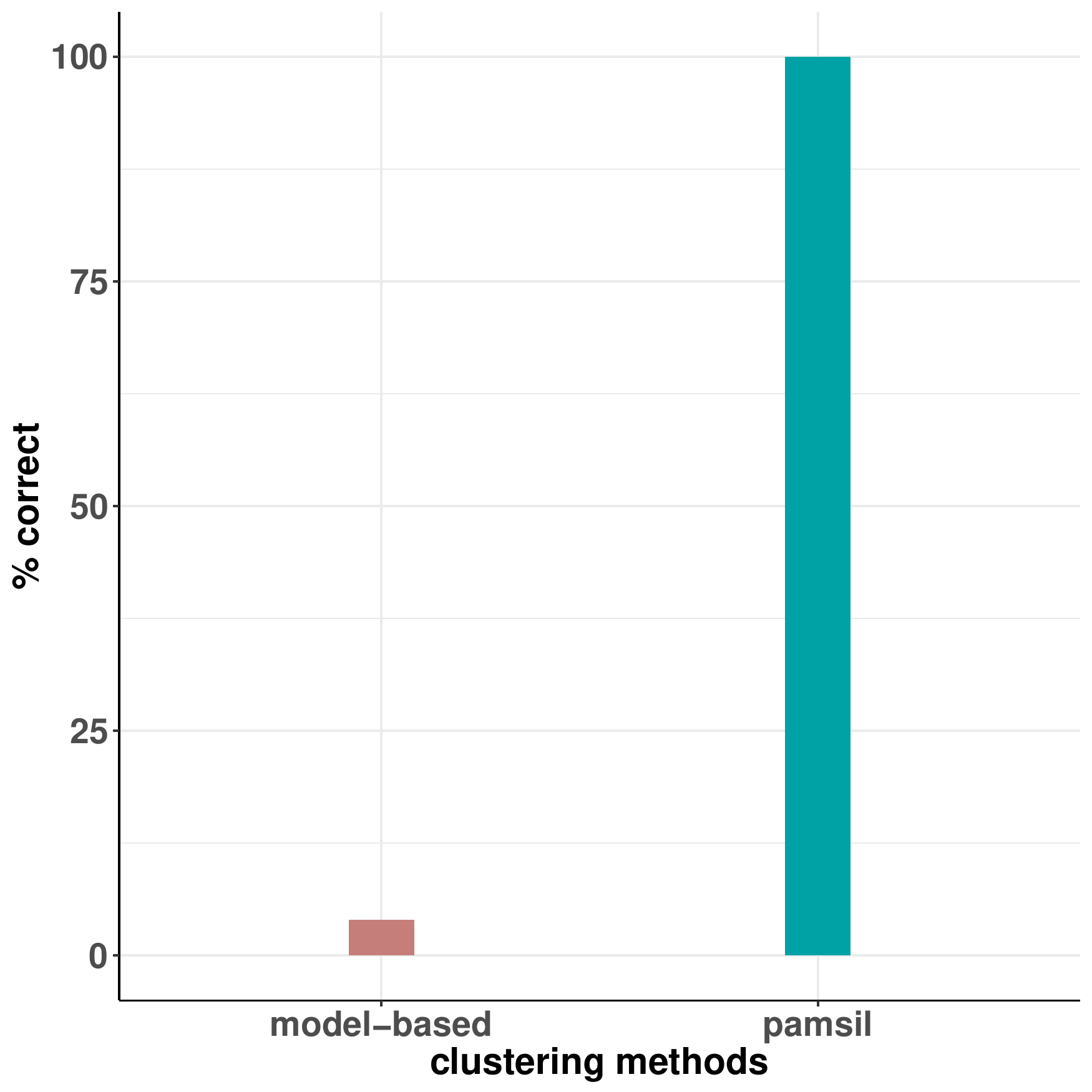}
}
\subfloat[ASW]{
  \includegraphics[width=35mm]{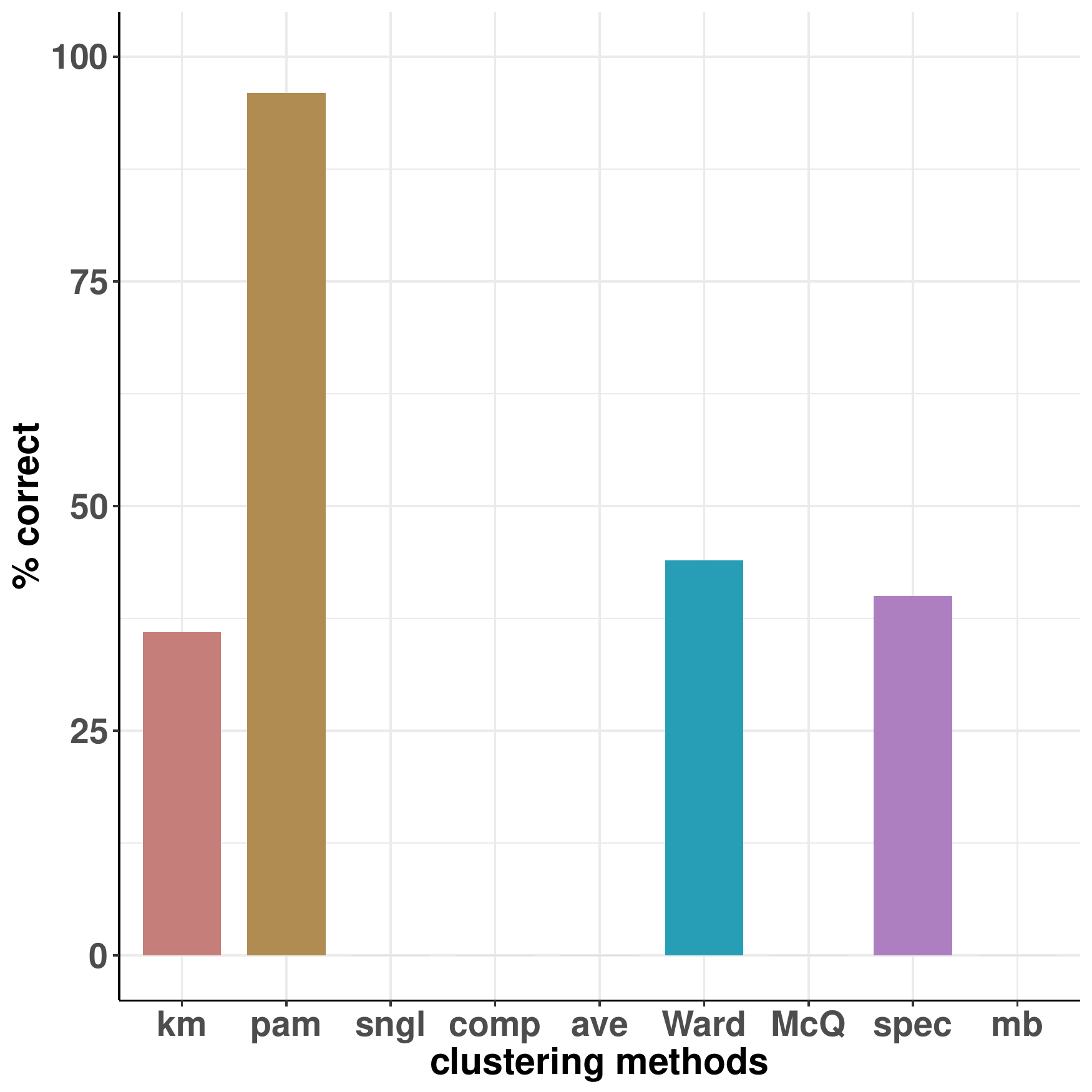}
}
\subfloat[OASW]{
  \includegraphics[width=35mm]{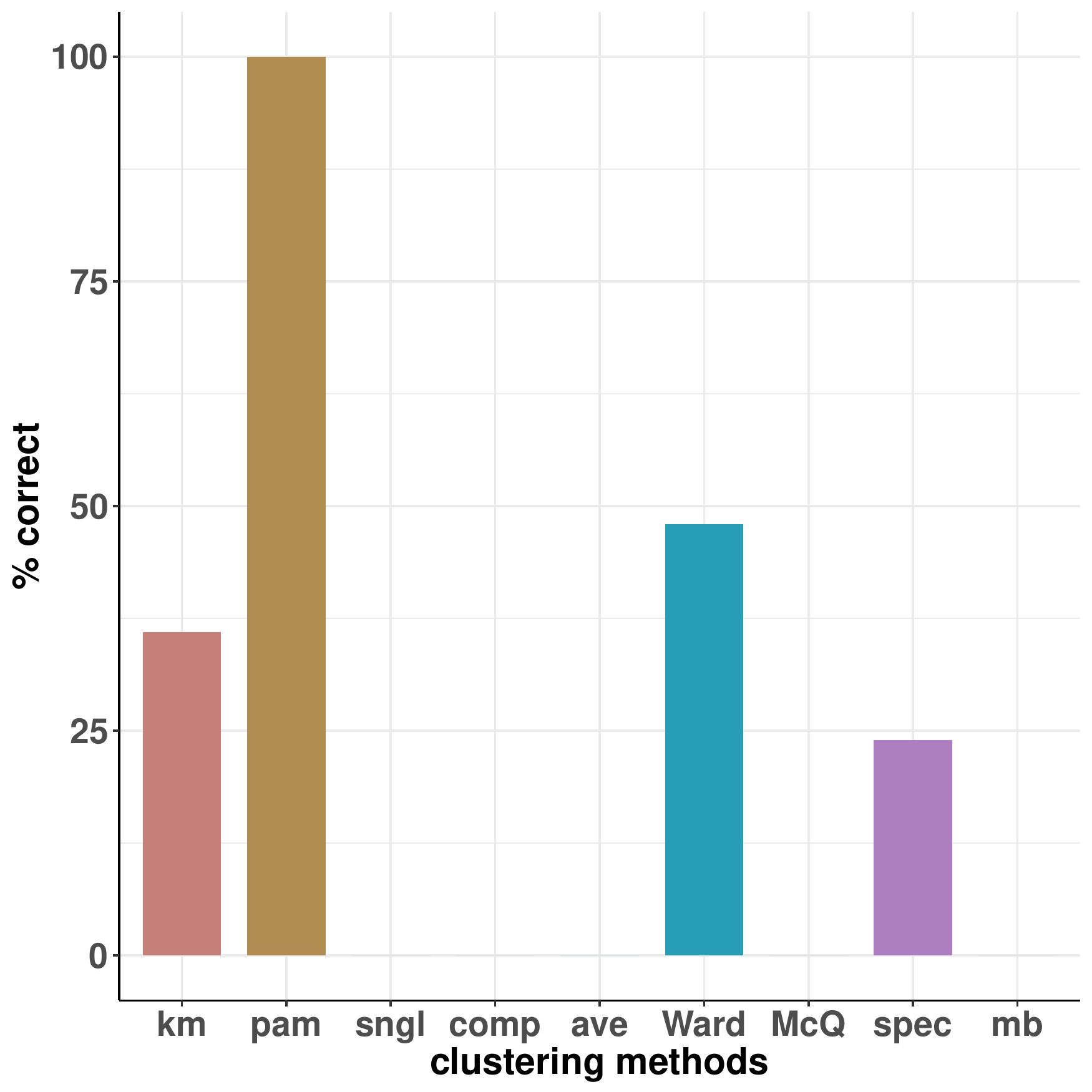}
}
\caption{Bar plots for the estimation of k for Model 5. The Jump method was never able to estimate correct number of clusters for Model 5. }
\label{appendix:estkmodelfive}
\end{figure}

\begin{figure}[!hbtp]
\centering
\subfloat[CH]{
  \includegraphics[width=35mm]{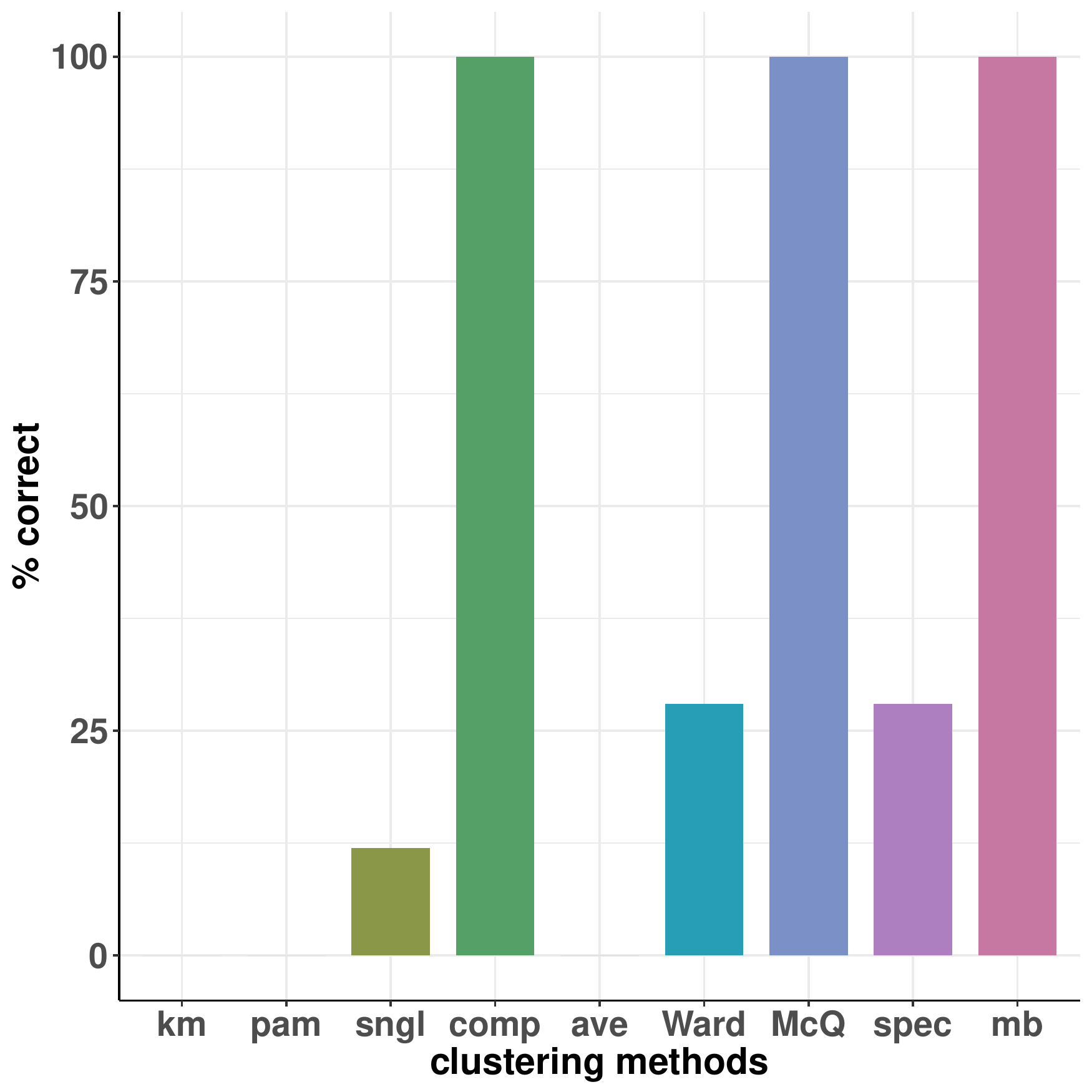}
}
\subfloat[H]{
  \includegraphics[width=35mm]{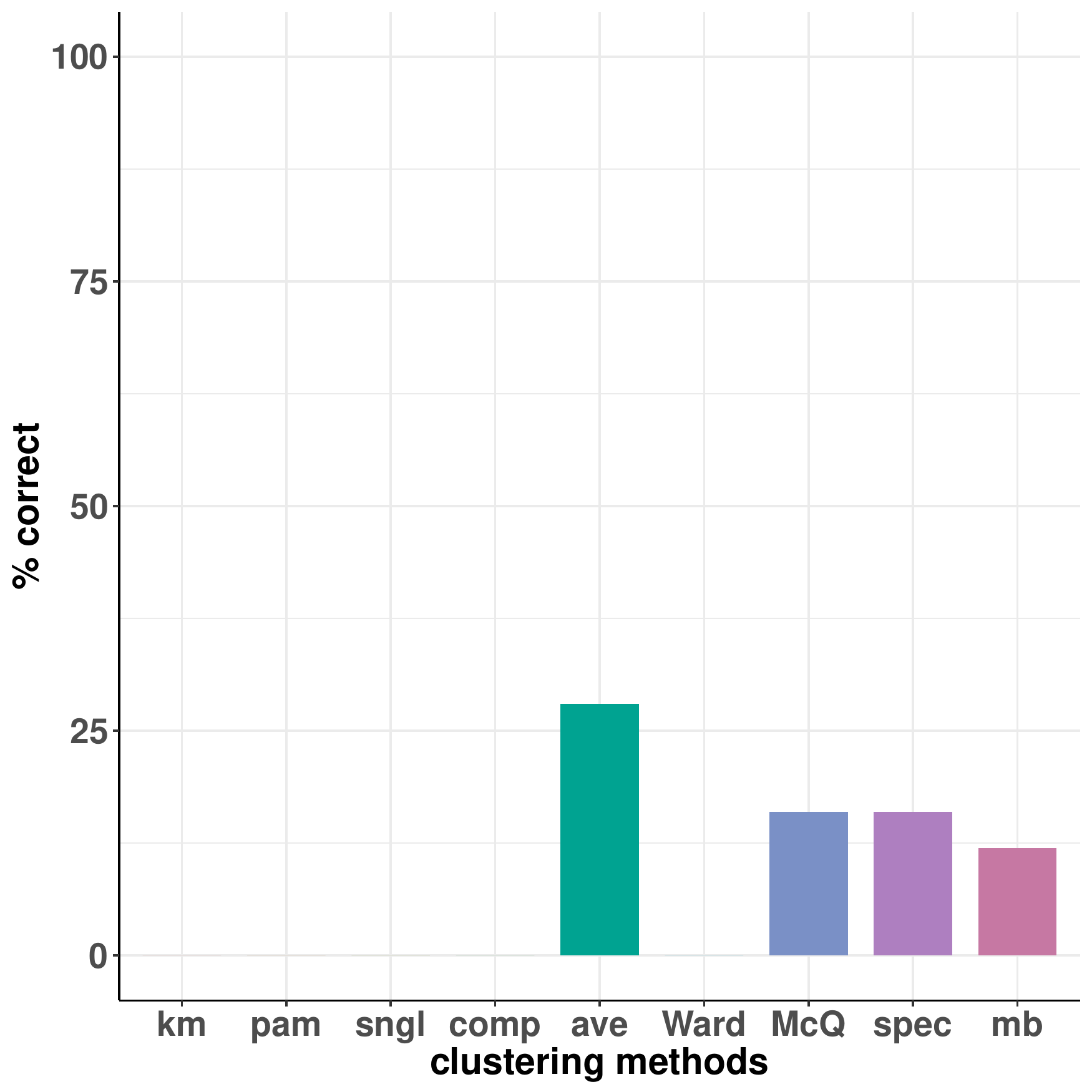}
}
\subfloat[KL]{
  \includegraphics[width=35mm]{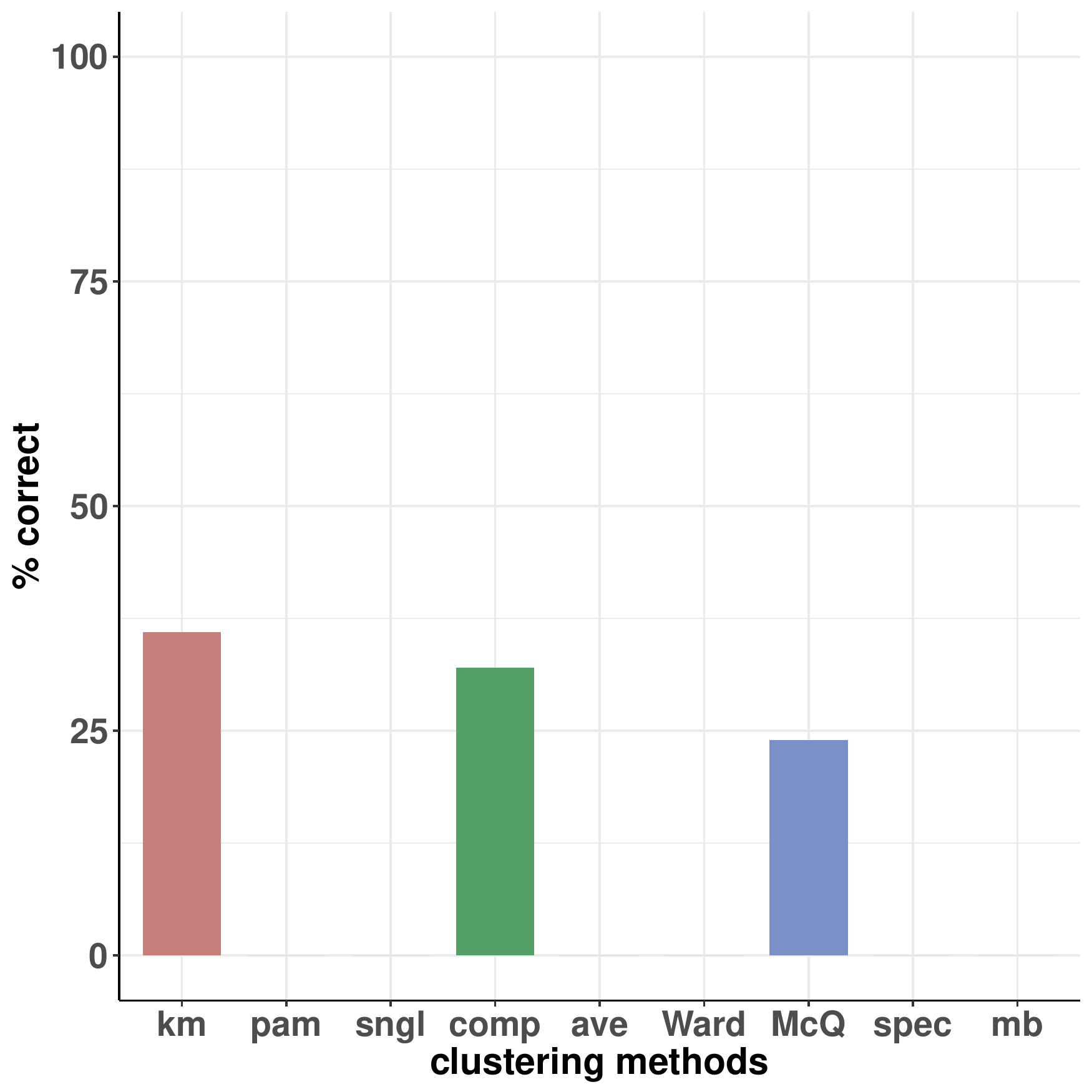}
}
\subfloat[gap]{
  \includegraphics[width=35mm]{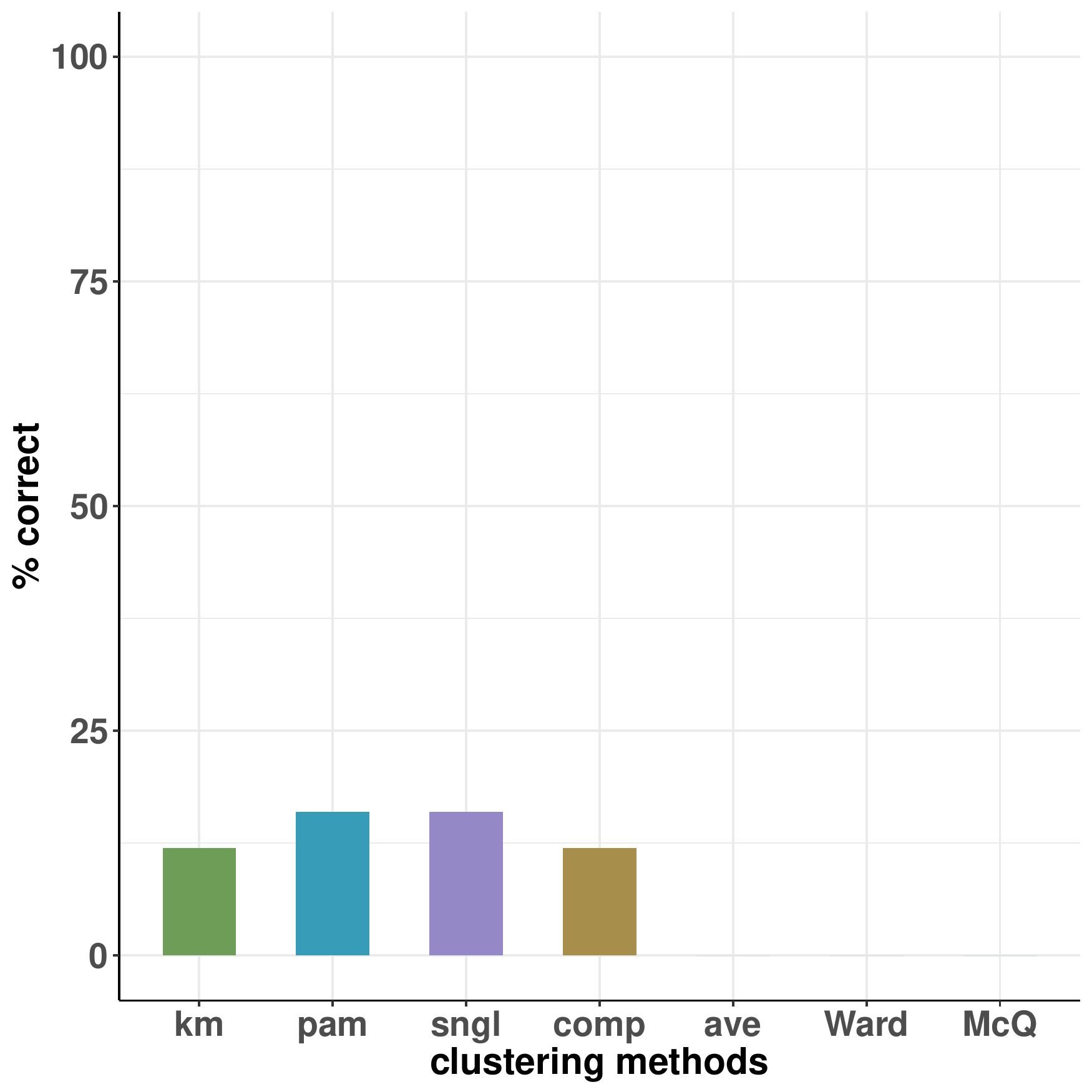}
}
\newline
\subfloat[jump]{
  \includegraphics[width=35mm]{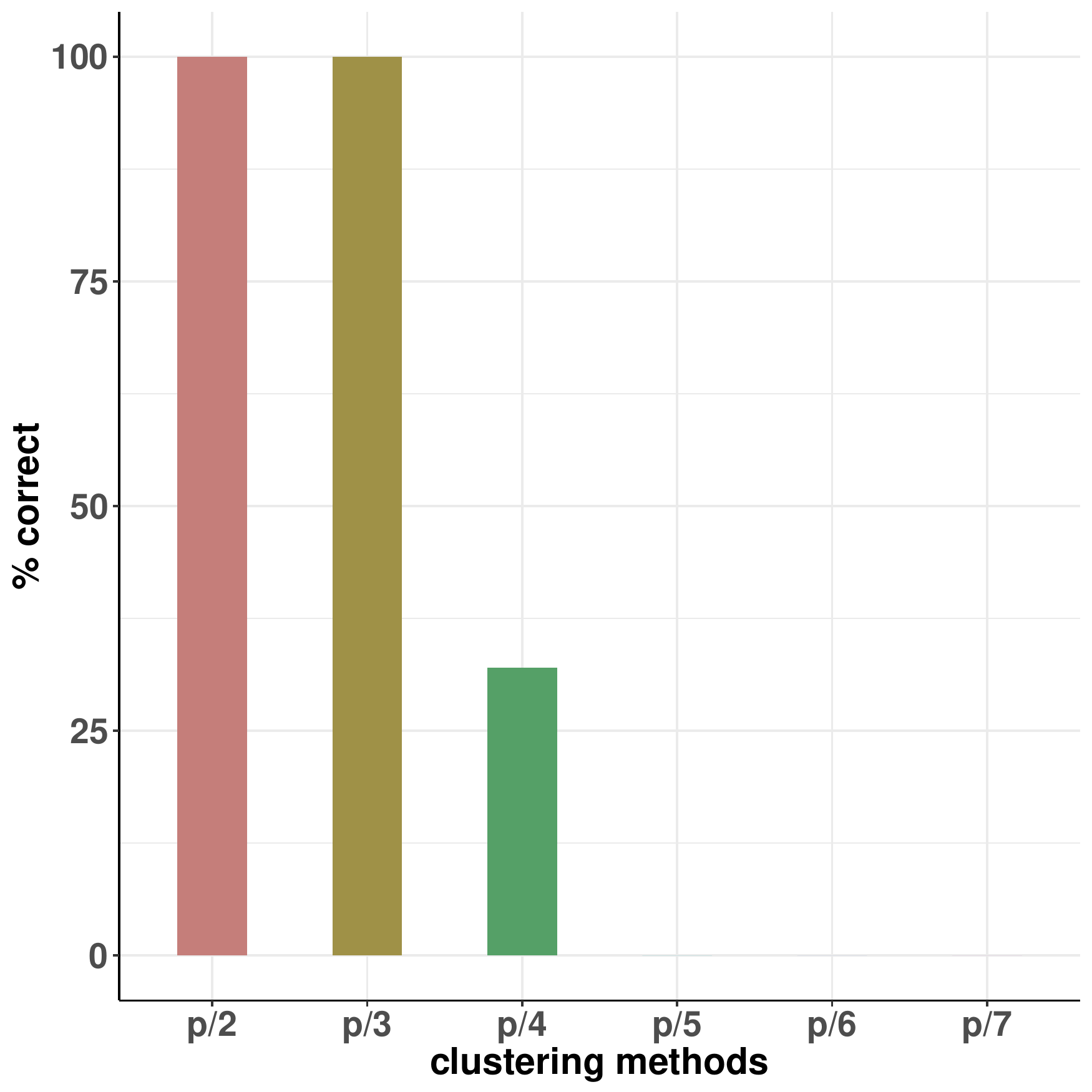}
}
\subfloat[PS]{
  \includegraphics[width=35mm]{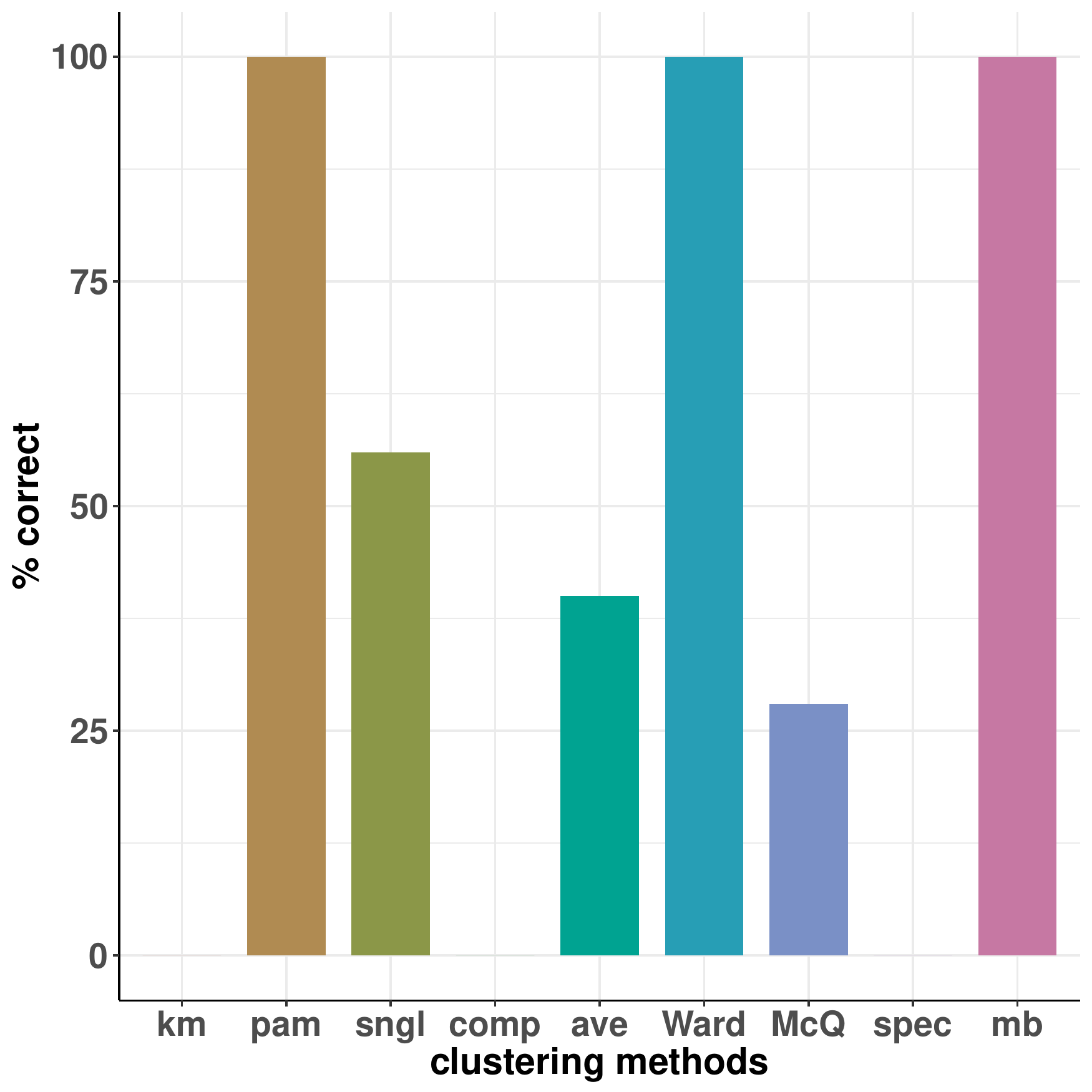}
}
\subfloat[CVNN]{
  \includegraphics[width=35mm]{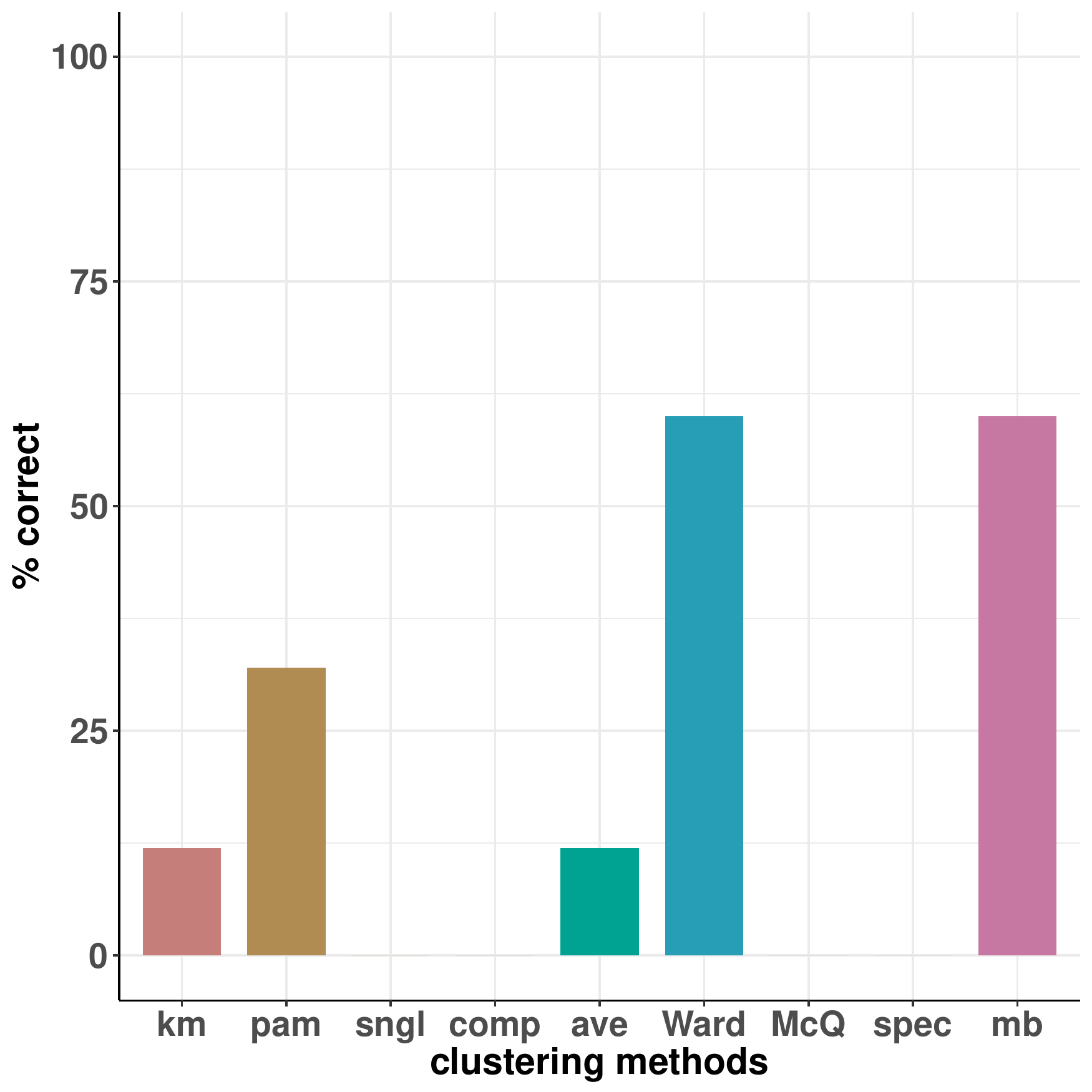}
}
\subfloat[BIC/PAMSIL]{
 \includegraphics[width=35mm]{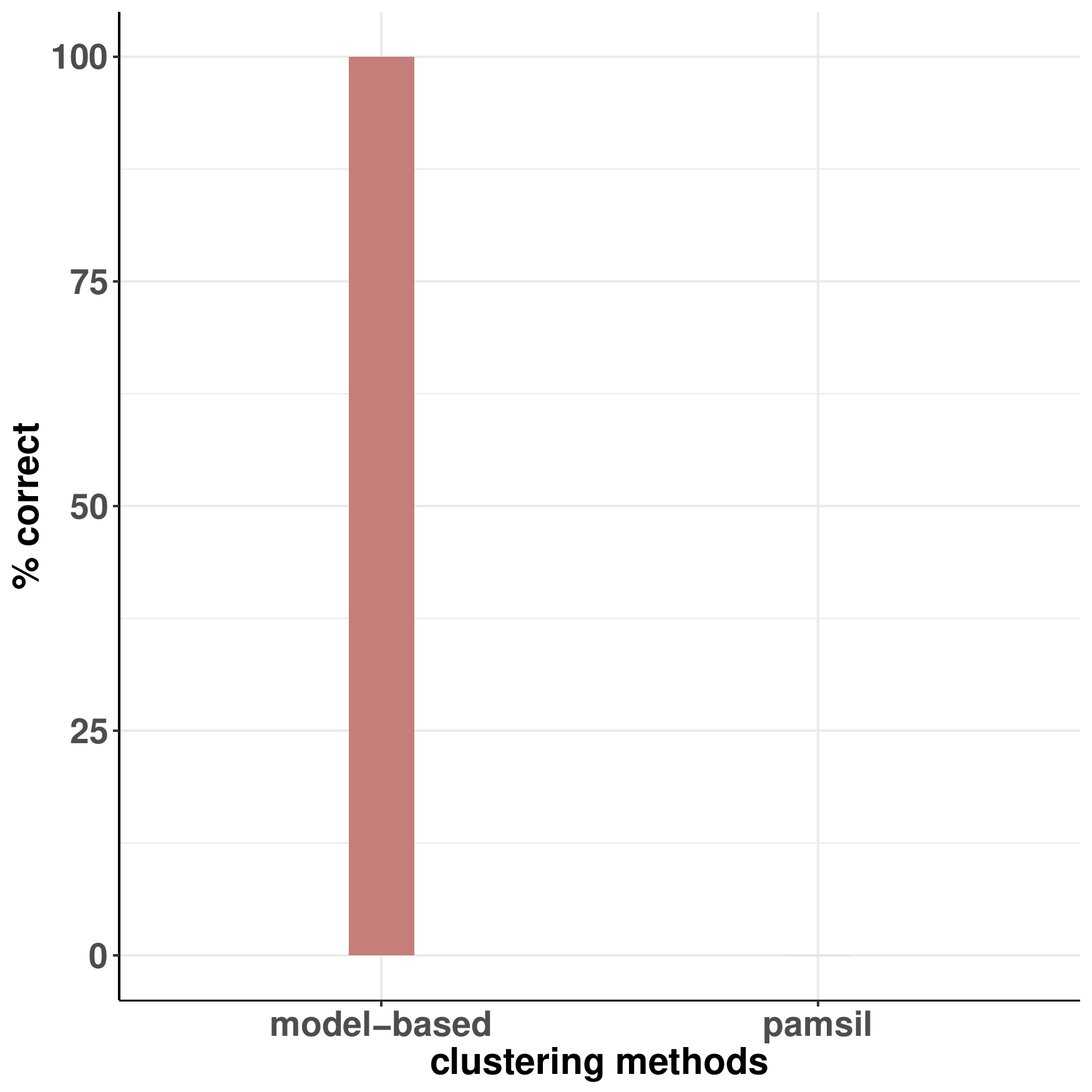}
 }
\newline
\rule{-60ex}{.2in}
\subfloat[OASW]{
  \includegraphics[width=35mm]{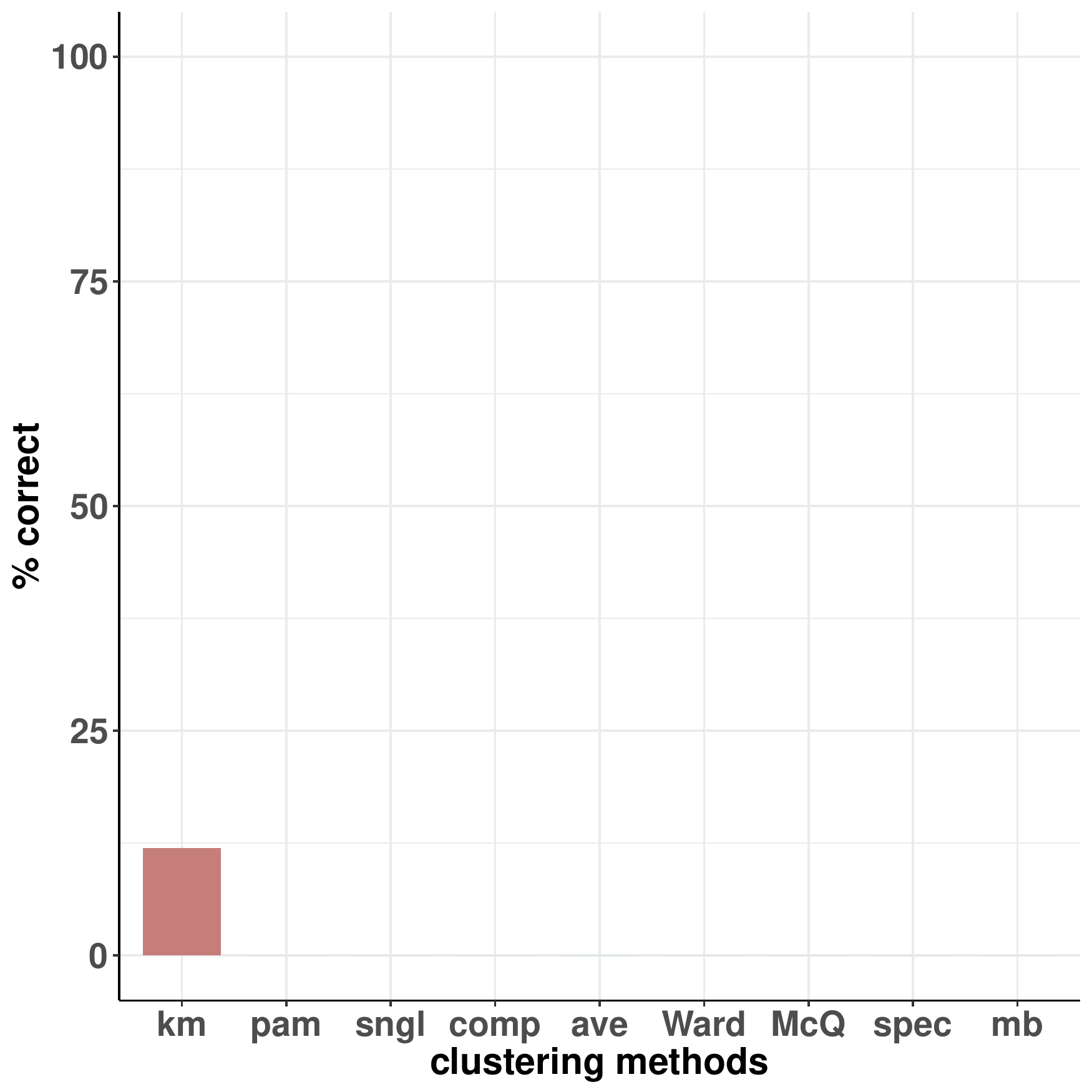}
}
\caption{Bar plots for the estimation of k for Model 6. The Gamma, C, BI, PAMSIL and ASW were never able to estimate correct number of clusters for Model 6.}
\label{appendix:estkmodelsix}
\end{figure}

\begin{figure}[H]
\centering
\subfloat[CH]{
  \includegraphics[width=35mm]{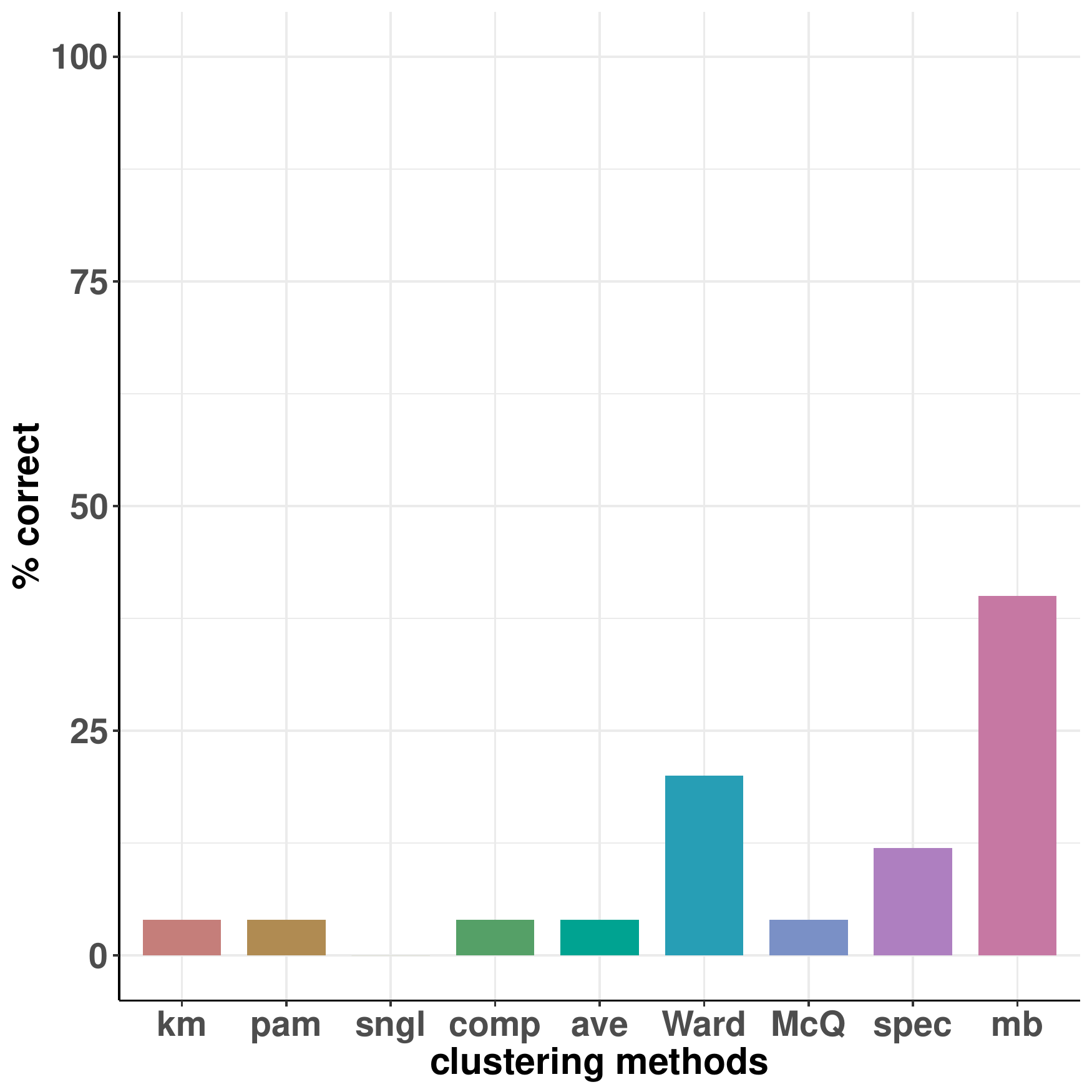}
}
\subfloat[H]{
  \includegraphics[width=35mm]{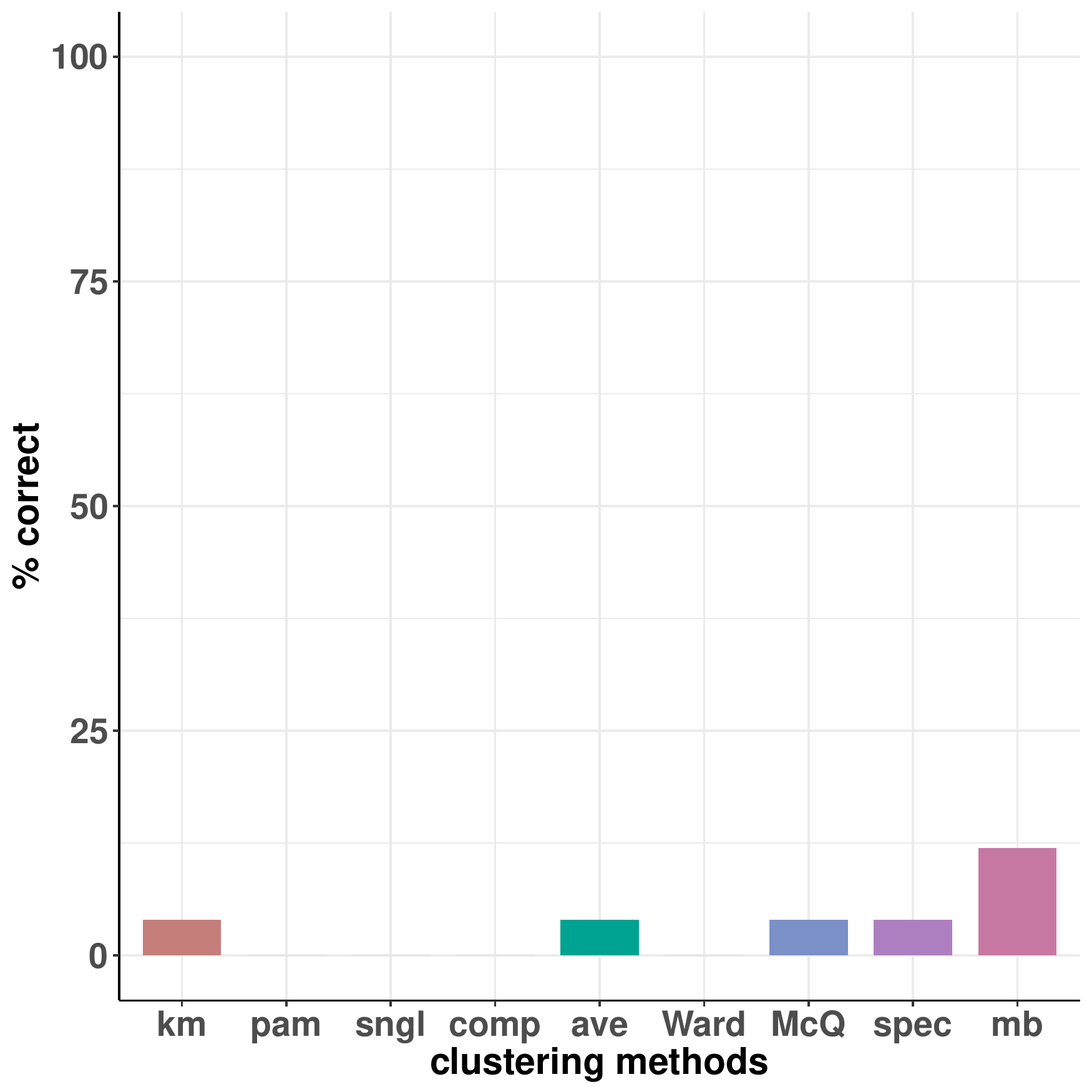}
}
  \subfloat[Gamma]{
  \includegraphics[width=35mm]{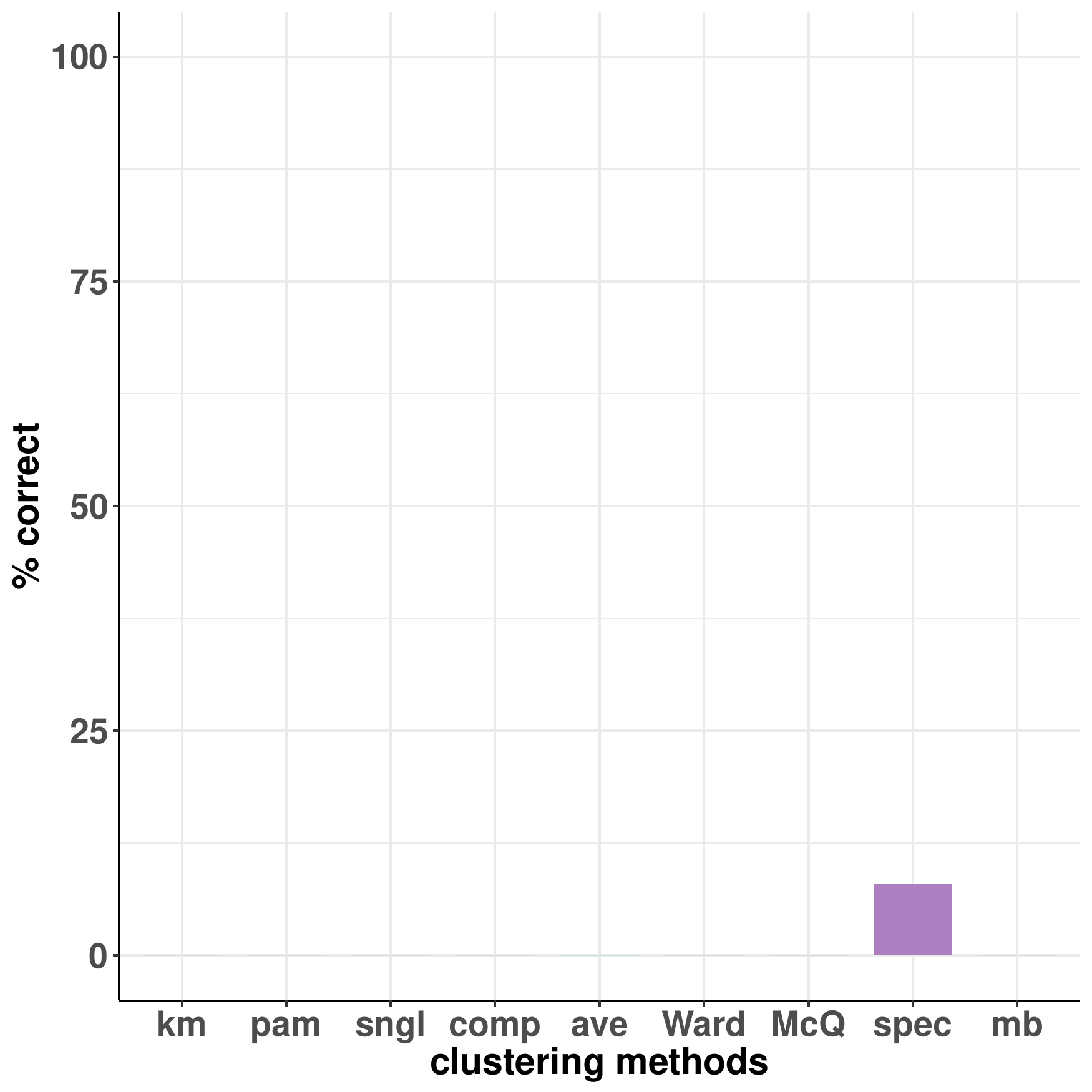}}
  \subfloat[C]{
  \includegraphics[width=35mm]{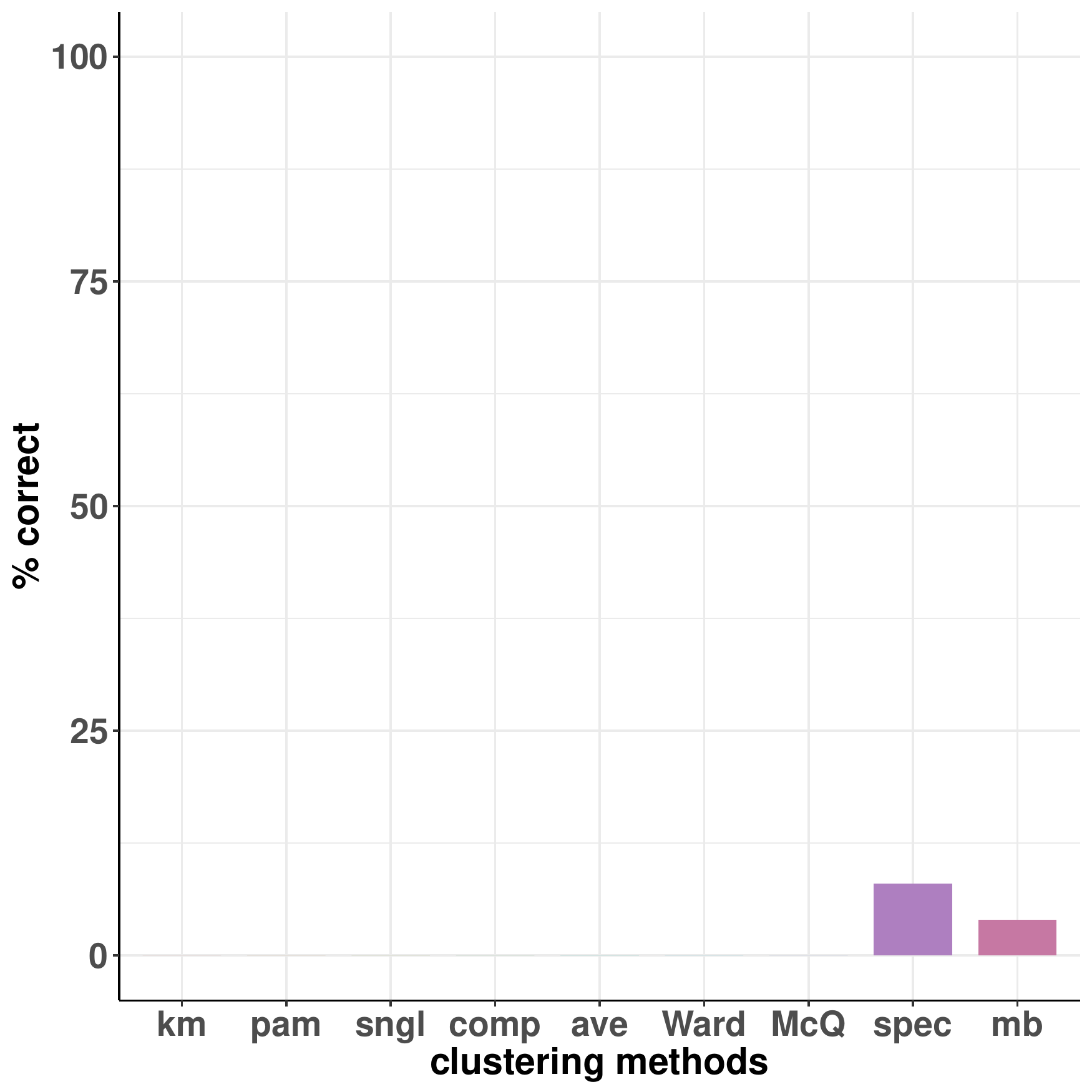}
}
\newline
\subfloat[KL]{
  \includegraphics[width=35mm]{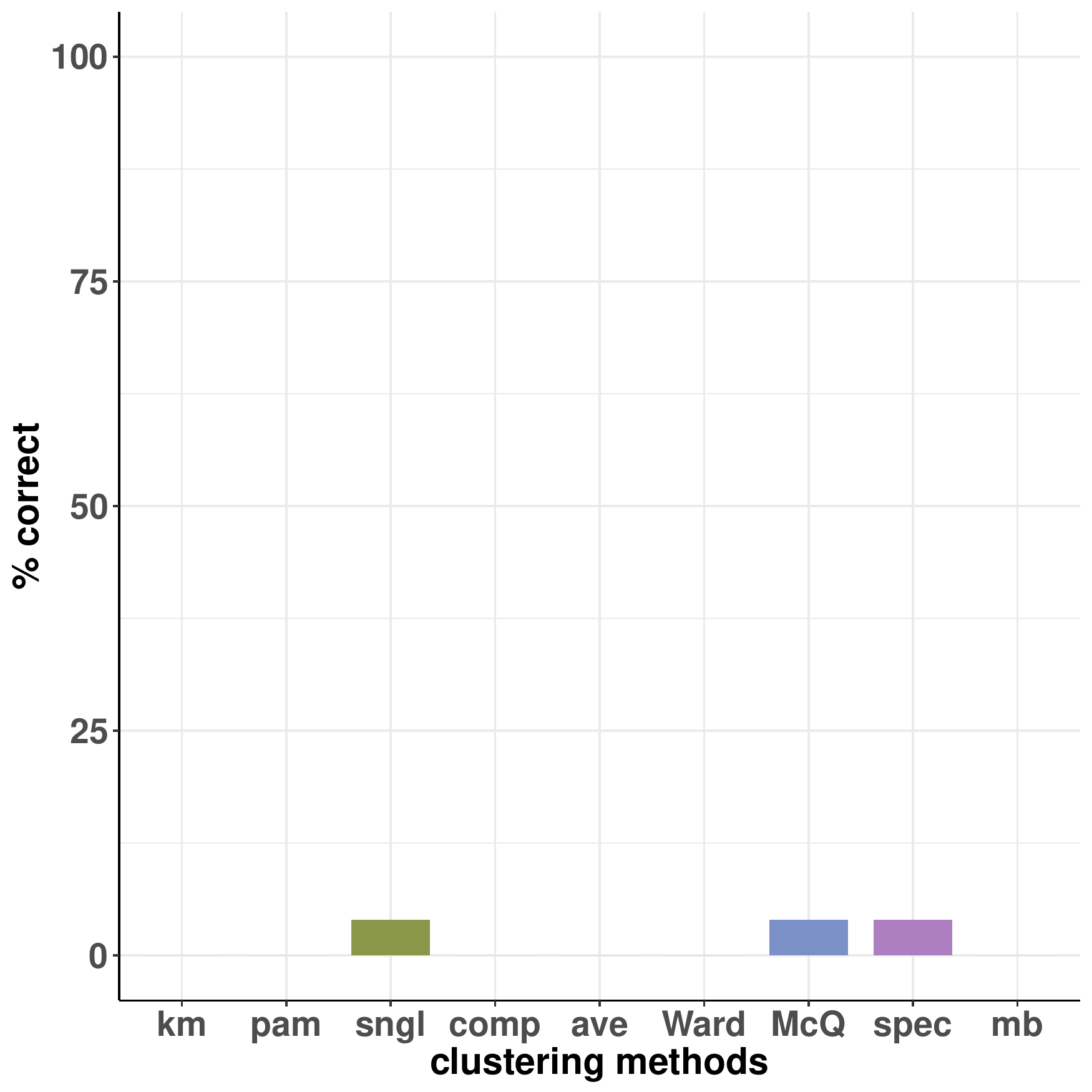}
}
\subfloat[gap]{
  \includegraphics[width=35mm]{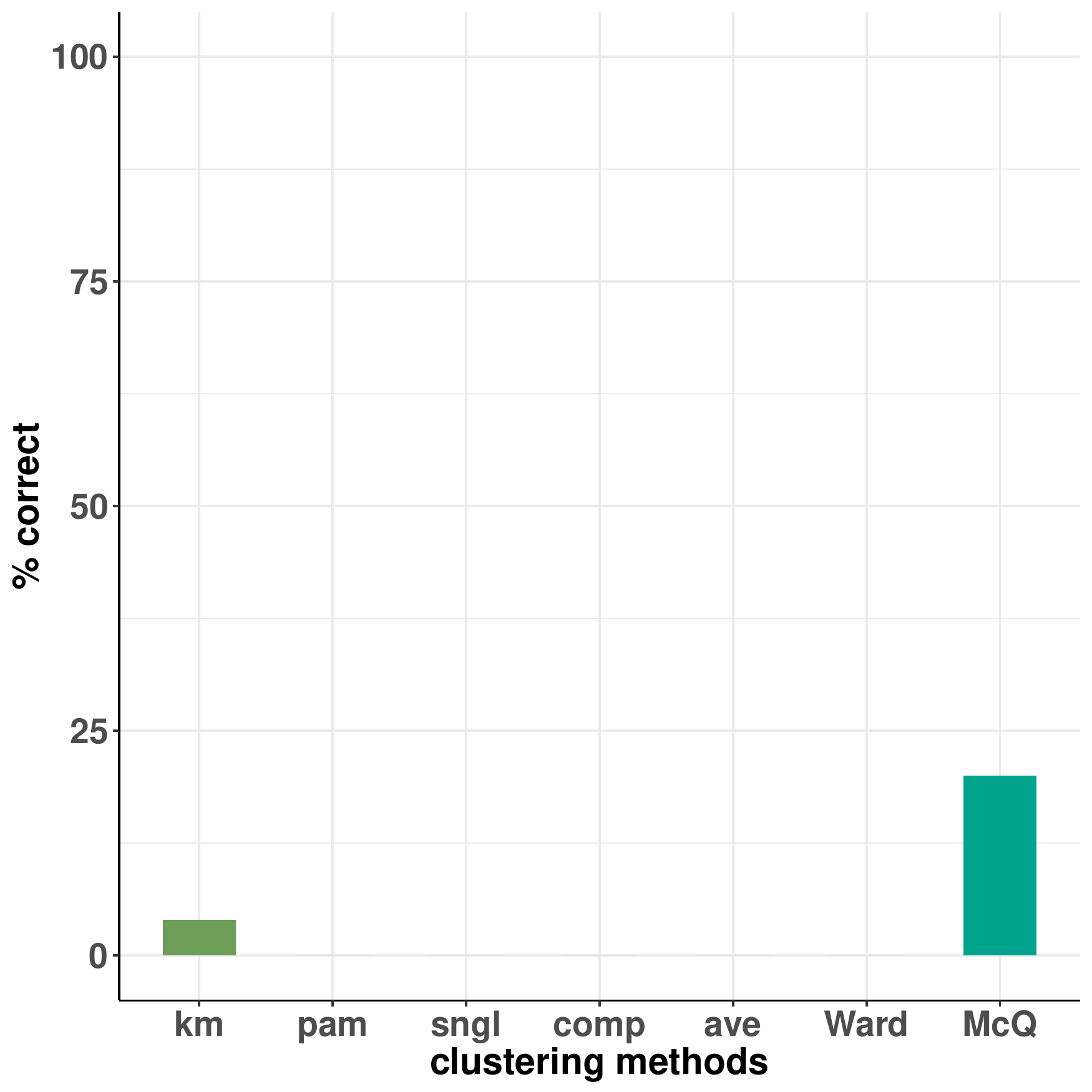}
}
\subfloat[PS]{
  \includegraphics[width=35mm]{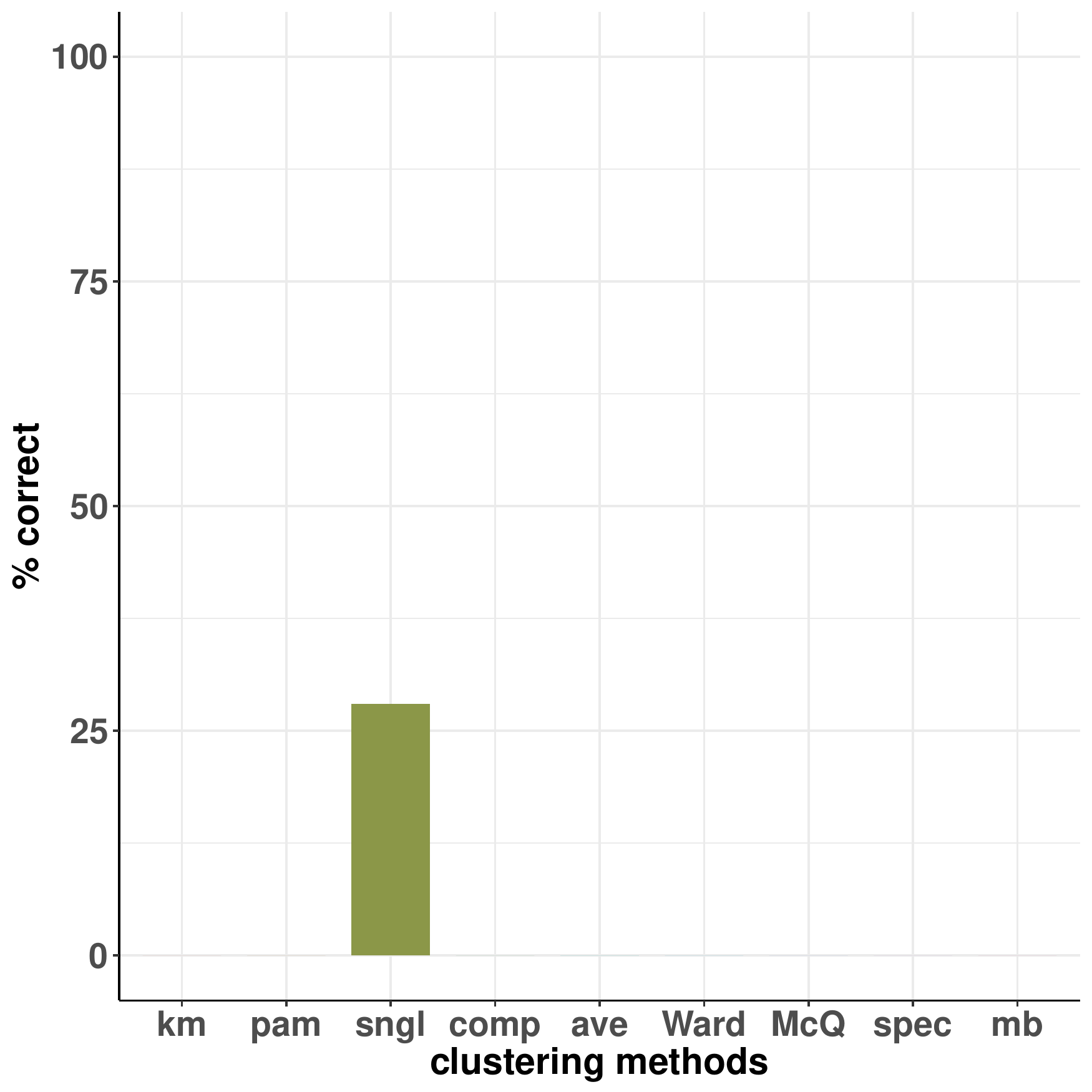}
}
\subfloat[BI]{
  \includegraphics[width=35mm]{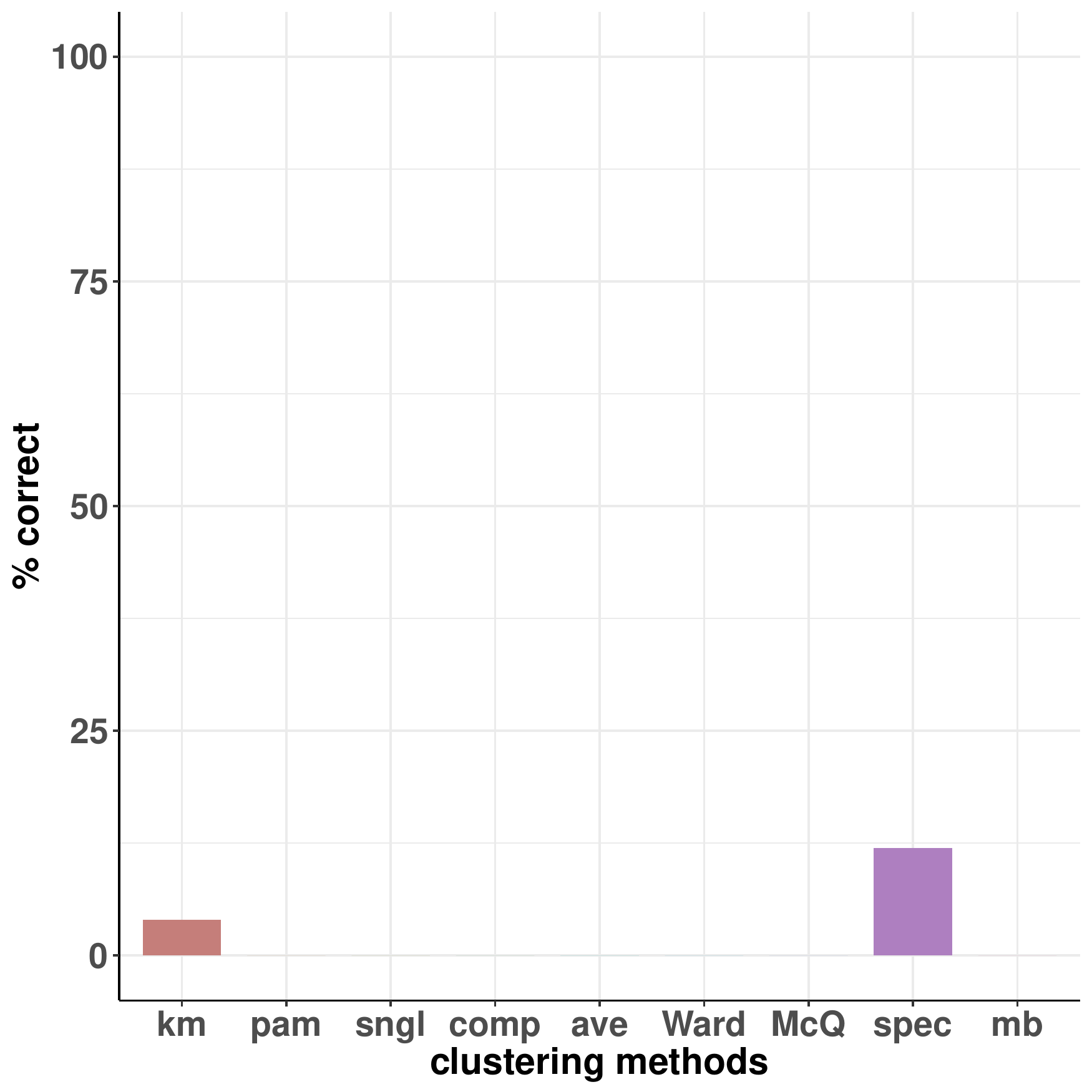}
}
\newline
\rule{-60ex}{.2in}
\subfloat[CVNN]{
  \includegraphics[width=35mm]{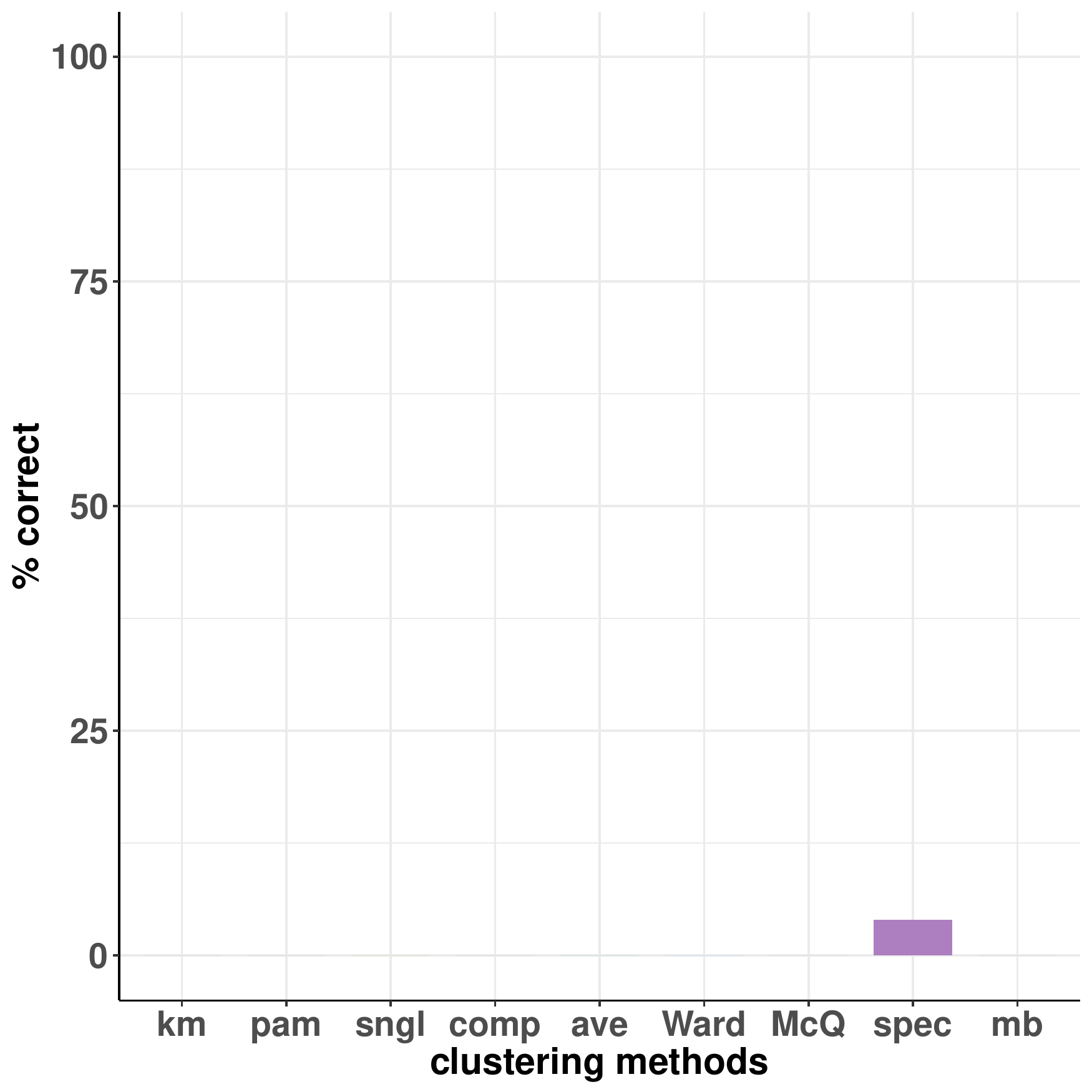}
}
\caption{Bar plots for the estimation of k for Model 7. The Jump, model-based clustering with BIC, PAMSIL, ASW and OSil were never able to estimate correct number of clusters for Model 7.}
\label{appendix:estkmodelseven}
\end{figure}

\begin{figure}[!hbtp]
\centering
\subfloat[CH]{
  \includegraphics[width=35mm]{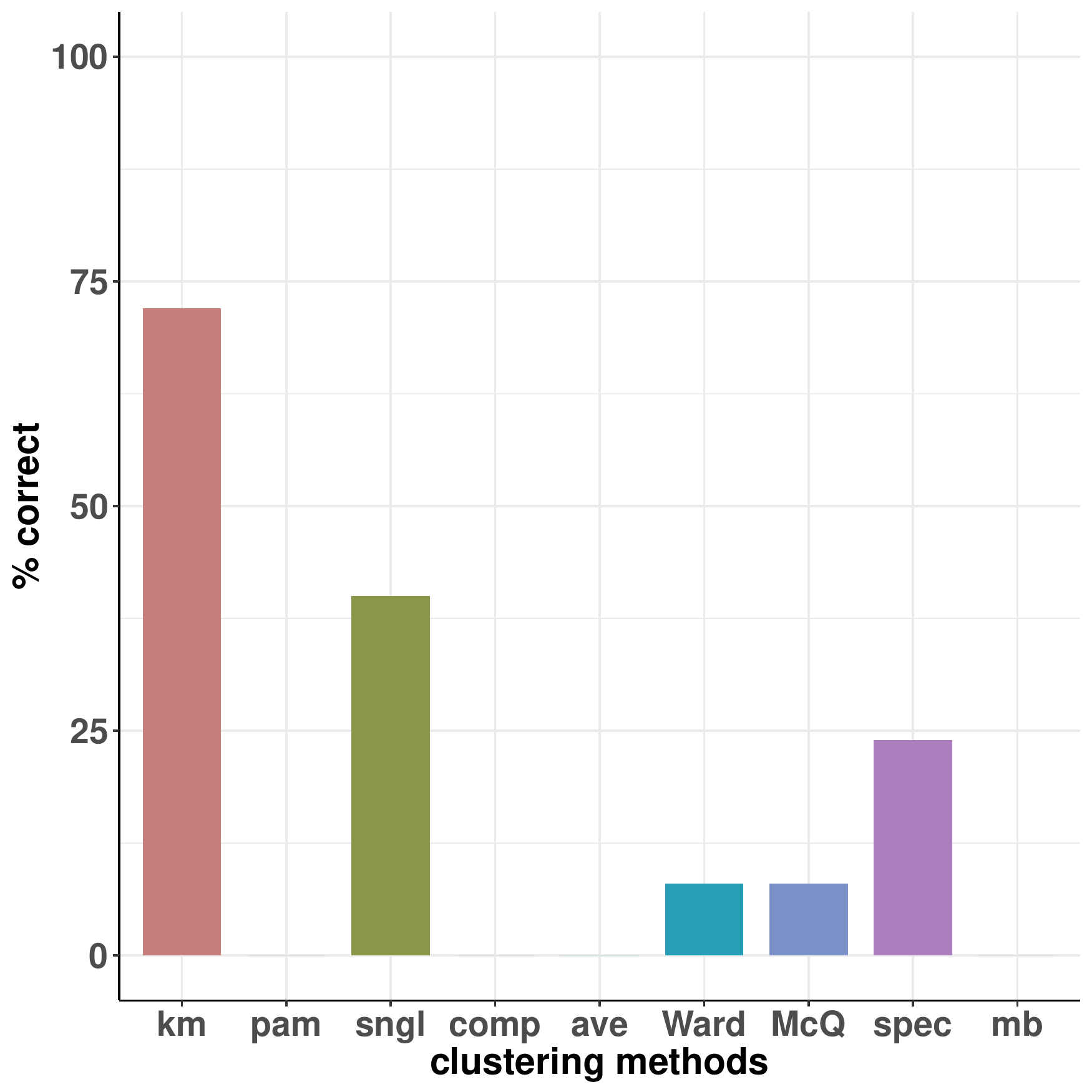}
}
\subfloat[H]{
  \includegraphics[width=35mm]{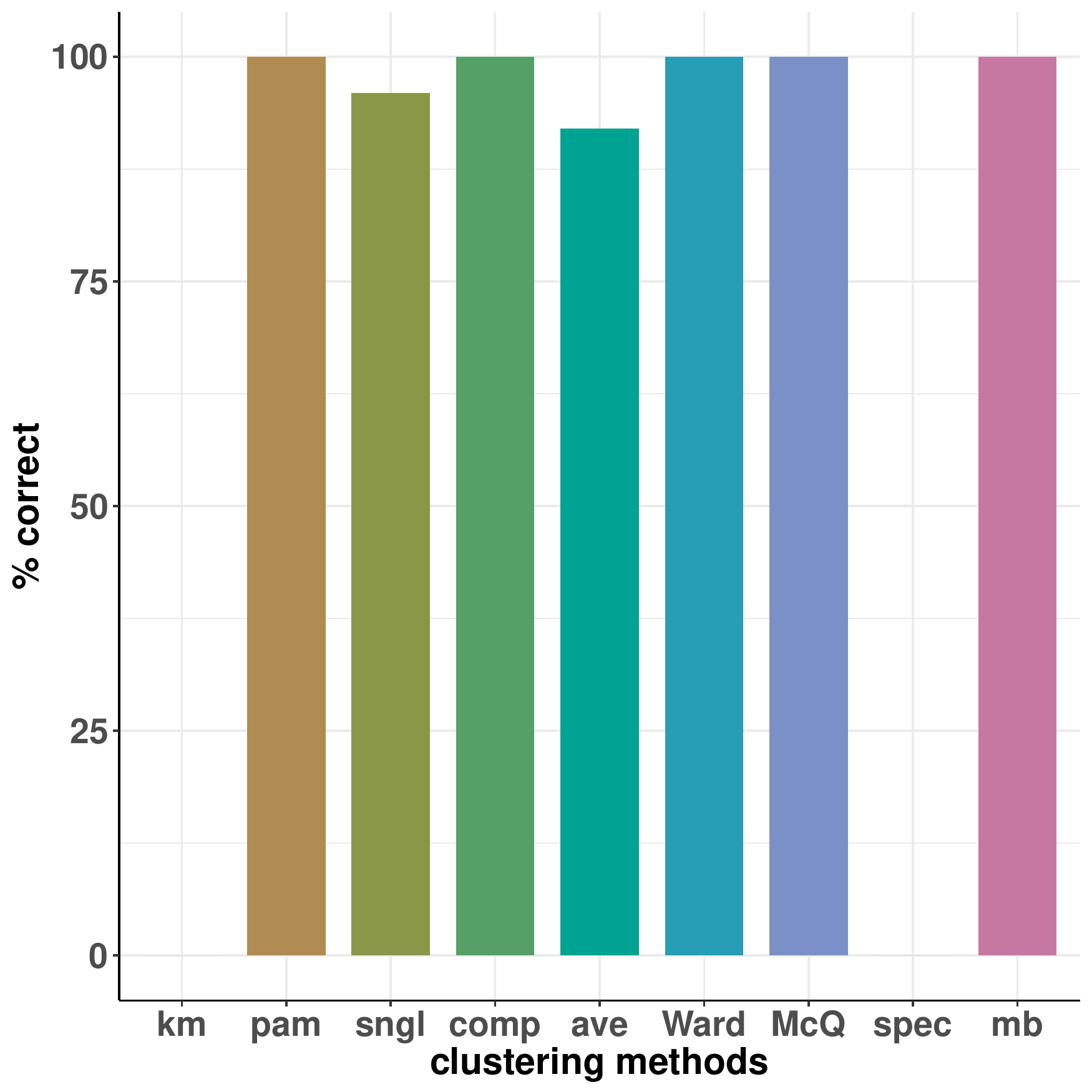}
}
\subfloat[Gamma]{
  \includegraphics[width=35mm]{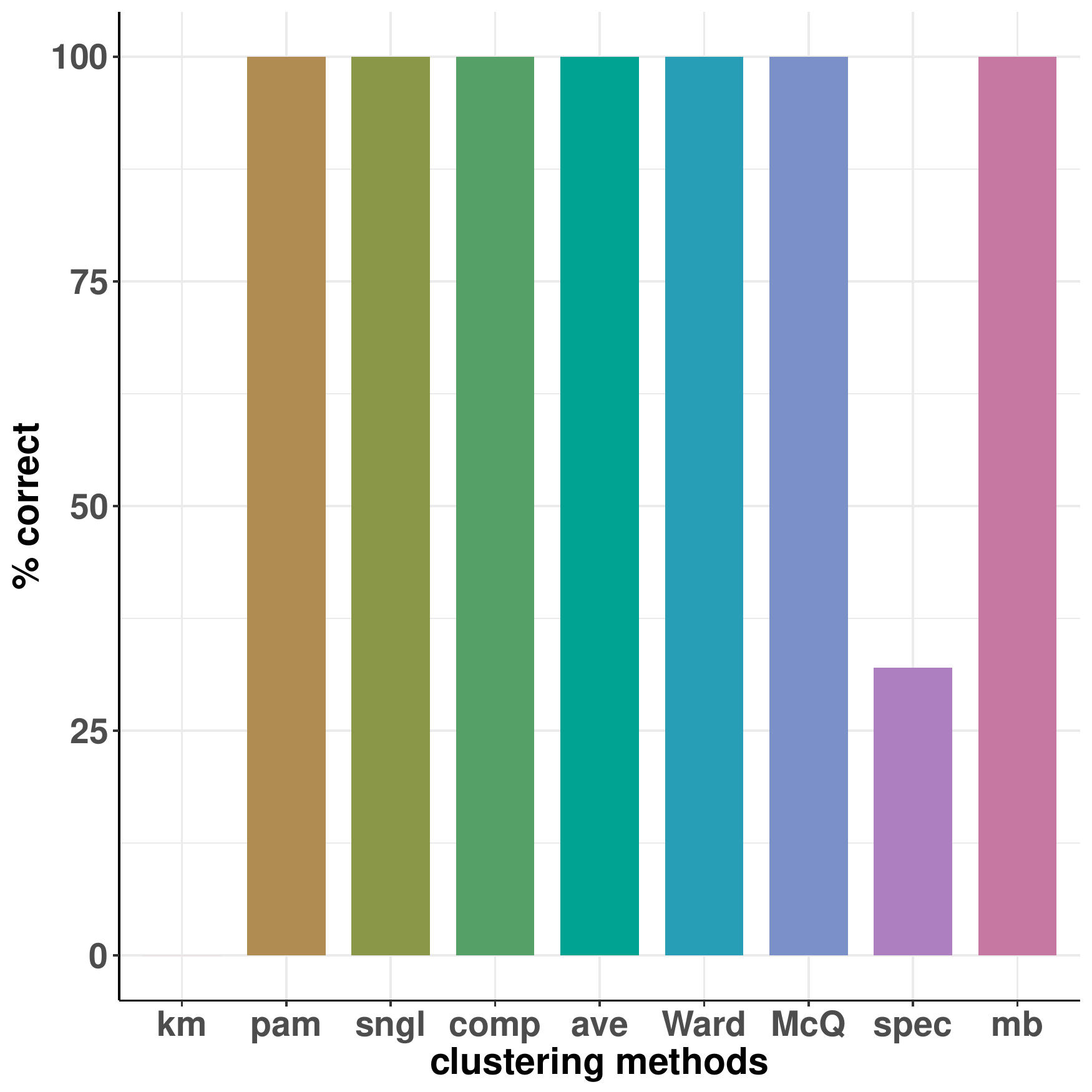}}
\subfloat[C]{
  \includegraphics[width=35mm]{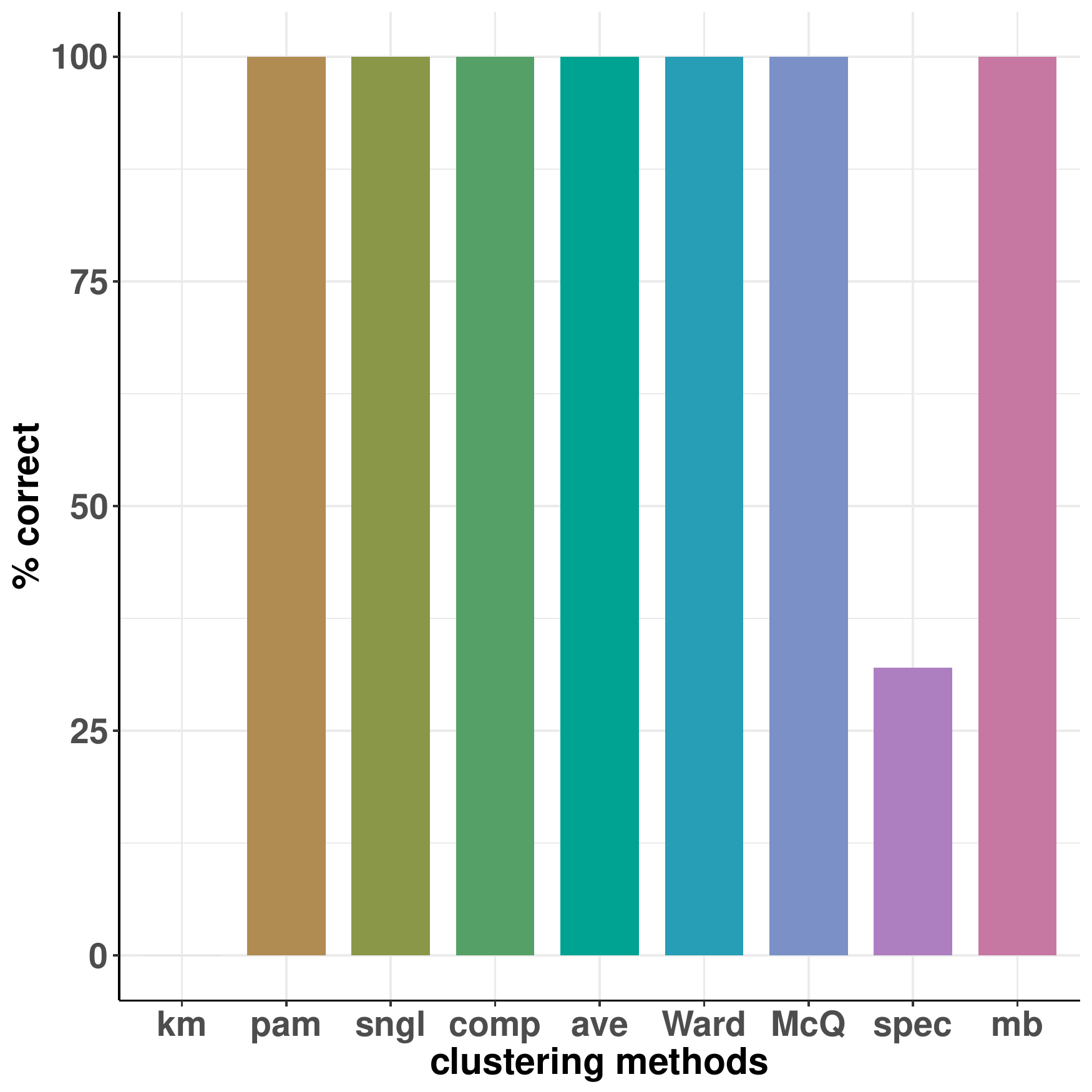}
}
\newline
\subfloat[PS]{
  \includegraphics[width=35mm]{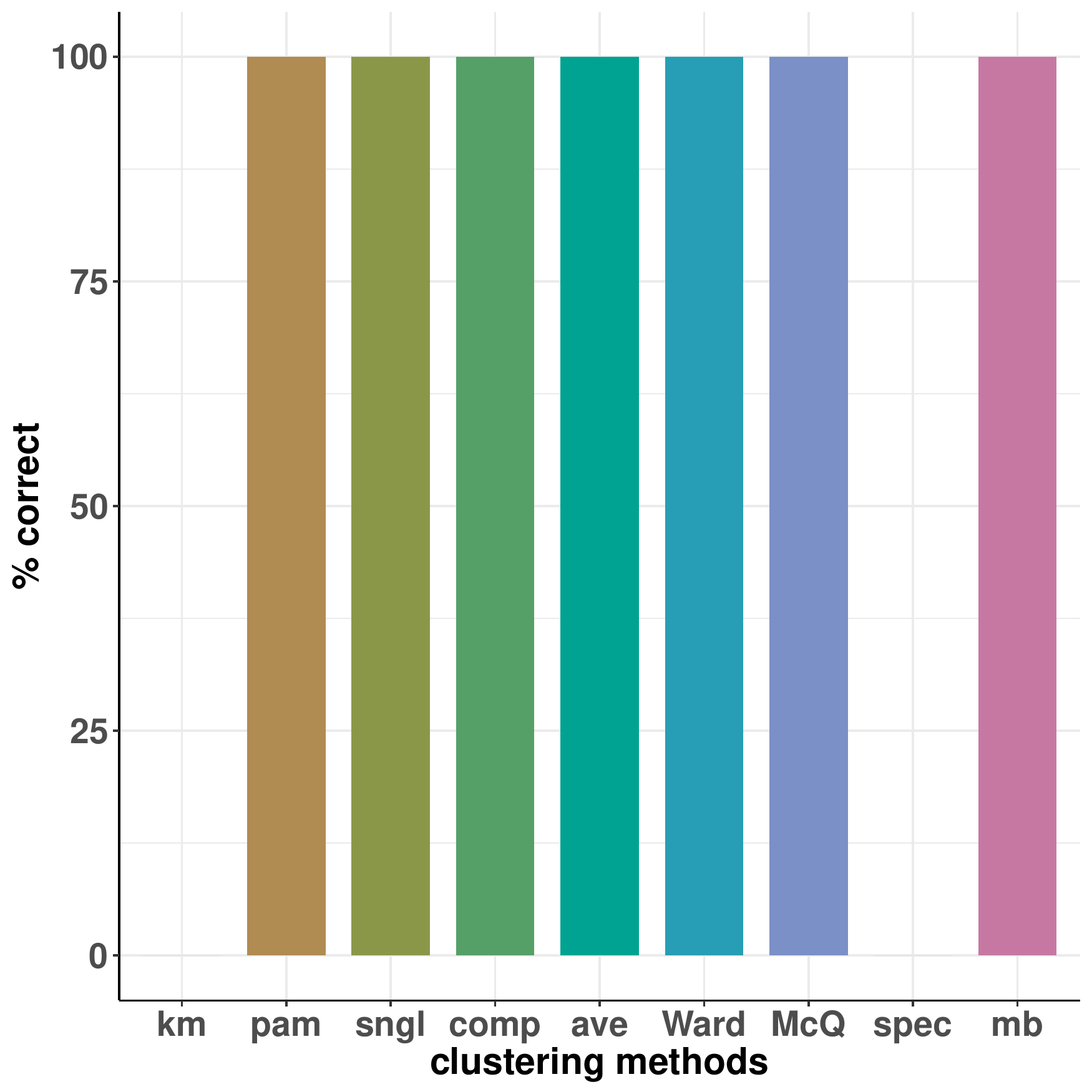}
}
\subfloat[BI]{
  \includegraphics[width=35mm]{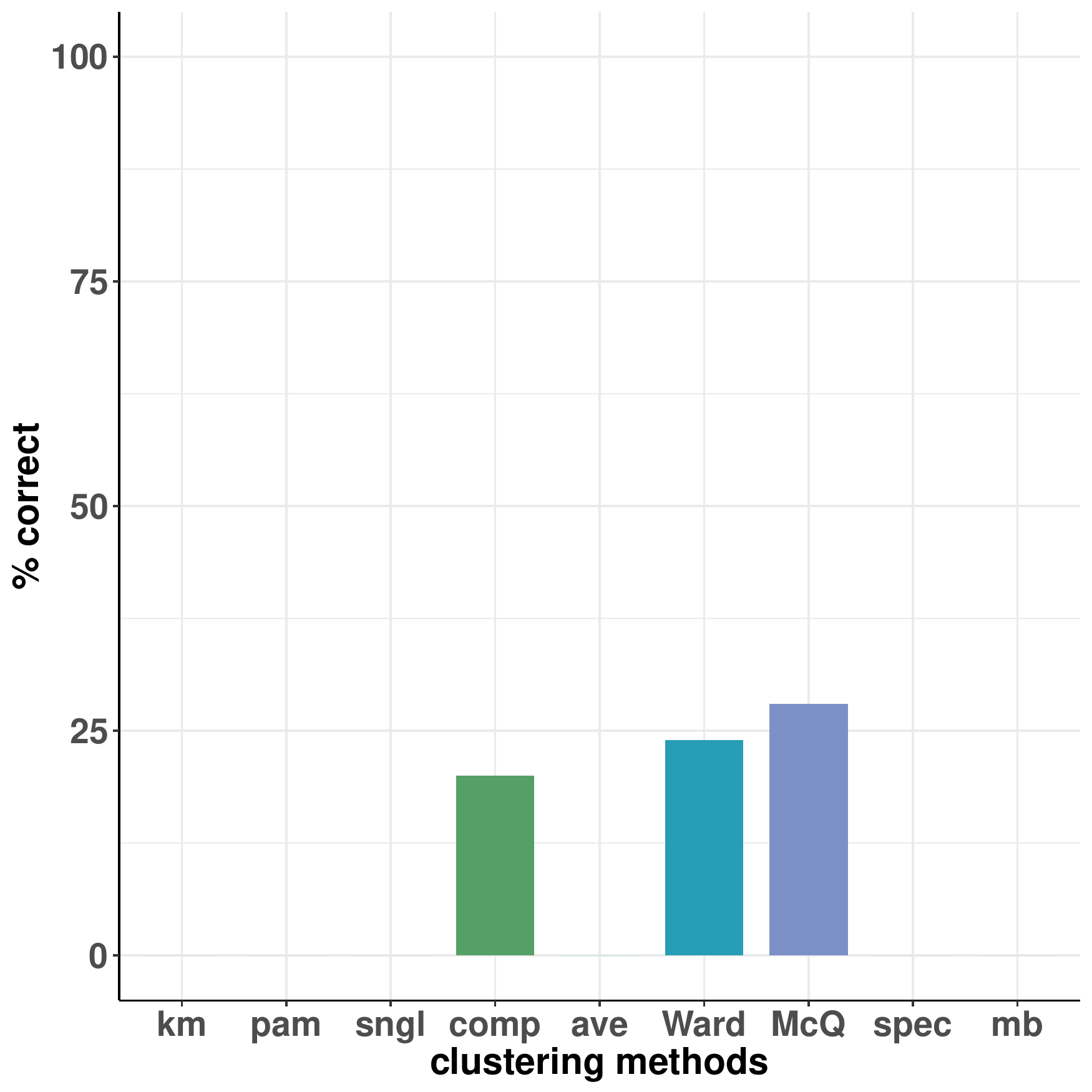}
}
\subfloat[BIC/PAMSIL]{
  \includegraphics[width=35mm]{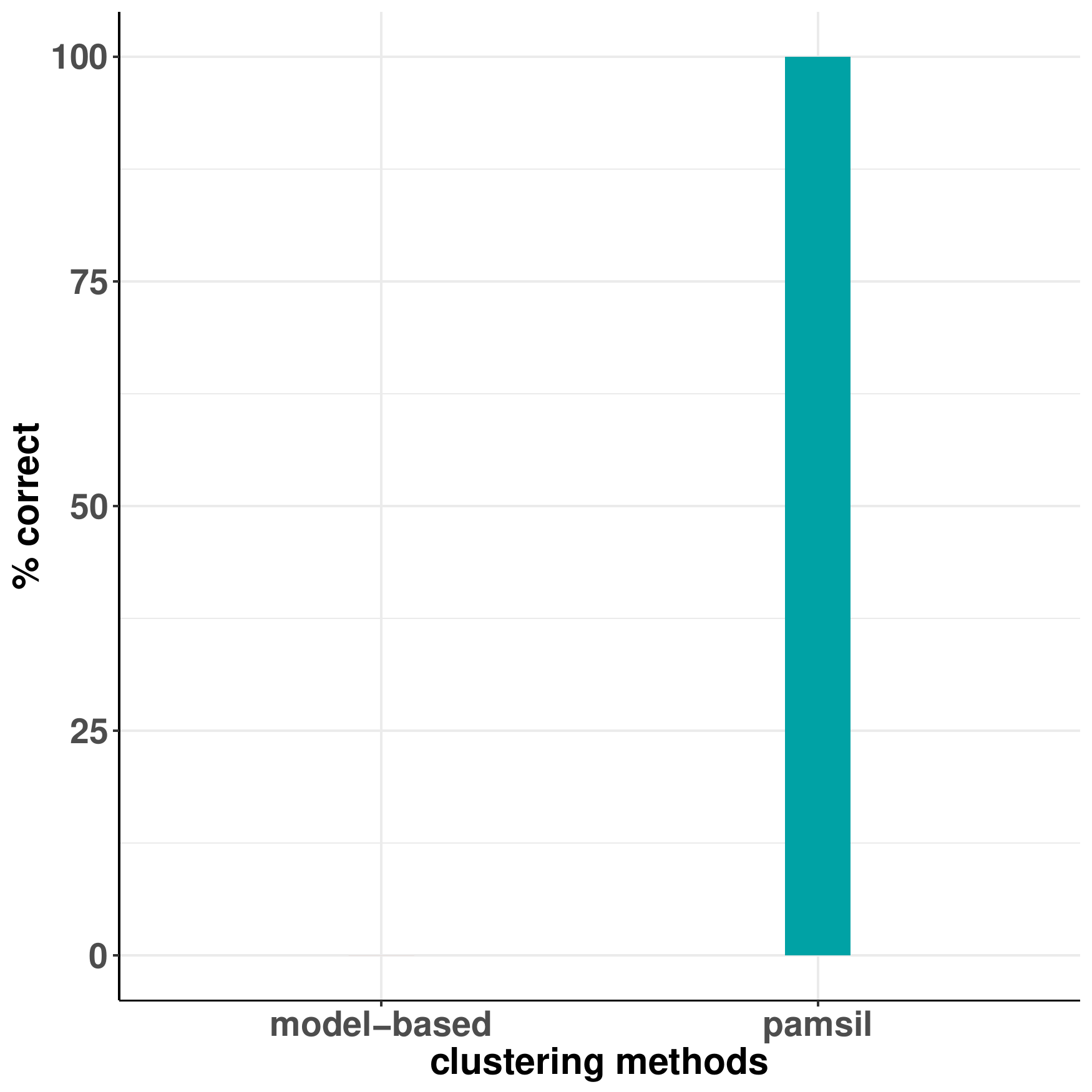}
}
\subfloat[ASW]{
  \includegraphics[width=35mm]{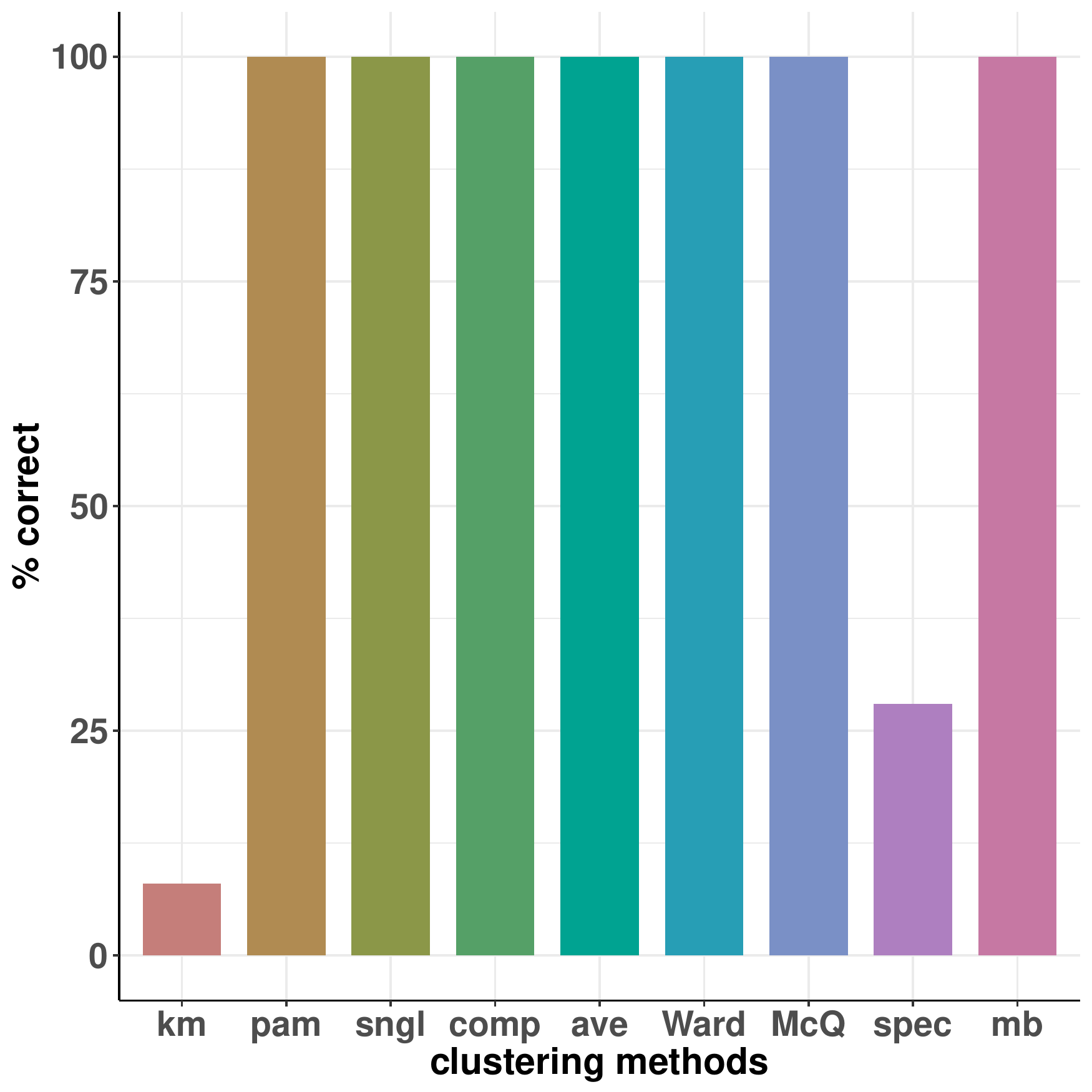}
}
\newline
\rule{-60ex}{.2in}
\subfloat[OASW]{
  \includegraphics[width=35mm]{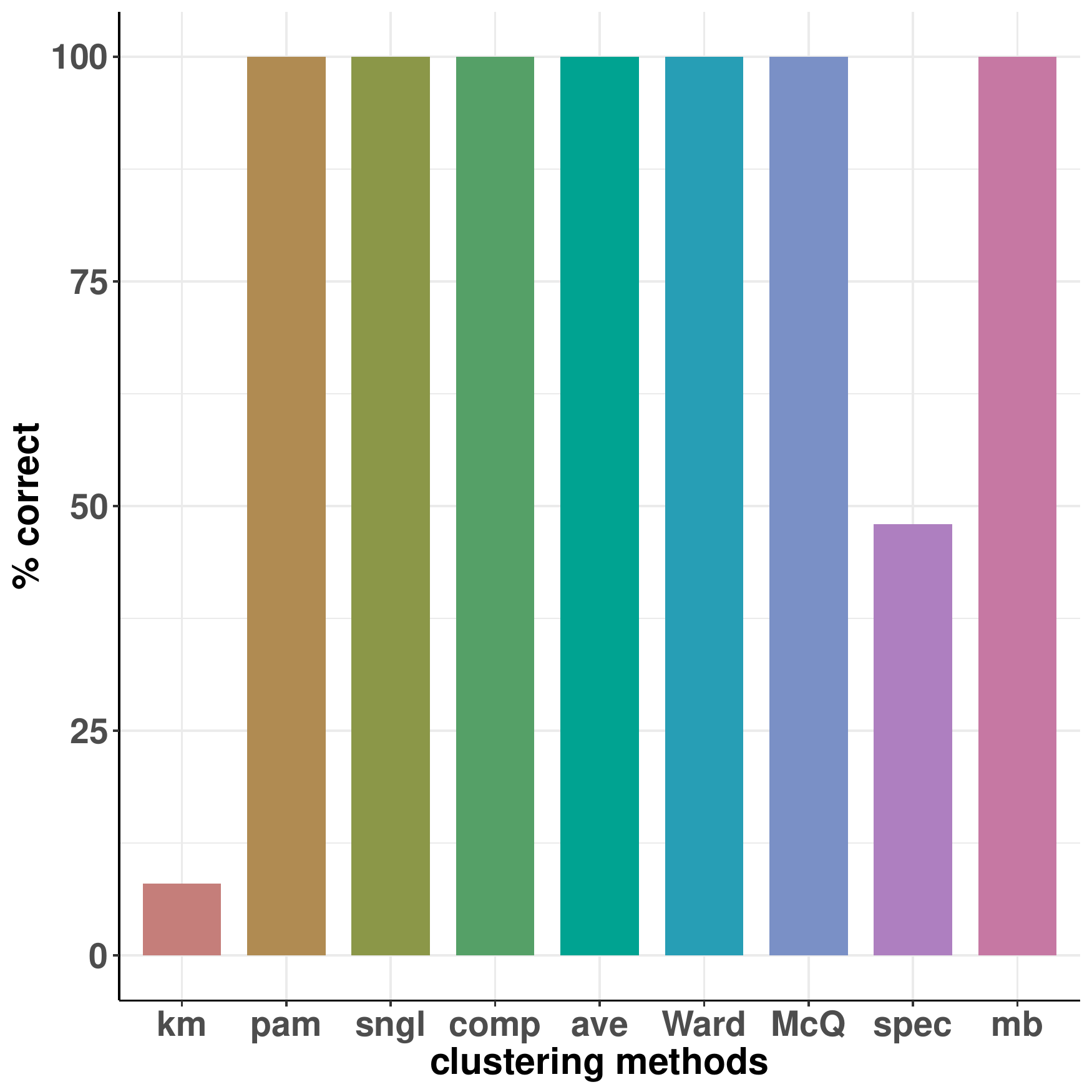}
}
\caption{Bar plots for the estimation of k for Model 8. The KL, Gap, Jump, CVNN were never able to estimate correct number of clusters for this model.}
\label{appendix:estkmodeleight}
\end{figure}

\begin{figure}[!hbtp]
\centering
\subfloat[CH]{
  \includegraphics[width=35mm]{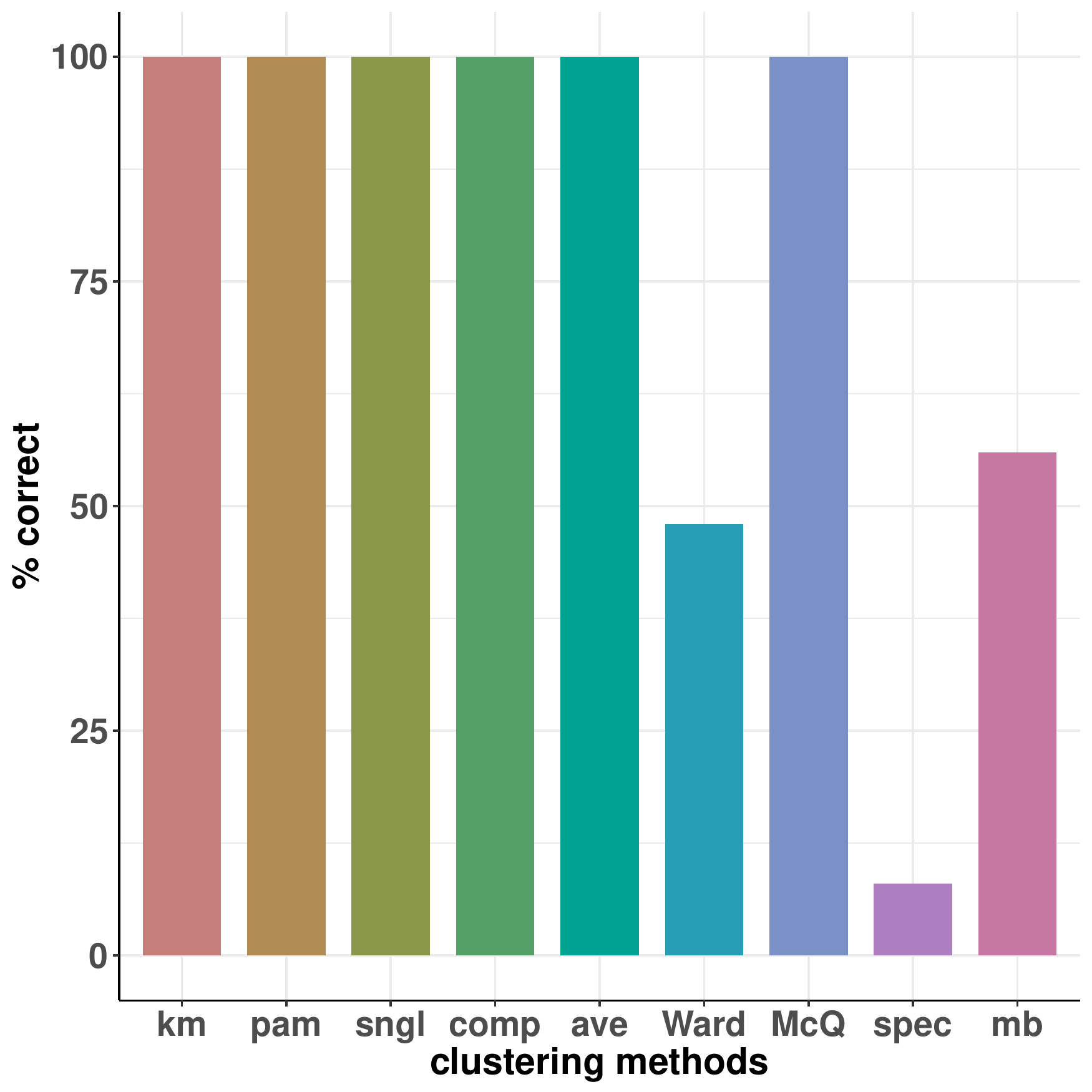}
}
\subfloat[H]{
  \includegraphics[width=35mm]{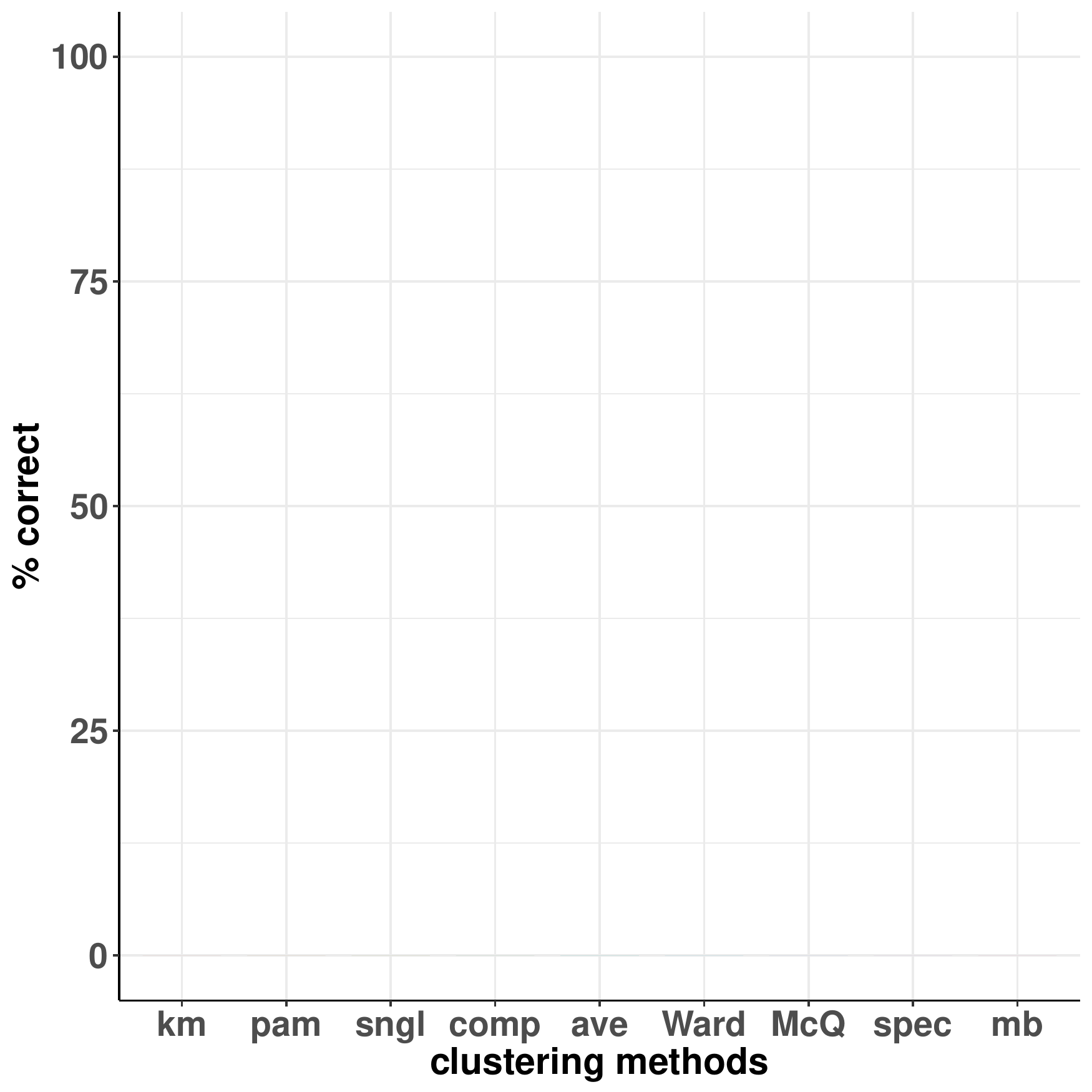}
}
\subfloat[Gamma]{
  \includegraphics[width=35mm]{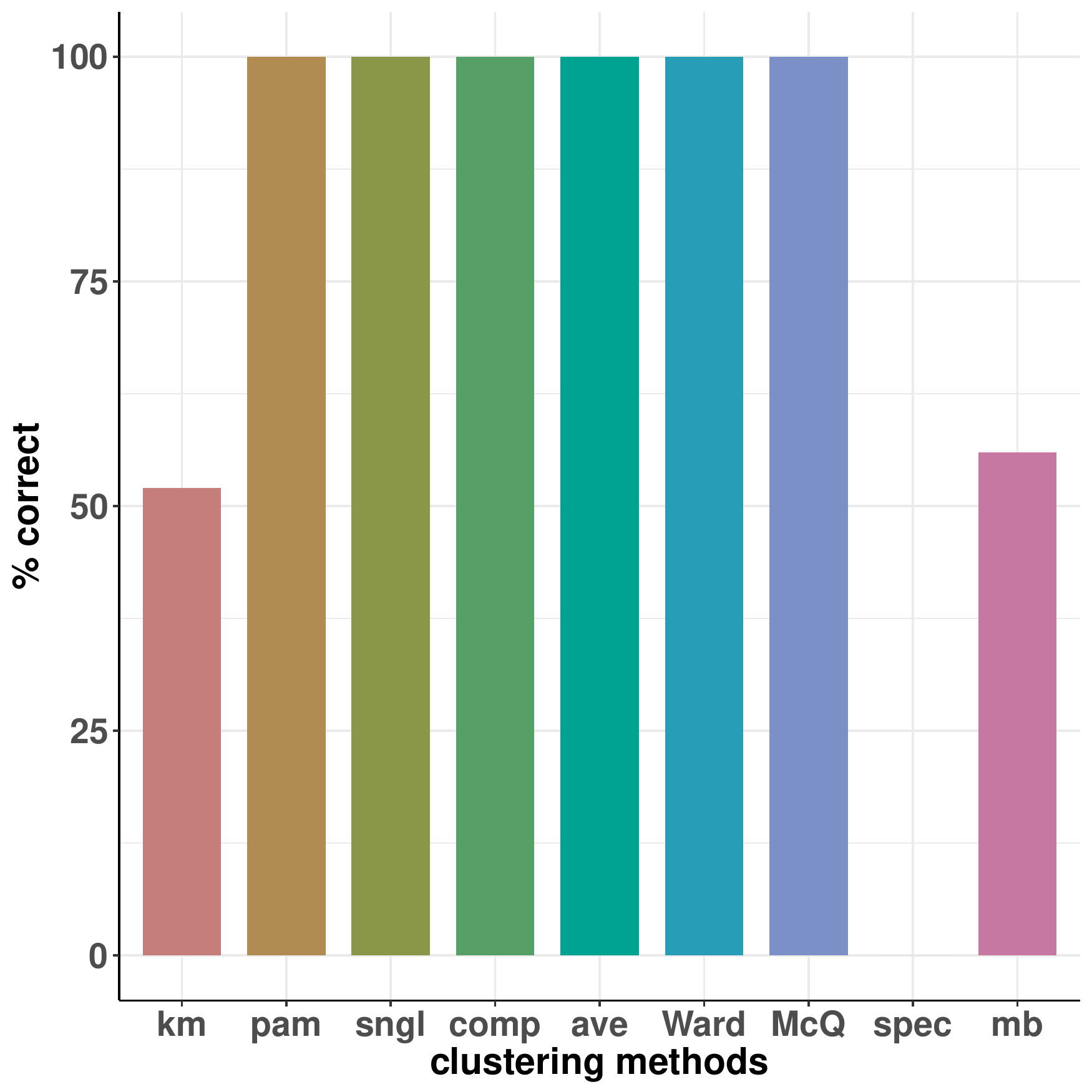}}
\subfloat[C]{
  \includegraphics[width=35mm]{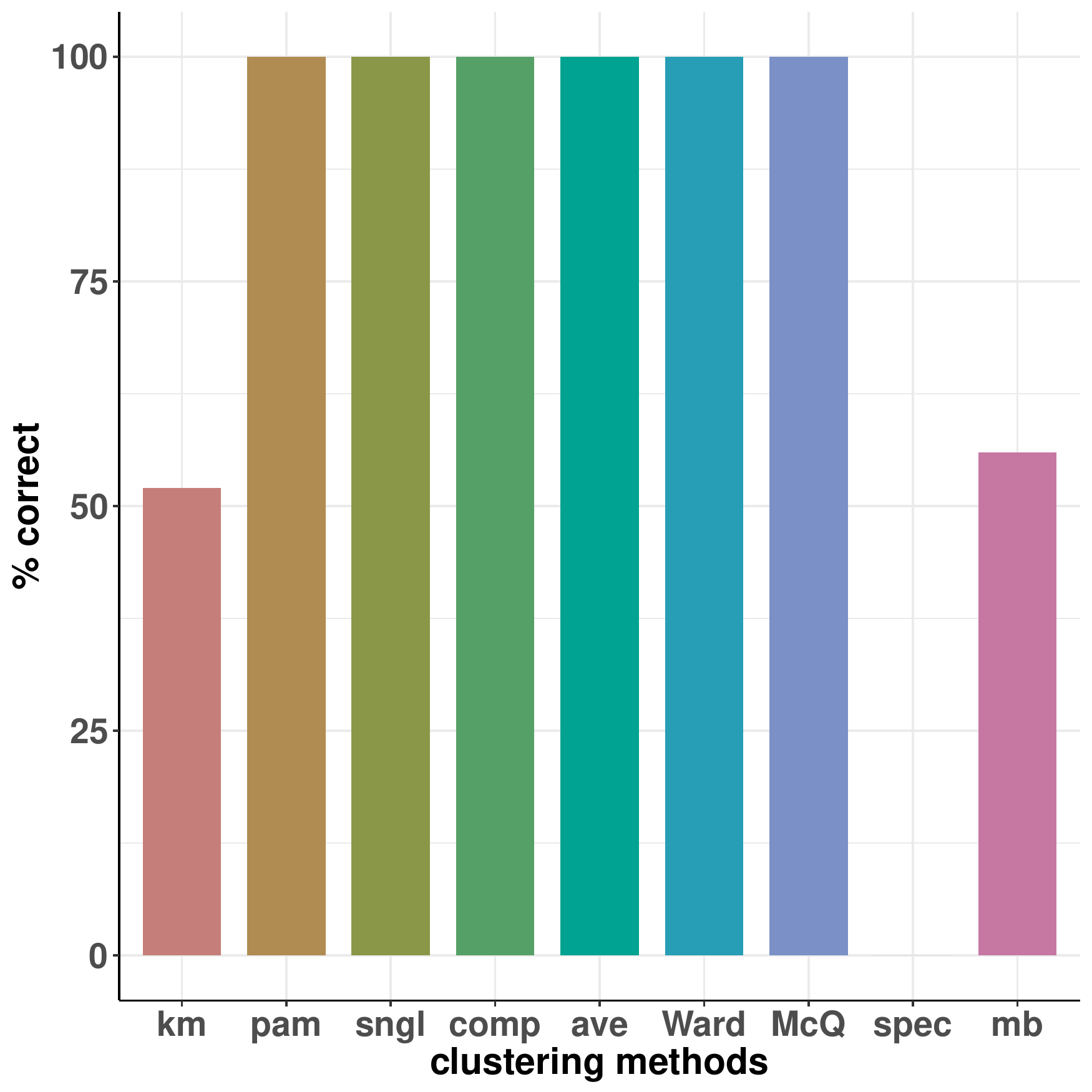}
}
\newline
\subfloat[KL]{
  \includegraphics[width=35mm]{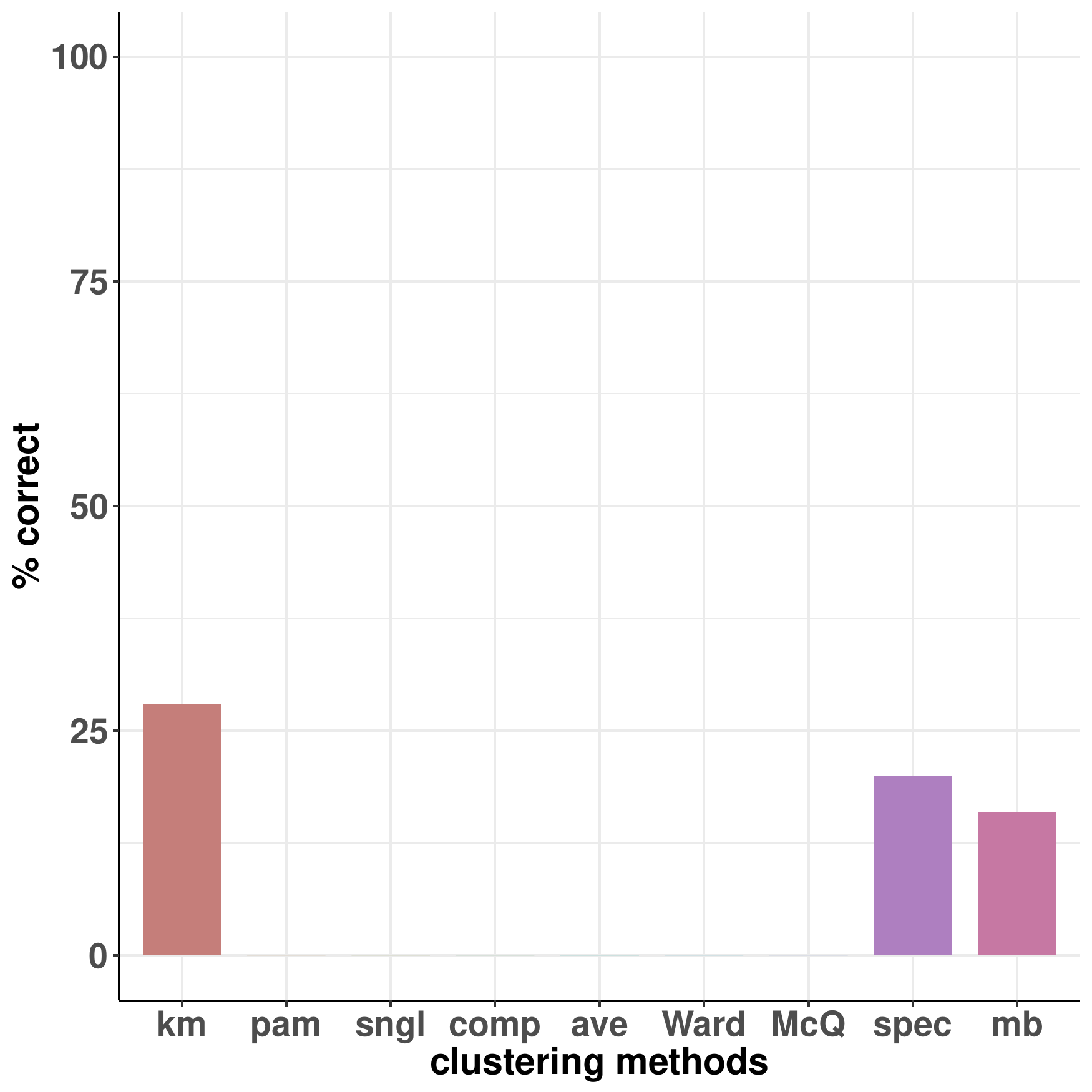}
}
\subfloat[gap]{
  \includegraphics[width=35mm]{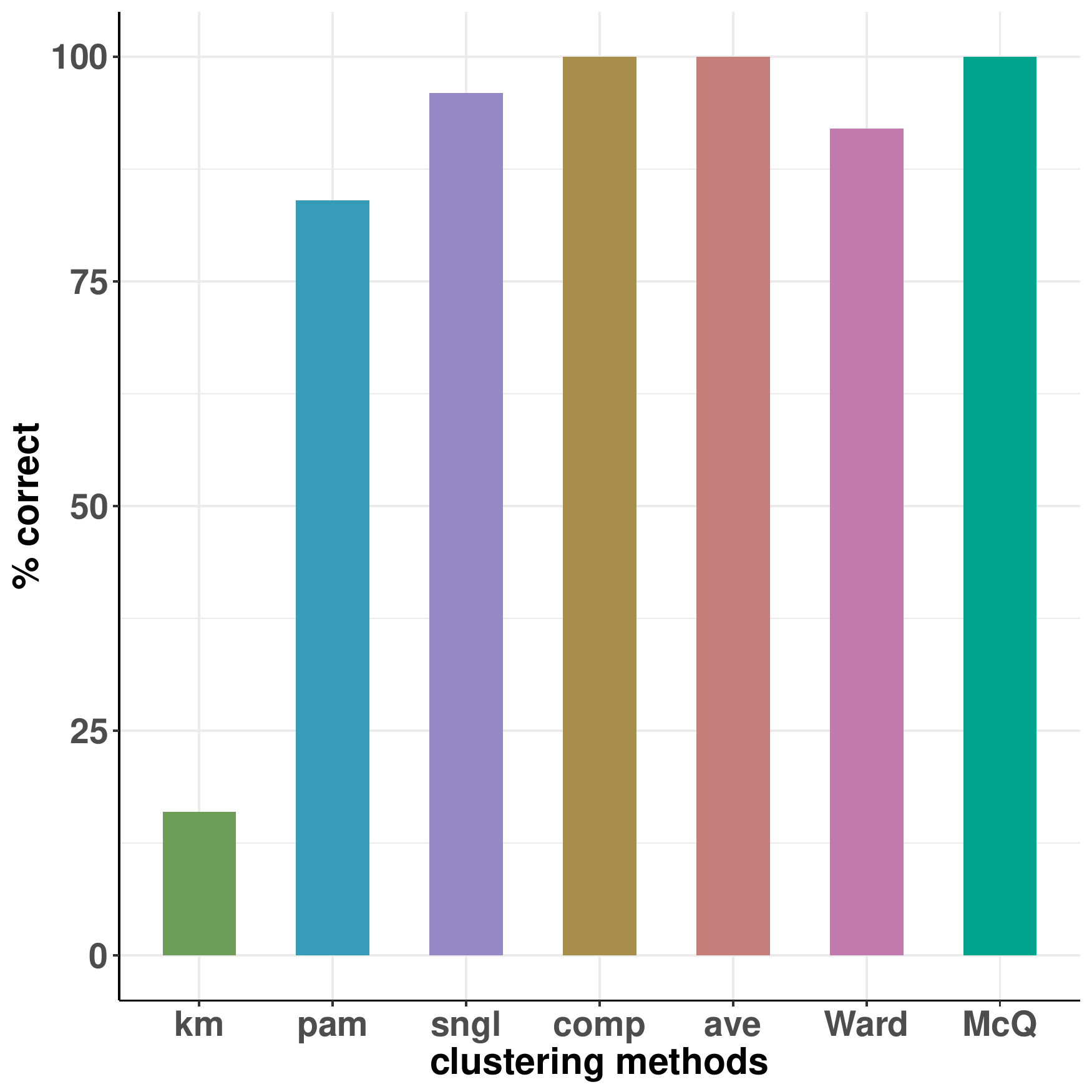}
}
\subfloat[jump]{
  \includegraphics[width=35mm]{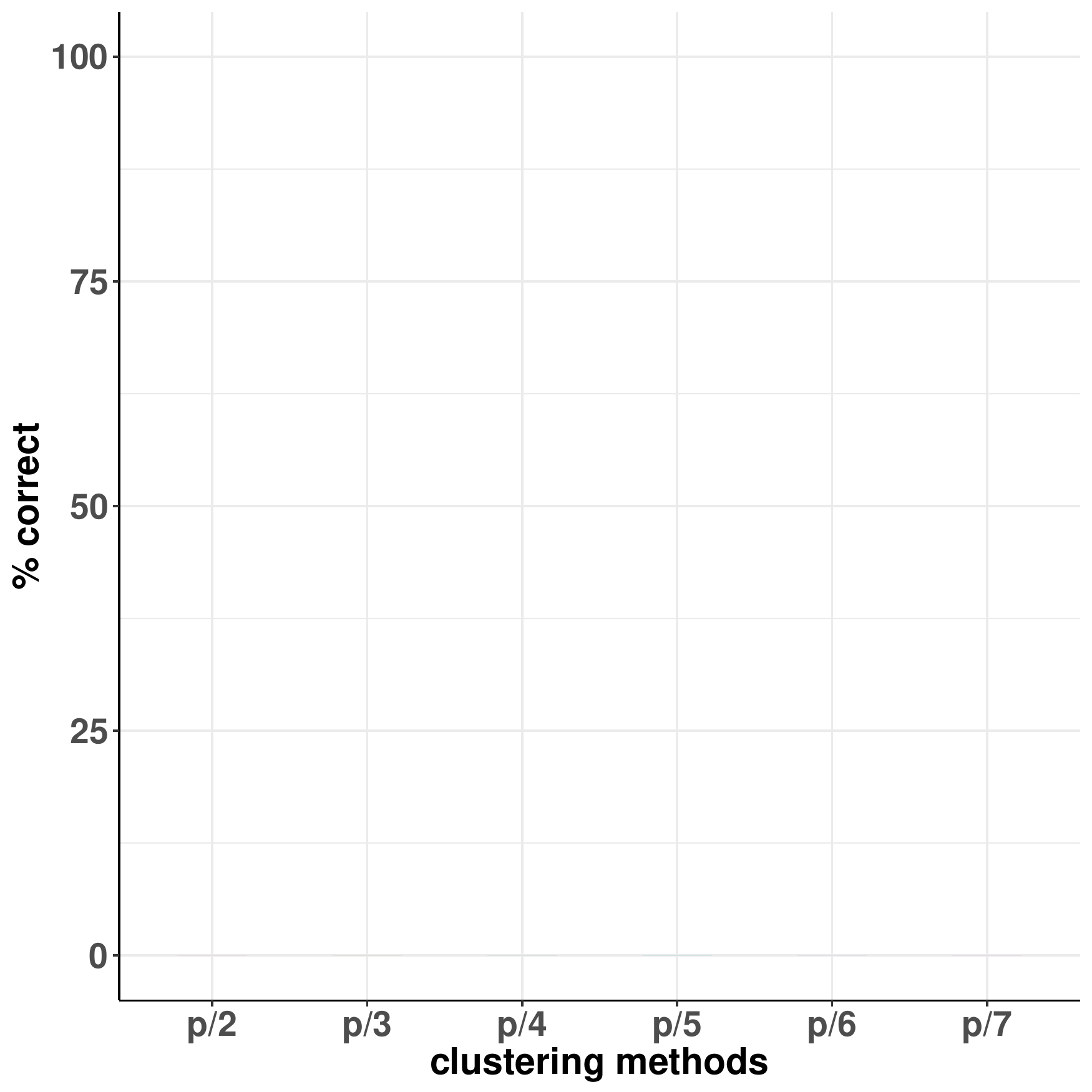}
}
\subfloat[PS]{
  \includegraphics[width=35mm]{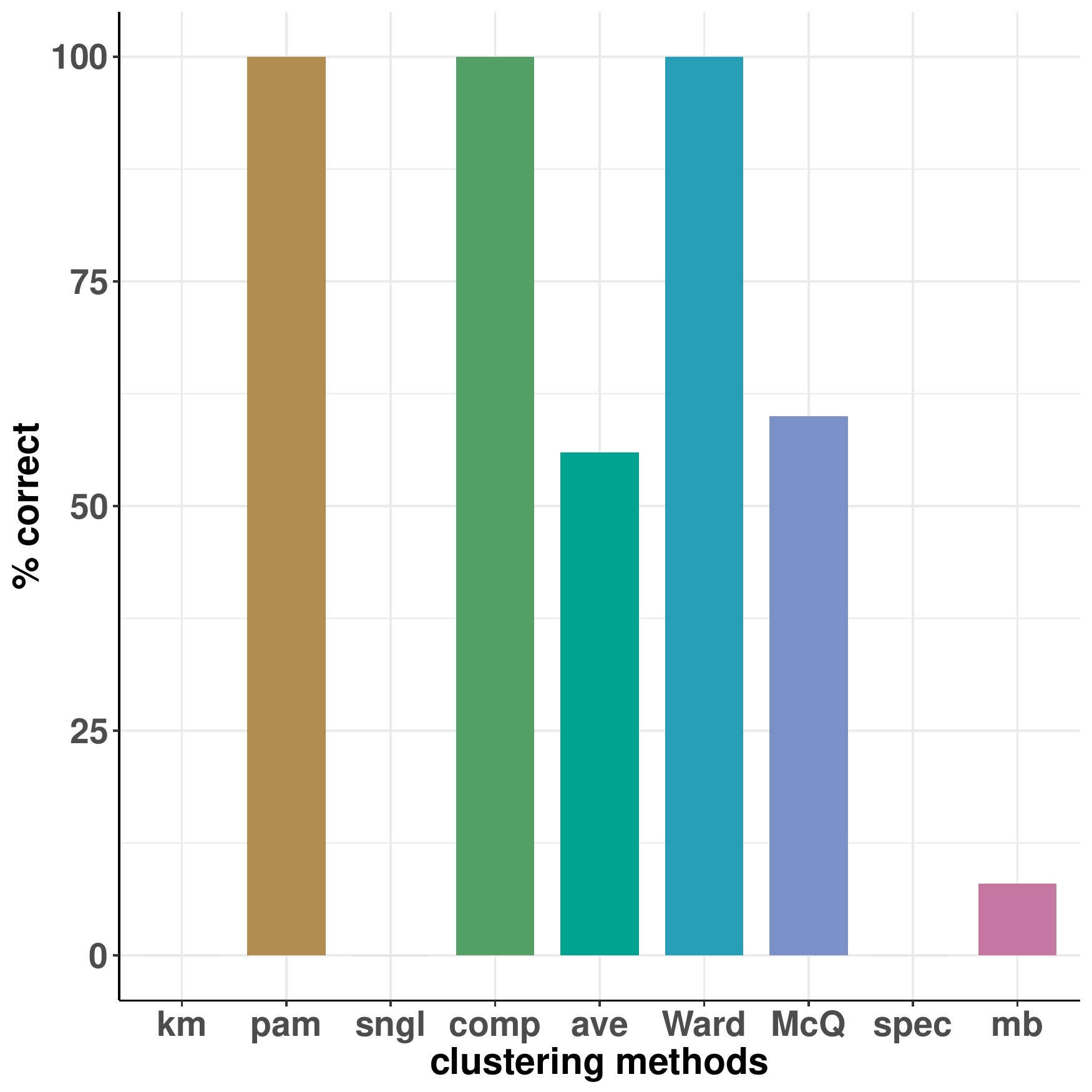}
}
\newline
\subfloat[BI]{
  \includegraphics[width=35mm]{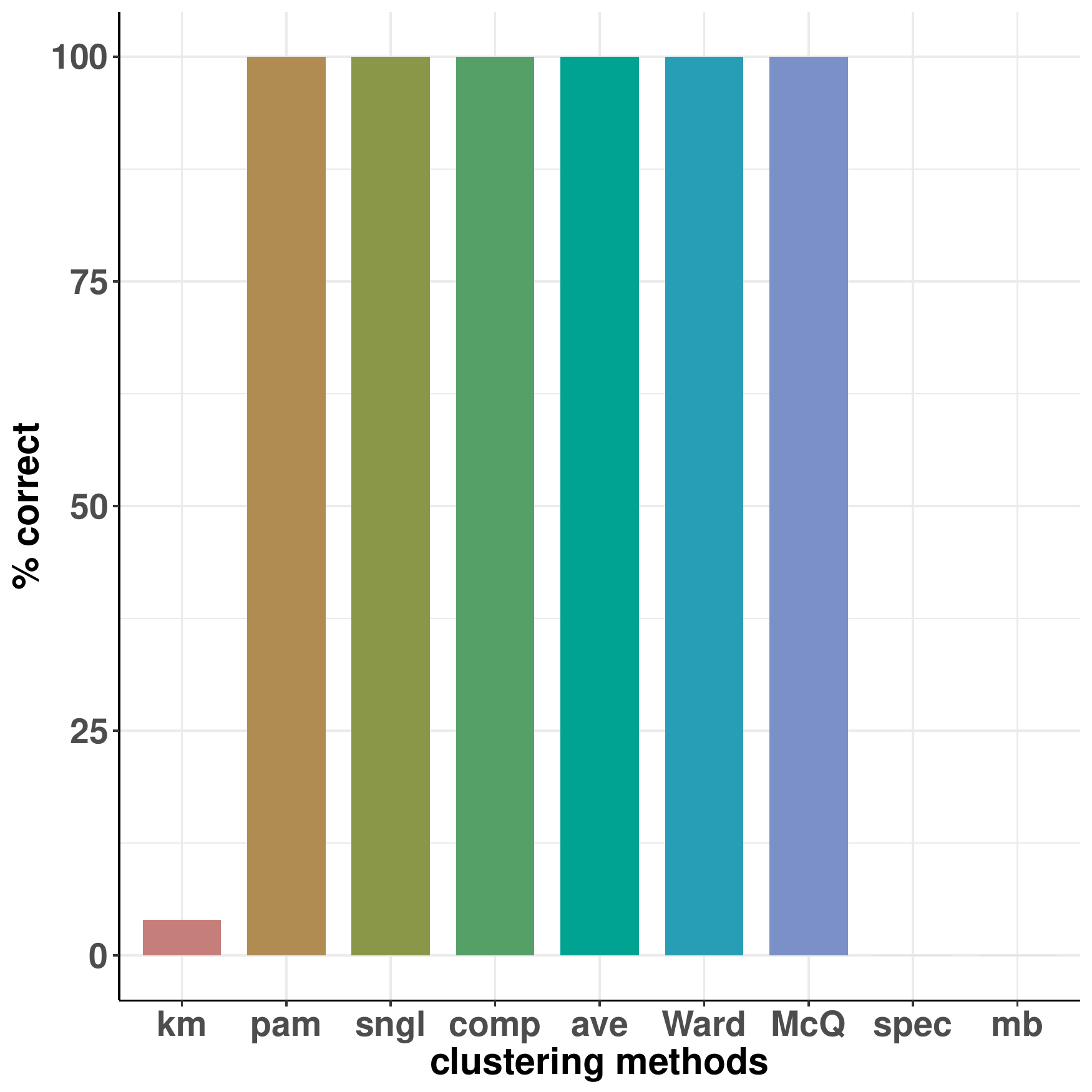}
}
\subfloat[CVNN]{
  \includegraphics[width=35mm]{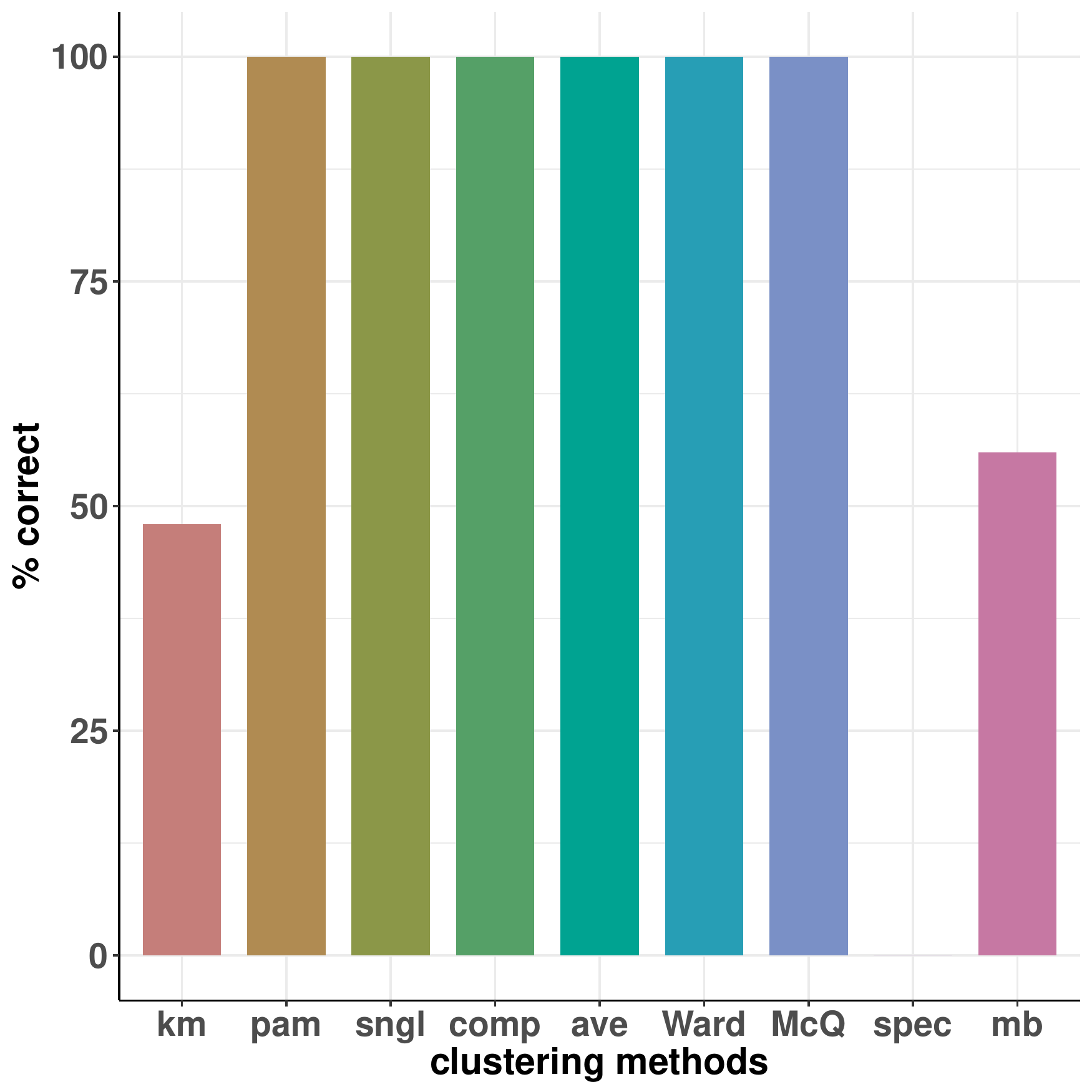}
}
\subfloat[BIC/PAMSIL]{
  \includegraphics[width=35mm]{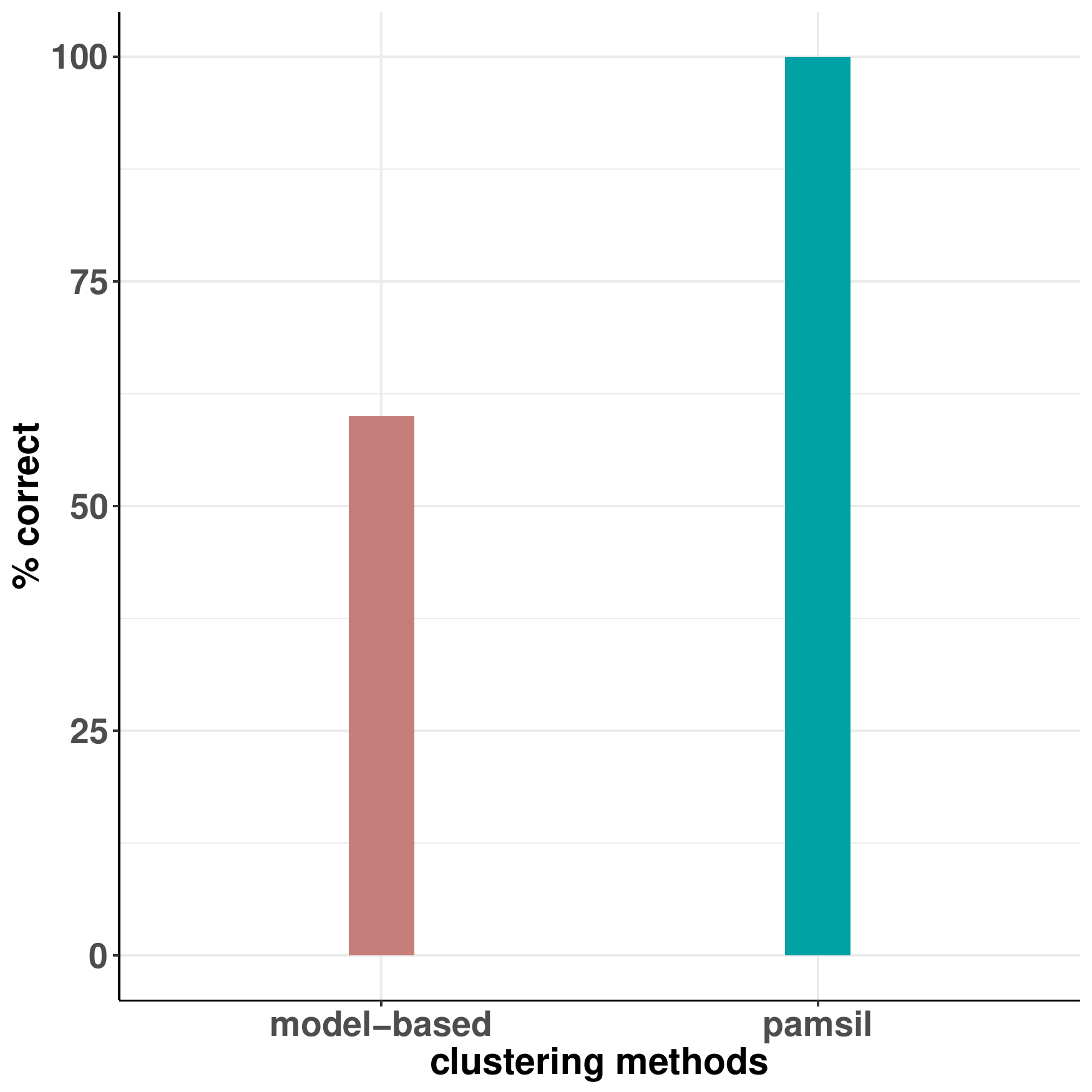}
}
\subfloat[ASW]{
  \includegraphics[width=35mm]{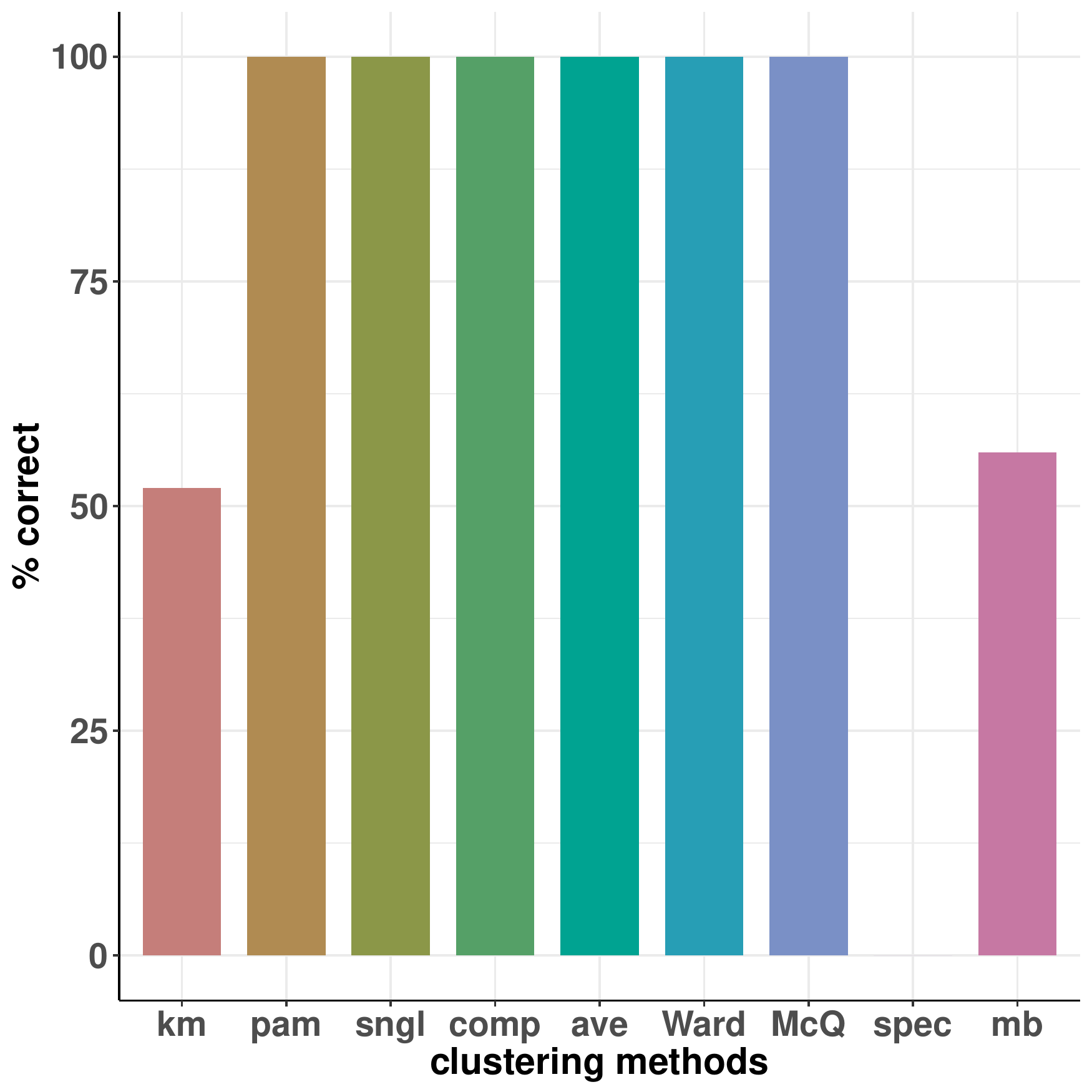}
}
\newline
\rule{-60ex}{.2in}
\subfloat[OASW]{
  \includegraphics[width=35mm]{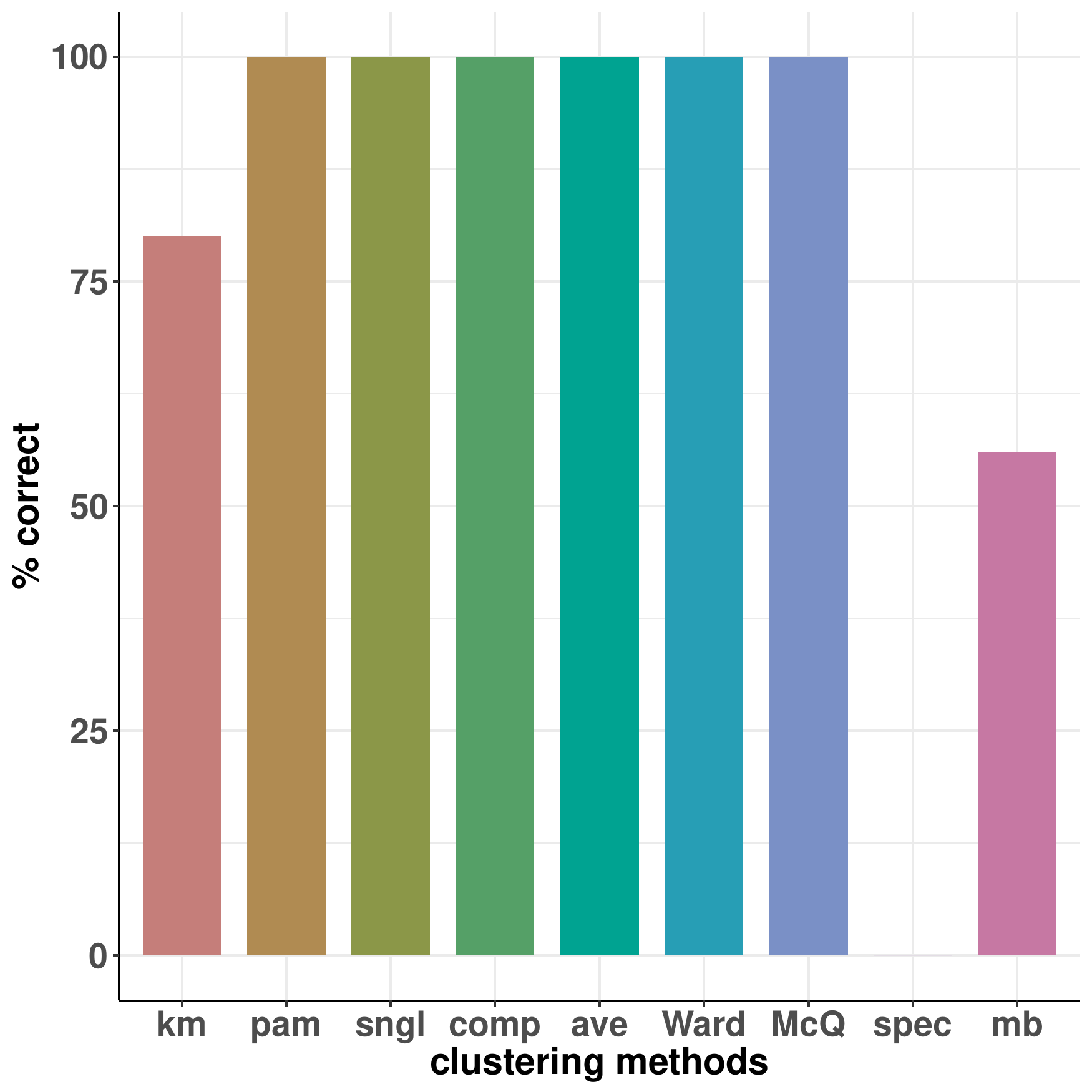}
}
\caption{Bar plots for the estimation of k for Model 10.}
\label{appendix:estkmodelten}
\end{figure}

\end{document}